\documentclass[floatfix,twocolumn,showpacs,preprintnumbers,amsmath,amssymb,pra,superscriptaddress,longbibliography]{revtex4-1}
\usepackage{color}
\usepackage[usenames,dvipsnames,svgnames,table]{xcolor}
\usepackage[colorlinks=true,linkcolor=blue,urlcolor=blue,citecolor=blue]{hyperref}
\usepackage{mathtools}
\usepackage{graphicx}
\usepackage{dcolumn}
\usepackage{array}
\usepackage{lipsum}
\usepackage{bm}
\usepackage{subfigure}
\usepackage{amssymb}
\usepackage{multirow}
\usepackage{tabularx}
\usepackage{amsmath}
\usepackage{braket}
\graphicspath{{plots/}}
 \usepackage{lipsum}
\usepackage{mathrsfs}

\renewcommand{\vec}[1]{\mathbf{#1}}

\begin{document}
\title{
Analytical representation of the Local Field Correction of the Uniform Electron Gas within the Effective Static Approximation
}

\author{Tobias Dornheim}
\email{t.dornheim@hzdr.de}

\affiliation{Center for Advanced Systems Understanding (CASUS), D-02826 G\"orlitz, Germany}
\affiliation{Helmholtz-Zentrum Dresden-Rossendorf (HZDR), D-01328 Dresden, Germany}

\author{Zhandos A.~Moldabekov}

\affiliation{Center for Advanced Systems Understanding (CASUS), D-02826 G\"orlitz, Germany}
\affiliation{Helmholtz-Zentrum Dresden-Rossendorf (HZDR), D-01328 Dresden, Germany}

\author{Panagiotis Tolias}
\affiliation{Space and Plasma Physics, Royal Institute of Technology, Stockholm, SE-100 44, Sweden}

\begin{abstract}
The description of electronic exchange--correlation effects is of paramount importance for many applications in physics, chemistry, and beyond. In a recent Letter, Dornheim \textit{et al.} [\textit{Phys.~Rev.~Lett.}~\textbf{125}, 235001 (2020)] have presented the \emph{effective static approximation} (ESA) to the local field correction (LFC), which allows for the highly accurate estimation of electronic properties such as the interaction energy and the static structure factor. In the present work, we give an analytical parametrization of the LFC within ESA that is valid for any wave number, and available for the entire range of densities ($0.7\leq r_s \leq20$) and temperatures ($0\leq \theta\leq 4$) that are relevant for applications both in the ground state and in the warm dense matter regime. A short implementation in Python is provided, which can easily be incorporated into existing codes.

In addition, we present an extensive analysis of the performance of ESA regarding the estimation of various quantities like the dynamic structure factor, static dielectric function,  the electronically screened ion-potential, and also stopping power in electronic medium. In summary, we find that the ESA gives an excellent description of all these quantities in the warm dense matter regime, and only becomes inaccurate when the electrons start to form a strongly correlated electron liquid ($r_s\sim20$). Moreover, we note that the exact incorporation of exact asymptotic limits often leads to a superior accuracy compared to the neural-net representation of the static LFC [\textit{J.~Chem.~Phys.}~\textbf{151}, 194104 (2019)]. 
\end{abstract}

\maketitle

\section{Introduction\label{sec:introduction}}


The accurate description of many-electron systems is of paramount importance for many applications in physics, quantum chemistry, material science, and related disciplines~\cite{quantum_theory,Foulkes_RevModPhys_2001}. In this regard, the uniform electron gas (UEG)~\cite{loos,review}, which is comprised of correlated electrons in a homogeneous, neutralizing positive background (also known as "jellium" or quantum one-component plasma), constitutes a fundamental model system. Indeed, our improved understanding of the UEG has facilitated many key insights like the quasi-particle picture of collective excitations~\cite{pines} and the Bardeen-Cooper-Schrieffer theory of superconductivity~\cite{Bardeen_PhysRev_1957}.

In the ground state, many properties of the UEG have been accurately determined on the basis of quantum Monte Carlo (QMC) simulations~\cite{Ceperley_UEG_1978,Ceperley_Alder_PRL_1980,bowen2,moroni,moroni2,Ortiz_Ballone_PRB_1994,Ortiz_Harris_Ballone_PRL_1999,Zong_Lin_Ceperley_PRE_2002,Shepherd_UEG_2012,Shepherd_UEG_PRB_2012,Spink_PRB_2013,Drummond_Wigner_2004,Fraser_Foulkes_PRB_1996}, which have subsequently been used as input for various parametrizations~\cite{Perdew_Zunger_PRB_1981,Perdew_Wang_PRB_1992,Perdew_Wang_PDF_1992,vwn,Gori_Giorgi_PRB_2000,cdop,Takada_PRB_2016}. These, in turn, have provided the basis of the possibly unrivaled success of density functional theory (DFT) regarding the description of real materials~\cite{PBE_1996,Burke_Perspective_JCP_2012,Jones_RevModPhys_2015}.

Over the last decade or so, there has emerged a remarkable interest in warm dense matter (WDM)--an exotic state with high temperatures and extreme densities. In nature, these conditions occur in various astrophysical objects such as giant planet interiors~\cite{saumon1,Militzer_2008,Guillot2018}, brown dwarfs~\cite{becker,saumon1}, and neutron star crusts~\cite{Daligault_2009}. On earth, WDM has been predicted to occur on the pathway towards inertial confinement fusion~\cite{hu_ICF}, and is relevant for the new field of hot-electron chemistry~\cite{Brongersma2015,Mukherjee2013}.

Consequently, WDM is nowadays routinely realized in large research facilities around the globe; see Ref.~\cite{falk_wdm} for a recent review of different experimental techniques. Further, we mention that there have been many remarkable experimental discoveries in this field, such as the observation of diamond formation by Kraus \textit{et al.}~\cite{Kraus2016,Kraus2017}, or the measurement of plasmons in aluminum by Sperling \textit{et al.}~\cite{Sperling_PRL_2015}.

At the same time, the theoretical description of WDM is notoriously difficult~\cite{wdm_book,new_POP} due to the complicated interplay of i) Coulomb coupling, ii) quantum degeneracy of the electrons, and iii) thermal excitations. Formally, these conditions are conveniently expressed by two characteristic parameters that are of the order of one simultaneously: the density parameter (Wigner-Seitz radius) $r_s=\overline{r}/a_\textnormal{B}$, where $\overline{r}$ and $a_\textnormal{B}$ are the average interparticle distance and Bohr radius, and the degeneracy temperature $\theta=k_\textnormal{B}T/E_\textnormal{F}$, with $E_\textnormal{F}$ being the usual Fermi energy~\cite{Ott2018,quantum_theory}. In particular, the high temperature rules out ground state approaches and thermal DFT~\cite{Mermin_DFT_1965} simulations, too, require as input an exchange--correlation (XC) functional that has been developed for finite temperature~\cite{kushal,karasiev_importance,Dharma-wardana_computation_2016,Sjostrom_PRB_2014}.

This challenge has resulted in a substantial progress regarding the development of electronic QMC simulations at WDM conditions~\cite{Driver_Militzer_PRL_2012,Blunt_PRB_2014,dornheim_POP,Brown_PRL_2013,Dornheim_NJP_2015, Schoof_PRL_2015,Malone_JCP_2015,Militzer_Driver_PRL_2015,Malone_PRL_2016,dornheim_prl,dornheim_cpp,groth_jcp,dornheim_pre,Driver_PRE_2018,Dornheim_PRL_2020,Dornheim_JCP_2020,lee2020phaseless,Rubenstein_auxiliary_finite_T,Yilmaz_JCP_2020}, which ultimately led to the first parametrizations of the XC-free energy $f_\textnormal{xc}$ of the UEG~\cite{groth_prl,ksdt}, allowing for thermal DFT calculations on the level of the local density approximation (LDA). At the same time, DFT approaches are being developed that deal efficiently with the drastic increase in the basis size for high temperatures~\cite{White_PRL_2013,Gao_PRB_2016,Zhang_POP_2016,Ding_PRL_2018,Mandy_highT_DFT_JCP_2020}, and even gradient corrections to the LDA have become available~\cite{Sjostrom_PRB_2014,Karasiev_PRL_2018}.

Of particular relevance for the further development of WDM theory is the response of the electrons to an external perturbation as it is described by the dynamic density response function $\chi(\mathbf{q},\omega)$, see Eq.~(\ref{eq:chi}) below, where $\mathbf{q}$ and $\omega$ denote the wave vector and frequency. Such information is vital for the interpretation of X-ray Thomson scattering experiments (XRTS)--a standard method of diagnostics for WDM which gives access to plasma parameters such as the electronic temperature~\cite{siegfried_review,kraus_xrts}. Furthermore, accurate knowledge of $\chi(\mathbf{q},\omega)$ would allow for the construction of advanced XC-functionals for DFT based on the adiabatic connection formula and the fluctuation-dissipation theorem, see Refs.~\cite{Lu_JCP_2014,Thygesen_JCP_2015,Goerling_PRB_2019,pribram} for details, or as the incorporation as the dynamic XC-kernel in time-dependent DFT~\cite{dynamic1,Baczewski_PRL_2016}. Finally, we mention the calculation of energy-loss properties like the stopping power~\cite{Moldabekov_PRE_2020}, the construction of effective ion-ion potentials~\cite{Ceperley_Potential_1996,Moldabekov_CPP_2017,zhandos1}, the description of electrical and thermal conductivities~\cite{Hamann_PRB_2020}, and the incorporation of electronic exchange--correlation effects into other theories such as quantum hydrodynamics~\cite{Diaw2017,zhandos_QHD} or average atom models~\cite{Sterne_average_atom_HEDP_2007}.

Being motivated by these applications, Dornheim and co-workers have recently presented a number of investigations of both the static and dynamic density response of the warm dense electron gas based on \textit{ab initio} path integral Monte Carlo (PIMC)~\cite{cep} simulations~\cite{dornheim_ML,dynamic_folgepaper,dornheim_dynamic,Dornheim_PRE_2020,Hamann_CPP_2020,Hamann_PRB_2020,dornheim_HEDP}. In particular, they have reported that often a static treatment of electronic XC-effects is sufficient for a highly accurate description of dynamic properties such as $\chi(\mathbf{q},\omega)$
or the dynamic structure factor (DSF) $S(\mathbf{q},\omega)$. Unfortunately, this \emph{static approximation} (see Sec.~\ref{sec:SA} below) leads to a substantial bias in frequency-averaged properties like the interaction energy $v$~\cite{dornheim_PRL_ESA_2020}.

To overcome this limitation, Dornheim \textit{et al.}~\cite{dornheim_PRL_ESA_2020} have presented the \emph{effective static approximation} (ESA), which entails a frequency-averaged description of electronic XC-effects by combining the neural-net representation of the static local field correction (LFC) from Ref.~\cite{dornheim_ML} with a consistent limit for large wave vectors based on QMC data for the pair distribution function evaluated at zero distance; see Ref.~\cite{Hunger_PRE_2021} for a recent investigation of this quantity. In particular, the ESA has been shown to give highly accurate results for different electronic properties such as the interaction energy and the static structure factor (SSF) $S(\mathbf{q})$ at the same computational cost as the random phase approximation (RPA). Furthermore, the value of the ESA for the interpretation of XRTS experiments has been demonstrated by re-evaluating the  study of aluminum by Sperling \textit{et al.}~\cite{Sperling_PRL_2015}.

The aim of the present work is two-fold: i) we introduce an accurate analytical parametrization of the LFC within ESA, which exactly reproduces the correct limits at both small and large wave numbers $q=|\mathbf{q}|$ and can be easily incorporated into existing codes without relying on the neural net from Ref.~\cite{dornheim_ML}; a short Python implementation is freely available online~\cite{code}; ii) we further analyze the performance of the ESA regarding the estimation of various electronic properties such as $S(q,\omega)$ and $\chi(q)$ over a large range of densities and temperatures.

The paper is organized as follows:
In Sec.~\ref{sec:theory}, we introduce the underlying theoretical background including the density response function, its relation to the dynamic structure factor, and the basic idea of the ESA scheme. Sec.~\ref{sec:analy} is devoted to our new analytical parametrization of the LFC within ESA (see Sec.~\ref{sec:fina} for the final result), which is analyzed in the subsequent Sec.~\ref{sec:results} regarding the estimation of numerous electronic properties. The paper is concluded by a brief summary and outlook in Sec.~\ref{sec:summary}.

\section{Theory\label{sec:theory}}

We assume Hartree atomic units throughout this work.

\subsection{Density response and local field correction}

The density response of an electron gas to an external harmonic perturbation~\cite{Dornheim_PRL_2020} of wave-number $q$ and frequency $\omega$ is---within linear response theory---fully described by the dynamic density response function $\chi(q,\omega)$. The latter is conveniently expressed as~\cite{quantum_theory,kugler1}
\begin{eqnarray}\label{eq:chi}
\chi(q,\omega) = \frac{\chi_0(q,\omega)}{1-\frac{4\pi}{q^2}\left[1-G(q,\omega)\right]\chi_0(q,\omega)}\ ,
\end{eqnarray}
where $\chi_0(q,\omega)$ denotes the density response function of an ideal Fermi gas known from theory and the full wave-number- and frequency-resolved information about exchange--correlation effects is contained in the dynamic local field correction $G(q,\omega)$. Hence, setting $G(q,\omega)=0$ in Eq.~(\ref{eq:chi}) leads to the well known RPA which entails only a mean-field description of the density response.

Naturally, the computation of accurate data for $G(q,\omega)$ constitutes a most formidable challenge, although first \textit{ab initio} results have become available recently at least for parts of the WDM regime~\cite{dornheim_dynamic,dynamic_folgepaper,Dornheim_PRE_2020,Hamann_PRB_2020,Hamann_CPP_2020}.

Let us next consider the static limit, i.e.,
\begin{eqnarray}
\chi(q) = \lim_{\omega\to0}\chi(q,\omega) \ .
\end{eqnarray}
In this limit, accurate data for Eq.~(\ref{eq:chi}) have been presented by Dornheim \textit{et al.}~\cite{dornheim_ML,dornheim_electron_liquid,dornheim_HEDP} based on the relation~\cite{bowen2}
\begin{eqnarray}\label{eq:static_chi}
\chi({q}) = -n\int_0^\beta \textnormal{d}\tau\ F({q},\tau) \quad ,
\end{eqnarray}
with the imaginary-time density--density correlation function being defined as
\begin{eqnarray}\label{eq:F}
F(q,\tau) = \frac{1}{N} \braket{\rho(q,\tau)\rho(-q,0)}\ .
\end{eqnarray}
We note that Eq.~(\ref{eq:F}) is the usual intermediate scattering function~\cite{siegfried_review}, but evaluated at an imaginary-time argument $\tau\in[0,\beta]$.
In addition, we note that it is straightforward to then use $\chi(q)$ to solve Eq.~(\ref{eq:chi}) for the static local field correction 
\begin{eqnarray}
G(q) &=& \lim_{\omega\to0}G(q,\omega) \nonumber\\ &=&
1 - \frac{q^2}{4\pi}\left( 
\frac{1}{\chi_0(q)} - \frac{1}{\chi(q)}
\right)\ .\label{eq:G_static}
\end{eqnarray}
Based on Eq.~(\ref{eq:G_static}), Dornheim \textit{et al.}~\cite{dornheim_ML} have obtained an extensive data set for $G(q)$ for $N_p\sim50$ different density--temperature combinations. These data---together with the parametrization of $G(q;r_s)$ at zero temperature by Corradini \textit{et al.}~\cite{cdop} based on ground-state QMC simulations~\cite{moroni,moroni2}---was then used to train a deep neural network that functions as an accurate representation $G(q;r_s,\theta)$ for $0\leq q\leq 5q_\textnormal{F}$, $0.7\leq r_s \leq20$ and $0\leq\theta\leq4$.

\subsection{Fluctuation--dissipation theorem}

The fluctuation--dissipation theorem~\cite{quantum_theory}
\begin{eqnarray}\label{eq:FDT}
S({q},\omega) = - \frac{ \textnormal{Im}\chi({q},\omega)  }{ \pi n (1-e^{-\beta\omega})}
\end{eqnarray}
relates Eq.~(\ref{eq:chi}) to the dynamic structure factor $S(q,\omega)$ and, thus, directly connects the LFC to different material properties. 
First and foremost, we mention that the DSF can be directly measured, e.g. with the XRTS technique~\cite{siegfried_review}, which means that the accurate prediction of $S(q,\omega)$ from theory is of key importance for the diagnostics of state-of-the-art WDM experiments~\cite{kraus_xrts}.

The static structure factor is defined as the normalization of the DSF
\begin{eqnarray}\label{eq:Sq}
S(q) = \int_{-\infty}^\infty \textnormal{d}\omega\ S(q,\omega)\ ,
\end{eqnarray}
and thus entails an averaging over the full frequency range. We stress that this is in contrast to the static density response function $\chi(q)$ introduced in the previous section, which is defined as the limit of $\omega\to0$.
The SSF, in turn, gives direct access to the interaction energy of the system, and for a uniform system it holds~\cite{review}
\begin{eqnarray}\label{eq:v}
v = \frac{1}{\pi} \int_0^\infty \textnormal{d}q\ \left[
S(q)-1
\right]\ .
\end{eqnarray}
Finally, we mention the adiabatic connection formula~\cite{review,groth_prl,ksdt}
\begin{eqnarray}
f_{xc}(r_s,\theta) = \frac{1}{r_s^2} \int_0^{r_s} \textnormal{d}\overline{r}_s\ v(\overline{r}_s,\theta)\overline{r}_s\ ,
\end{eqnarray}
which implies that the free energy (and, equivalently the partition function $Z$) can be inferred if the dynamic density response function---the only unknown part of which is the dynamic LFC $G(q,\omega)$---of a system is known for all wave numbers and frequencies, and for different values of the coupling parameter $r_s$. This idea is at the heart of the construction of advanced exchange--correlation functionals for DFT calculations within the ACFDT formulation; see, e.g., Refs.~\cite{Thygesen_JCP_2015,Lu_JCP_2014,pribram,Goerling_PRB_2019} for more details.

\subsection{The static approximation\label{sec:SA}}

Since the full frequency-dependence of $G(q,\omega)$ remains to this date unknown for most parts of the WDM regime (and also in the ground-state), one might neglect dynamic effects and simply substitute $G(q)$ in Eq.~(\ref{eq:chi}). This leads to the dynamic density response function within the \emph{static approximation}~\cite{dornheim_dynamic,Hamann_PRB_2020},
\begin{eqnarray}\label{eq:static_approximation}
\chi_\textnormal{stat}(q,\omega) = \frac{\chi_0(q,\omega)}{1-\frac{4\pi}{q^2}\left[1-G(q)\right]\chi_0(q,\omega)}\ ,
\end{eqnarray}
which entails the frequency-dependence on an RPA level, but exchange-correlation effects are incorporated statically.
Indeed, it was recently shown that Eq.~(\ref{eq:static_approximation}) allows to obtain nearly exact results for $\chi(q,\omega)$, $S(q,\omega)$, and related quantities for $r_s\lesssim5$ and $\theta\gtrsim1$.

Yet, while results for individual wave numbers are relatively good, the \emph{static approximation} is problematic for quantities that require an integration over $q$, such as the interaction energy $v$~\cite{dornheim_PRL_ESA_2020}. More specifically, it can be shown that neglecting the frequency dependence in the LFC (LFCs that are explicitly defined without a frequency dependence are hereafter denoted as $\overline{G}(q)$) leads to the relation~\cite{stls}
\begin{eqnarray}\label{eq:onTop}
\lim_{q\to\infty} \overline{G}(q) = 1 - g(0)\ ,
\end{eqnarray}
where $g(0)$ denotes the pair distribution function (PDF) $g(r)$ evaluated at zero distance, sometimes also called the on-top PDF or contact probability. Yet, is has been shown both in the ground state~\cite{holas_limit,farid} and at finite temperature~\cite{dornheim_ML,dornheim_electron_liquid} that the exact static limit of the dynamic LFC diverges towards either positive or negative infinity in the $q\to\infty$ limit. Eq.~(\ref{eq:onTop}) thus implies that using $G(q)$ as $\overline{G}(q)$ in Eq.~(\ref{eq:static_approximation}) leads to a diverging on-top PDF, which is, of course, unphysical. This, too, is the reason for spurious contributions to wave-number integrated quantities like $v$ at large $q$.

\subsection{The Effective Static Approximation\label{sec:ESA}}

To overcome these limitation of the \emph{static approximation}, Dornheim \textit{et al.}~\cite{dornheim_PRL_ESA_2020} have proposed to define an effectively frequency-averaged theory that combines the good performance of Eq.~(\ref{eq:static_approximation}) for $q\lesssim3q_\textnormal{F}$ with the consistent limit of $\overline{G}(q)$ from Eq.~(\ref{eq:onTop}).

More specifically, this so-called effective static approximation is constructed as~\cite{dornheim_PRL_ESA_2020}
\begin{eqnarray}\label{eq:ESA}
\overline{G}_\textnormal{ESA}(q;r_s,\theta) &=& G_\textnormal{nn}(q;r_s,\theta)\left(1-A(x)\right) \\ \nonumber & &+ \left(1-g(0;r_s,\theta)\right)A(x) \ ,
\end{eqnarray}
with $x=q/q_\textnormal{F}$, and where $G_\textnormal{nn}(q;t_s,\theta)$ is the neural-net representation of PIMC data for the exact static limit $G(q)=G(q,0)$ of the UEG~\cite{dornheim_ML}, and $g(0;r_s,\theta)$ denotes the on-top pair distribution function that was parametrized in Ref.~\cite{dornheim_PRL_ESA_2020} on the basis of restricted PIMC data by Brown \textit{et al.}~\cite{Brown_PRL_2013}. Further, $A(x)$ denotes the activation function
\begin{eqnarray}\label{eq:activation}
A(x) = A(x,x_m,\eta) = \frac{1}{2}\left[
1+\text{tanh}\left(\eta(x-x_m)\right)
\right] \
\end{eqnarray}
resulting in a smooth transition between $G_\textnormal{nn}$ and Eq.~(\ref{eq:onTop}) for large $q$. Here the parameters $x_m$ and $\eta$ can be used to tune the position and width of the activation. In practice, the performance of the ESA only weakly depends on $\eta$ and we always use $\eta=3$ throughout this work. The appropriate choice of the position $x_m$ is less trivial and is discussed below.

\begin{figure}\centering
\includegraphics[width=0.55\textwidth]{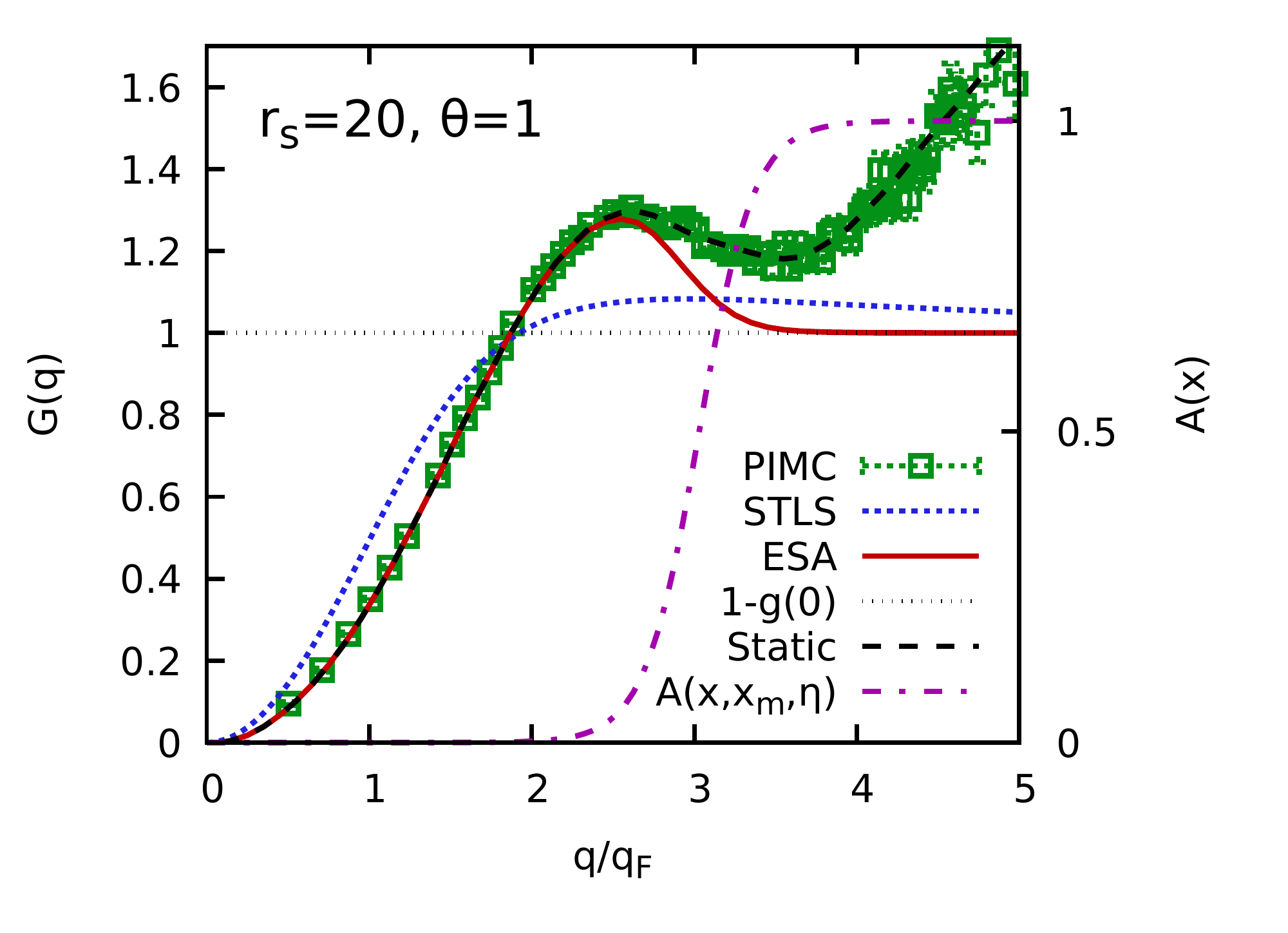}\\\vspace*{-1cm}\hspace*{-0.02\textwidth}
\includegraphics[width=0.47\textwidth]{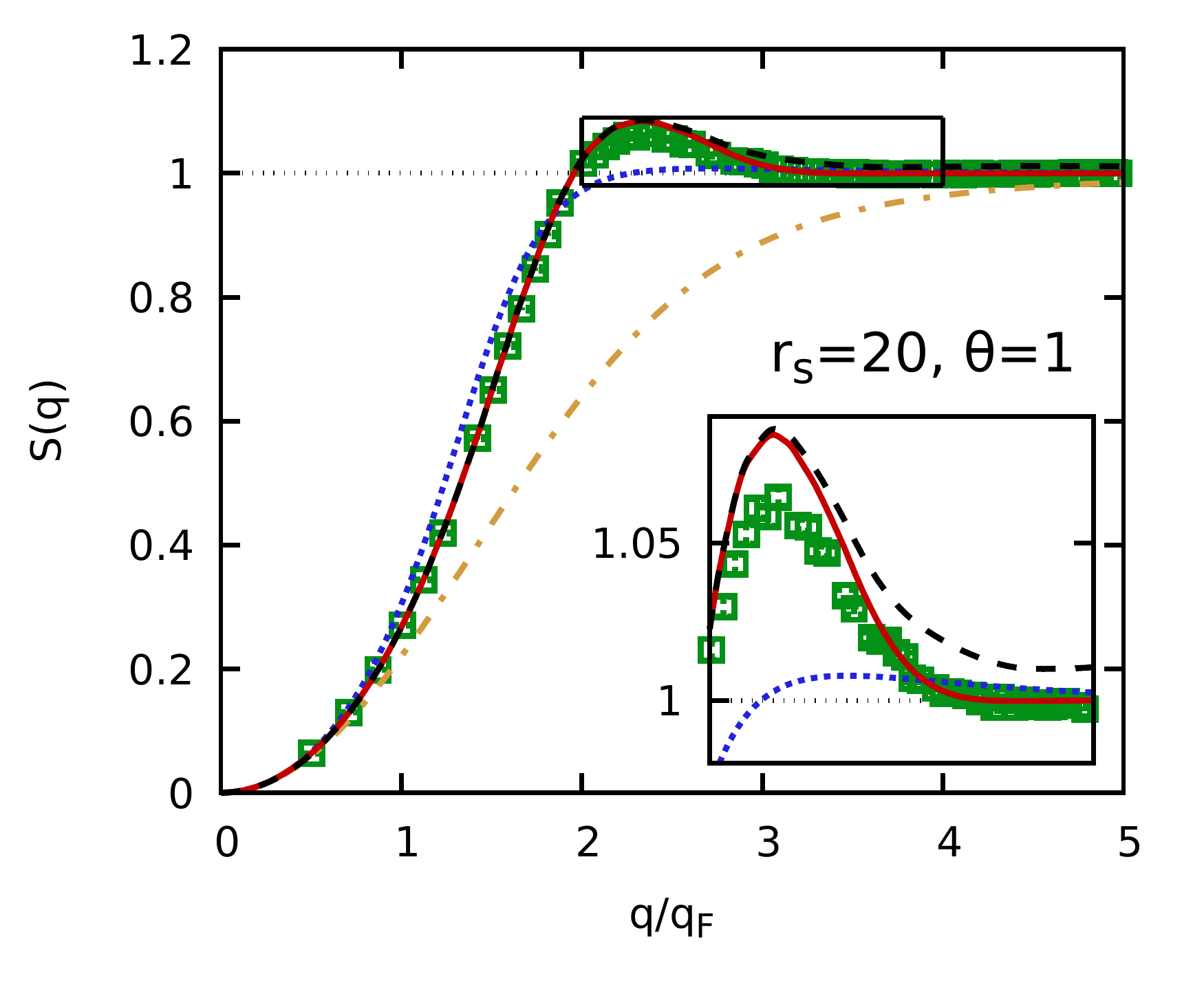}
\caption{\label{fig:ESA}
Illustration of the effective static approximation (ESA)~\cite{dornheim_PRL_ESA_2020} for $r_s=20$ and $\theta=1$. Top panel: Static LFC. Green squares are exact PIMC data for $G(q;r_s,\theta)$ taken from Ref.~\cite{dornheim_electron_liquid}, and dashed black line the neural net representation from Ref.~\cite{dornheim_ML}. The solid red curve shows the frequency-averaged LFC $\overline{G}(q;r_s,\theta)$ within ESA [Eq.~(\ref{eq:ESA})] and the dotted blue curve the same quantity within STLS~\cite{stls2,stls}. The purple dash-dotted line shows the activation function $A(x,x_m,\eta)$ [for $x_m=\eta=3$, see Eq.~(\ref{eq:activation})] and corresponds to the right $y$-axis.
Top panel: Static structure factor $S(q)$ from the same methods, and in RPA (dash-dotted yellow).
}
\end{figure}

An example for the construction of the ESA is shown in Fig.~\ref{fig:ESA} for the UEG at $r_s=20$ and $\theta=1$. In the top panel, we show the wave-number dependence of the static LFC $G(q)$, with the green squares depicting exact PIMC data for $N=66$ taken from Ref.~\cite{dornheim_electron_liquid} and the black dashed curve the neural-net representation from Ref.~\cite{dornheim_ML}. Observe the positively increasing tail at large $q$ from both data sets, which is consistent to the positive value of the exchange-correlation contribution to the kinetic energy at these conditions~\cite{holas_limit,Militzer_PRL_2002}.

The solid red line corresponds to the ESA and is indistinguishable from the neural net for $q\lesssim2 q_\textnormal{F}$. Further, it smoothly goes over into Eq.~(\ref{eq:onTop}) for larger $q$ and attains this limit for $q\gtrsim3.5q_\textnormal{F}$.
The purple dash-dotted curve shows the corresponding activation function $A(x)$ [using $x_m=3$] on the right $y$-axis and illustrates the shape of the switchover between the two limits.
As a reference, we have also included $\overline{G}(q)$ computed within the finite-temperature version~\cite{stls,stls2} of the STLS approximation~\cite{stls_original}, see the dotted blue curve. First and foremost, we note that STLS constitutes a purely static theory for the LFC and, thus, exactly fulfills Eq.~(\ref{eq:onTop}), i.e., it attains a constant value in the limit of large wave numbers, although for significantly larger values of $q$. In addition, STLS is well known to violate the exact compressibility sum-rule~\cite{stls2} (see Eq.~(\ref{eq:CSR}) below) and deviates from the other curves even in the small-$q$ limit. Finally, we note that it does not reproduce the peak of both the neural net and ESA around $q=2.5q_\textnormal{F}$.

The bottom panel of Fig.~\ref{fig:ESA} shows the corresponding results for the static structure factor $S(q)$, with the green crosses again being the exact PIMC results from Ref.~\cite{dornheim_electron_liquid}. At this point, we feel that a note of caution is pertinent. On the one hand, the PIMC method is limited to simulations in the static limit, as dynamic simulations are afflicted with an exponentially hard phase problem~\cite{Segal_PRB_2010} in addition to the usual fermion sign problem~\cite{dornheim_sign_problem}. Therefore, PIMC results for both $\chi(q,\omega)$ and $G(q,\omega)$ are only available for $\omega=0$. Yet, the PIMC method is also capable to give exact results for frequency-averaged quantities like $S(q)$, as the frequency integration is carried out in the imaginary time~\cite{cep}. Thus, the green squares do correspond to the results one would obtain if the correct, dynamic LFC $G(q,\omega)$ was inserted into Eq.~(\ref{eq:chi}).

This is in contrast to the black dashed curve, that has been obtained on the basis of the \emph{static approximation}, Eq.~(\ref{eq:static_approximation}), using as input the neural-net representation~\cite{dornheim_ML} of the exact static limit $G(q)$. Evidently, the static treatment of exchange--correlation effects is well justified for $q\lesssim2q_\textnormal{F}$, but there appear systematic deviations for larger $q$; see also the inset showing a magnified segment around the maximum of $S(q)$. In particular, $S(q)$ does not decay to $1$, and, while being small for each individual $q$, the error accumulates under the integral in Eq.~(\ref{eq:v}).

The solid red curve has been obtained by inserting $\overline{G}(q)$ within the ESA into Eq.~(\ref{eq:static_approximation}). Plainly, the inclusion of the on-top PDF via Eq.~(\ref{eq:ESA}) removes the spurious effects from the \emph{static approximation}, and the ESA curve is strikingly accurate over the entire $q$-range.

The dotted blue curve has been computed using $\overline{G}(q)$ within the STLS approximation. For small $q$, it, too obeys the correct parabolic limit~\cite{dornheim_prl,kugler_bounds}, which is the consequence of \emph{perfect screening} in the UEG~\cite{quantum_theory}. For larger $q$, there appear systematic deviations, and the correlation-induced peak of $S(q)$ around $q\sim2.2q_\textnormal{F}$ is not reproduced by this theory; see also Ref.~\cite{dornheim_electron_liquid} for an extensive analysis including even stronger values of the coupling strength $r_s$.

Finally, the dash-dotted yellow curve has been computed within the RPA. Clearly, neglecting exchange--correlation effects in Eq.~(\ref{eq:chi}) leads to an insufficient description of the SSF, and we find systematic deviations of up to $\sim30\%$.

\section{Analytical representation of the ESA\label{sec:analy}}

\subsection{Choice of the activation function}

The ESA as it has been defined in Eq.~(\ref{eq:ESA}) has, in principle, two free parameters, which have to be defined/parametrized before an analytical representation of $\overline{G}(q;r_s,\theta)$ can be introduced. More specifically, these are the transition wave number $x_m$ and scaling parameter $\eta$ from the activation function $A(x;x_m,\eta)$; see Eq.~(\ref{eq:activation}).

\textbf{Scaling parameter $\eta$:} We choose $\eta(r_s,\theta)=3=\textnormal{const}$, as $\overline{G}_\textnormal{ESA}(q;r_s,\theta)$ only weakly depends on this parameter; see Ref.~\cite{dornheim_PRL_ESA_2020} for an example.

\begin{figure}\centering
\includegraphics[width=0.475\textwidth]{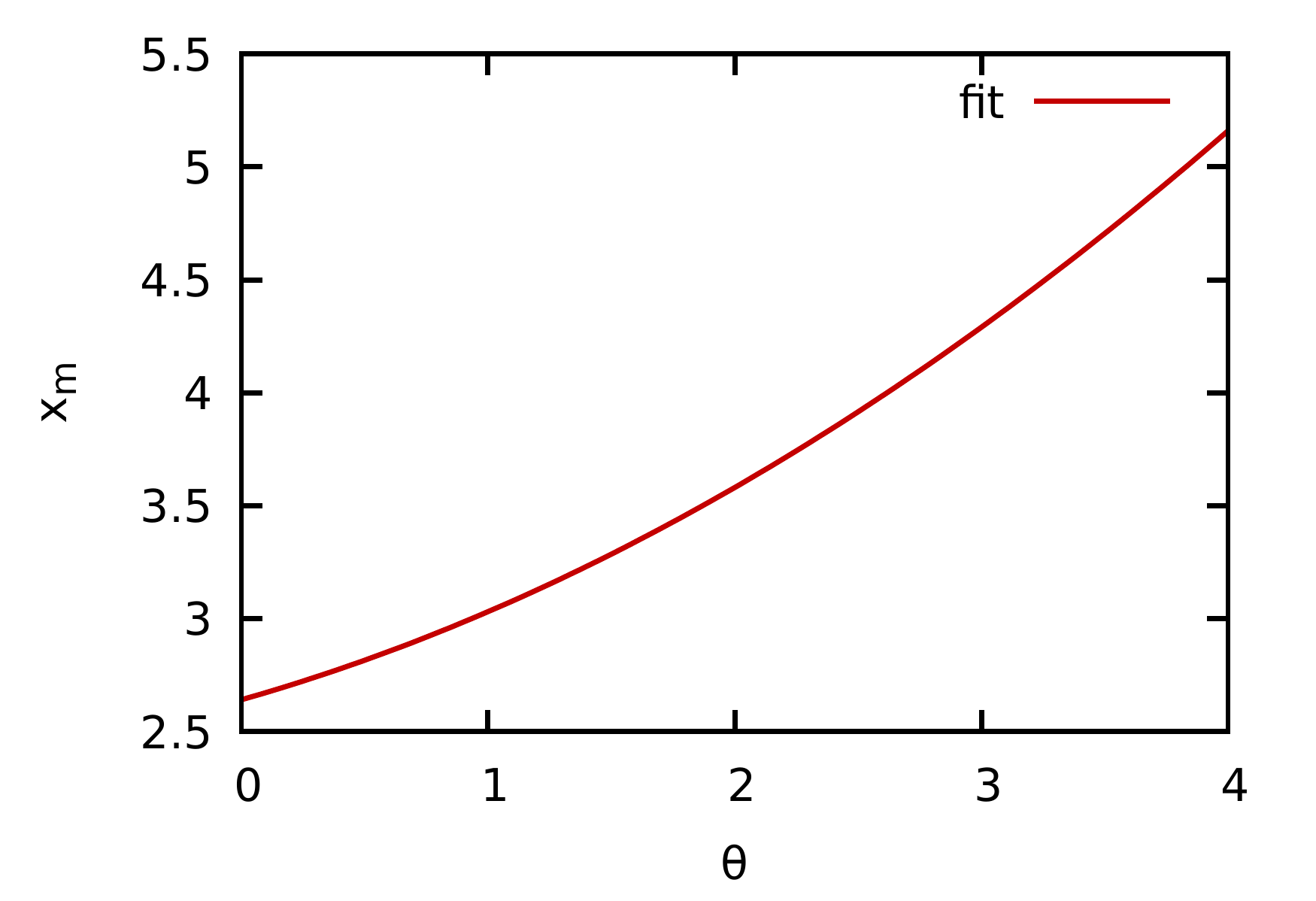}
\caption{\label{fig:activation_xm}
Temperature dependence of the transition wave-number $x_m$ from Eq.~(\ref{eq:activation_xm}).
}
\end{figure}

\textbf{Transition wave-number $x_m$:} The choice of a reasonable wave-number of the transition between the neural-net and Eq.~(\ref{eq:onTop}) is less trivial. What we need is a transition around $x_m\sim2.5q_\textnormal{F}$ for $\theta\lesssim1$, whereas it should move to larger wave-number for higher temperatures.
The dependence on the density parameter $r_s$, on the other hand, is less pronounced and can be neglected.
We thus construct the function 
\begin{equation}\label{eq:activation_xm}
    x_m(\theta) = A_x + B_x \theta + C_x \theta^2\ ,
\end{equation}
with $A_x$, $B_x$, and $C_x$ being free parameters that we determine empirically. In particular, we find $A_x=2.64$, $B_x=0.31$, and $C_x=0.08$.
A graphical depiction of Eq.~(\ref{eq:activation_xm}) is shown in Fig.~\ref{fig:activation_xm}

\begin{figure}\centering
\includegraphics[width=0.475\textwidth]{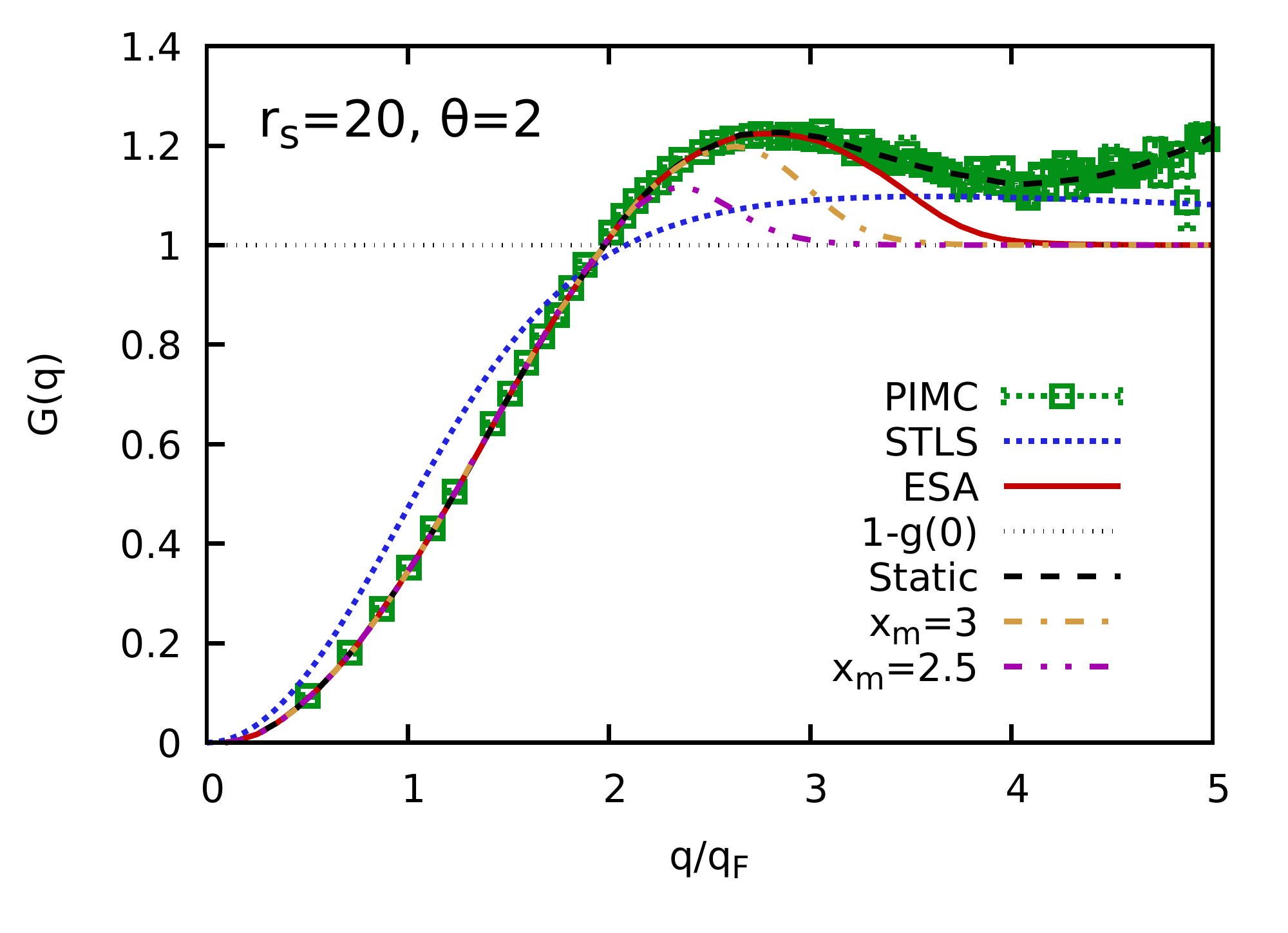}\\\vspace*{-1.1cm}\hspace*{-0.014\textwidth}
\includegraphics[width=0.495\textwidth]{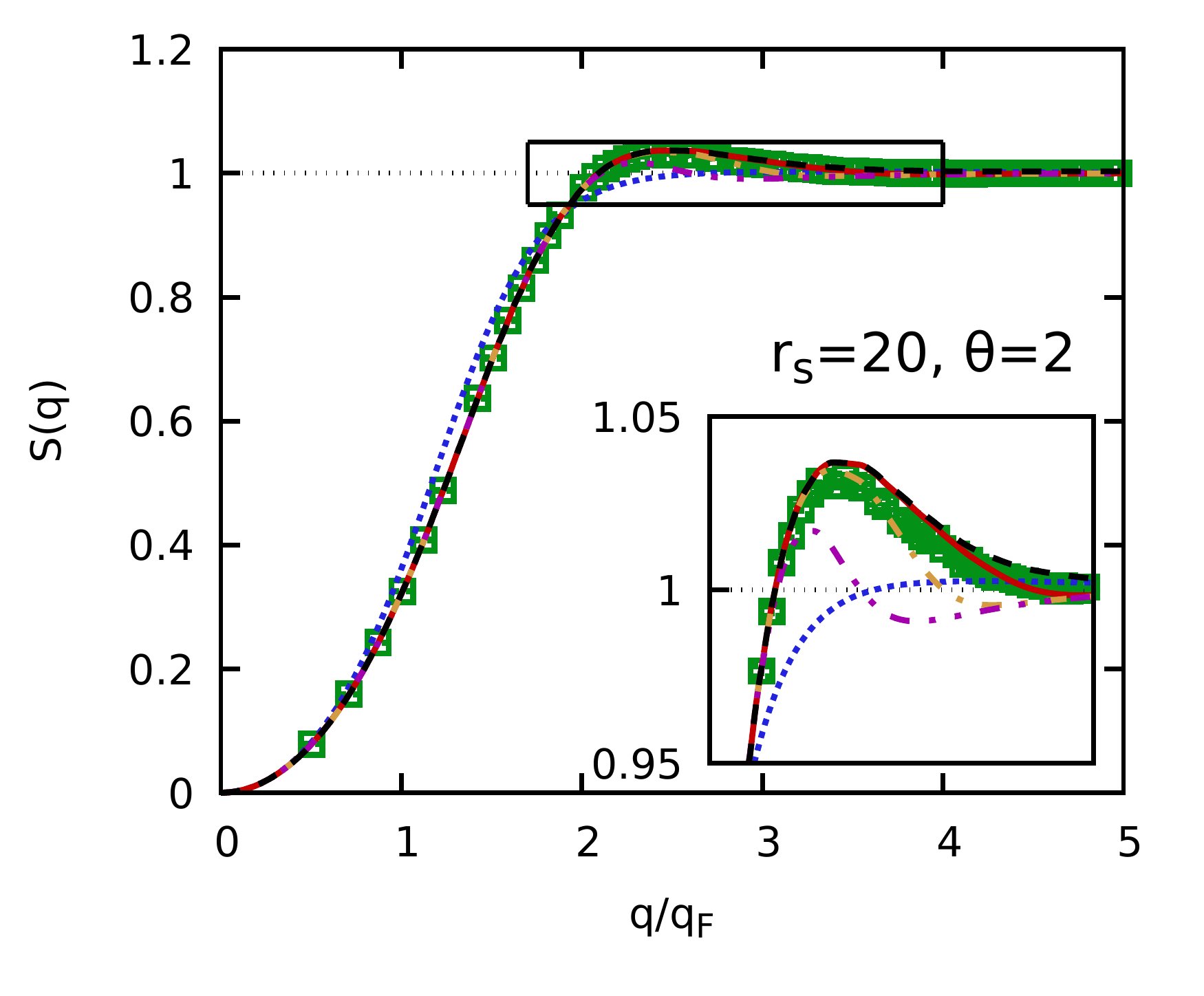}
\caption{\label{fig:Gq_rs20_theta2}
Top panel: Static local field correction for $r_s=20$ and $\theta=2$. Green squares are PIMC data for $G(q)$ from Ref.~\cite{dornheim_electron_liquid}, and dashed black line the neural-net representation from Ref.~\cite{dornheim_ML}. The solid red line shows $\overline{G}(q)$ within the ESA using Eq.~(\ref{eq:activation_xm}) [i.e., $x_m=3.58$], and the dash-dotted yellow and dash-double-dotted purple line show the ESA for $x_m=3$ and $x_m=2.5$. The dotted blue line shows $\overline{G}(q)$ from STLS, and the light grey line the analytical limit from Eq.~(\ref{eq:onTop}).
Bottom panel: Corresponding results for the static structure factor $S(q)$.
}
\end{figure}

An example for the impact of $x_m$ on both $\overline{G}(q)$ and the corresponding SSF is shown in Fig.~\ref{fig:Gq_rs20_theta2}. The top panel shows the LFC, and we observe an overall similar trend as for $\theta=1$ depicted in Fig.~\ref{fig:ESA}. The main differences both in the PIMC data and the neural net results for $G(q)$ are i) the comparably reduced height of the maximum, ii) the increased width of the maximum regarding $q$, and iii) the decreased slope of the positive tail at large wave numbers.
The red curve shows the ESA results for $\overline{G}(q)$ using the transition wave-number obtained from Eq.~(\ref{eq:activation_xm}), i.e., $x_m\approx3.58$. In particular, the red curve reproduces the peak structure of the exact static limit $G(q)$, and subsequently approaches the large-$q$ limit from Eq.~(\ref{eq:onTop}) [light dotted grey line].
In contrast, the dash-dotted yellow and dashed-double-dotted purple lines are ESA results for $x_m=3$ and $x_m=2.5$, respectively, and start to significantly deviate from $G(q)$ before the peak.
Finally, the dotted blue curve shows $\overline{G}(q)$ from STLS, and has been included as a reference.

Regarding $S(q)$, the solid red curve shows the best agreement to the PIMC data, whereas the \emph{static approximation} again exhibits the spurious behaviour for large $q$, albeit less pronounced than for $\theta=1$ shown above.
The ESA results for $x_m=3$, too, is in good agreement to the PIMC data, although there appears an unphysical minimum around $q=3q_\textnormal{F}$. The ESA curve for $x_m=2.5$, on the other hand, does not reproduce the maximum in $S(q)$ from the other data sets. Finally, the STLS curve does not provide an accurate description of the physical behaviour and systematically deviates from the exact results except in the limits of large and small $q$.

\subsection{Analytical representation}

Let us start this discussion by introducing a suitable functional form for the $q$-dependence of $\overline{G}_\textnormal{ESA}$ when $r_s$ and $\theta$ are fixed. First and foremost, we note that our parametrization is always constructed from Eq.~(\ref{eq:ESA}), which means that the task at hand is to find an appropriate representation of $G_\textnormal{nn}(q;r_s,\theta)$ that is sufficiently accurate in the wave-number regime where the neural net contributes to the ESA. The correct limit for large $q$, on the other hand, is built in automatically.

In addition, we would like to incorporate the exact long-wavelength limit of the static LFC that is given by the compressibility sum-rule~\cite{dornheim_ML,stls2} (CSR)
\begin{eqnarray}\label{eq:CSR}
\lim_{q\to0} G(q;r_s,\theta) &=& G_\textnormal{CSR}(q;r_s,\theta)\\ &=&  - \frac{q^2}{4\pi} \frac{\partial^2}{\partial n^2} \left( n f_\textnormal{xc} \right) \quad . \nonumber
\end{eqnarray}
This is achieved by the ansatz
\begin{eqnarray}\label{eq:const_fit}
G^{r_s,\theta}_\textnormal{nn,fit}(q) &=& G_\textnormal{CSR}(q;r_s,\theta)\\\nonumber & &\times\left[ 
\frac{1+\alpha^{r_s,\theta} x + \beta^{r_s,\theta}\sqrt{x}}{1+\gamma^{r_s,\theta}x + \delta^{r_s,\theta}x^{1.25} + G_\textnormal{CSR}(q;r_s,\theta)}
\right]\ ,
\end{eqnarray}
where $x=q/q_\textnormal{F}$ is the reduced wave-number and the super-scripts in the four free parameters $\alpha^{r_s,\theta}$, $\beta^{r_s,\theta}$, $\gamma^{r_s,\theta}$, and $\delta^{r_s,\theta}$ indicate that they are obtained for fixed values of $\theta$ and $r_s$. We note that the $G_\textnormal{CSR}(q;r_s,\theta)$ term in the denominator of the square brackets compensates the equal pre-factor for large $q$.

\begin{figure}\centering
\includegraphics[width=0.475\textwidth]{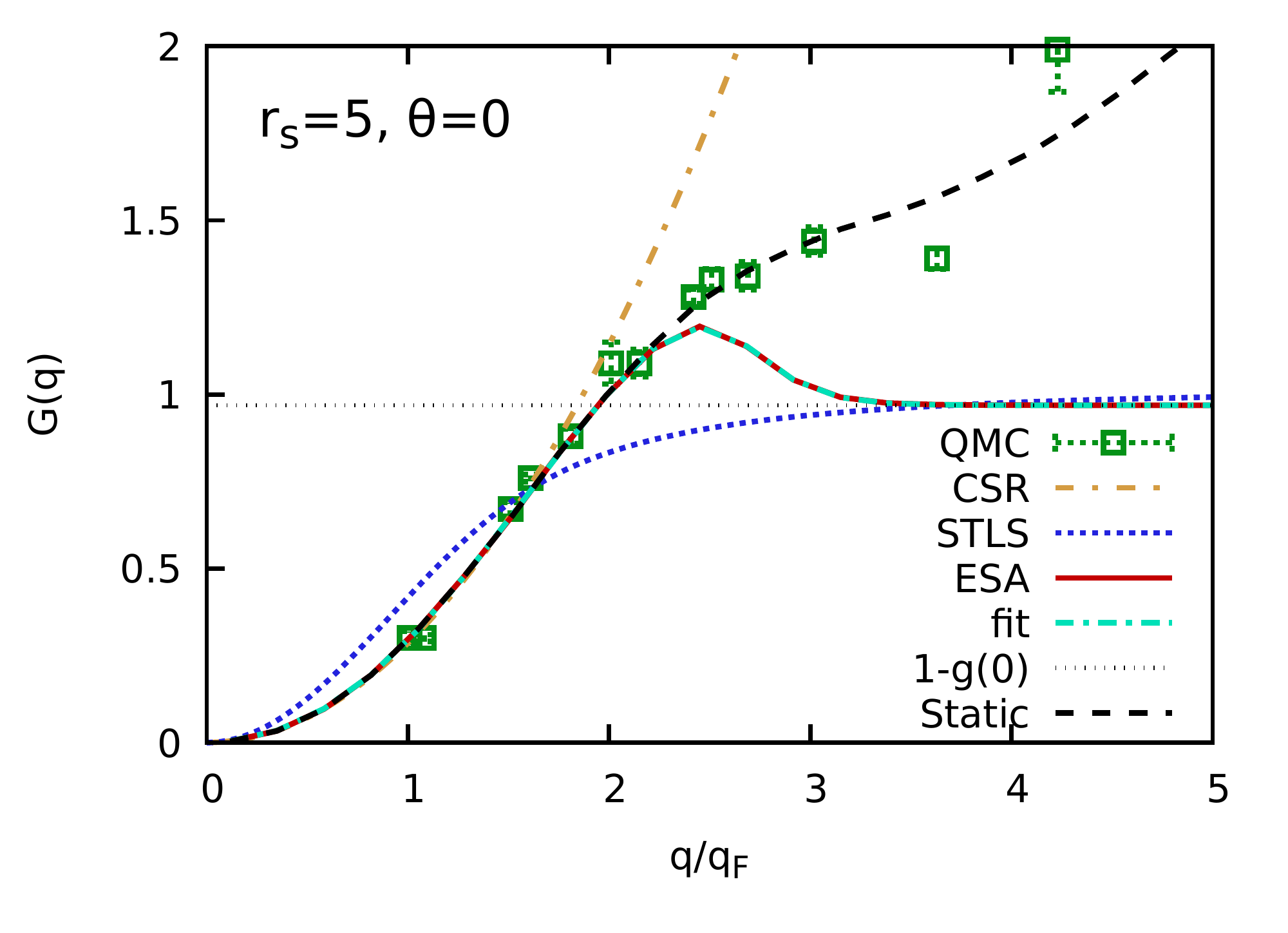}\\\vspace*{-1.1cm}
\includegraphics[width=0.475\textwidth]{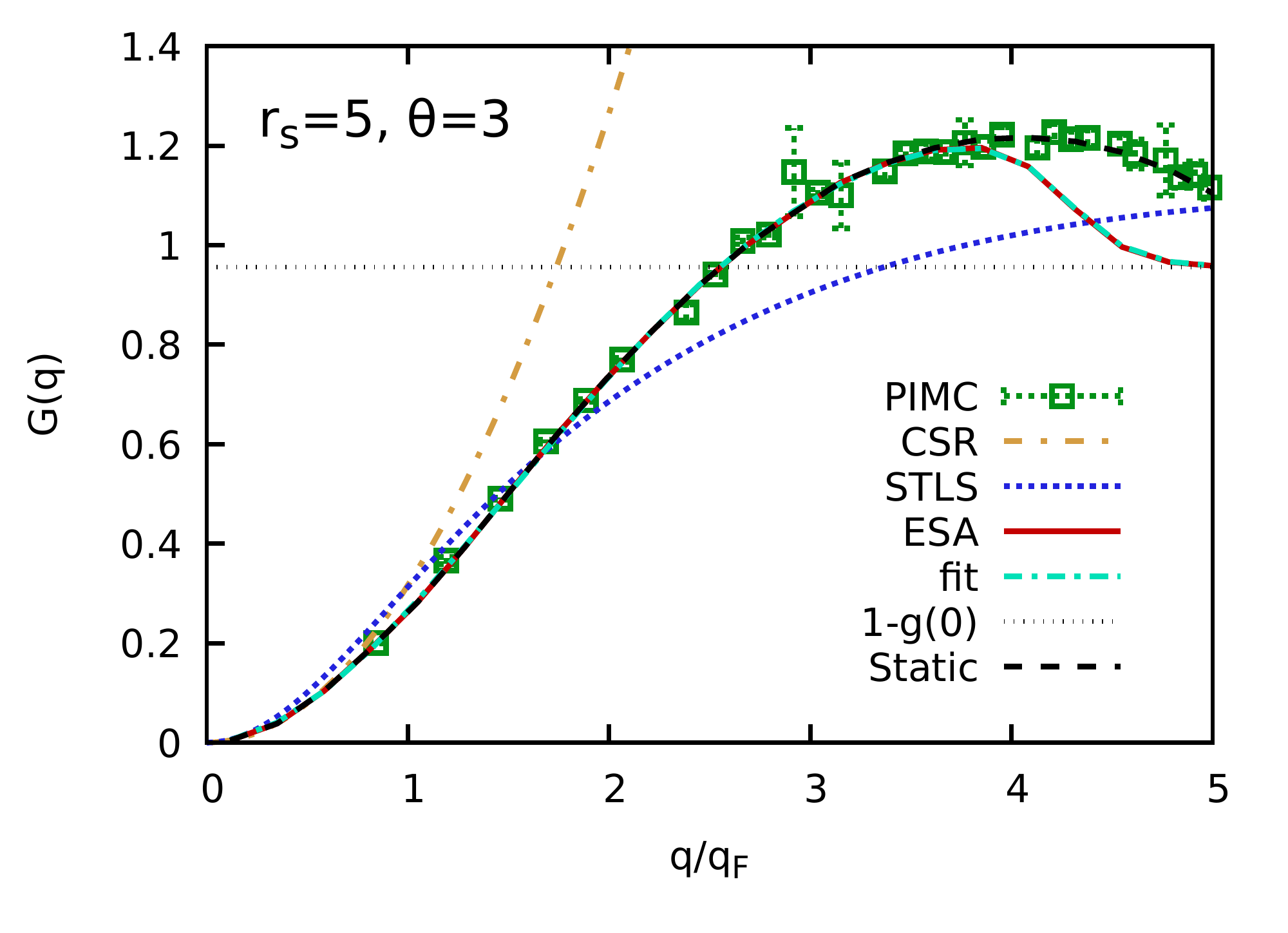}
\caption{\label{fig:Gq_rs5}
Static local field correction for $r_s=5$ and $\theta=0$ (top) and $\theta=3$ (bottom). Green squares are ground-state QMC data from Ref.~\cite{moroni2} (PIMC data for $G(q)$ from Ref.~\cite{dornheim_ML}) for $\theta=0$ ($\theta=3$), and dashed black lines the neural-net representation from Ref.~\cite{dornheim_ML}. The solid red line shows $\overline{G}(q)$ within the ESA using Eq.~(\ref{eq:activation_xm}), and the light blue dash-dotted curve the corresponding fit from Eq.~(\ref{eq:const_fit}).
The dotted blue line shows $\overline{G}(q)$ from STLS, and the light grey line the analytical limit from Eq.~(\ref{eq:onTop}).
}
\end{figure}

Two examples for the application of Eq.~(\ref{eq:const_fit}) are shown in Fig.~\ref{fig:Gq_rs5}, where the local field correction is shown for $r_s=5$ and $\theta=0$ (top) and $\theta=3$ (bottom). The red curve shows $\overline{G}(q)$ within the ESA, and the light blue dash-dotted curve a fit to these data using Eq.~(\ref{eq:const_fit}) as a functional form for $\theta$ and $r_s$ being constant.
First and foremost, we note that the fit perfectly reproduces the ESA, and no fitting error can be resolved with the naked eye.

The dash-dotted yellow curves show the CSR [Eq.~(\ref{eq:CSR})], which has been included into Eq.~(\ref{eq:const_fit}).
In the ground state, we indeed find good agreement between the CSR, the QMC data, the neural net, and also the ESA for $q\lesssim2q_\textnormal{F}$.
This is somewhat changed for $\theta=3$, where the yellow curve exhibits more pronounced deviations from the PIMC data and all other curves. Still, we note that the functional form from Eq.~(\ref{eq:const_fit}) is capable to accommodate this finding, and attains the small-wave number limit only for small $q$ in this case.

We thus conclude that Eq.~(\ref{eq:const_fit}) constitutes a suitable basis for the desired analytical representation $\overline{G}_\textnormal{ESA}(q;r_s,\theta)$. As a next step, we make Eq.~(\ref{eq:const_fit}) dependent on the density parameter $r_s$. To achieve this goal, we parametrize the free parameters as:
\begin{eqnarray}\label{eq:kappa_rs_dependence}
\kappa^\theta(r_s) = \frac{a_\kappa^\theta + b_\kappa^\theta r_s}{1 + c_\kappa^\theta r_s}\ ,
\end{eqnarray}
with $\kappa\in\{\alpha,\beta,\gamma,\delta\}$. Thus, the characterization of the $r_s$-dependence for a single isotherm requires the determination of $12$ free parameters.
This results in the isothermic representation of the LFC of the form
\begin{eqnarray}\label{eq:isotherm_fit}
G^{\theta}_\textnormal{nn,fit}(&q&;r_s) = G_\textnormal{CSR}(q;r_s,\theta)\\\nonumber  &\times&\left[ 
\frac{1+\alpha^{\theta}(r_s) x + \beta^{\theta}(r_s)\sqrt{x}}{1+\gamma^{\theta}(r_s)x + \delta^{\theta}(r_s)x^{1.25} + G_\textnormal{CSR}(q;r_s,\theta)}
\right]\ ,
\end{eqnarray}

\begin{figure*}\centering
\includegraphics[width=0.475\textwidth]{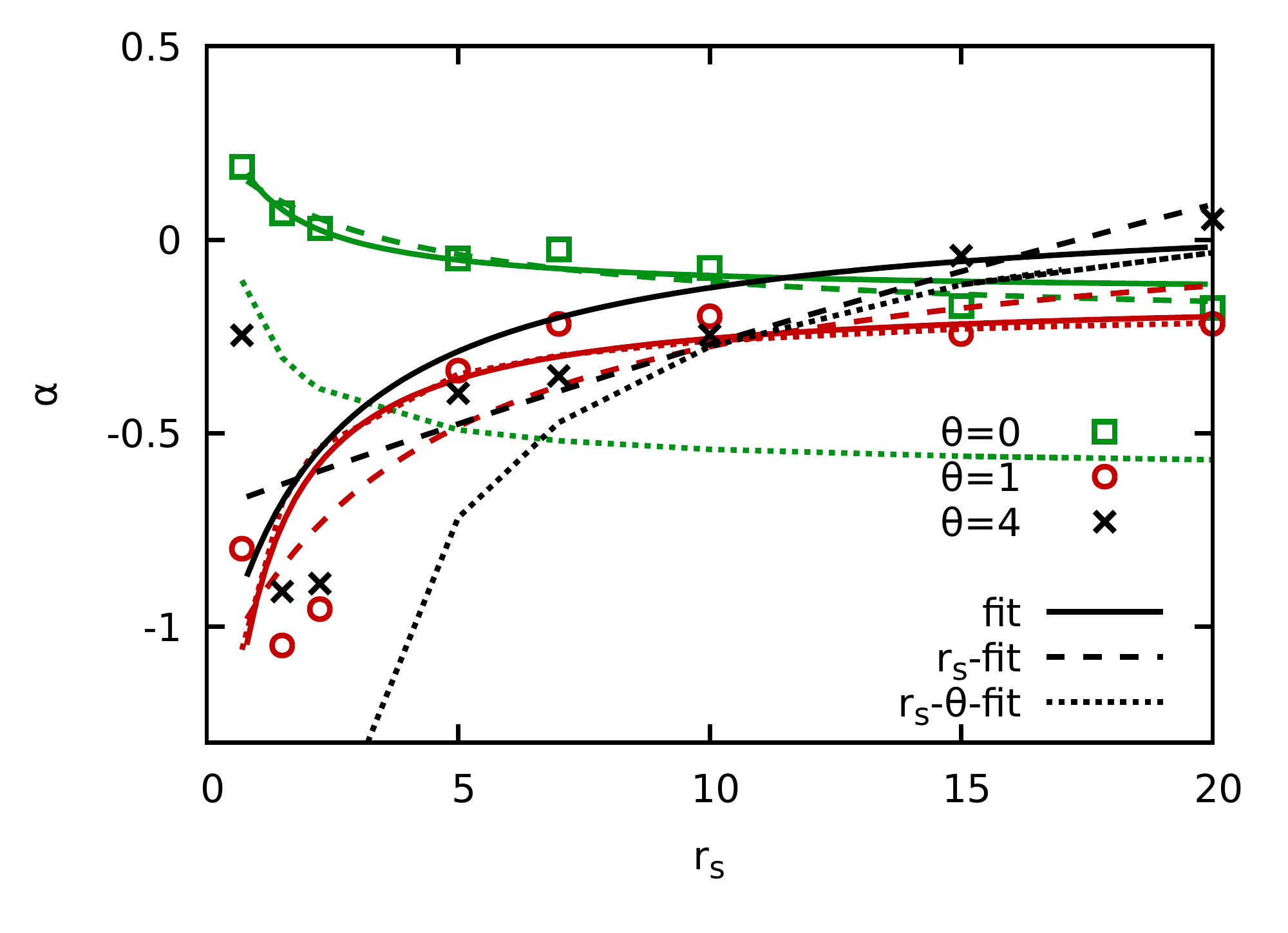}\includegraphics[width=0.475\textwidth]{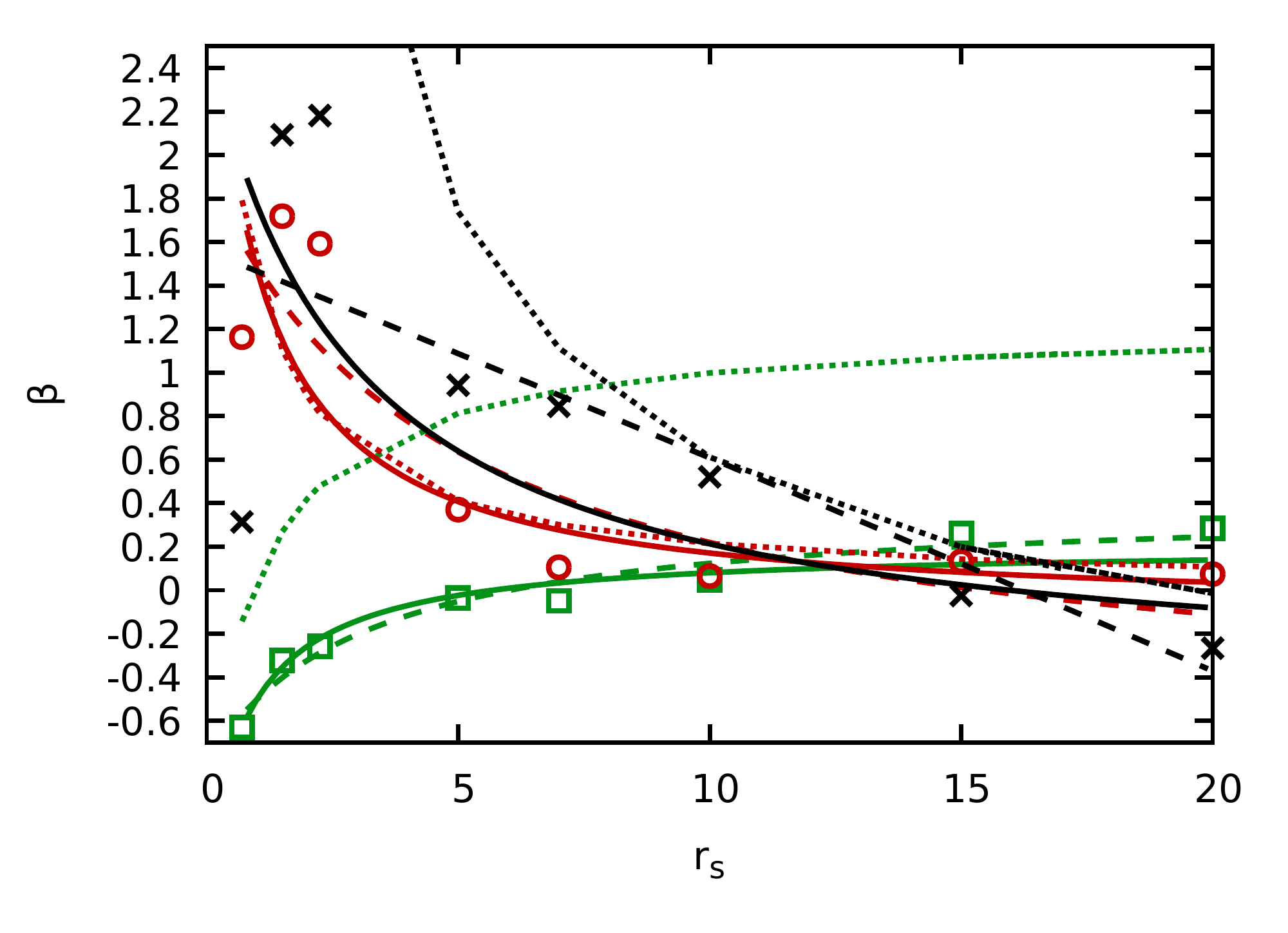}\\\vspace*{-1.1cm}\includegraphics[width=0.475\textwidth]{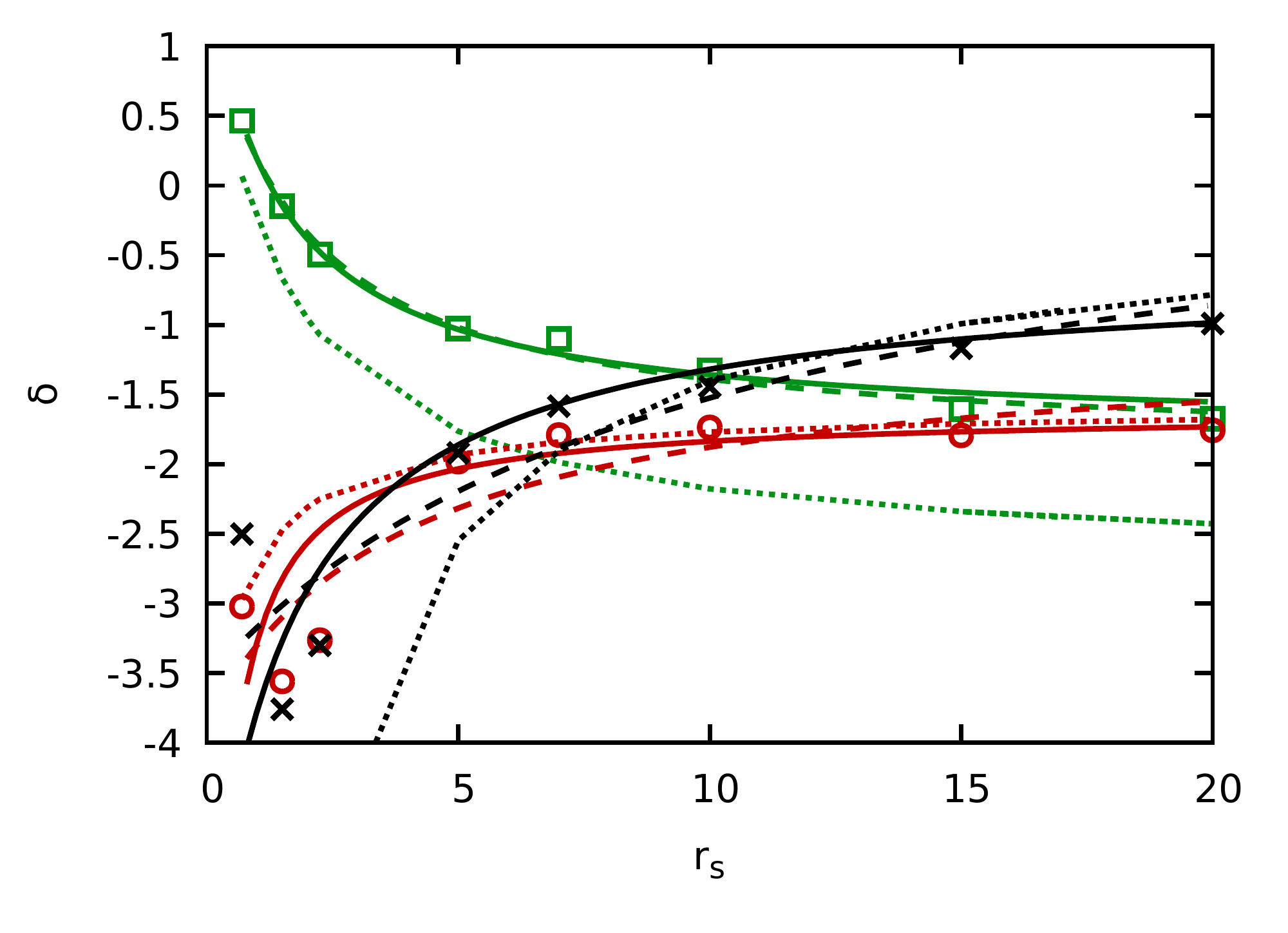}\includegraphics[width=0.475\textwidth]{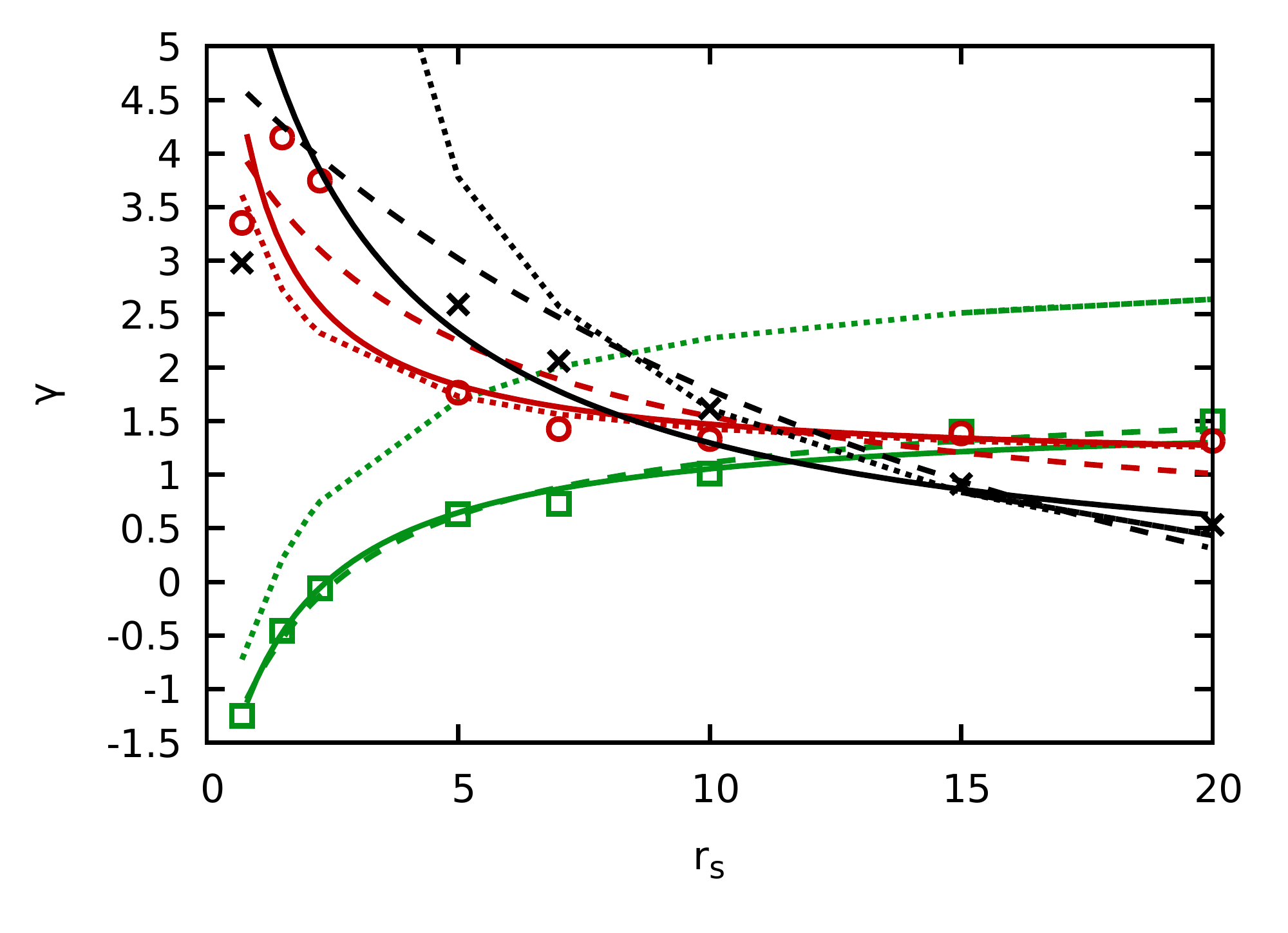}
\caption{\label{fig:ABCD}
Dependence of the fit parameters $\alpha-\delta$ (clockwise) from Eq.~(\ref{eq:const_fit} on the density parameter $r_s$ for $\theta=0$ (green), $\theta=1$ (red), and $\theta=4$ (black). The symbols have been obtained by fitting Eq.~(\ref{eq:const_fit}) to ESA data for individual constant values of both $r_s$ and $\theta$, and the solid lines have been fitted to these data using Eq.~(\ref{eq:kappa_rs_dependence}) as a functional form. The dashed lines have been obtained from isothermal fits over the full $r_s$-range using Eq.~(\ref{eq:isotherm_fit}), and the dotted curves from the final fit over the full $r_s$-$\theta$-$q$-dependence, see~Eq.~(\ref{eq:final_fit}).
}
\end{figure*}

This isothermic representation is illustrated in Fig.~\ref{fig:ABCD}, where we show the full $r_s$-dependence of the four free parameter $\alpha-\delta$ (clockwise) for $\theta=0$ (green), $\theta=1$ (red), and $\theta=4$ (black). The symbols have been obtained by fitting Eq.~(\ref{eq:const_fit}) to ESA data for $\overline{G}(q)$ for constant values of $r_s$ and $\theta$. The solid lines have been subsequently obtained by fitting the representation of Eq.~(\ref{eq:kappa_rs_dependence}) to these data over the entire $r_s$-range. The resulting curves are indeed smooth and qualitatively capture the main trends from the data points. 
Finally, the dashed curves have been computed by fitting Eq.~(\ref{eq:isotherm_fit}) to ESA data over the entire $r_s$-range, but for constant values of $\theta$.
Interestingly, this final optimization step results in qualitative change of the description of all four parameters for $\theta=4$, but only mildly changes the results for both $\theta=1$ and $\theta=0$.

\begin{figure}\centering
\includegraphics[width=0.475\textwidth]{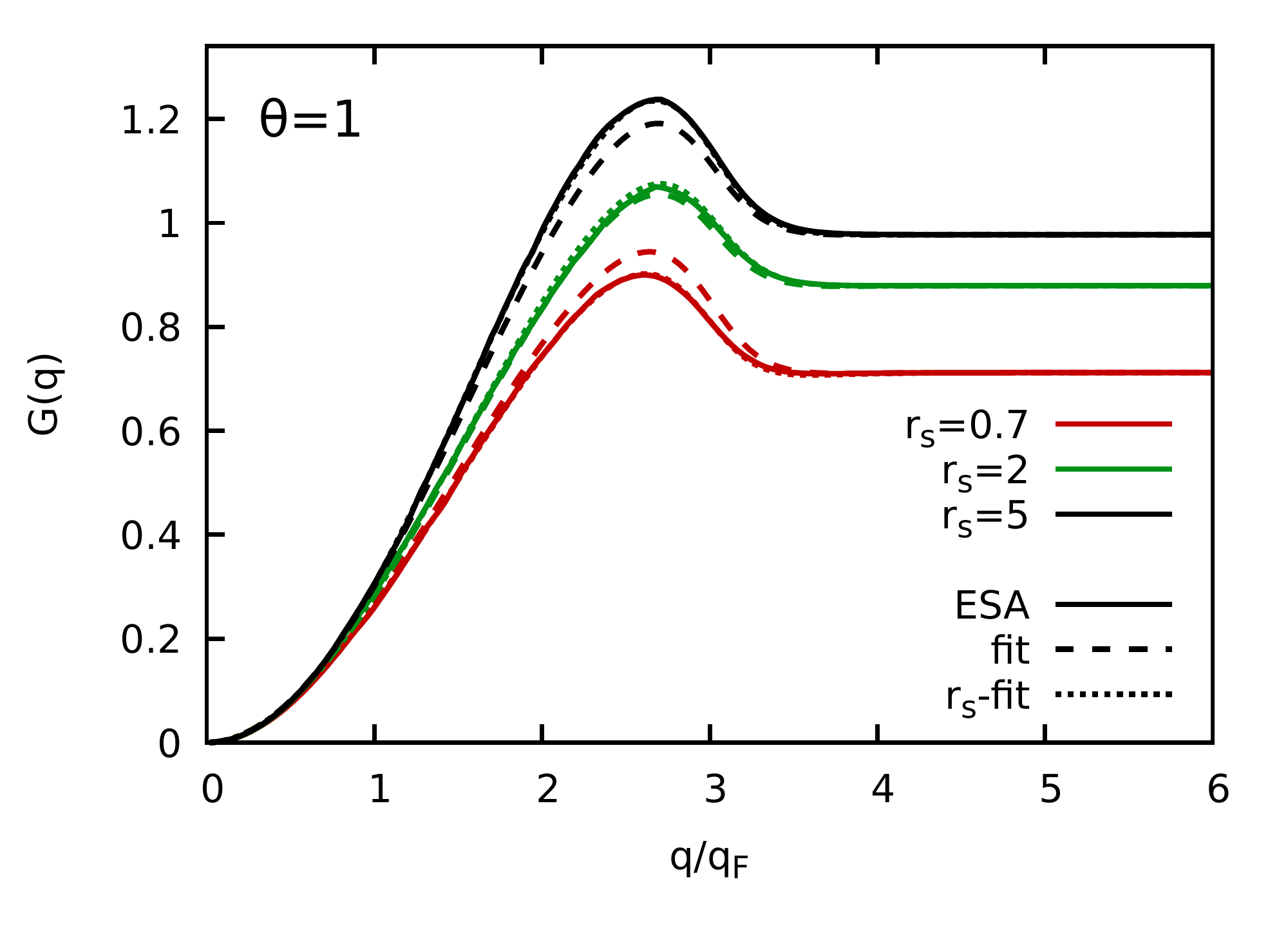}\\\vspace*{-1.1cm}
\includegraphics[width=0.475\textwidth]{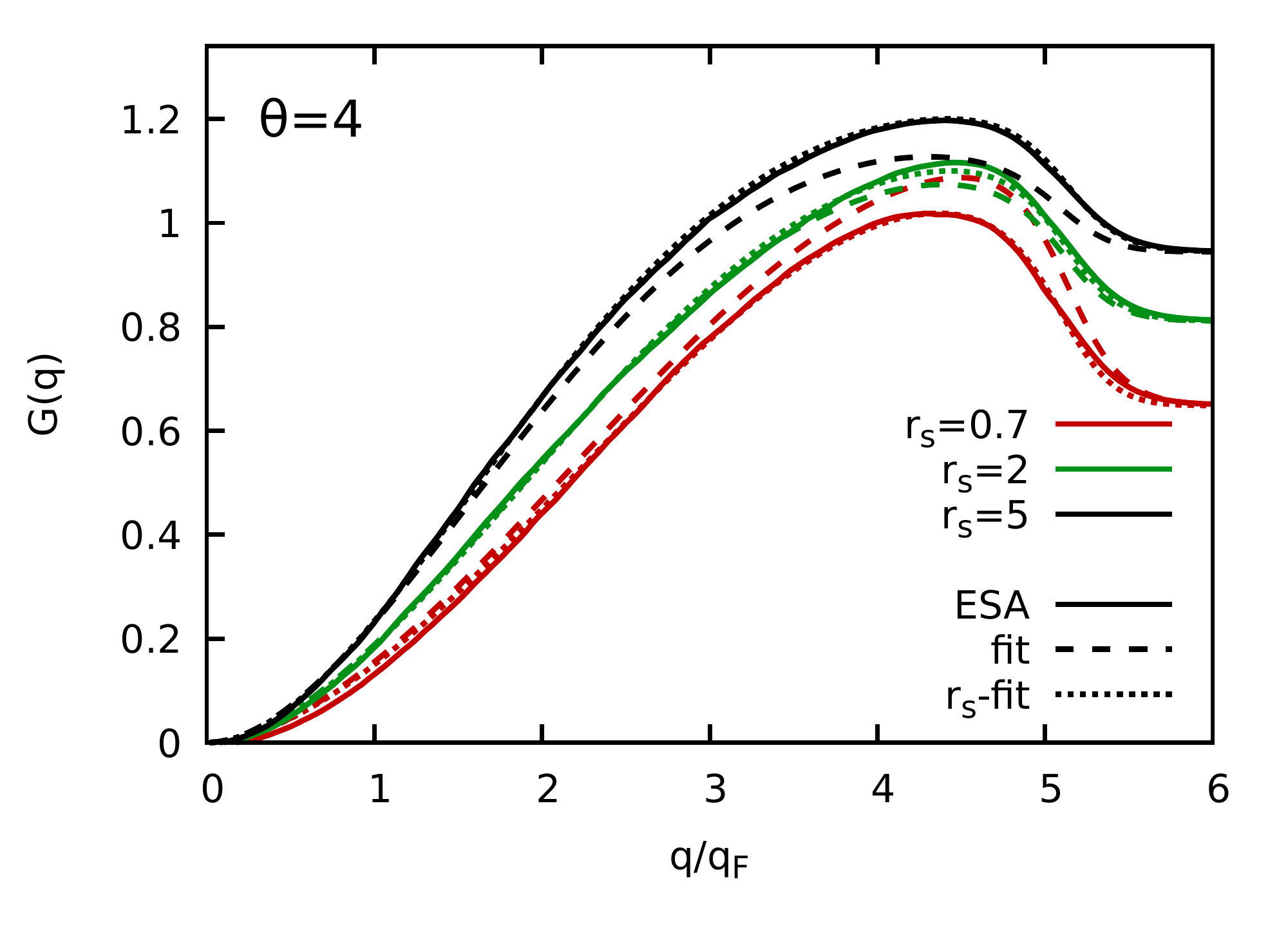}
\caption{\label{fig:Gq_isotherm}
Illustration of the isothermic fit function of the local field correction $\overline{G}_\textnormal{ESA}(q)$ for $\theta=1$ (top) and $\theta=4$ (bottom).
The red, green, and black curves depict different results for $r_s=0.7$, $r_s=2$, and $r_s=5$, respectively. Solid: ESA; dashed: fitted $r_s$-dependence of the individual coefficients $\alpha-\delta$ from Eq.~(\ref{eq:const_fit}) according to Eq.~(\ref{eq:kappa_rs_dependence}); dotted: full isothermic fits of $\overline{G}_\textnormal{ESA}^\theta(q)$ via Eq.~(\ref{eq:isotherm_fit}).
}
\end{figure}

Let us for now postpone the discussion of the dotted curve in Fig.~\ref{fig:ABCD}, and consider Fig.~\ref{fig:Gq_isotherm} instead. In particular, we show the results of the isothermic fitting procedure for $\theta=1$ (top) and $\theta=4$ (bottom), with the red, green, and black curves corresponding to different data sets for $r_s=0.7$, $r_s=2$, and $r_s=5$, respectively.
More specifically, the solid lines show the ESA reference data for $\overline{G}(q)$, and the dashed curves have been obtained by fitting the data points for $\alpha-\delta$ shown in Fig.~\ref{fig:ABCD} via Eq.~(\ref{eq:kappa_rs_dependence}). For $\theta=1$, this simple procedure alone leads to an excellent representation of $\overline{G}^\theta(q;r_s)$. The dotted curve has been obtained by performing the full isothermic fits, i.e., by fitting Eq.~(\ref{eq:isotherm_fit}) to ESA data over the entire $r_s$-range, but with $\theta$ being constant. Indeed, we find only minor deviations between the dashed and the dotted curve.

For $\theta=4$, on the other hand, the simple representation of the fit parameters from Eq.~(\ref{eq:const_fit}) results in a substantially less accurate representation of $\overline{G}^\theta_\textnormal{ESA}(q;r_s)$, and the systematic error is most pronounced at high density, $r_s=0.7$.
This shortcoming can be remedied by performing the full isothermic fit of the entire $q$-$r_s$-dependence, and the dotted curves are in excellent agreement to the original ESA data everywhere.
We thus conclude that the functional form of Eq.~(\ref{eq:isotherm_fit}) constitutes an adequate representation of $\overline{G}^\theta_\textnormal{ESA}(q;r_s)$.

\subsection{\label{sec:fina}Final representation of $\overline{G}_\textnormal{ESA}(q;r_s,\theta)$}

The final step is then given by the construction of
an analytical representation of the full $r_s$-$\theta$-$q$-dependence by expressing the parameters $a_\kappa^\theta$, $b_\kappa^\theta$, and $c_\kappa^\theta$ in Eq.~(\ref{eq:kappa_rs_dependence}) as a function of $\theta$,
\begin{eqnarray}\label{eq:f_kappa}
f_\kappa(\theta) = a_f + b_f \theta + c_f \theta^{1.5}\ .
\end{eqnarray}
This results in three free parameters for each of the $12$ coefficients required for the characterization of the $r_s$-dependence, i.e., a total of $36$ parameters that have to be determined by the fitting procedure.

The full three-dimensional fit-function is then given by
\begin{eqnarray}\label{eq:final_fit}
G_\textnormal{nn,fit}(&q&;r_s,\theta) = G_\textnormal{CSR}(q;r_s,\theta)\\ &\times&\left[ 
\frac{1+\alpha(r_s,\theta) x + \beta(r_s,\theta)\sqrt{x}}{1+\gamma(r_s,\theta)x + \delta(r_s,\theta)x^{1.25}+G_\textnormal{CSR}(q;r_s,\theta)}
\right]\ ,\nonumber
\end{eqnarray}
where the functions $\kappa(r_s,\theta)$ [with $\kappa\in\{\alpha,\beta,\gamma,\delta\}$] are given by
\begin{eqnarray}
\kappa(r_s,\theta) = \frac{a_\kappa(\theta) + b_\kappa(\theta) r_s}{1 + c_\kappa(\theta) r_s}\ ,
\end{eqnarray}
and the $\theta$-dependent coefficients follow Eq.~(\ref{eq:f_kappa}).

\begin{table*}\caption{\label{tab:tab} Fit parameters for the analytic parametrization of $\overline{G}_\textnormal{ESA}(q;r_s,\theta)$ from Eq.~(\ref{eq:final_fit}). For each of the coefficients $a_\alpha, b_\alpha, \dots, c_\gamma$, we give the three free parameters from Eq.~(\ref{eq:f_kappa}), $a_f$, $b_f$, and $c_f$. A short python implementation is freely available online~\cite{code}.}
\begin{ruledtabular}\begin{tabular}{c|c|rrr}
 \hspace*{1.5cm}&  $a_\alpha$& $0.66477593$ & $-4.59280227$ &  $1.24649624$ \\
 $\alpha$ & $b_\alpha$ & $-1.27089927$ &
  $1.26706839$ &  $-0.4327608$ \\ 
  & $c_\alpha$ & $2.09717766$ &  $1.15424724$ &
  $-0.65356955$ \\ \hline 
  & $a_\beta$ & $-1.0206202$ & $5.16041218$ & $-0.23880981$ \\
  $\beta$ & $b_\beta$ &   $1.07356921$ & $-1.67311761$ &  $0.58928105$ \\
  & $c_\beta$ & $0.8469662$ &
 $1.54029035$ &  $-0.71145445$ \\ \hline
 & $a_\gamma$ & $-2.31252076$ &  $5.83181391$ &
 $2.29489749$\\
 $\gamma$ & $b_\gamma$ &  $1.76614589$ & $-0.09710839$ &  $-0.33180686$ \\
  & $c_\gamma$ &  $0.56560236$ & $1.10948188$ &  $-0.43213648$ \\ \hline 
  & $a_\delta$ &  $1.3742155$ &
 $-4.01393906$ &  $-1.65187145$ \\
 $\delta$ & $b_\delta$ & $-1.75381153$ &  $-1.17022854$ &
  $0.76772906$ \\
 & $c_\delta$ & $0.63867766$ & $1.07863273$ &  $-0.35630091$
\end{tabular}\end{ruledtabular}\end{table*}

Our final analytical representation of the LFC within the effective static approximation immediately follows from plugging Eq.~(\ref{eq:final_fit}) into Eq.~(\ref{eq:ESA}),
\begin{eqnarray}\label{eq:analytical}
\overline{G}_\textnormal{ESA,fit}(q;r_s,\theta) &=& G_\textnormal{nn,fit}(q;r_s,\theta)\left(1-A(x)\right) \\ \nonumber & &+ \left(1-g(0;r_s,\theta)\right)A(x) \ .
\end{eqnarray}
The thus fitted coefficients are given in Tab.~\ref{tab:tab}, and a corresponding python implementation is freely available online~\cite{code}.

\begin{figure}\centering
\includegraphics[width=0.475\textwidth]{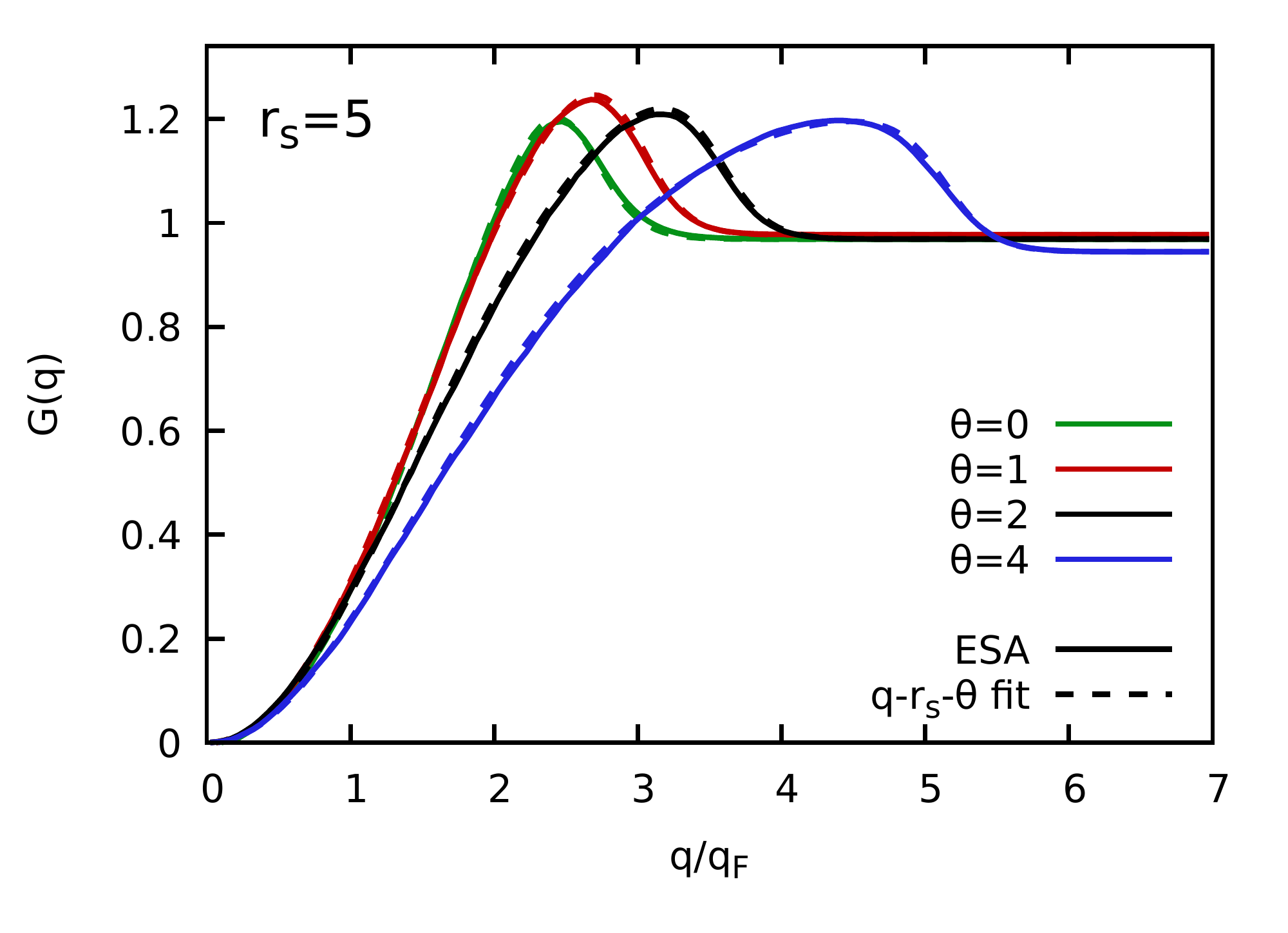}\\\vspace*{-1.1cm}
\includegraphics[width=0.475\textwidth]{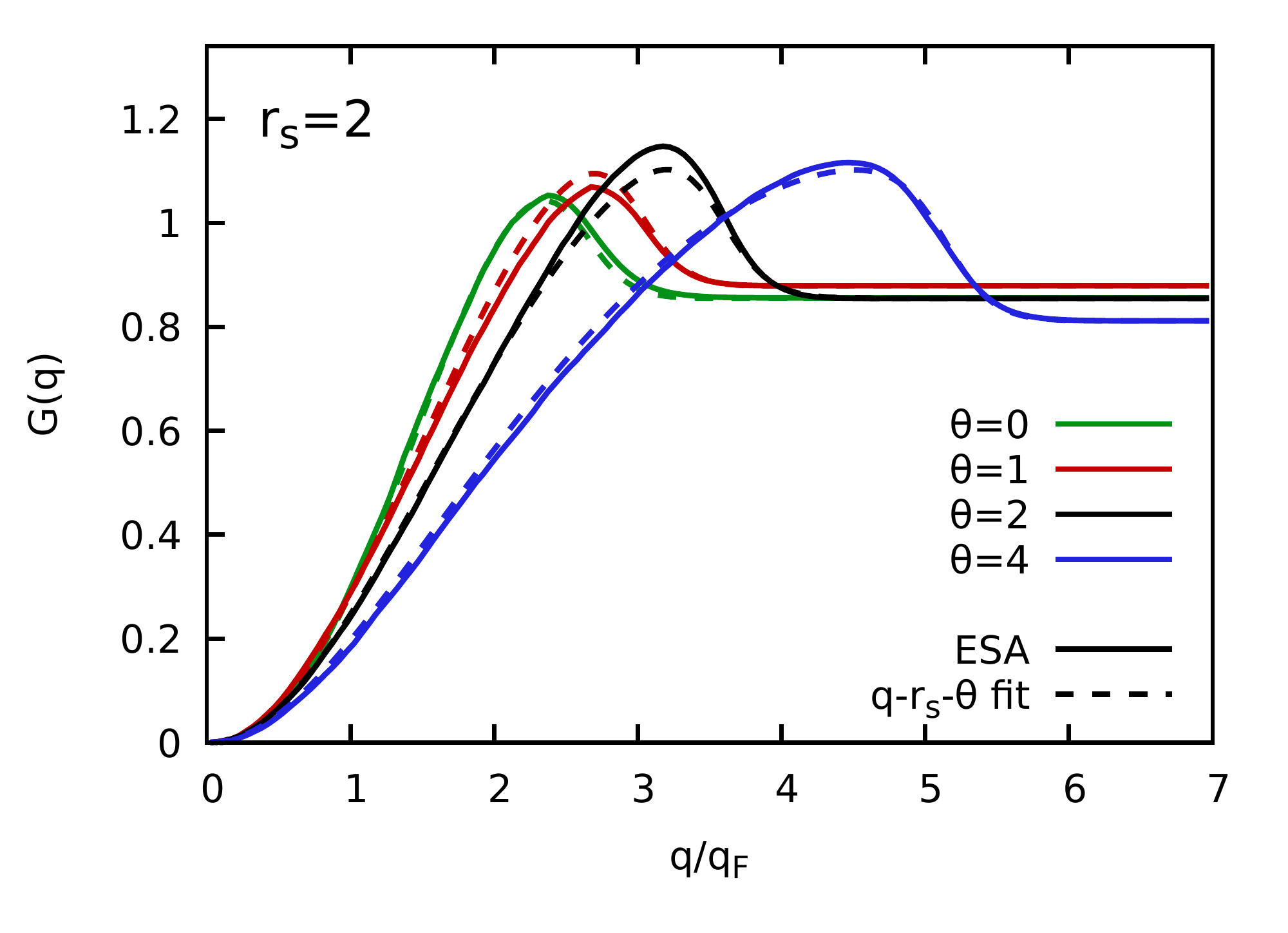}
\caption{\label{fig:Gq_3D}
Analytical representation of $\overline{G}_\textnormal{ESA}(q;r_s,\theta)$: Shown are ESA results (solid lines) and our final analytical representation, Eq.~(\ref{eq:analytical}).
}
\end{figure}

The resulting analytical representation $\overline{G}_\textnormal{ESA}(q;r_s,\theta)$ is illustrated in Fig.~\ref{fig:Gq_3D}, where we compare it (dashed lines) to the original ESA data at $r_s=5$ (top) and $r_s=2$, i.e., two metallic densities that are of high interest in the context of WDM research.

More specifically, $r_s=5$ corresponds to a strongly coupled system, where an accurate treatment of electronic exchange--correlation effects is paramount~\cite{low_density1}.
These conditions can be realized experimentally in hydrogen jets~\cite{Zastrau} and evaporation experiments~\cite{benage,karasiev_importance,low_density1,low_density2}. 
The green, red, black, and blue curves show results for $\theta=0$, $\theta=1$, $\theta=2$, and $\theta=4$, respectively, and we find that our new analytical representation of $\overline{G}_\textnormal{ESA}(q;r_s,\theta)$ is in excellent agreement to the ESA input data everywhere.

The bottom panel corresponds to $r_s=2$, which is relevant e.g. for the investigation of aluminum~\cite{Sperling_PRL_2015,Ramakrishna_PRB_2021}. Here, too, we find excellent agreement between the fitted function and the ESA input data for $\theta=0$ and $\theta=4$, while small, yet significant deviations appear at intermediate wave numbers for $\theta=2$ and $\theta=1$. Still, it is important to note that these deviations do not exceed the statistical uncertainty of the original PIMC input data for $G(q)$ on which the neural net from Ref.~\cite{dornheim_ML} and the ESA are based.

We thus conclude that our analytical representation of $\overline{G}_\textnormal{ESA}(q;r_s,\theta)$ provides a highly accurate description of electronic--exchange correlation effects over the entire relevant parameter range. The application of this representation for the computation of other material properties like the static structure factor $S(q)$, interaction energy $v$, or dielectric function $\epsilon(q)$ is discussed in detail in Sec.~\ref{sec:results}.

\section{Results\label{sec:results}}

\subsection{The static local field correction}

\begin{figure}\centering
\includegraphics[width=0.475\textwidth]{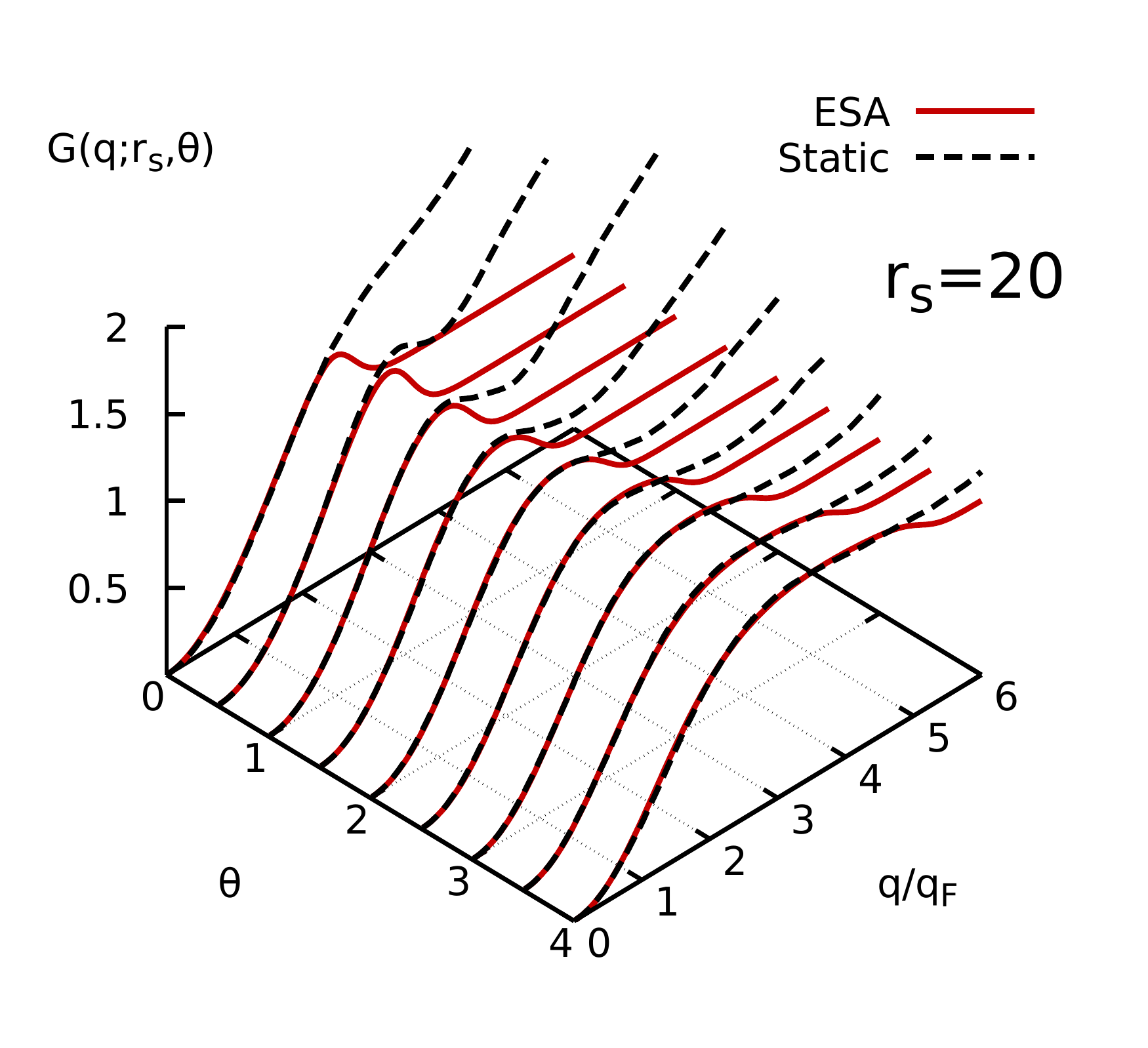}\\\vspace*{-0.65cm}
\includegraphics[width=0.475\textwidth]{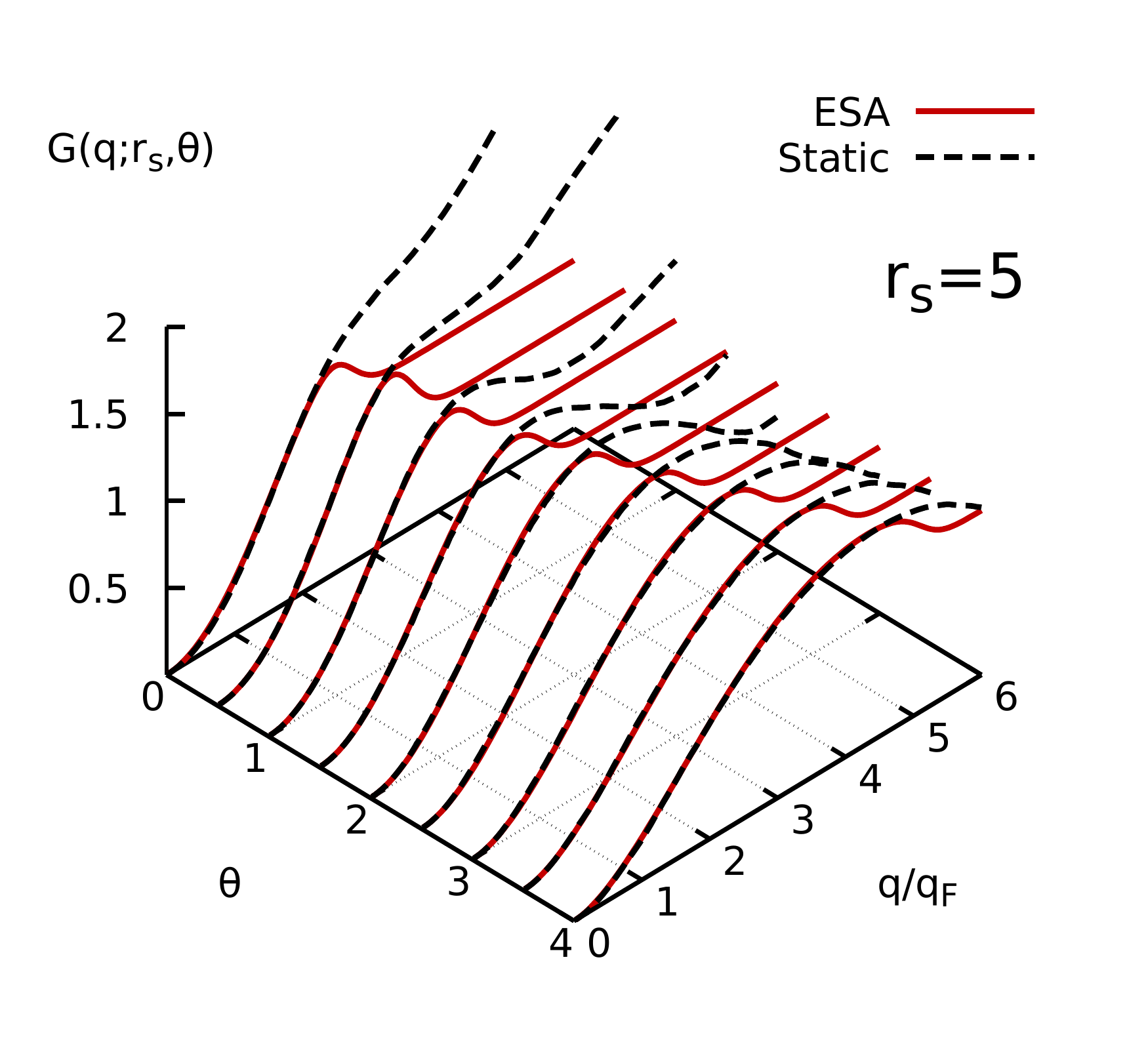}
\caption{\label{fig:Gq_3D_theta}
The local field correction in the $\theta$-$q$-plane: The solid red and dashed black curves show our analytical representation of $\overline{G}_\textnormal{ESA}(q;r_s,\theta)$ and the neural-net representation of the exact static limit $G(q)$ from Ref.~\cite{dornheim_ML} for $r_s=20$ (top) and $r_s=5$ (bottom).
}
\end{figure}

Let us begin the investigation of the results that can be obtained within the ESA by briefly recapitulating a few important properties of $\overline{G}_\textnormal{ESA}(q;r_s,\theta)$ itself. To this end, we show the LFC in the $\theta$-$q$-plane for $r_s=20$ (top) and $r_s=5$ (bottom) in Fig.~\ref{fig:Gq_3D_theta}. More specifically, the dashed black lines show the neural-net results for $G(q)$ from Ref.~\cite{dornheim_ML}, and the solid red lines the corresponding data for our analytical representation of $\overline{G}_\textnormal{ESA}(q;r_s,\theta)$. First and foremost, we note that the temperature dependence is qualitatively similar for both values of the density parameter; a more detailed analysis of the $r_s$-dependence of the LFC is presented in Fig.~\ref{fig:Gq_3D_rs} below.
As usual, $G(q)$ exhibits a non-constant behaviour for large wave numbers, whereas the ESA converges towards Eq.~(\ref{eq:onTop}). In addition, our parametrization nicely reproduces the neural-net for $x<x_m(\theta)$, which further illustrates the high quality of the representation.
Finally, we find that the exact static limit of the LFC, too, becomes increasingly flat at large $q$ for high temperatures, which can be seen particularly well for $r_s=20$. In fact, simultaneously considering large values of $r_s$ and $\theta$ brings us to the classical limit, where $G(q)$ converges towards one for large wave numbers~\cite{ICHIMARU198791},
\begin{eqnarray}
\lim_{q\to\infty} G_\textnormal{classical}(q) = 1\ .
\end{eqnarray}
Moreover, the ESA and $G(q)$ converge in this regime as the static structure factor can always be computed from the static LFC only via the exact relation~\cite{ICHIMARU198791,Mithen_PRE_2012}
\begin{eqnarray}\label{eq:S_from_G}
S_\textnormal{classical}(q) = \frac{1}{1-\frac{4\pi}{q^2}\left[G_\textnormal{classical}(q)-1\right]\beta n}\ .
\end{eqnarray}
In other words, the spurious effects due to the \emph{static approximation} and the need for the ESA in WDM applications are a direct consequence of quantum effects on electronic exchange--correlation effects, which only vanish in the classical limit.

\begin{figure}\centering
\includegraphics[width=0.475\textwidth]{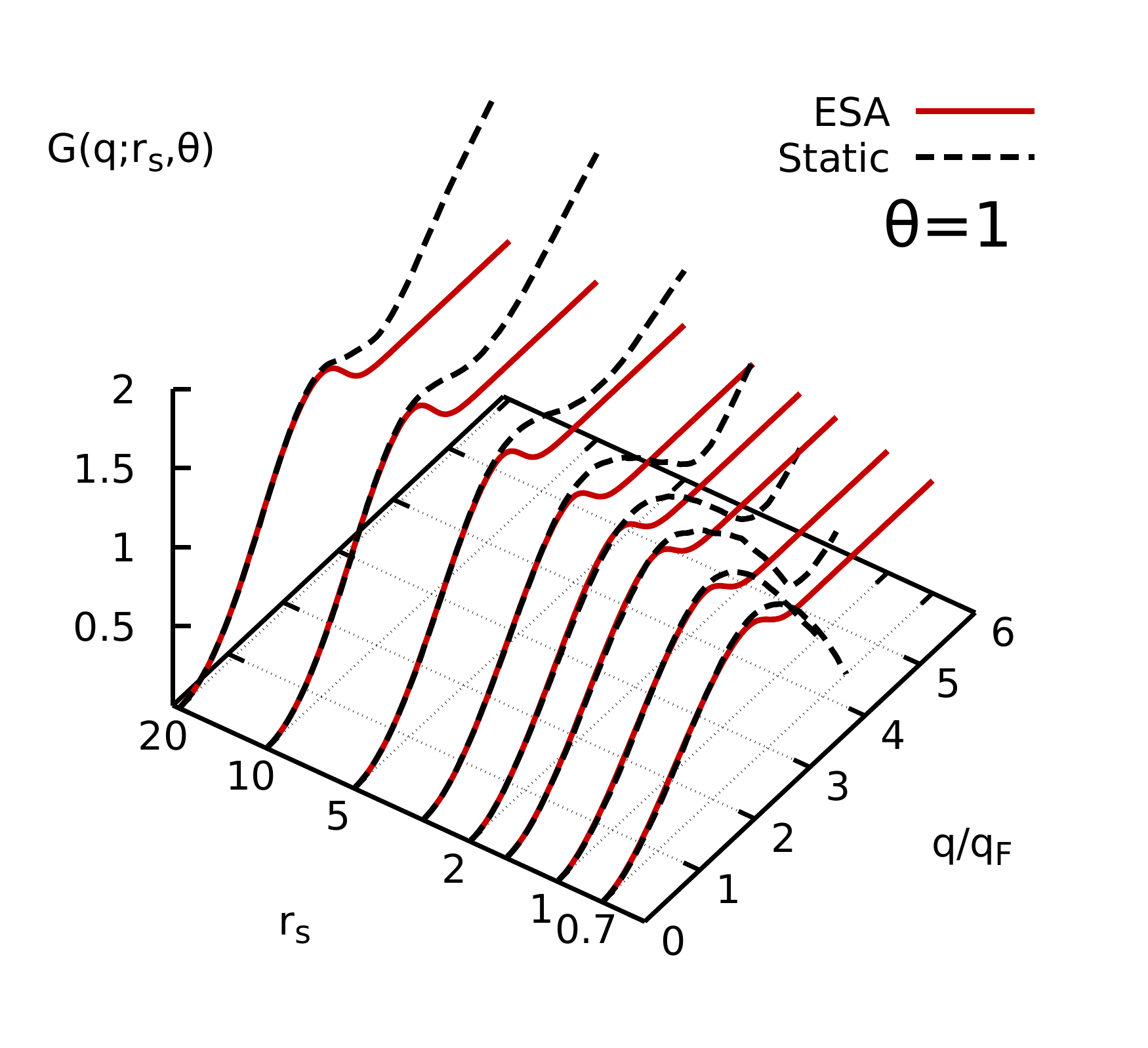}
\caption{\label{fig:Gq_3D_rs}
The local field correction in the $r_s$-$q$-plane: The solid red and dashed black curves show our analytical representation of $\overline{G}_\textnormal{ESA}(q;r_s,\theta)$ and the neural-net representation of the exact static limit $G(q)$ from Ref.~\cite{dornheim_ML} for $\theta=1$. Note the logarithmic scale of the $r_s$-axis.
}
\end{figure}

Let us next consider the dependence of the LFC on the density parameter $r_s$, which is shown in Fig.~\ref{fig:Gq_3D_rs} for $\theta=1$. For strong coupling, we observe a positive tail in the neural-net results for $G(q)$ which begins at smaller values of $x=q/q_\textnormal{F}$ for larger $r_s$. Between $r_s=2$ and $r_s=1$, i.e., in the middle of the WDM regime, this behaviour changes and we find instead a negative slope, which ultimately even leads to negative values of $G(q)$.
From a physical perspective, the long wave-number limit is dominated by single-particle effects and the sign of the slope follows from the exchange--correlation contribution to the kinetic energy $K$~\cite{holas_limit,farid}, which changes its sign at these conditions~\cite{Militzer_PRL_2002,Hunger_PRE_2021}.

The ESA, on the other hand, is invariant to this effect and, as usually, attains the consistent limit for $\overline{G}(q)$ given by Eq.~(\ref{eq:onTop}) for all values of $r_s$.


\begin{figure}\centering\includegraphics[width=0.475\textwidth]{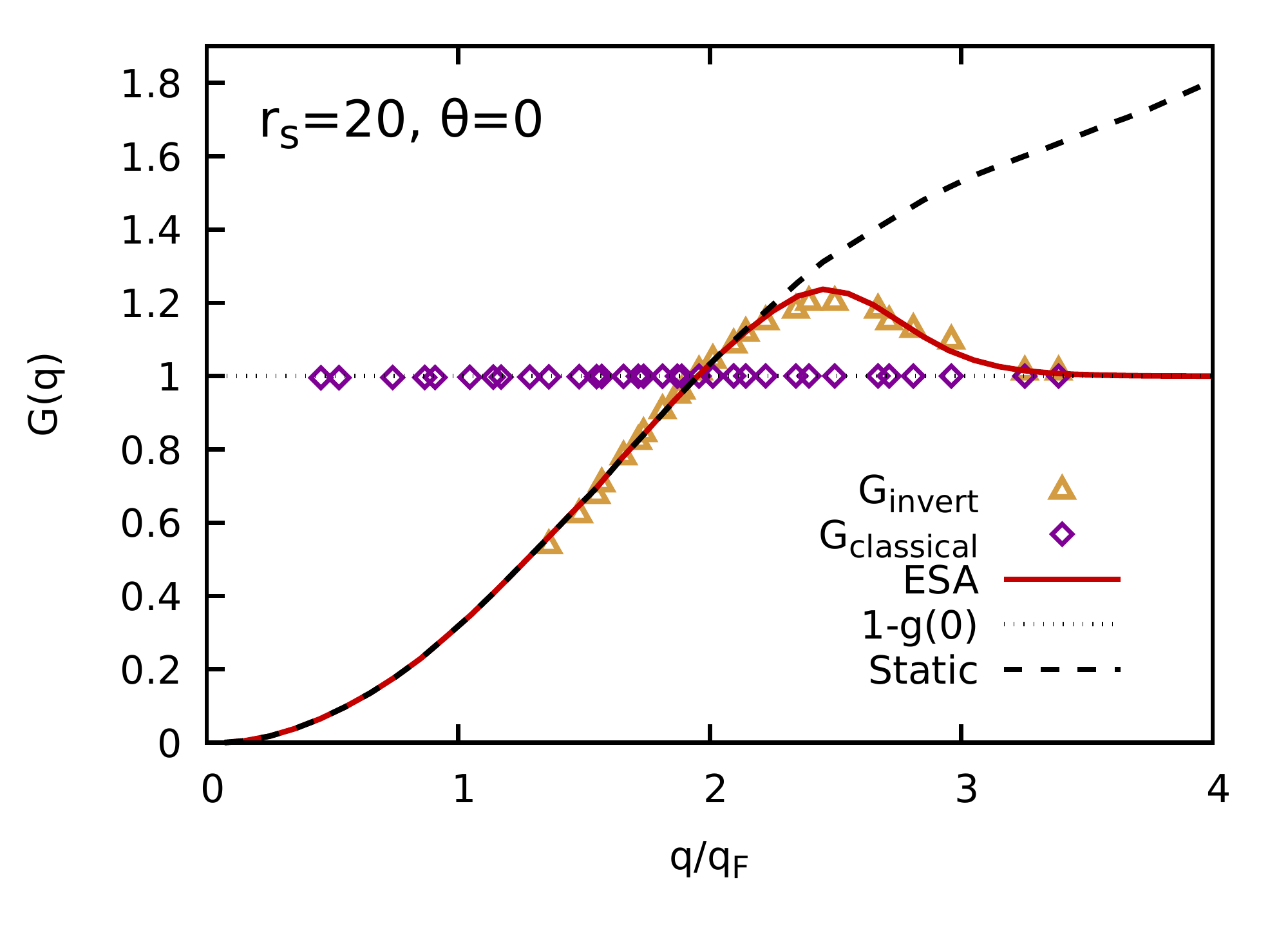}\\\vspace*{-1.1cm}
\includegraphics[width=0.475\textwidth]{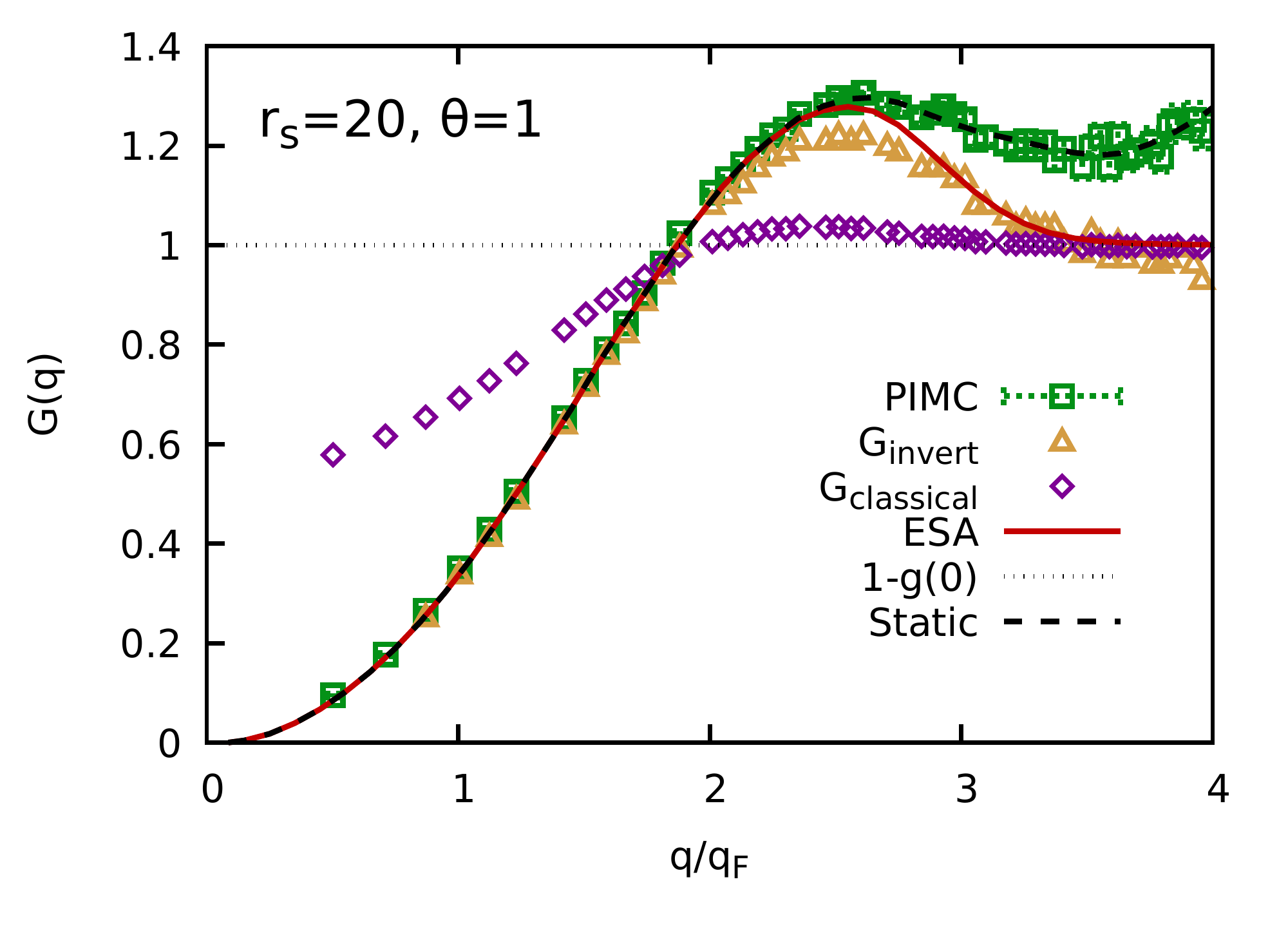}\\\vspace*{-0.7cm}
\includegraphics[width=0.475\textwidth]{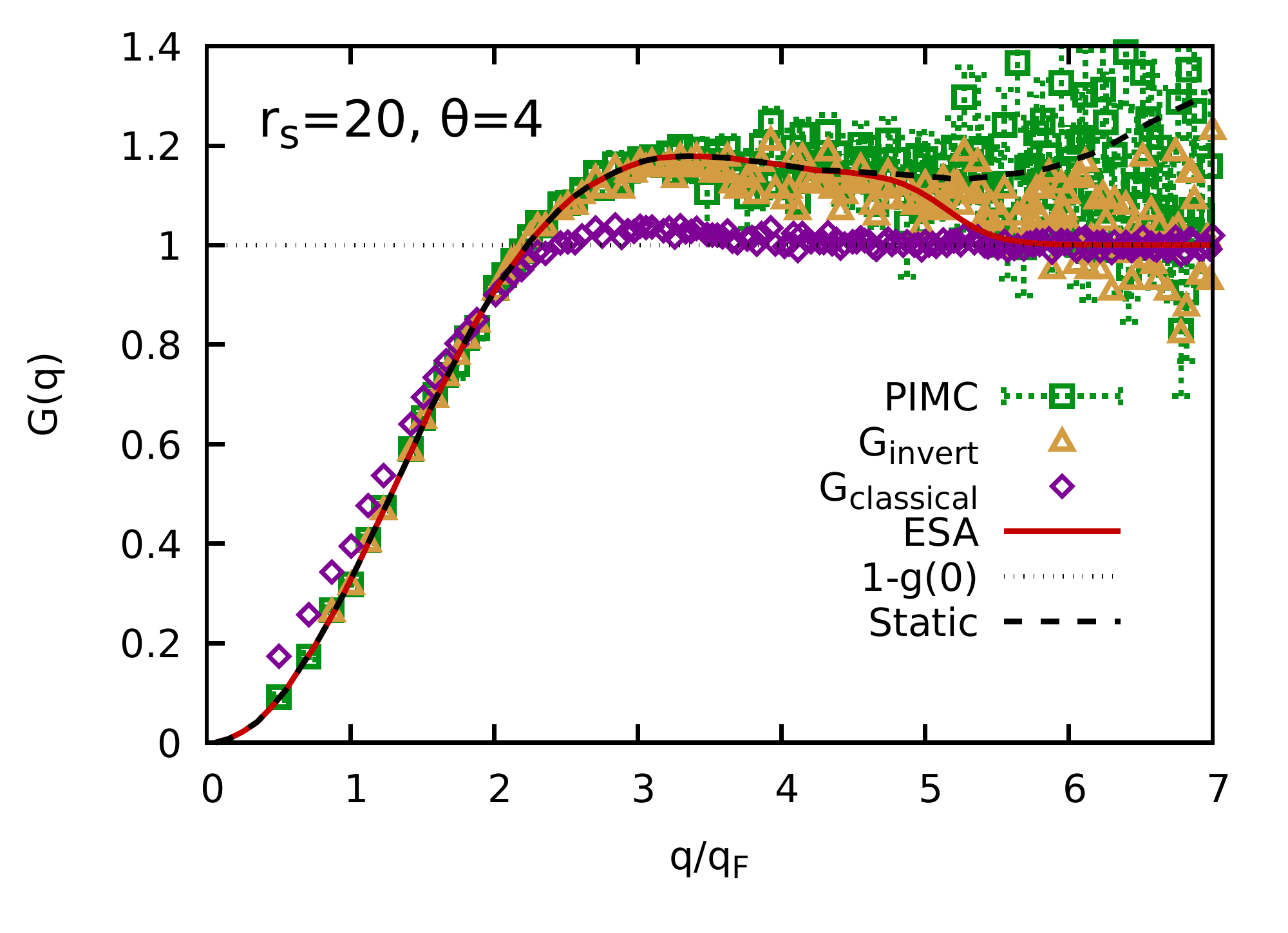}
\caption{\label{fig:Invert_rs20}
Inverted local field correction at finite temperature: The for $r_s=20$ at $\theta=0$ (top), $\theta=1$ (center) and $\theta=4$ (bottom). Green squares and black dashed line: PIMC results for $G(q)$ from Ref.~\cite{dornheim_electron_liquid} and corresponding neural-net results~\cite{dornheim_ML}. Solid red and dotted grey: ESA and large-$q$ limit, Eq.~(\ref{eq:onTop}). Yellow triangles: inverted LFC $\overline{G}_\textnormal{classical}(q)$, see Eq.~(\ref{eq:invert}). Purple diamonds: LFC from the classical relation Eq.~(\ref{eq:G_from_S}).
}
\end{figure}

As a further motivation for our ESA scheme, we consider an effective local field correction $\overline{G}_\textnormal{invert}(q)$, which, by definition, exactly reproduces QMC data for $S(q)$ where they are available. More specifically, such a quantity can be defined as
\begin{eqnarray}\label{eq:invert}
\overline{G}_\textnormal{invert}(q) = \textnormal{min}_{\overline{G}}\left(
\left|
S^{\overline{G}}(q) - S(q)
\right|
\right)\ ,
\end{eqnarray}
where $S^{\overline{G}}(q)$ denotes the SSF computed with respect to some trial static LFC $\overline{G}$. In practice, we solve Eq.~(\ref{eq:invert}) by scanning over a dense $\overline{G}$-grid for each $q$-point and search for the minimum deviation in the SSF. In this way, we have effectively inverted $S(q)$ for the LFC $\overline{G}$, even though the relation between the two quantities is not straightforward when quantum mechanical effects cannot be neglected.

The results for this procedure are depicted in Fig.~\ref{fig:Invert_rs20}, where we show different LFCs at $r_s=20$. The top and center panels corresponds to $\theta=0$ and $\theta=1$, and both $G(q)$ and $\overline{G}_\textnormal{ESA}(q)$ exhibit the familiar behaviour that has been discussed in the context of Fig.~\ref{fig:ESA} above. The yellow triangles show the inverted results for Eq.~(\ref{eq:invert}) and are in remarkably good agreement to both $G(q)$ and $\overline{G}_\textnormal{ESA}(q)$ for $q\lesssim2q_\textnormal{F}$. For larger $q$, $\overline{G}_\textnormal{invert}(q)$ follows $\overline{G}_\textnormal{ESA}(q)$ and attains the same finite limit instead of diverging like the exact static limit of the LFC. In fact, the curves can hardly be distinguished within the given level of accuracy (in particular at $\theta=0$), which further substantiates the simple construction of the ESA, Eq.~(\ref{eq:ESA}). 

Let us briefly postpone the discussion of the purple diamonds and instead consider the bottom panel of Fig.~\ref{fig:Invert_rs20} showing results for $\theta=4$. At these conditions, $G(q)$ and $\overline{G}_\textnormal{ESA}(q)$ only start to noticeably deviate for $q\gtrsim 5q_\textnormal{F}$, and the PIMC data, too, appear to remain nearly constant for large $q$. In addition, the black dashed curve is only reliable for $q\leq5q_\textnormal{F}$ as data for larger wave numbers had not been included into the training of the neural net, see Ref.~\cite{dornheim_ML} for details.  

Unsurprisingly, the inverted data for $\overline{G}_\textnormal{inverted}(q)$ closely follow $\overline{G}_\textnormal{ESA}(q)$ over the entire $q$-range, and both ESA and the \emph{static approximation} give highly accurate results for $S(q)$ and $v$.

Let us next more closely examine the connection between the ESA and the classical limit, where $G(q)$ is sufficient to compute exact results for $S(q)$, see Eq.~(\ref{eq:S_from_G}) above. In particular, Eq.~(\ref{eq:S_from_G}) can be straightforwardly solved for $G(q)$, which gives the relation
\begin{eqnarray}\label{eq:G_from_S}
G_\textnormal{classical}(q) = 1 - \frac{q^2}{4\pi}\left( \frac{1}{S(q)}-1\right)\frac{1}{\beta n}\ ,
\end{eqnarray}
which, too, is exact in the classical limit.

At the same time, it is interesting to evaluate Eq.~(\ref{eq:G_from_S}) for a quantum system to gauge the impact of quantum effects on exchange--correlation effects at different wave numbers $q$. The results are depicted by the purple diamonds in Fig.~\ref{fig:Invert_rs20}. In the ground state, i.e., $\beta\to\infty$, it holds $G_\textnormal{classical}(q)=1$ for all $q$, as the second term is proportional to $T$ and, hence, vanishes.
For $\theta=1$, $G_\textnormal{classical}(q)$ does depend on $q$, but is still qualitatively wrong over the entire depicted wave-number range. In particular, it strongly violates the compressibility sum-rule Eq.~(\ref{eq:CSR}) and does not even decay to zero in the limit of small $q$.
Finally, $G_\textnormal{classical}(q)$ does more closely resemble the other curves at $\theta=4$, but still substantially deviates everywhere. We thus conclude that quantum effects are paramount even at $\theta=4$ and $r_s=20$, and can only be neglected at significantly higher temperatures.


\subsection{The static structure factor}

\begin{figure*}\centering\includegraphics[width=0.475\textwidth]{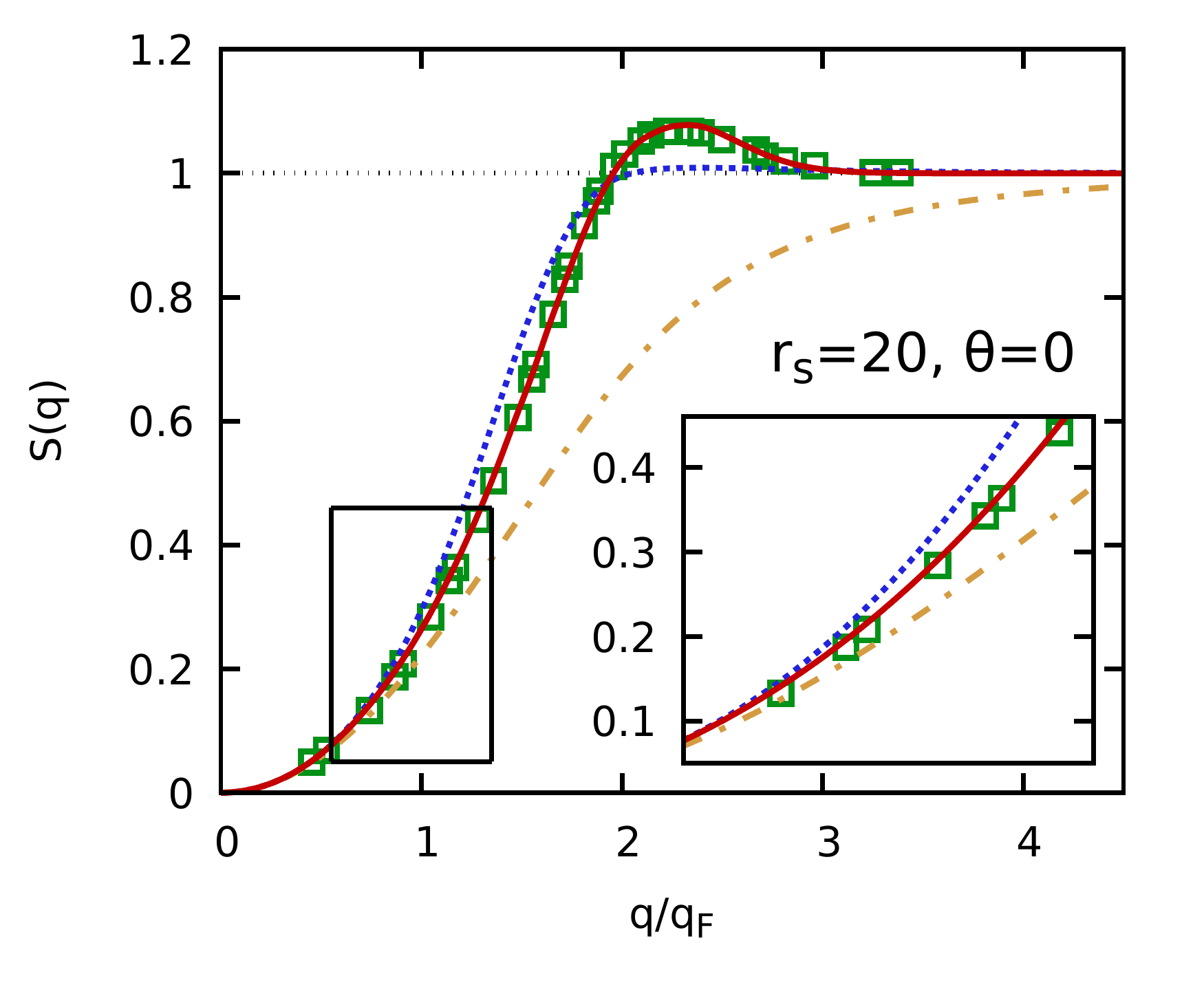}\includegraphics[width=0.475\textwidth]{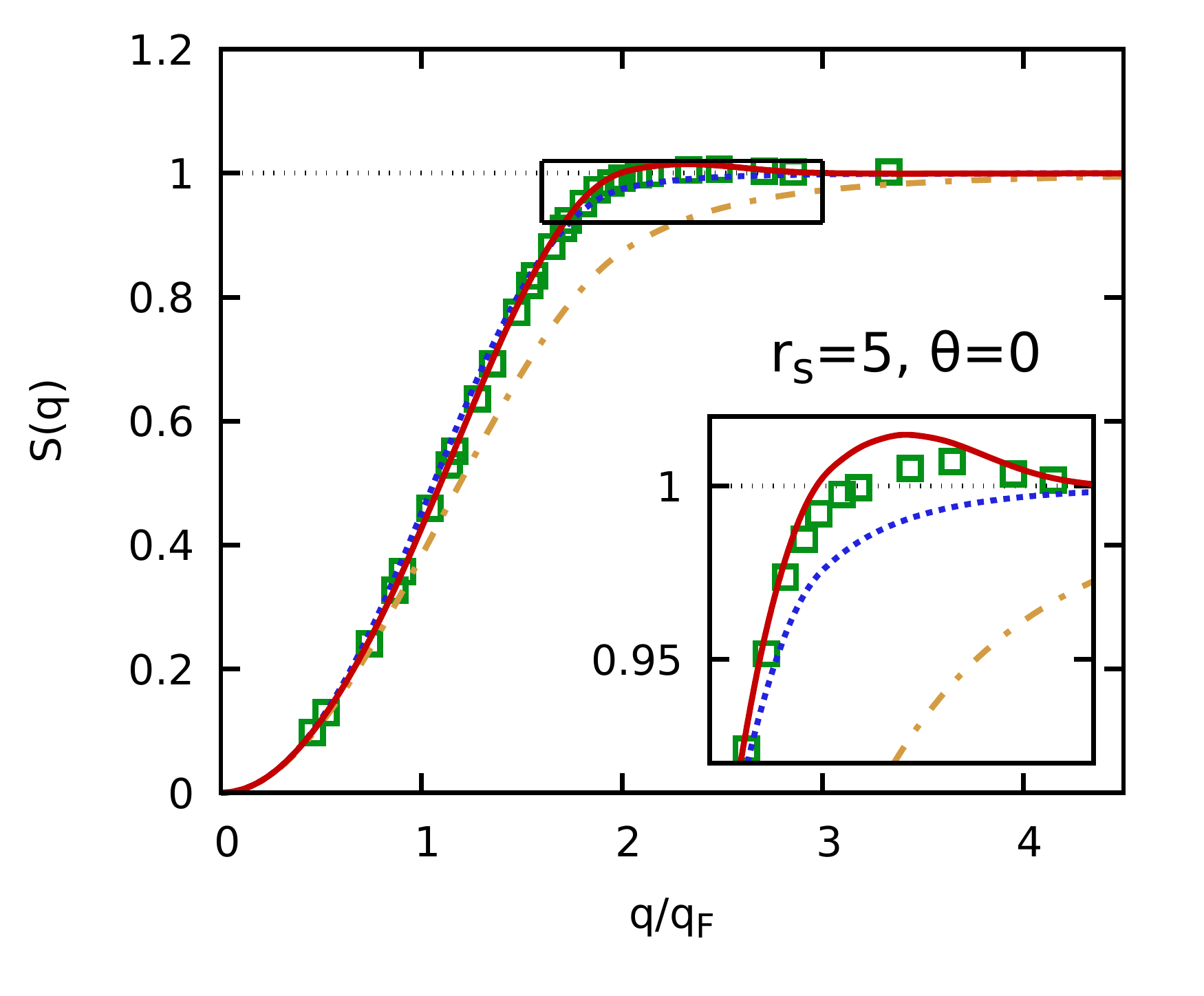}\\\vspace*{-1.35cm}
\includegraphics[width=0.475\textwidth]{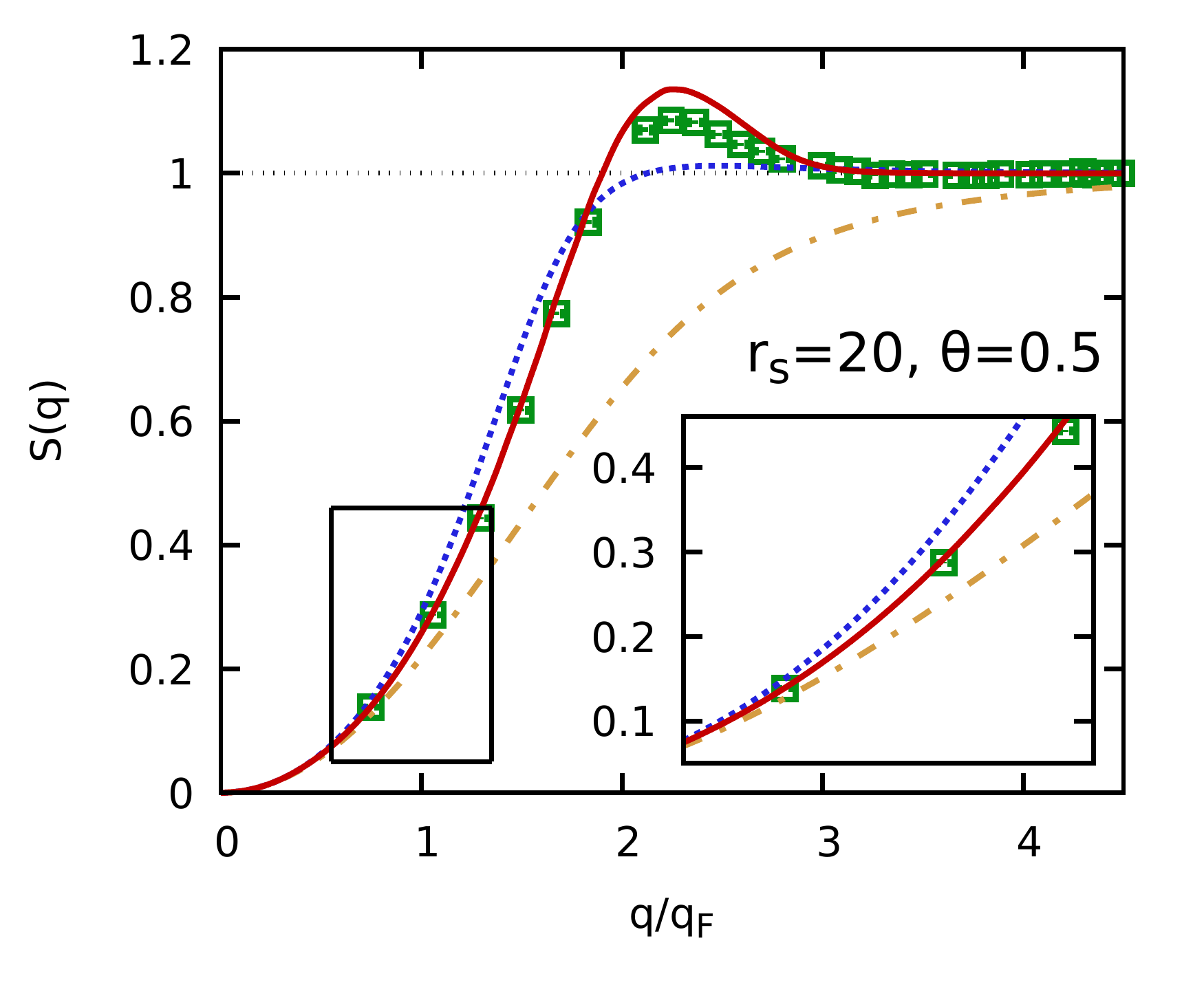}\includegraphics[width=0.475\textwidth]{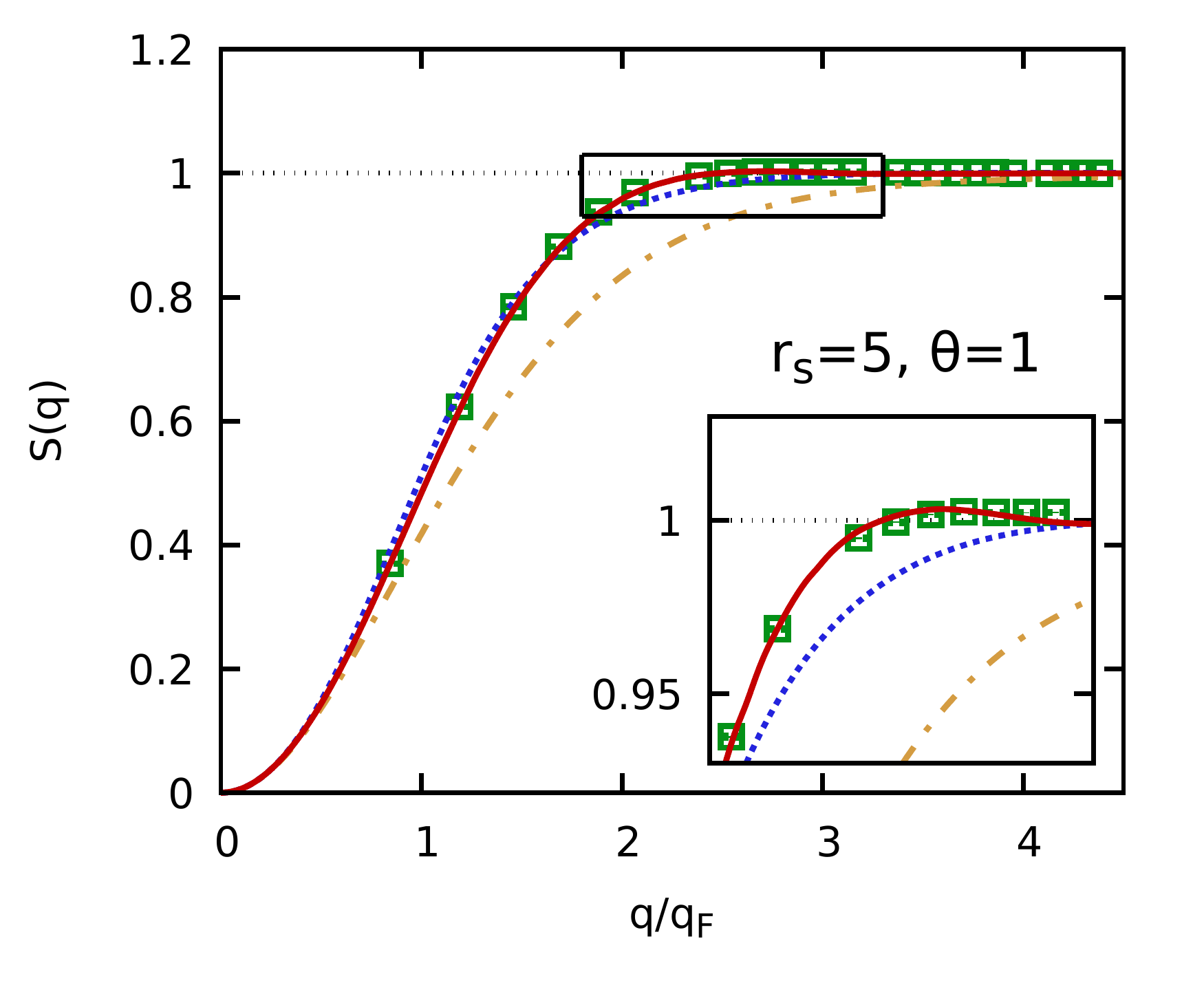}\\\vspace*{-1.35cm}
\includegraphics[width=0.475\textwidth]{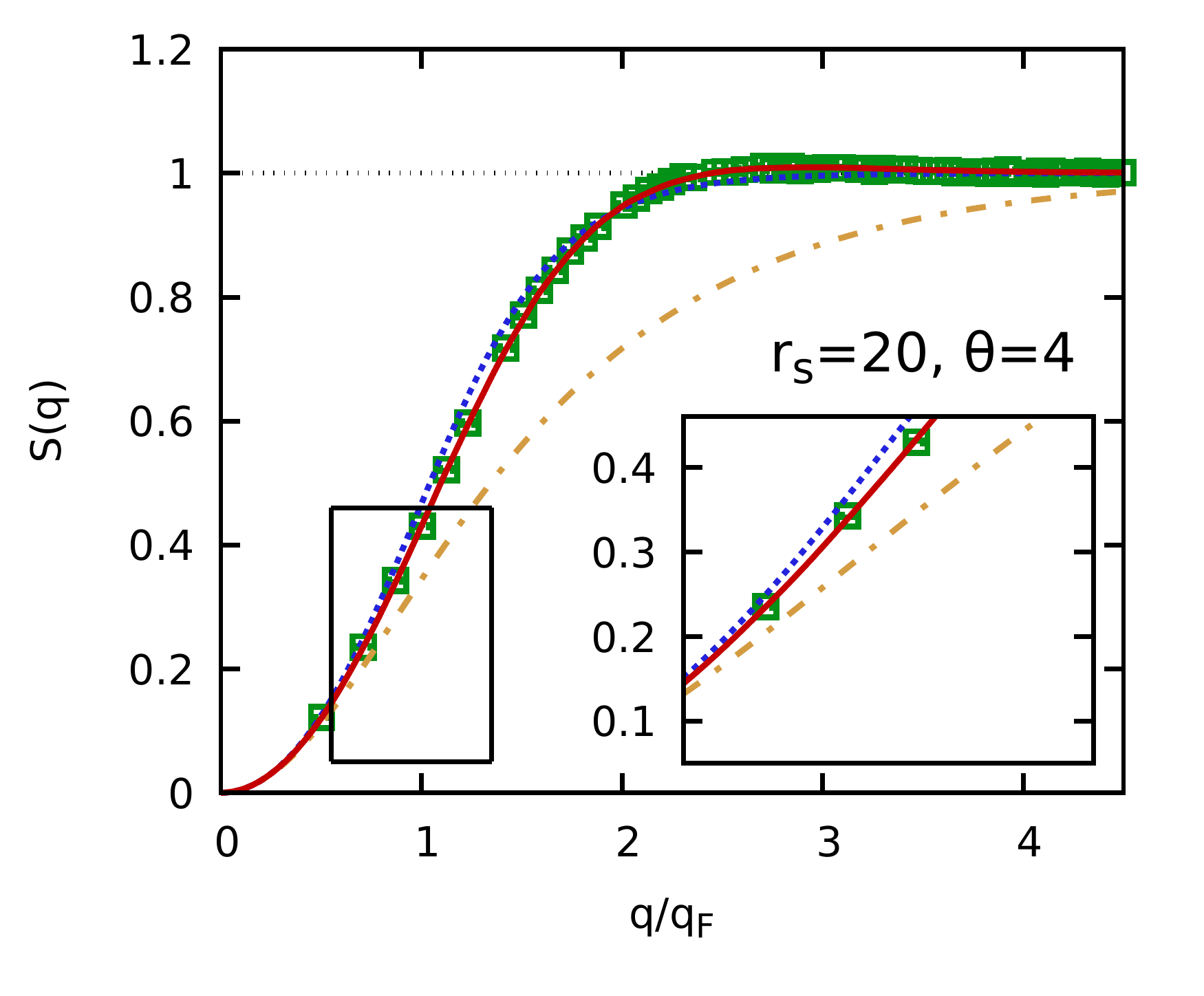}\includegraphics[width=0.475\textwidth]{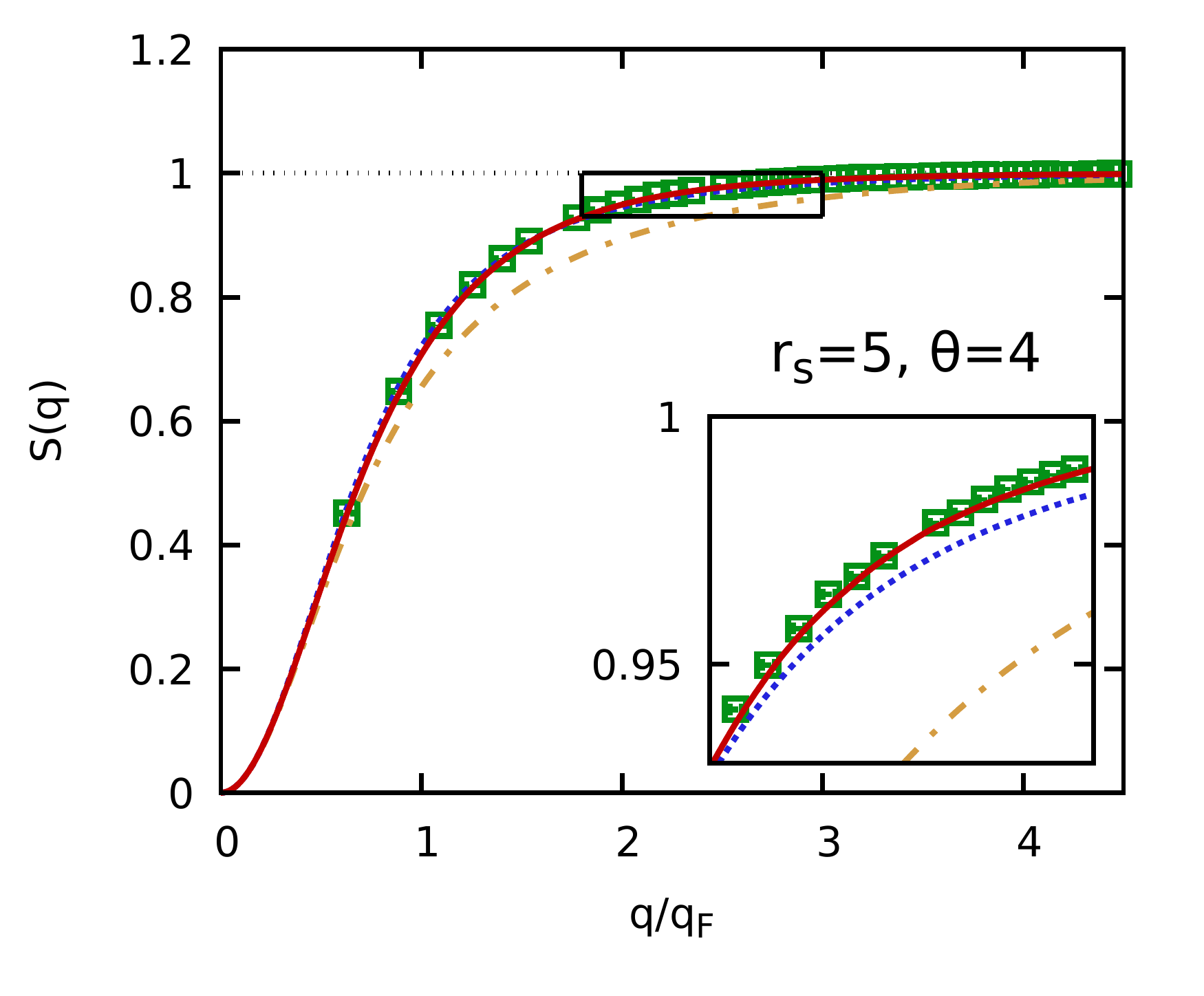}
\caption{\label{fig:Panel_Sq_rs20}
Static structure factor at $r_s=20$ (left) and $r_s=5$ (right) at different $\theta$. Green squares: $T=0$ QMC data~\cite{Spink_PRB_2013} and finite-$T$ PIMC data~\cite{dornheim_electron_liquid}; Solid red: ESA; dotted blue: STLS~\cite{stls_original,stls,stls2}; dash-dotted yellow: RPA.
}
\end{figure*}

The next quantity to be investigated with the ESA scheme is the static structure factor $S(q)$, which we show in Fig.~\ref{fig:Panel_Sq_rs20}.
The left column corresponds to $r_s=20$ and, thus, constitutes the most challenging case for the ESA due to the dominant character of exchange--correlation effects at these conditions.

Let us start with the top panel, showing results for the ground state. The green squares are state-of-the-art diffusion Monte Carlo results by Spink \textit{et al.}~\cite{Spink_PRB_2013} and constitute the gold standard for benchmarks. The solid red curve has been obtained using $\overline{G}_\textnormal{ESA}(q)$ and is in remarkable agreement for all $q$, even in the vicinity of the peak of $S(q)$ around $q\approx2.25q_\textnormal{F}$. In contrast, the blue dotted STLS curve does not capture this feature and exhibits pronounced systematic deviations except in the limits of small and large wave numbers.

The center panel in the left column has been obtained for $\theta=0.5$, and the green squares are finite-$T$ PIMC data taken from Dornheim \textit{et al.}~\cite{dornheim_electron_liquid}. Again, the ESA gives a very good description of $S(q)$, although the peak height is somewhat overestimated. Still, the description is strikingly improved compared to the STLS approximation.

Lastly, the bottom panel has been obtained for $\theta=4$, where ESA cannot be distinguished from the PIMC reference data within the given Monte Carlo error bars. STLS, too, is quite accurate in this regime, although there remain systematic deviations at intermediate $q$.

Finally, we mention the dash-dotted yellow curve in all three panels, that have been obtained within RPA. Evidently, this mean field description is unsuitable at such low densities even at relatively high values of the reduced temperature $\theta$.

The right column of Fig.~\ref{fig:Panel_Sq_rs20} has been obtained for a density that is of prime interest to WDM research, $r_s=5$. Again, the top panel corresponds to the ground-state and shows relatively good agreement between diffusion Monte Carlo, ESA, and STLS, although the latter does not capture the small correlation induced peak in $S(q)$. The RPA, on the other hand, remains inaccurate despite the reduced coupling strength compared to the left panel.

At $\theta=1$ (center panel), the situation is quite similar, with the ESA being nearly indistinguishable to the PIMC data over the entire $q$-range, whereas STLS is too large for small and too small for large wave numbers.

Finally, the bottom panel corresponds to $\theta=4$. Here, too, only the ESA is capable to reproduce the PIMC data, whereas STLS and in particular RPA exhibit systematic errors.

\subsection{Interaction energy}

\begin{figure*}\centering\includegraphics[width=0.475\textwidth]{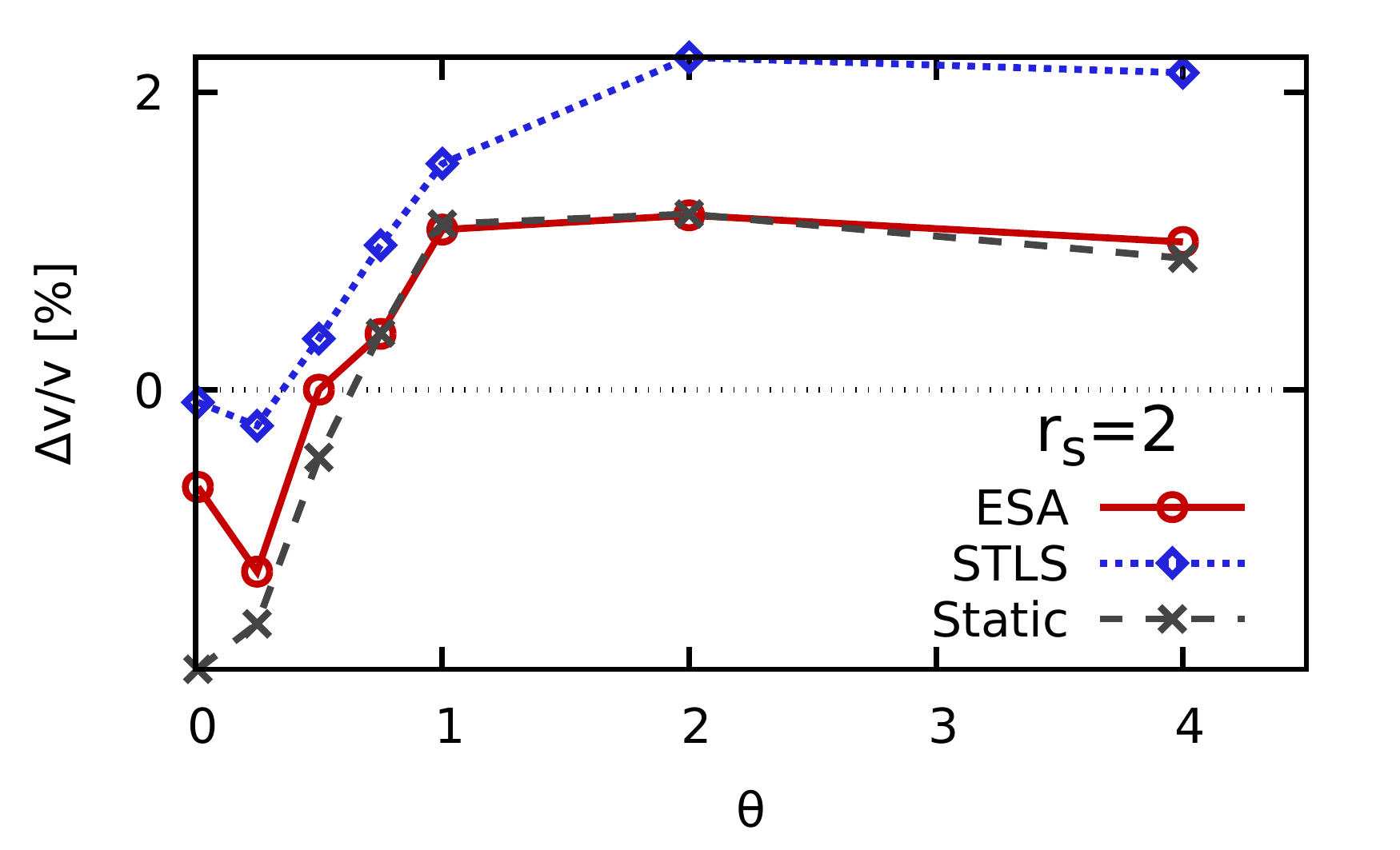}\includegraphics[width=0.475\textwidth]{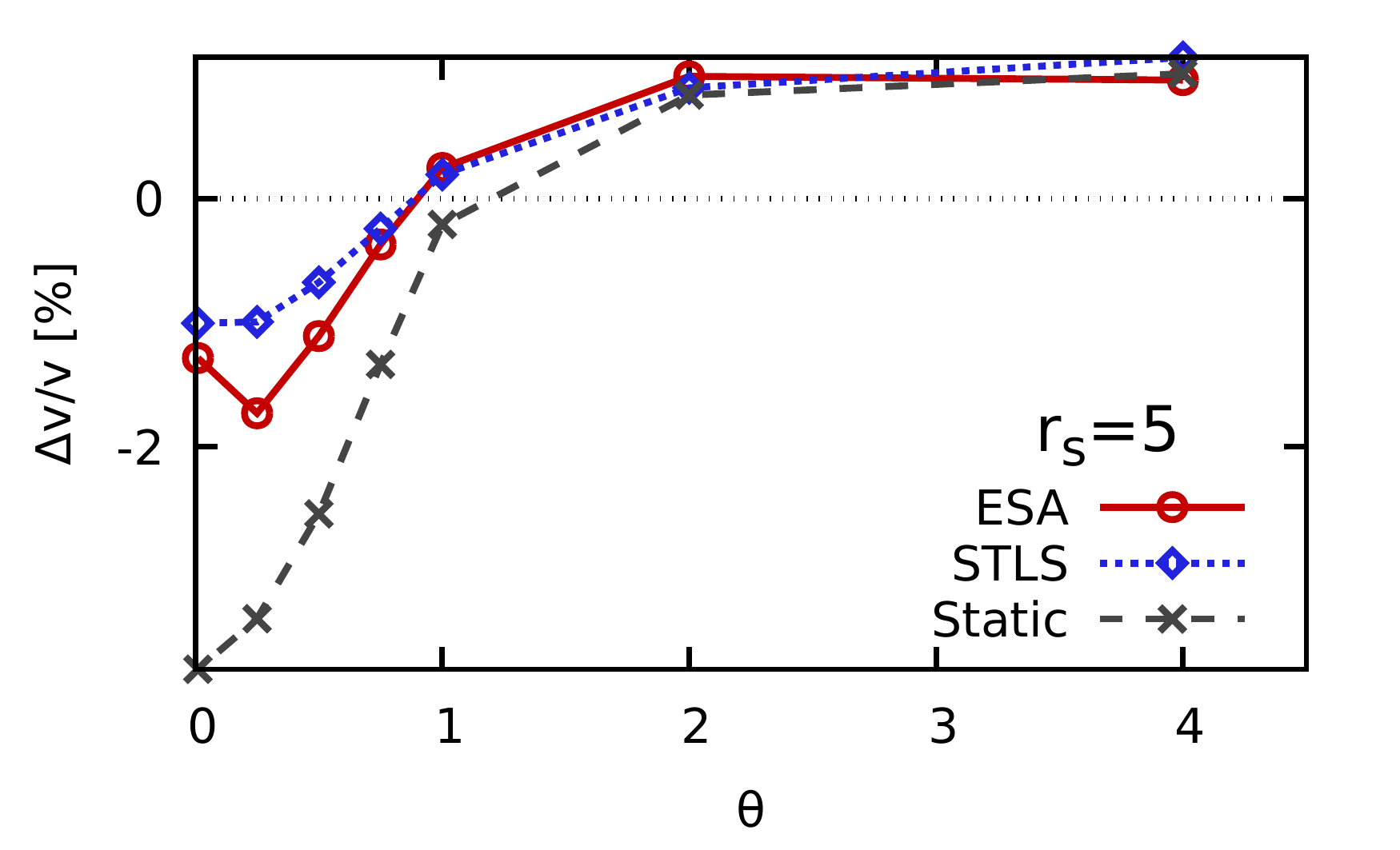}\\\vspace*{-1.06cm}
\includegraphics[width=0.475\textwidth]{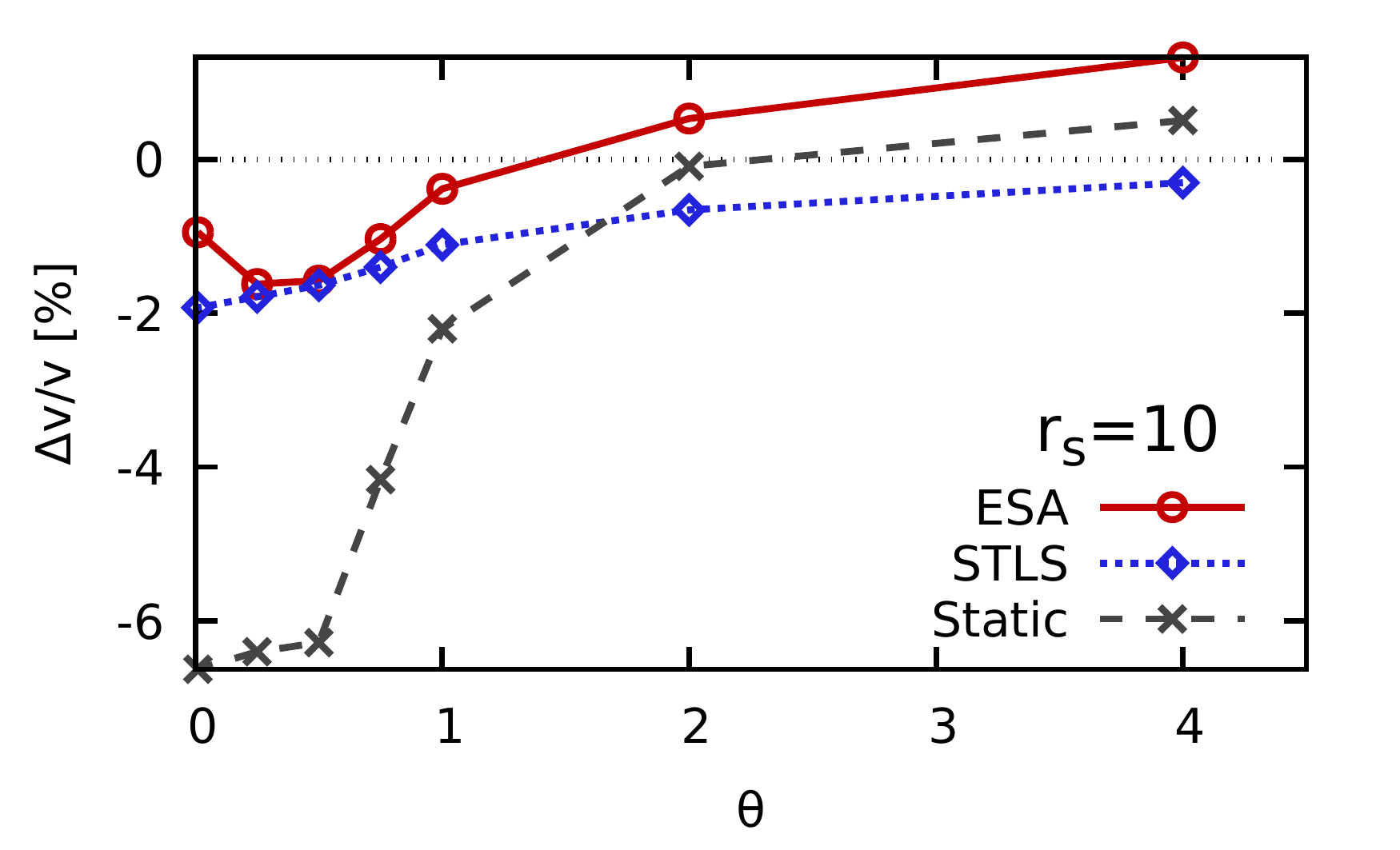}\hspace*{-0.013\textwidth}\includegraphics[width=0.488\textwidth]{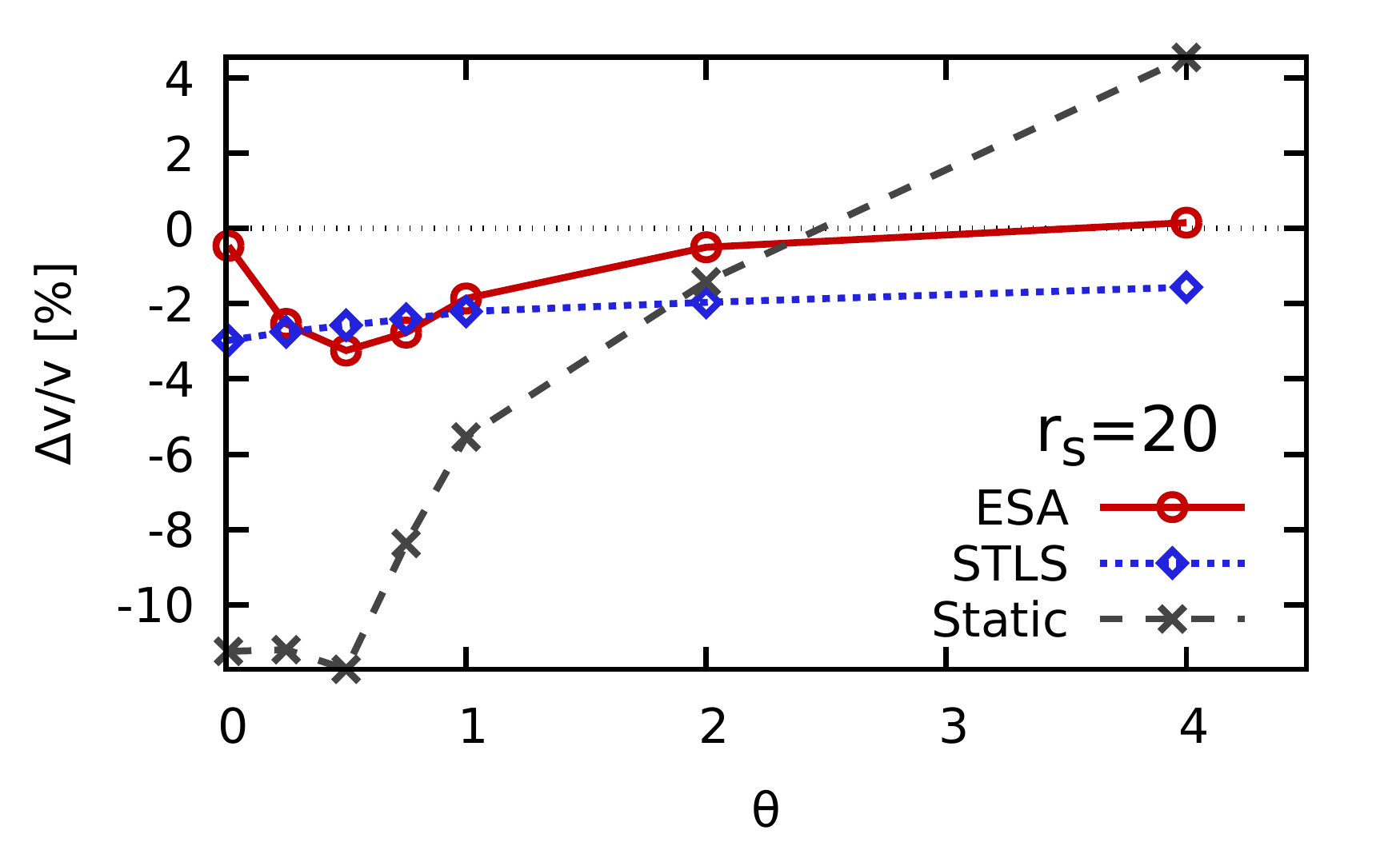}
\caption{\label{fig:Panel_v}
Relative deviation in the interaction energy $v$ [see Eq.~(\ref{eq:v})] compared to the parametrization by Groth \textit{et al.}~\cite{groth_prl}. Solid red circles: ESA; dotted blue diamonds: STLS~\cite{stls_original,stls,stls2}; dashed grey crosses: \emph{static approximation} using the neural-net representation from Ref.~\cite{dornheim_ML}.}
\end{figure*}

The next important quantity to be investigated in this work is the interaction energy $v$, which, in the case of a uniform electron gas, is simply given by a one-dimensional integral over the static structure factor $S(q)$ [see Eq.~(\ref{eq:v})] that we evaluate numerically. The results are shown in Fig.~\ref{fig:Panel_v}, where we depict the $\theta$-dependence of $v$ for four relevant values of the density parameter $r_s$.

More specifically, the top left panel corresponds to $r_s=2$, i.e., a metallic density that is typical for WDM experiments using various materials, and we plot the relative deviation in $v$ compared to the accurate parametrization of the UEG by Groth \textit{et al.}~\cite{groth_prl}. At these conditions, both the ESA (solid red) and the \emph{static approximation} (dashed grey) are very accurate over the entire $\theta$-range, with a maximum deviation of $\Delta v/v\sim1\%$. The STLS approximation (dotted blue), too, is capable to provide accurate results for $v$, with a maximum deviation of $\sim2\%$.

Let us proceed to the top right panel corresponding to $r_s=5$, a relatively sparse density that can be realized e.g. in experiments with hydrogen jets, see above. First and foremost, we note that both the ESA and STLS provide a remarkably good description of the interaction energy, and the systematic error never exceeds $2\%$. Somewhat surprisingly, STLS even gives slightly more accurate dara for small values of $\theta$ compared to ESA. Yet, this is due to a fortunate cancellation of errors in $S(q)$ under the integral in Eq.~(\ref{eq:v}) [$S(q)$ is too large for small $q$ and too small for large $q$, which roughly balances out]~\cite{dornheim_PRL_ESA_2020,review}, since the static structure factor $S(q)$ is comparatively much better in ESA than in STLS, cf.~Fig.~\ref{fig:Panel_Sq_rs20}. In addition, we note that the \emph{static approximation} performs substantially worse for low temperatures, which is due to the unphysically slow convergence of $S(q)$ towards $1$ for large $q$, see Secs.~\ref{sec:SA} and \ref{sec:ESA} above.

The bottom left panel shows the same analysis for $r_s=10$, and even for this strong coupling strength that constitutes the boundary of the electron liquid regime~\cite{dornheim_dynamic}, the error in ESA does not exceed $2\%$. In addition, the STLS exhibits a comparable accuracy in $v$, whereas the \emph{static approximation} fails at low $\theta$ as it is expected.

Finally, the bottom right panel shows results for very strong coupling, $r_s=20$. Overall, the ESA gives the most accurate data for $v$ of all depicted approximations, and is particularly good both at large temperature and in the ground state. In contrast, the STLS approximation for $\overline{G}(q)$ results in a relatively constant relative deviation of $\sim2-3\%$, whereas the \emph{static approximation} cannot reasonably used for this values of the density parameter.

\subsection{Density response function}

\begin{figure}\centering\hspace*{0.01\textwidth}\includegraphics[width=0.462\textwidth]{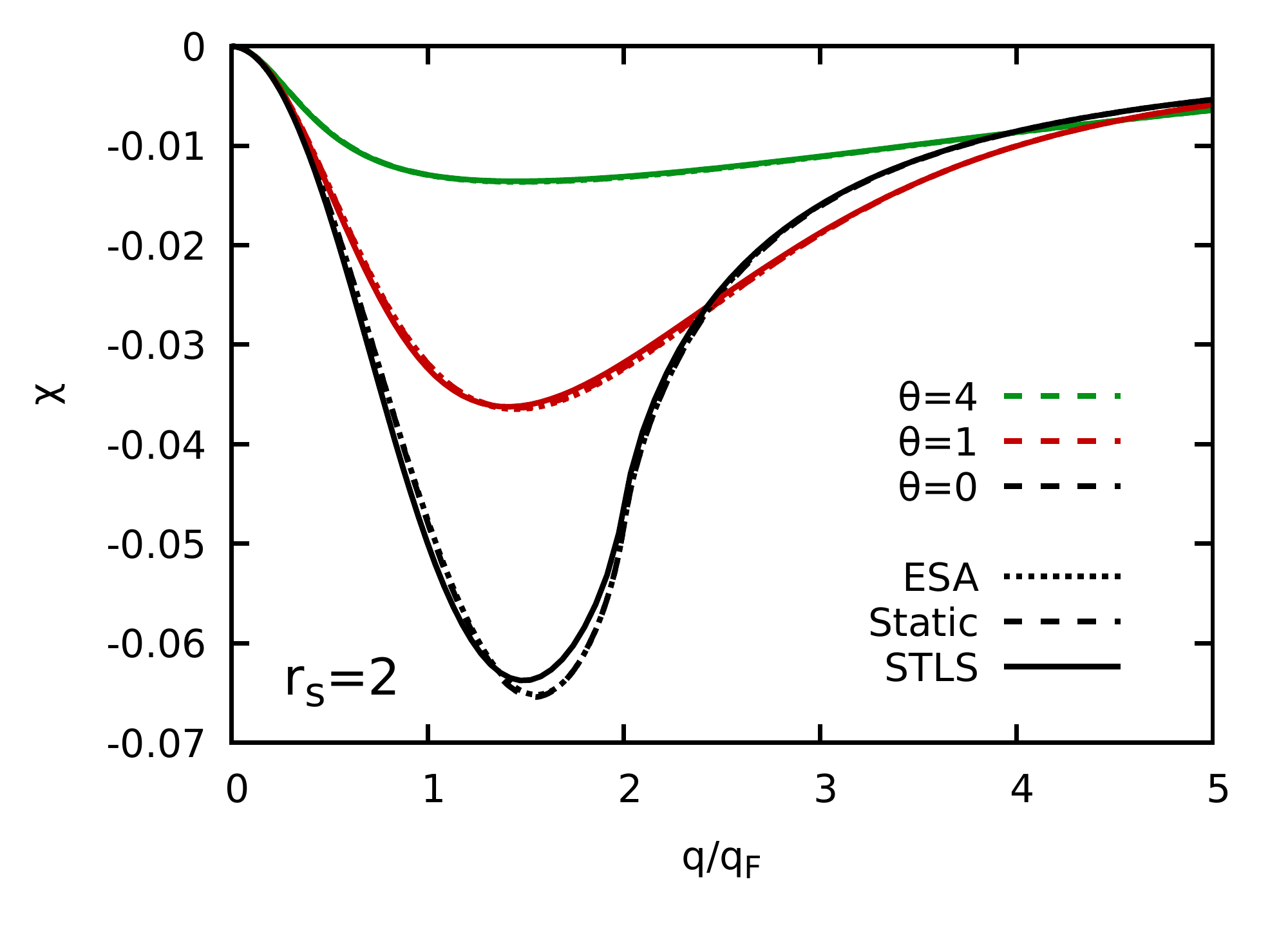}\\\vspace*{-1.1cm}
\includegraphics[width=0.475\textwidth]{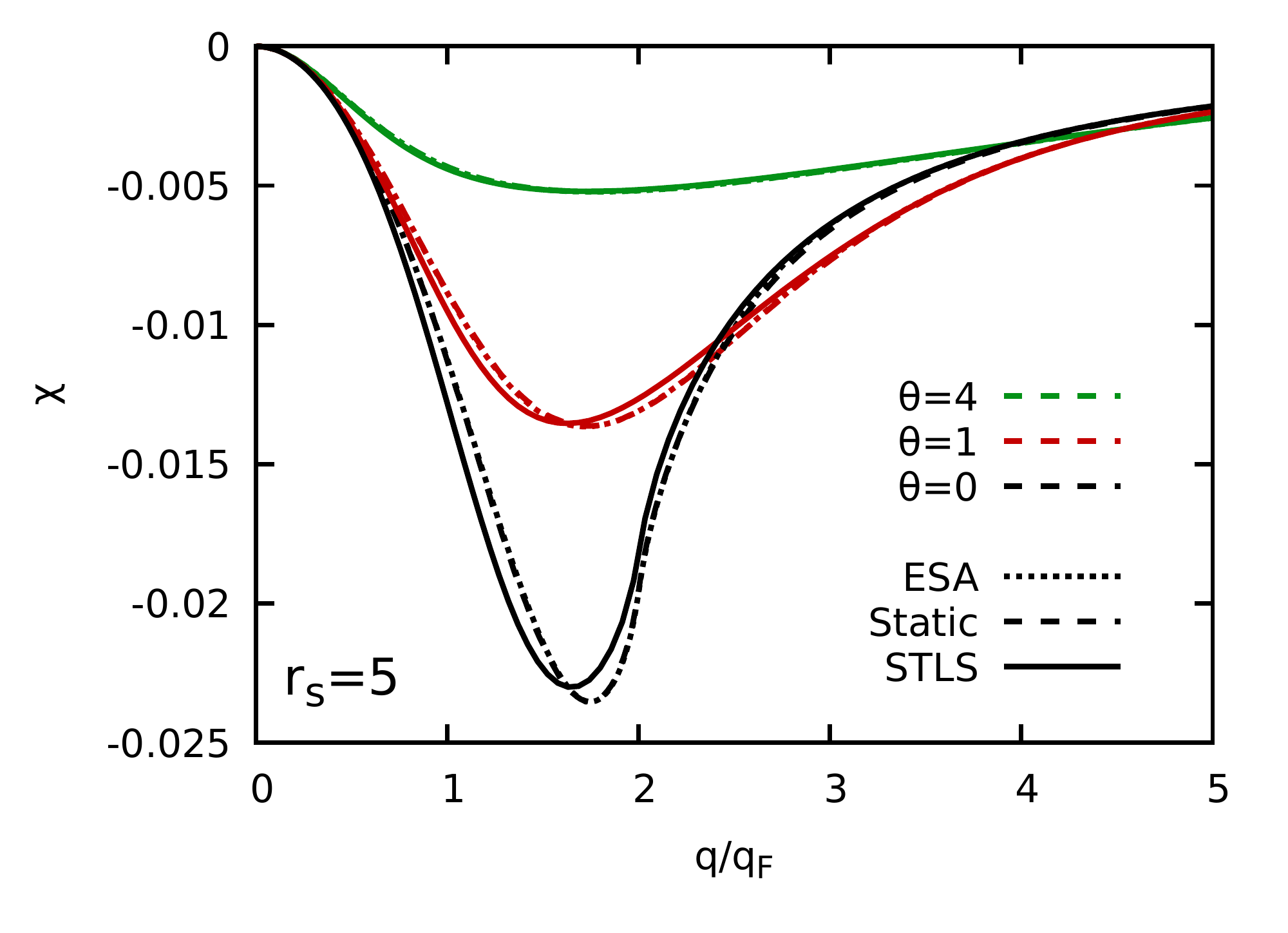}\\\vspace*{-1.1cm}
\includegraphics[width=0.475\textwidth]{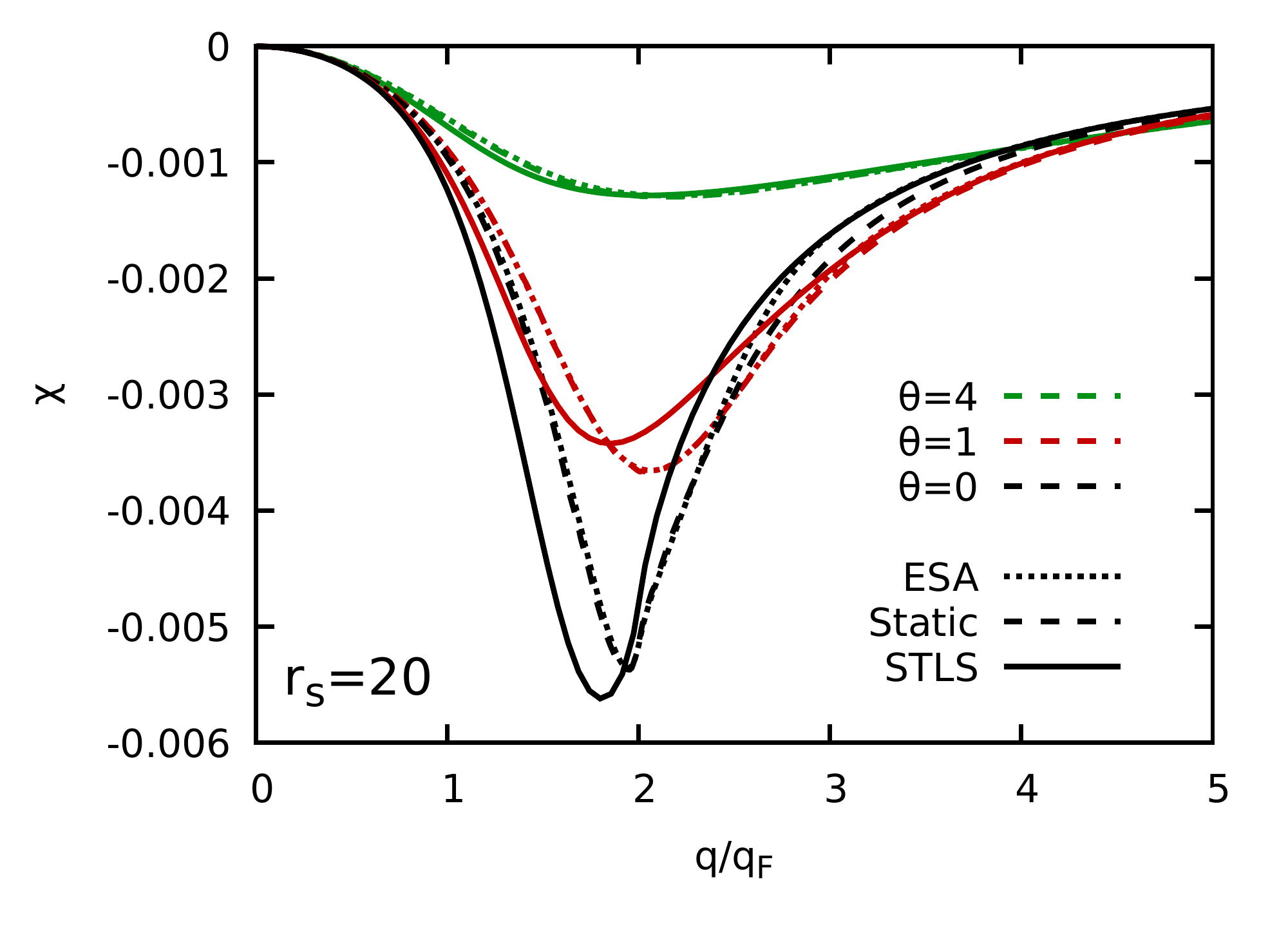}
\caption{\label{fig:Chi}
Static density response function $\chi(q)$ for $r_s=2$ (top), $r_s=5$, and $r_s=20$ (bottom). The dotted, solid, and dashed lines have been obtained by inserting into Eq.~(\ref{eq:static_approximation}) $\overline{G}_\textnormal{ESA}(q)$, $\overline{G}_\textnormal{STLS}(q)$, and the neural-net representation of $G(q,0)$ from Ref.~\cite{dornheim_ML}, respectively. Green curves: $\theta=4$; red: $\theta=1$; black: $\theta=0$.
}
\end{figure}

This section is devoted to a discussion of the suitability of frequency-averaged LFCs for the determination of the exact static limit of the density response function $\chi(q)$. In this case, the previously discussed \emph{static approximation}, i.e., using the neural-net representation of $G(q,0)$ from Ref.~\cite{dornheim_ML}, is exact, and the large-$q$ limit of frequency-independent theories $\overline{G}(q)$ given by Eq.~(\ref{eq:onTop}) is spurious. 
On the other hand, we might expect that the impact of the LFC decreases for large $q$, such that $\overline{G}_\textnormal{ESA}(q)$ and $G(q)$ could potentially give similar results.

To resolve this question, we show $\chi(q)$ in Fig.~\ref{fig:Chi} for three representative values of the density parameter $r_s$, with the green, red, and black sets of curves corresponding to $\theta=4$, $\theta=1$, and $\theta=0$, respectively. 
Let us start with the top panel showing results for a metallic density, $r_s=2$, with the dotted, dashed, and solid curves corresponding to ESA, the exact static limit, and STLS, respectively.
Firstly, we note that all three curves exhibit the correct parabolic shape for small wave-numbers~\cite{kugler_bounds},
\begin{eqnarray}\label{eq:chi_limit}
\lim_{q\to0}\chi(q) = -\frac{4\pi}{q^2}\ .
\end{eqnarray}
In particular, Eq.~(\ref{eq:chi_limit}) is a direct consequence of the $4\pi/q^2$ pre-factor in front of the LFC in Eqs.~(\ref{eq:chi}) and (\ref{eq:static_approximation}), which means that its impact vanishes for small $q$.
With increasing wave numbers, $\chi(q)$ exhibits a broad peak around $q\approx1.5q_\textnormal{F}$, which is also well reproduced by all curves. Moreover, the ESA is virtually indistinguishable from the exact result for all three temperatures, whereas STLS noticeably deviates, in particular at $\theta=0$.

The center panel shows the same analysis for $r_s=5$. As discussed above, the increased coupling strength means that the impact of the LFC is more pronounced in this case, and the STLS curve substantially deviates at intermediate wave numbers, except for the highest temperature $\theta=4$. In stark contrast, the ESA is in excellent agreement to the exact curve everywhere, and we find only minor deviations for $2q_\textnormal{F}\lesssim q \lesssim 3q_\textnormal{F}$.
In this sense, the ESA combines the best from two worlds, by giving excellent results both for frequency-averaged quantities like $S(q)$, and really static properties like $\chi(q,0)$ over the entire WDM regime.

This nice feature of the ESA is only lost when entering the strongly coupled electron liquid regime, as it is demonstrated in the bottom panel of Fig.~\ref{fig:Chi} for $r_s=20$. In this case, the static density response function is more sharply peaked at low temperature and exhibits a nontrivial shape that is difficult to resolve. 
Therefore, the STLS approximation is not capable to give a reasonable description of either the peak position or the shape, see Ref.~\cite{dornheim_electron_liquid} for a more extensive analysis on this point including even larger values of the density parameter $r_s$.
The ESA, on the other hand, is strikingly accurate for both $\theta=4$ and $\theta=1$, but substantially deviates from the exact curve for $2q_\textnormal{F}\lesssim q \lesssim 4q_\textnormal{F}$ in the ground state.

\subsection{Dielectric function\label{sec:Diel}}

\begin{figure*}\centering\hspace*{0.01\textwidth}\includegraphics[width=0.462\textwidth]{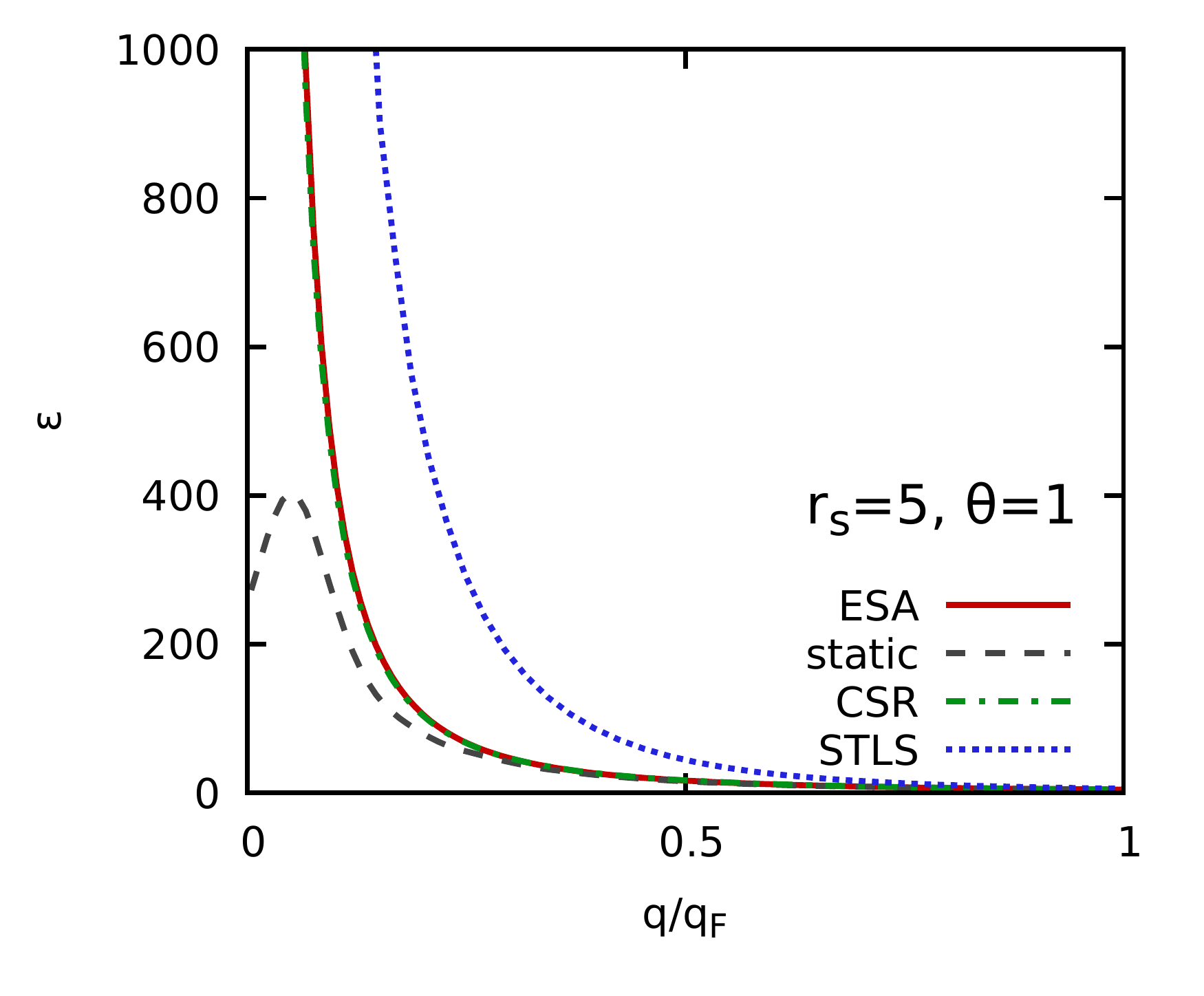}
\includegraphics[width=0.462\textwidth]{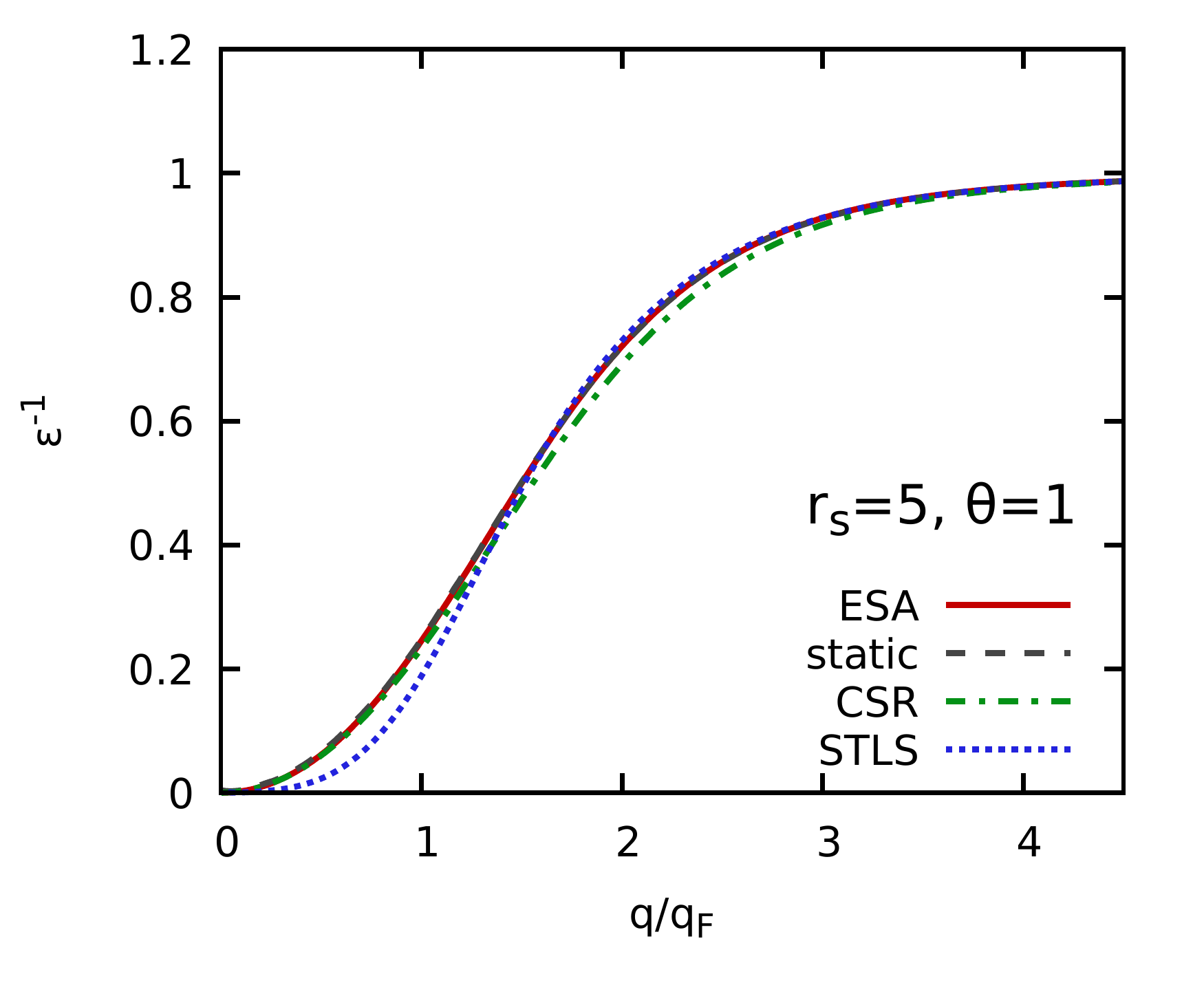}
\caption{\label{fig:Dielectric_rs5}
Left: Static dielectric function $\epsilon(q)$ for $r_s=5$ and $\theta=1$. Solid red: ESA; dashed grey: exact static limit using the neural-net from Ref.~\cite{dornheim_ML}; dash-dotted green: CSR, Eq.~(\ref{eq:CSR}); dotted-blue: STLS~\cite{stls2,stls,stls_original}. Right: Same data for the inverse dielectric function $\epsilon^{-1}(q)$.
}
\end{figure*}

The dynamic dielectric function $\epsilon(q,\omega)$ is defined as
\begin{eqnarray}\label{eq:epsilon}
\epsilon(q,\omega) = 1 - \frac{\chi(q,\omega)}{\frac{q^2}{4\pi}+\chi(q,\omega)}\ ,
\end{eqnarray}
and is important in both classical and quantum electrodynamics, in particular for the description of plasma oscillations~\cite{bonitz_book,alexandrov,Hamann_CPP_2020}. 
Since a more detailed analysis of this quantity has been presented elsewhere~\cite{Hamann_PRB_2020,Hamann_CPP_2020}, here we restrict ourselves to a brief discussion of ESA results for the static limit of Eq.~(\ref{eq:epsilon}), $\epsilon(q)$.

The results are shown in Fig.~\ref{fig:Dielectric_rs5}, where the left panel shows the dielectric function for $r_s=5$ and $\theta=1$. Remarkably, we find substantial disagreement between the different results for small wave numbers $q$, which is in striking contrast to linear response properties like $\chi(q)$ and also the SSF $S(q)$. For the latter quantities, the impact of the LFC vanishes for small $q$ as it has been explained above, such that even the mean-field description within the RPA becomes exact in this limit.
The dielectric function, on the other hand, always diverges for small $q$, and this divergence is connected to the CSR for the static LFC [Eq.~(\ref{eq:CSR})]~\cite{stls2,Hamann_PRB_2020},
\begin{eqnarray}\label{eq:duality}
\lim_{q\to0}\epsilon(q) = - \frac{4\pi\chi_0(q)}{q^2\left[1 + 4\pi C\chi_0(q)\right]}\ ,
\end{eqnarray}
where $C$ is the pre-factor to the parabola in Eq.~(\ref{eq:CSR}),
\begin{eqnarray}
C = - \frac{1}{4\pi} \frac{\partial^2}{\partial n^2} \left( n f_\textnormal{xc} \right)\ .
\end{eqnarray}
In principle, exact knowledge of the static LFC as it is encoded in the neural-net representation from Ref.~\cite{dornheim_ML} gives access to the exact static dielectric function depicted in Fig.~\ref{fig:Dielectric_rs5}. Yet, while the exact relation Eq.~(\ref{eq:CSR}) was indeed incorporated into the training procedure of the neural net, it was not strictly enforced and, thus, is only fulfilled by the \emph{static} (grey dashed) curve with a finite accuracy.
Therefore, this curve violates Eq.~(\ref{eq:duality}) and attains a finite value in the limit of $q\to0$, which is unphysical.

Our new analytical representation of $\overline{G}_\textnormal{ESA}(q)$, in contrast, exactly incorporates the CSR, which means that the solid red curve exhibits the correct asymptotic behaviour (depicted as the dash-dotted green curve).
Finally, the dotted blue curve has been obtained on the basis of the approximate $\overline{G}_\textnormal{STLS}(q)$, and starkly deviates from the exact asymptotic limit. Indeed, the violation of the CSR is a well-known shortcoming of the STLS approach~\cite{stls2}, which has ultimately led to the development of the approach by Vashista and Singwi~\cite{vs_original,stolzmann}.

The right panel of Fig.~\ref{fig:Dielectric_rs5} shows the corresponding data for the inverse dielectric function $\epsilon^{-1}(q)$. Here the \emph{static} and ESA curves are in excellent agreement over the entire $q$-range, which, again, highlights the value of the analytical parametrization which is capable to accurately describe both $\epsilon(q)$ and $\epsilon^{-1}(q)$ at the same time.

\begin{figure*}\centering\hspace*{0.01\textwidth}\includegraphics[width=0.462\textwidth]{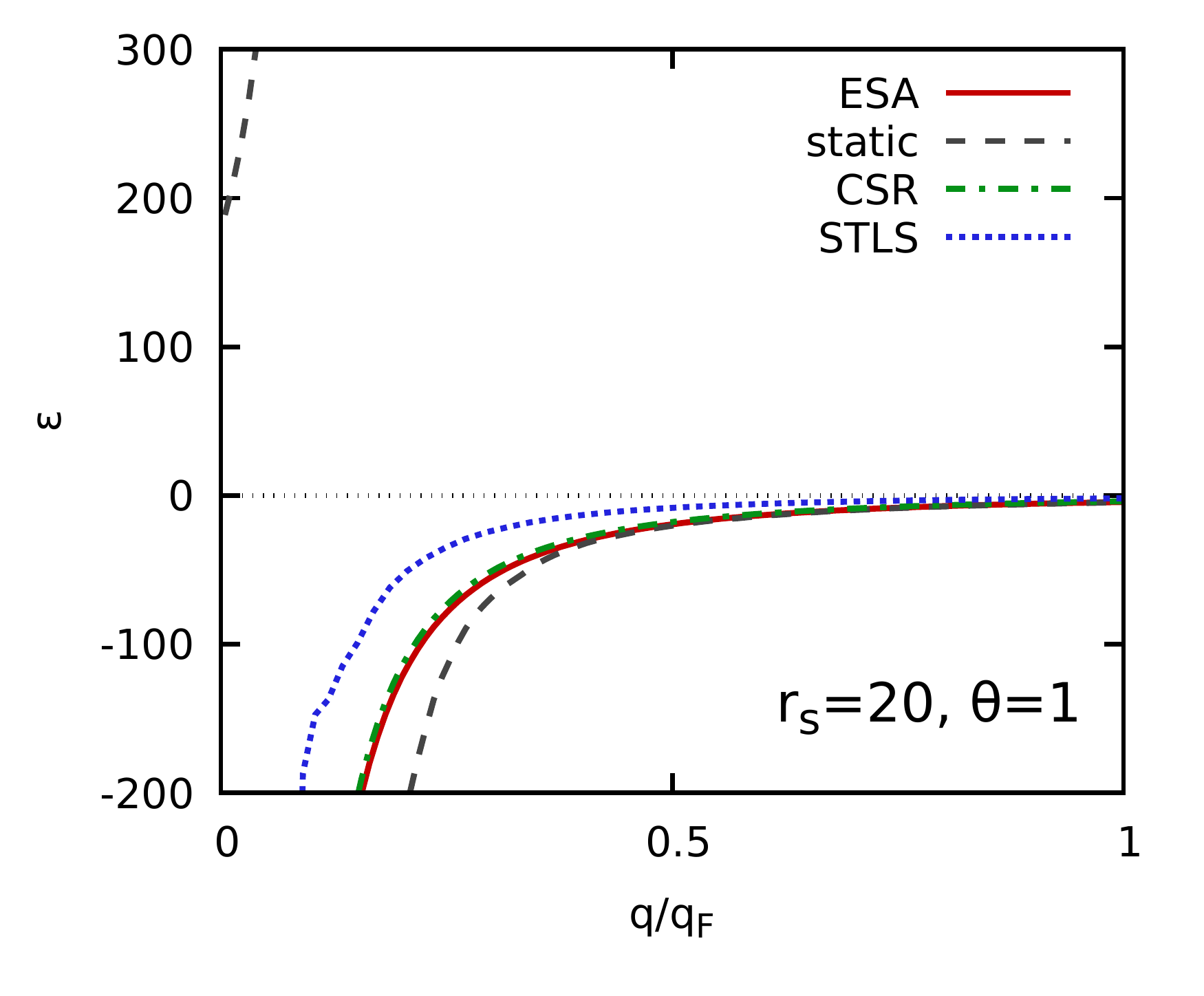}
\includegraphics[width=0.462\textwidth]{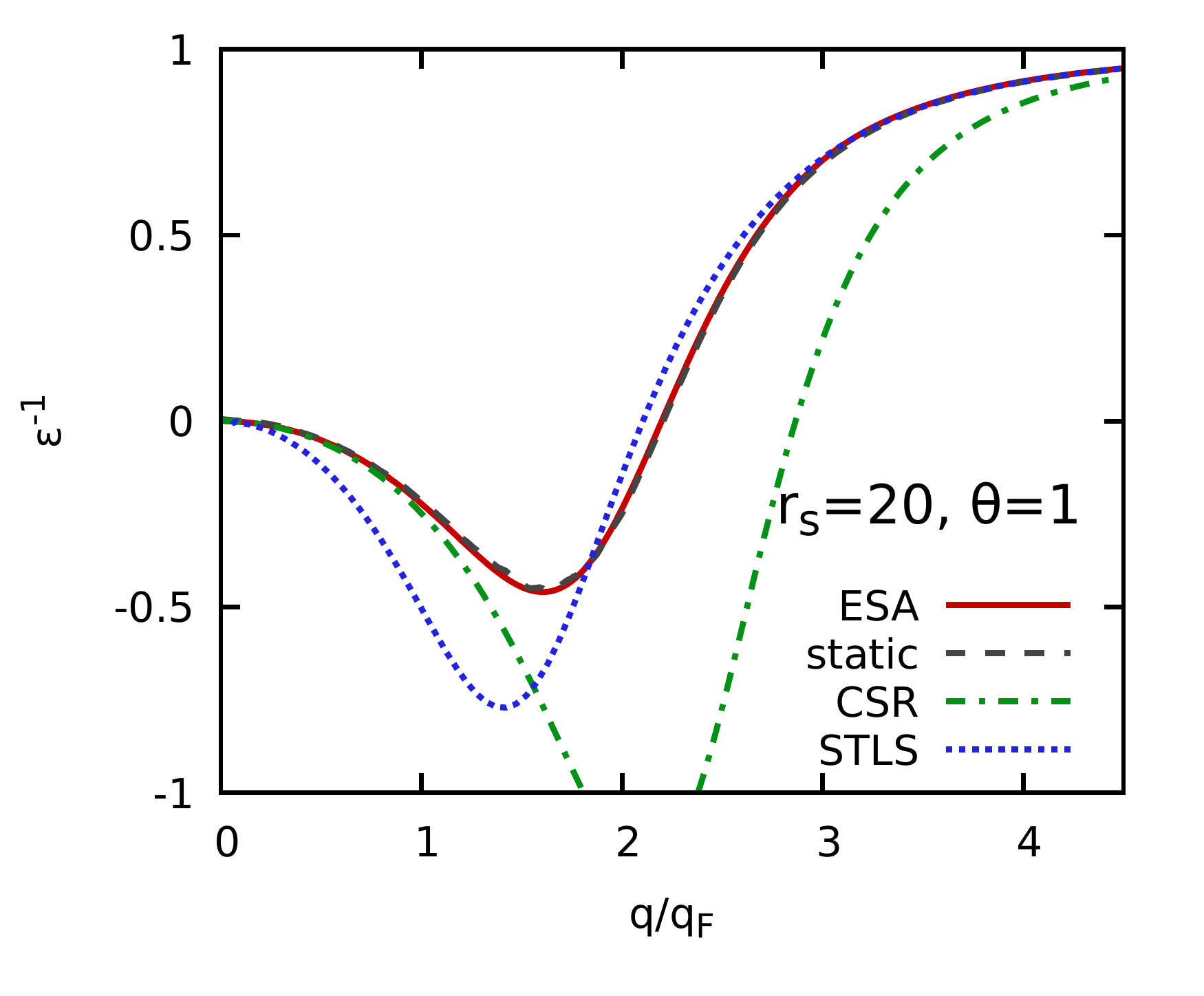}
\caption{\label{fig:Dielectric_rs20}
Left: Static dielectric function $\epsilon(q)$ for $r_s=20$ and $\theta=1$. Solid red: ESA; dashed grey: exact static limit using the neural-net from Ref.~\cite{dornheim_ML}; dash-dotted green: CSR, Eq.~(\ref{eq:CSR}); dotted-blue: STLS~\cite{stls2,stls,stls_original}. Right: Same data for the inverse dielectric function $\epsilon^{-1}(q)$.
}
\end{figure*}

Let us conclude this section with an example at strong coupling, $r_s=20$ and $\theta=1$, depicted in Fig.~\ref{fig:Dielectric_rs20}. Firstly, we note that here the ESA and CSR curves for $\epsilon(q)$ diverge towards negative infinity, which is the result of a negative compressibility at these conditions, see also Refs.~\cite{Hamann_PRB_2020,stls2}. For completeness, we note that this is a necessary, but not sufficient condition for instability~\cite{quantum_theory}, and, thus, not problematic. The STLS curve, too, diverges towards negative infinity, although with a substantially different slope. Finally, the \emph{static} curve becomes increasingly inaccurate for small $q$ and again attains a finite value for $q=0$.

Regarding the inverse dielectric function (right panel), the negative compressibility is reflected by a nontrivial shape of this quantity, with a minimum around $q\approx1.8q_\textnormal{F}$. Here, too, we note that ESA and the \emph{static} curve are in excellent agreement everywhere, whereas the STLS approximation gives a substantially wrong prediction of both the location and the depth of the minimum in $\epsilon^{-1}(q)$.

\subsection{Dynamic structure factor}

\begin{figure*}\centering\hspace*{-0.05\textwidth}\includegraphics[width=0.562\textwidth]{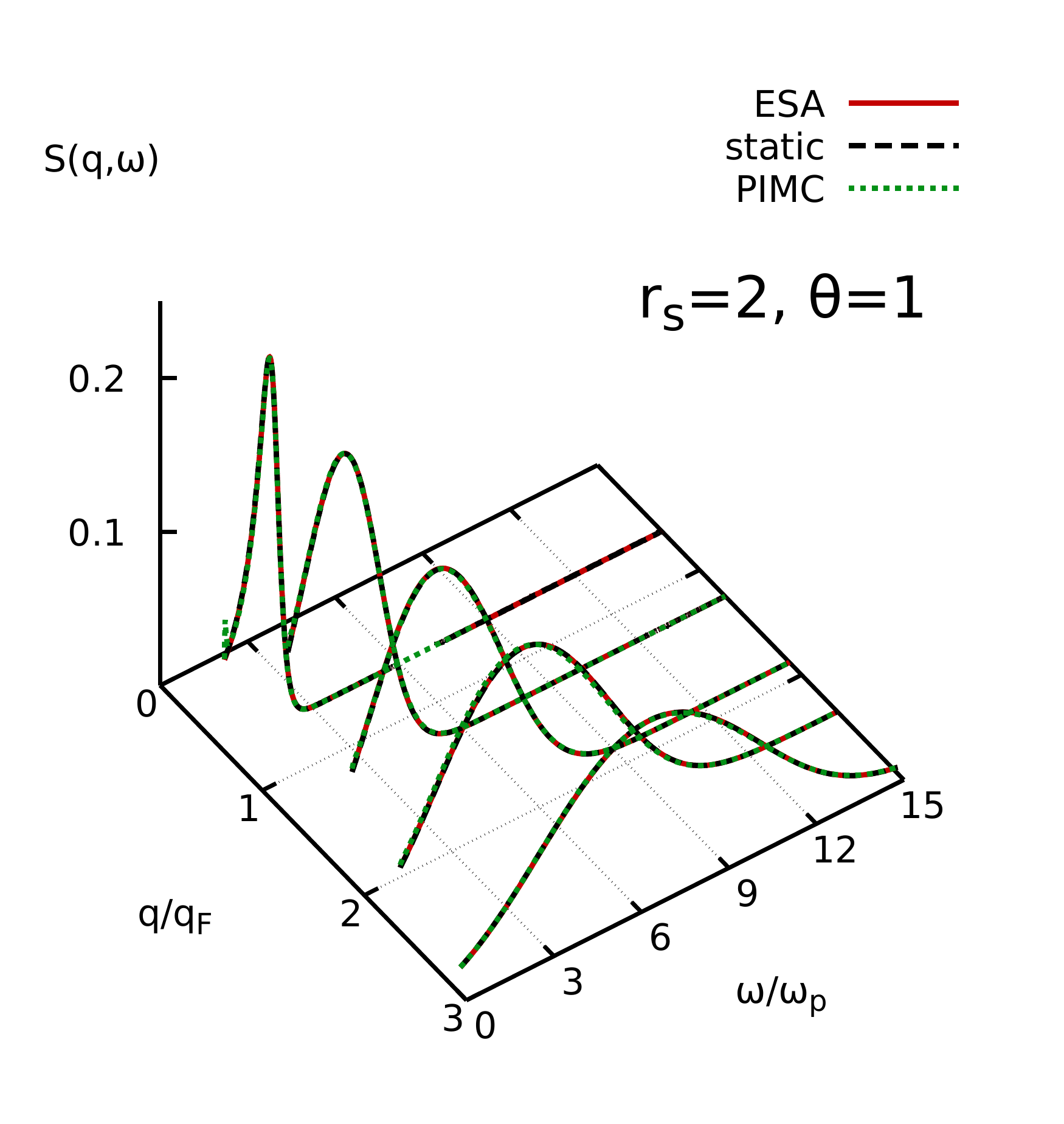}\hspace*{-0.05\textwidth}
\includegraphics[width=0.562\textwidth]{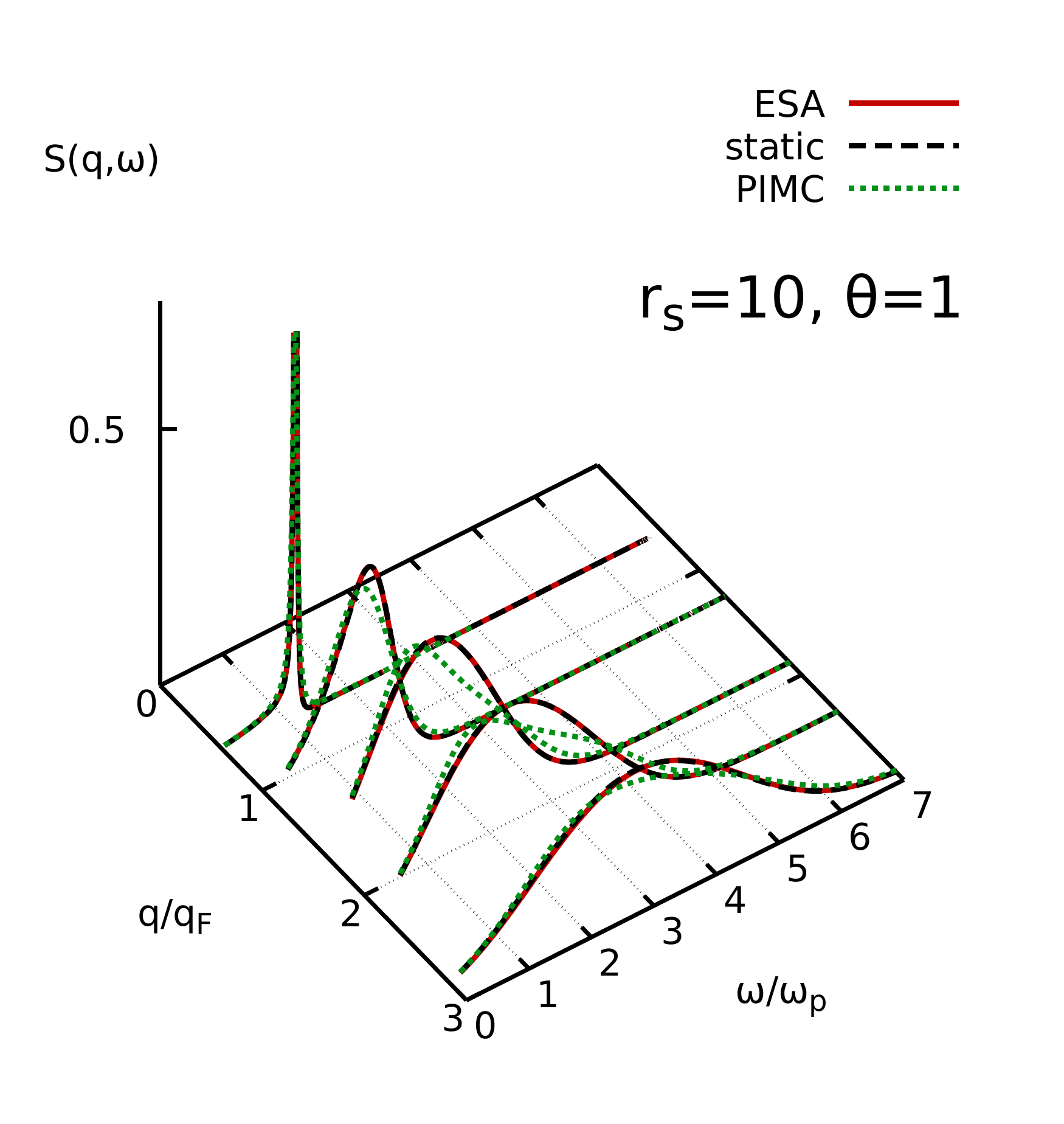}\\\vspace*{-0.5cm}
\caption{\label{fig:DSF}
Dynamic structure factor of the uniform electron gas at $\theta=1$ for $r_s=2$ (left) and $r_s=10$ (right). Solid red: ESA; dashed black: \emph{static approximation}; dotted green: \textit{ab initio} reconstructed PIMC results using a stochastically sampled dynamic LFC, taken from Ref.~\cite{dornheim_dynamic}.
}
\end{figure*}

The final property of the UEG to be investigated in this work is the dynamic structure factor $S(q,\omega)$, which is shown in Fig.~\ref{fig:DSF} for $\theta=1$.
The left panel corresponds to the usual metallic density, $r_s=2$, and the dotted green curves are \textit{ab initio} PIMC results taken from Ref.~\cite{dornheim_dynamic} that have been obtained by stochastically sampling the dynamic LFC $G(q,\omega)$. In addition, the solid red and dashed black curves have been obtained by using the ESA and the \emph{static approximation}, and are in virtually perfect agreement to the PIMC data everywhere. This illustrates that a static description of the LFC is fully sufficient to describe the dynamic density response of electrons at these conditions, see also Refs.~\cite{dornheim_dynamic,dynamic_folgepaper,Hamann_PRB_2020,Dornheim_PRE_2020} for more details.

The right panel corresponds to a stronger coupling strength, $r_s=10$, which is located at the margins of the electron liquid regime. While the ESA and \emph{static approximation} here, too, basically give the same results, both curves exhibit systematic deviations towards the exact PIMC data. This is a direct consequence of the increased impact of the frequency-dependence of electronic exchange--correlation effects expressed via the dynamic LFC at these conditions~\cite{dornheim_dynamic}.

\begin{figure}\centering\includegraphics[width=0.462\textwidth]{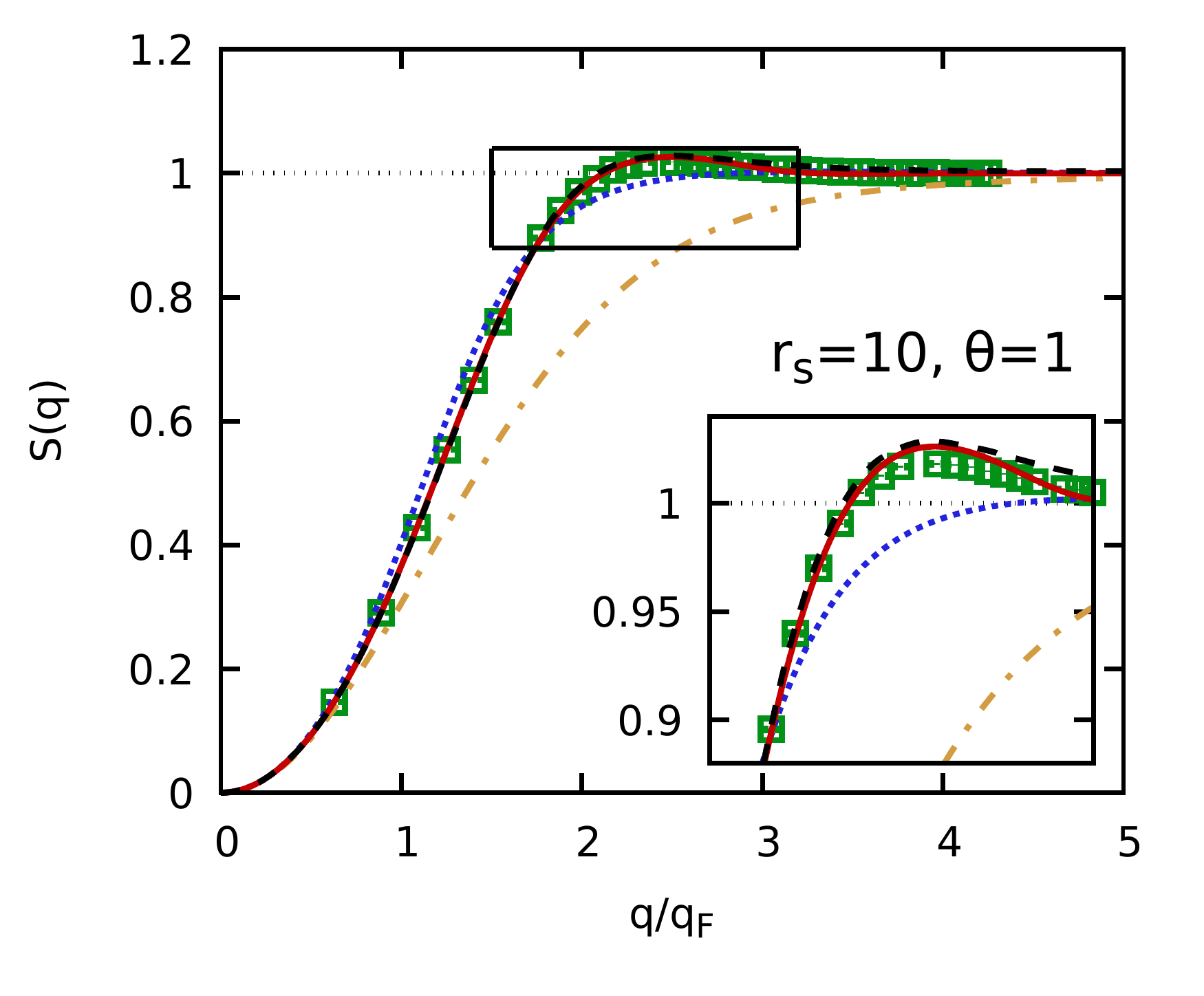}
\caption{\label{fig:Sq_rs10}
Static structure factor of the UEG for $r_s=10$ and $\theta=1$ (cf.~right panel of Fig.~\ref{fig:DSF}). Green squares: PIMC data taken from Ref.~\cite{dornheim_dynamic}; solid red: ESA; dashed black: \emph{static approximation}; dotted blue: STLS~\cite{stls_original,stls,stls2}; dash-dotted yellow: RPA.
}
\end{figure} 

Interestingly, the impact of the dynamic LFC only manifests in a pronounced way in the shape of $S(q,\omega)$, whereas its normalization [i.e., the SSF, see Eq.~(\ref{eq:Sq})] is hardly affected. This is demonstrated in Fig.~\ref{fig:Sq_rs10}, where we show the corresponding $S(q)$ for the same conditions. For example, for both $q=1.25q_\textnormal{F}$ and $q=1.88q_\textnormal{F}$, the shape of the PIMC data for $S(q,\omega)$ significantly deviates from the other curves, whereas the SSF is nearly perfectly reproduced by both the ESA and the \emph{static approximation}.

For larger $q$, the results for the SSF of $G(q)$ and $\overline{G}_\textnormal{ESA}(q)$ do start to deviate, but this has no pronounced impact on $S(q,\omega)$ itself.

We thus conclude that both the usual \emph{static approximation} and our new ESA scheme~\cite{dornheim_PRL_ESA_2020} are equally well suited for the description of dynamic properties at WDM conditions, but are not suited for a qualitative description of the dynamic density response of the strongly coupled electron liquid regime, for which a fully dynamic local field correction has been shown to be indispensable.

\subsection{Test charge screening.}

According to linear response theory, the screened potential of an ion (with charge $Ze$) can be  computed using the static dielectric function  as \cite{Hansen, zhandos1}:
 \begin{equation} \label{eq:pot_stat}
 \Phi(\vec r)   = \int\!\frac{\mathrm{d}^3q}{(2 \pi)^3 }~ \frac{4\pi Ze}{q^2}\frac{e^{i \vec q \cdot \vec r}}{\epsilon(q)},
\end{equation}
which is valid for the weak electron-ion coupling. The latter condition is satisfied at large distances from the ion \cite{zhandos_cpp21}. 

\begin{figure}\vspace{0.5cm}\centering\includegraphics[width=0.462\textwidth]{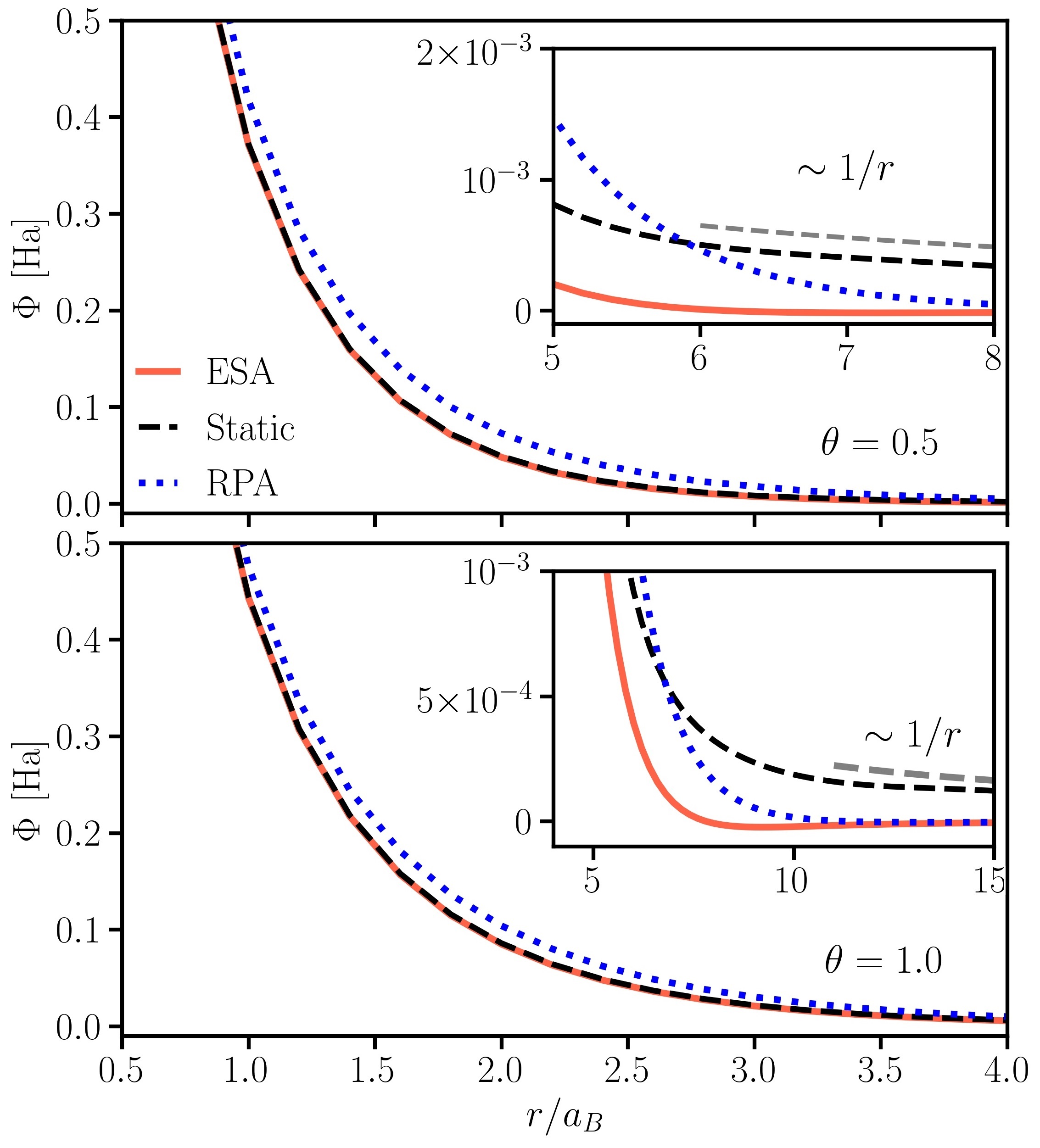}
\caption{\label{fig:potential}
Screened ion potential at $r_s=2$, $\theta=0.5$ and $\theta=1$. Solid red: the data computed using the analytical representation of the ESA Eq.~(\ref{eq:analytical}); dashed black: static approximation computed using the neural-net representation of the static local field correction from Ref.~\cite{dornheim_ML}; dotted blue:  RPA result; dashed grey line shows $\sim 1/r$ behavior of the  neural-net representation based data at large distances. 
}
\end{figure}

As discussed in Sec.~\ref{sec:Diel} above, the violation of the exact limit Eq.~(\ref{eq:duality}) leads to the  unphysical behavior of the static dielectric function computed using the neural-net representation of the LFC from Ref.~\cite{dornheim_ML}. This results in incomplete screening when the corresponding static dielectric function is used to compute the screened potential.   To illustrate this, we show the screened ion potential (with $Z=1$) for $r_s=2$, $\theta=0.5$ and $\theta=1.0$ in Fig.~\ref{fig:potential}, where the screened ion potential is computed using ESA given by Eq.~(\ref{eq:analytical}), the neural-net representation of the LFC from Ref.~\cite{dornheim_ML}, and RPA.

From Fig.~\ref{fig:potential}, it is clearly seen that the neural-net representation based result for the screened potential exhibits an $\sim 1/r$ asymptotic behavior at large distances. In contrast, the screened potential obtained using  the analytical representation $\overline{G}_\textnormal{ESA}(q;r_s,\theta)$ correctly reproduces complete screening like RPA based data, with a Yukawa type exponential screening  at large distances \cite{zhandos_cpp21}.  Finally, we note that electronic exchange--correlation effects, taken into account by using the LFC, lead to a stronger screening of the ion potential compared to the RPA result \cite{zhandos1, zhandos_cpp21, zhandos_cpp17}. 

\subsection{Stopping power}
A further example for the application of the LFC is the calculation of the  stopping power, i.e. the mean energy loss of a projectile (an ion) per unit path length, and related quantities such as the penetration length, straggling rate etc. These energy dissipation characteristics are of paramount importance for such applications as ICF and laboratory astrophysics~\cite{GRABOWSKI2020100905, MRE2018}.   
A linear response expression based on the dynamic dielectric function that describes the stopping power for a low-Z projectile when the ion--electron coupling is weak~\cite{PhysRevA.23.1898, Zwicknagel}  is given by~\cite{PhysRevA.23.1898}: 
\begin{equation}\label{eq:sp}
    S(v)=\frac{2Z^2e^2}{\pi v^2}\int_0^{\infty} \, \frac{{\rm d}k}{k} \, \int_0^{kv} {\rm d} \omega ~\omega ~{\rm Im} \left[\frac{-1}{\epsilon(k,\omega)}\right],
\end{equation}
where  $v$ is the ion velocity.   

\begin{figure}\vspace{0.25cm}\centering\includegraphics[width=0.48\textwidth]{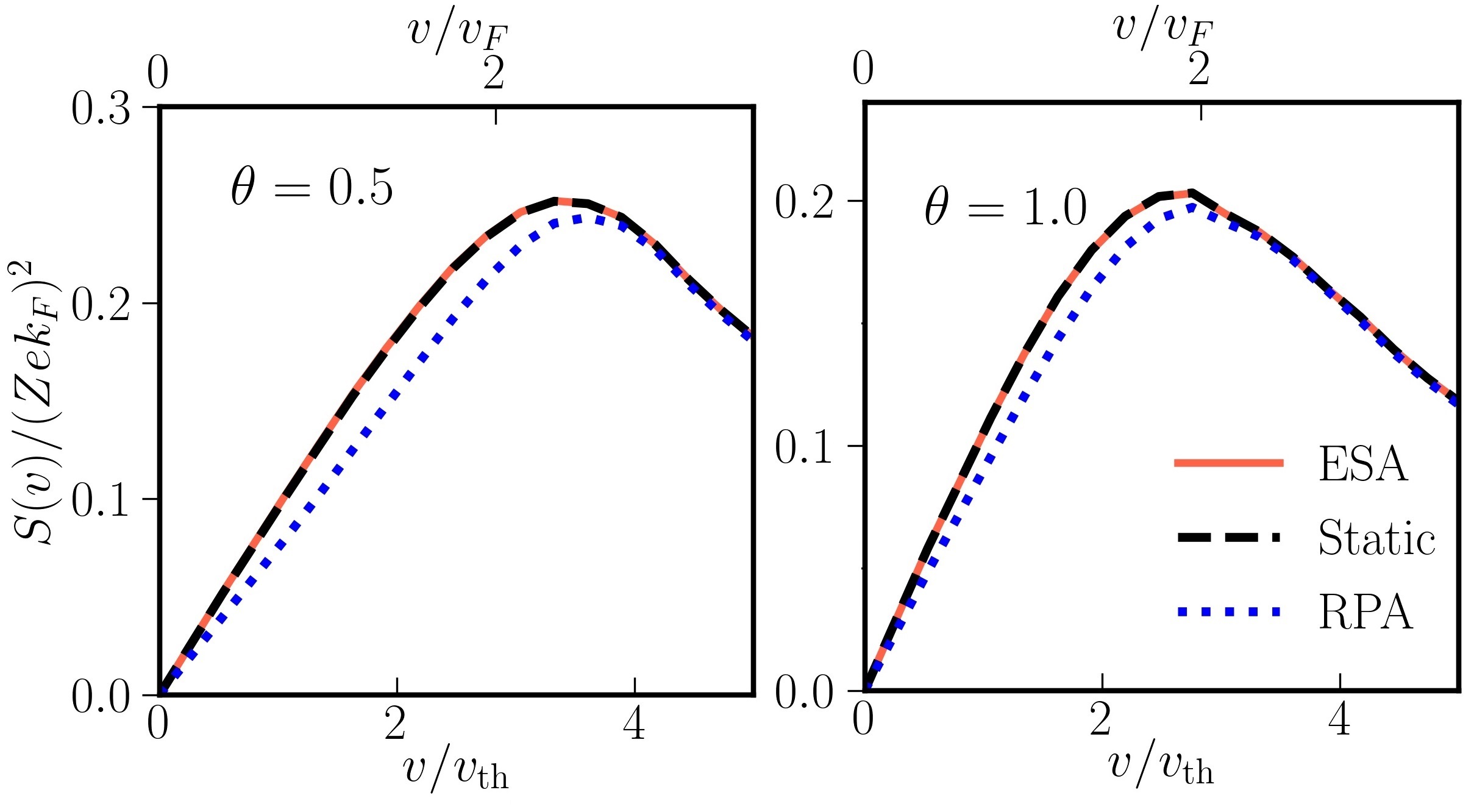}
\caption{\label{fig:sp}
Stopping power at $r_s=2$, $\theta=0.5$ and $\theta=1$.Solid red: data computed using the analytical representation of the ESA Eq.~(\ref{eq:analytical}); dashed black: \emph{static approximation} computed using the neural-net representation of the static local field correction from Ref.~\cite{dornheim_ML}; dotted blue:  RPA result. The lower $x$ axis corresponds to $v/v_{\rm th}$ and the upper $x$ axis to $v/v_F$, with $v_{\rm th}$ and  $v_F$ being the thermal velocity and Fermi velocity of electrons, respectively. }
\end{figure}

Recently,  using Eq.~(\ref{eq:sp}), the neural-net representation of the LFC \cite{dornheim_ML} was used to study the ion energy-loss characteristics and friction in a free-electron gas at warm dense matter conditions  \cite{Moldabekov_PRE_2020}.  
Therefore, it is required to check whether  the discussed  unphysical behavior of certain quantities based on  the neural-net representation of the LFC \cite{dornheim_ML} also manifests in the stopping power.
The comparison of the ESA ~(\ref{eq:analytical}) based data for the stopping power  to the results obtained using  the neural-net representation of the LFC \cite{dornheim_ML} is shown in Fig.~\ref{fig:sp} for $r_s=2$, $\theta=0.5$ and $\theta=1.0$. From Fig.~\ref{fig:sp} we see that the ESA  and the neural-net representation based results for the stopping power are in agreement  with a high accuracy. Additionally, a comparison to the RPA based data shows that electronic exchange-correlation effects are significant at projectile velocities $v\lesssim v_F$. We refer an interested reader to Ref.~\cite{Moldabekov_PRE_2020}  for a more detailed study in a wider parameter range. 

\section{Summary and Discussion\label{sec:summary}}
\subsection{Summary}
The first main achievement of this work is the construction of an accurate analytical representation of the \emph{effective static approximation} for the local field correction $\overline{G}_\textnormal{ESA}(q;r_s,\theta)$ covering all wave-numbers and the entire relevant range of densities ($0.7\leq r_s\leq20$) and temperatures ($0\leq\theta\leq4$). Our fit formula [Eq.~(\ref{eq:final_fit})] well reproduces the original ESA scheme presented in Ref.~\cite{dornheim_PRL_ESA_2020} while exactly incorporating the CSR in the limit of small wave numbers, and without the need for the evaluation of the neural-net from Ref.~\cite{dornheim_ML}. A short implementation of Eq.~(\ref{eq:final_fit}) in Python is freely available online~\cite{code} and can easily be incorporated into existing codes; see the next section for a short list of potential applications.

The second aim of this paper is the further analysis of the ESA in general and our fit formula in particular regarding the estimation of various electronic properties. Here one finding of considerable interest has been the estimation of an effective static LFC $\overline{G}_\textnormal{invert}(q)$ that, when being inserted into Eq.~(\ref{eq:chi}), exactly reproduces the static structure factor $S(q)$ known from QMC calculation both in the ground state and at finite temperature. Remarkably, $\overline{G}_\textnormal{invert}(q)$ almost exactly follows $\overline{G}_\textnormal{ESA}(q)$ for all wave numbers, which further substantiates the quality of the relatively simple idea behind the ESA. As it is expected, the latter gives very accurate results both for $S(q)$ and the interaction energy $v$, in particular at metallic densities where we find relative deviations to PIMC data not exceeding $1\%$.

A further point of interest is the utility of the ESA regarding the estimation of the static density response function $\chi(q)$ and the directly related dielectric function $\epsilon(q)$. More specifically, the neural-net representation of the exact static LFC $G(q;r_s,\theta)$ should give exact result for this quantities, whereas the definition of $\overline{G}_\textnormal{ESA}(q;r_s,\theta)$ as a frequency-averaged LFC could potentially introduce a bias in this limit. Yet, we find that the ESA gives virtually exact results over the entire WDM regime (even in the ground-state), whereas said bias only manifests in $\chi(q)$ for the strongly coupled electron liquid regime, $r_s=20$. In addition, the exact incorporation of the CSR for small $q$ in our parametrization of $\overline{G}_\textnormal{ESA}(q;r_s,\theta)$ means that the present results for the dielectric function $\epsilon(q)$ are even superior to the corresponding prediction by the neural net, where the CSR is only fulfilled approximately, i.e., with finite accuracy. In particular, the ESA gives the correct divergence behaviour of $\epsilon(q)$ in the limit of small $q$, whereas the neural-net predicts a finite value for $q=0$, which is unphysical~\cite{quantum_theory,Hamann_PRB_2020}.

A third item of our analysis is the application of the ESA for the estimation of the dynamic structure factor $S(q,\omega)$, where we find no difference to the usual \emph{static approximation}~\cite{dornheim_dynamic,dynamic_folgepaper,Hamann_PRB_2020}. More specifically, both $G(q;r_s,\theta)$ and $\overline{G}_\textnormal{ESA}(q;r_s,\theta)$ are highly accurate at WDM densities, but cannot reproduce the nontrivial shape of $S(q,\omega)$ associated with the predicted incipient excitonic mode~\cite{Takada_PRB_2016,Higuchi_Japan_2000} in the electron liquid regime.

Furthermore, we have compared our parametrization of $\overline{G}_\textnormal{ESA}(q;r_s,\theta)$ and the neural-net representation of $G(q;r_s,\theta)$ regarding the construction of an electronically screened ionic potential $\Phi(r)$. While the resulting potentials are in excellent agreement for small to intermediate distances $r$, the aforementioned inaccuracies of the neural net at small $q$ lead to a spuriously slow convergence of $\Phi(r)$ at large ionic separations $r$.

Finally,  the stopping power calculation results show that the ESA and the neural-net representation of the LFC  are equivalent for this application.  Therefore,  both the presented analytical fit formula for the ESA  and the neural-net representation of the LFC can be used to study ion energy-loss in WDM and hot dense matter.

\subsection{Discussion and outlook}

The ESA scheme has been shown to give a highly reliable description of electronic XC-effects and, in our opinion, constitutes the method of choice for many applications both in the context of WDM research and solid state physics in the ground state.

Due to its definition as a frequency-averaged LFC, the ESA is particularly suited for the construction of advanced XC-functionals for DFT simulations based on the adiabatic connection and the fluctuation dissipation theorem~\cite{Lu_JCP_2014,Thygesen_JCP_2015,pribram,Goerling_PRB_2019}. This is a highly desirable project, as the predictive capability of DFT for WDM calculations is still limited~\cite{Ramakrishna_PRB_2021}.

Secondly, we mention the interpretation of XRTS experiments~\cite{siegfried_review,kraus_xrts} within the Chihara decomposition~\cite{Chihara_1987} where electronic correlations are often treated insufficiently. In this regard, the remarkable degree of accuracy provided by both ESA and the \emph{static approximation}, and the promising results for aluminum shown in Ref.~\cite{dornheim_PRL_ESA_2020} give us hope that an improved description of XRTS signals can be achieved with hardly any additional effort.

Thirdly, the ESA can be used to incorporate electronic XC-effects into many effective theories in a straightforward way. Here examples include quantum hydrodynamics~\cite{zhandos_QHD,Diaw2017, zhandos_POP15}, average atom models~\cite{Sterne_average_atom_HEDP_2007}, electronically screened ionic potentials~\cite{PhysRevE.91.023102, ctpp.201500137, zhandos_cpp17}, and dynamic electronic phase-field crystal methods \cite{PhysRevB.100.235116}.

Finally, we mention the value of the LFC in general and the ESA in particular for the estimation of a multitude of material properties like the electronic stopping power~\cite{Moldabekov_PRE_2020}, thermal and electrical conductivities~\cite{Hamann_PRB_2020}, and energy relaxation rates~\cite{transfer1,transfer2, PhysRevE.97.013205}.

From a theoretical perspective, the main open challenge is given by the estimation of the full frequency-dependence of the LFC $G(q,\omega)$, which is currently only possible for certain parameters~\cite{dornheim_dynamic,dynamic_folgepaper,Hamann_PRB_2020}. One way towards this goal would be the development of new fermionic QMC approaches at finite temperature, to estimate the imaginary-time density--density correlation function $F(q,\tau)$--the crucial ingredient for the reconstruction of both $S(q,\omega)$ and $G(q,\omega)$. Here the phaseless auxiliary-field QMC method constitutes a promising candidate~\cite{lee2020phaseless}.

A second topic for future research is given by the comparison of $\overline{G}_\textnormal{ESA}(q;r_s,\theta)$ to different dielectric theories~\cite{tanaka_hnc,stls,stls2,Tanaka_CPP_2017,stolzmann,Panholzer_PRL_2018}, in particular the
recent scheme by Tanaka~\cite{tanaka_hnc} and the frequency-dependent version of STLS~\cite{arora,schweng,dynamic_ii}.

\section*{Acknowledgments}
We thank Jan Vorberger for helpful comments.
This work was partly funded by the Center for Advanced Systems Understanding (CASUS) which is financed by Germany's Federal Ministry of Education and Research (BMBF) and by the Saxon Ministry for Science, Culture and Tourism (SMWK) with tax funds on the basis of the budget approved by the Saxon State Parliament.
We gratefully acknowledge CPU-time at the Norddeutscher Verbund f\"ur Hoch- und H\"ochstleistungsrechnen (HLRN) under grant shp00026 and on a Bull Cluster at the Center for Information Services and High Performace Computing (ZIH) at Technische Universit\"at Dresden.

\bibliography{bibliography}

\begin{thebibliography}{146}%
\makeatletter
\providecommand \@ifxundefined [1]{%
 \@ifx{#1\undefined}
}%
\providecommand \@ifnum [1]{%
 \ifnum #1\expandafter \@firstoftwo
 \else \expandafter \@secondoftwo
 \fi
}%
\providecommand \@ifx [1]{%
 \ifx #1\expandafter \@firstoftwo
 \else \expandafter \@secondoftwo
 \fi
}%
\providecommand \natexlab [1]{#1}%
\providecommand \enquote  [1]{``#1''}%
\providecommand \bibnamefont  [1]{#1}%
\providecommand \bibfnamefont [1]{#1}%
\providecommand \citenamefont [1]{#1}%
\providecommand \href@noop [0]{\@secondoftwo}%
\providecommand \href [0]{\begingroup \@sanitize@url \@href}%
\providecommand \@href[1]{\@@startlink{#1}\@@href}%
\providecommand \@@href[1]{\endgroup#1\@@endlink}%
\providecommand \@sanitize@url [0]{\catcode `\\12\catcode `\$12\catcode
  `\&12\catcode `\#12\catcode `\^12\catcode `\_12\catcode `\%12\relax}%
\providecommand \@@startlink[1]{}%
\providecommand \@@endlink[0]{}%
\providecommand \url  [0]{\begingroup\@sanitize@url \@url }%
\providecommand \@url [1]{\endgroup\@href {#1}{\urlprefix }}%
\providecommand \urlprefix  [0]{URL }%
\providecommand \Eprint [0]{\href }%
\providecommand \doibase [0]{http://dx.doi.org/}%
\providecommand \selectlanguage [0]{\@gobble}%
\providecommand \bibinfo  [0]{\@secondoftwo}%
\providecommand \bibfield  [0]{\@secondoftwo}%
\providecommand \translation [1]{[#1]}%
\providecommand \BibitemOpen [0]{}%
\providecommand \bibitemStop [0]{}%
\providecommand \bibitemNoStop [0]{.\EOS\space}%
\providecommand \EOS [0]{\spacefactor3000\relax}%
\providecommand \BibitemShut  [1]{\csname bibitem#1\endcsname}%
\let\auto@bib@innerbib\@empty
\bibitem [{\citenamefont {Giuliani}\ and\ \citenamefont
  {Vignale}(2008)}]{quantum_theory}%
  \BibitemOpen
  \bibfield  {author} {\bibinfo {author} {\bibfnamefont {G.}~\bibnamefont
  {Giuliani}}\ and\ \bibinfo {author} {\bibfnamefont {G.}~\bibnamefont
  {Vignale}},\ }\href@noop {} {\emph {\bibinfo {title} {Quantum Theory of the
  Electron Liquid}}}\ (\bibinfo  {publisher} {Cambridge University Press},\
  \bibinfo {address} {Cambridge},\ \bibinfo {year} {2008})\BibitemShut
  {NoStop}%
\bibitem [{\citenamefont {Foulkes}\ \emph {et~al.}(2001)\citenamefont
  {Foulkes}, \citenamefont {Mitas}, \citenamefont {Needs},\ and\ \citenamefont
  {Rajagopal}}]{Foulkes_RevModPhys_2001}%
  \BibitemOpen
  \bibfield  {author} {\bibinfo {author} {\bibfnamefont {W.~M.~C.}\
  \bibnamefont {Foulkes}}, \bibinfo {author} {\bibfnamefont {L.}~\bibnamefont
  {Mitas}}, \bibinfo {author} {\bibfnamefont {R.~J.}\ \bibnamefont {Needs}}, \
  and\ \bibinfo {author} {\bibfnamefont {G.}~\bibnamefont {Rajagopal}},\
  }\bibfield  {title} {\enquote {\bibinfo {title} {Quantum monte carlo
  simulations of solids},}\ }\href {\doibase 10.1103/RevModPhys.73.33}
  {\bibfield  {journal} {\bibinfo  {journal} {Rev. Mod. Phys.}\ }\textbf
  {\bibinfo {volume} {73}},\ \bibinfo {pages} {33--83} (\bibinfo {year}
  {2001})}\BibitemShut {NoStop}%
\bibitem [{\citenamefont {Loos}\ and\ \citenamefont {Gill}(2016)}]{loos}%
  \BibitemOpen
  \bibfield  {author} {\bibinfo {author} {\bibfnamefont {P.-F.}\ \bibnamefont
  {Loos}}\ and\ \bibinfo {author} {\bibfnamefont {P.~M.~W.}\ \bibnamefont
  {Gill}},\ }\bibfield  {title} {\enquote {\bibinfo {title} {The uniform
  electron gas},}\ }\href
  {http://onlinelibrary.wiley.com/doi/10.1002/wcms.1257/abstract} {\bibfield
  {journal} {\bibinfo  {journal} {Comput. Mol. Sci}\ }\textbf {\bibinfo
  {volume} {6}},\ \bibinfo {pages} {410--429} (\bibinfo {year}
  {2016})}\BibitemShut {NoStop}%
\bibitem [{\citenamefont {Dornheim}\ \emph
  {et~al.}(2018{\natexlab{a}})\citenamefont {Dornheim}, \citenamefont {Groth},\
  and\ \citenamefont {Bonitz}}]{review}%
  \BibitemOpen
  \bibfield  {author} {\bibinfo {author} {\bibfnamefont {T.}~\bibnamefont
  {Dornheim}}, \bibinfo {author} {\bibfnamefont {S.}~\bibnamefont {Groth}}, \
  and\ \bibinfo {author} {\bibfnamefont {M.}~\bibnamefont {Bonitz}},\
  }\bibfield  {title} {\enquote {\bibinfo {title} {The uniform electron gas at
  warm dense matter conditions},}\ }\href
  {https://www.sciencedirect.com/science/article/abs/pii/S0370157318300516}
  {\bibfield  {journal} {\bibinfo  {journal} {Phys. Reports}\ }\textbf
  {\bibinfo {volume} {744}},\ \bibinfo {pages} {1--86} (\bibinfo {year}
  {2018}{\natexlab{a}})}\BibitemShut {NoStop}%
\bibitem [{\citenamefont {Bohm}\ and\ \citenamefont {D.~Pines}(1952)}]{pines}%
  \BibitemOpen
  \bibfield  {author} {\bibinfo {author} {\bibfnamefont {D.}~\bibnamefont
  {Bohm}}\ and\ \bibinfo {author} {\bibfnamefont {A}~\bibnamefont {D.~Pines}},\
  }\bibfield  {title} {\enquote {\bibinfo {title} {Collective description of
  electron interactions: Ii. collective vs individual particle aspects of the
  interactions},}\ }\href
  {https://journals.aps.org/pr/abstract/10.1103/PhysRev.85.338} {\bibfield
  {journal} {\bibinfo  {journal} {Phys. Rev.}\ }\textbf {\bibinfo {volume}
  {85}},\ \bibinfo {pages} {338} (\bibinfo {year} {1952})}\BibitemShut
  {NoStop}%
\bibitem [{\citenamefont {Bardeen}\ \emph {et~al.}(1957)\citenamefont
  {Bardeen}, \citenamefont {Cooper},\ and\ \citenamefont
  {Schrieffer}}]{Bardeen_PhysRev_1957}%
  \BibitemOpen
  \bibfield  {author} {\bibinfo {author} {\bibfnamefont {J.}~\bibnamefont
  {Bardeen}}, \bibinfo {author} {\bibfnamefont {L.~N.}\ \bibnamefont {Cooper}},
  \ and\ \bibinfo {author} {\bibfnamefont {J.~R.}\ \bibnamefont {Schrieffer}},\
  }\bibfield  {title} {\enquote {\bibinfo {title} {Theory of
  superconductivity},}\ }\href {\doibase 10.1103/PhysRev.108.1175} {\bibfield
  {journal} {\bibinfo  {journal} {Phys. Rev.}\ }\textbf {\bibinfo {volume}
  {108}},\ \bibinfo {pages} {1175--1204} (\bibinfo {year} {1957})}\BibitemShut
  {NoStop}%
\bibitem [{\citenamefont {Ceperley}(1978)}]{Ceperley_UEG_1978}%
  \BibitemOpen
  \bibfield  {author} {\bibinfo {author} {\bibfnamefont {D.}~\bibnamefont
  {Ceperley}},\ }\bibfield  {title} {\enquote {\bibinfo {title} {Ground state
  of the fermion one-component plasma: A monte carlo study in two and three
  dimensions},}\ }\href {\doibase 10.1103/PhysRevB.18.3126} {\bibfield
  {journal} {\bibinfo  {journal} {Phys. Rev. B}\ }\textbf {\bibinfo {volume}
  {18}},\ \bibinfo {pages} {3126--3138} (\bibinfo {year} {1978})}\BibitemShut
  {NoStop}%
\bibitem [{\citenamefont {Ceperley}\ and\ \citenamefont
  {Alder}(1980)}]{Ceperley_Alder_PRL_1980}%
  \BibitemOpen
  \bibfield  {author} {\bibinfo {author} {\bibfnamefont {D.~M.}\ \bibnamefont
  {Ceperley}}\ and\ \bibinfo {author} {\bibfnamefont {B.~J.}\ \bibnamefont
  {Alder}},\ }\bibfield  {title} {\enquote {\bibinfo {title} {Ground state of
  the electron gas by a stochastic method},}\ }\href {\doibase
  10.1103/PhysRevLett.45.566} {\bibfield  {journal} {\bibinfo  {journal} {Phys.
  Rev. Lett.}\ }\textbf {\bibinfo {volume} {45}},\ \bibinfo {pages} {566--569}
  (\bibinfo {year} {1980})}\BibitemShut {NoStop}%
\bibitem [{\citenamefont {Bowen}\ \emph {et~al.}(1994)\citenamefont {Bowen},
  \citenamefont {Sugiyama},\ and\ \citenamefont {Alder}}]{bowen2}%
  \BibitemOpen
  \bibfield  {author} {\bibinfo {author} {\bibfnamefont {C.}~\bibnamefont
  {Bowen}}, \bibinfo {author} {\bibfnamefont {G.}~\bibnamefont {Sugiyama}}, \
  and\ \bibinfo {author} {\bibfnamefont {B.~J.}\ \bibnamefont {Alder}},\
  }\bibfield  {title} {\enquote {\bibinfo {title} {Static dielectric response
  of the electron gas},}\ }\href
  {http://link.aps.org/doi/10.1103/PhysRevB.50.14838} {\bibfield  {journal}
  {\bibinfo  {journal} {Phys. Rev. B}\ }\textbf {\bibinfo {volume} {50}},\
  \bibinfo {pages} {14838} (\bibinfo {year} {1994})}\BibitemShut {NoStop}%
\bibitem [{\citenamefont {Moroni}\ \emph {et~al.}(1992)\citenamefont {Moroni},
  \citenamefont {Ceperley},\ and\ \citenamefont {Senatore}}]{moroni}%
  \BibitemOpen
  \bibfield  {author} {\bibinfo {author} {\bibfnamefont {S.}~\bibnamefont
  {Moroni}}, \bibinfo {author} {\bibfnamefont {D.~M.}\ \bibnamefont
  {Ceperley}}, \ and\ \bibinfo {author} {\bibfnamefont {G.}~\bibnamefont
  {Senatore}},\ }\bibfield  {title} {\enquote {\bibinfo {title} {Static
  response from quantum {M}onte {C}arlo calculations},}\ }\href
  {https://journals.aps.org/prl/abstract/10.1103/PhysRevLett.69.1837}
  {\bibfield  {journal} {\bibinfo  {journal} {Phys. Rev. Lett}\ }\textbf
  {\bibinfo {volume} {69}},\ \bibinfo {pages} {1837} (\bibinfo {year}
  {1992})}\BibitemShut {NoStop}%
\bibitem [{\citenamefont {Moroni}\ \emph {et~al.}(1995)\citenamefont {Moroni},
  \citenamefont {Ceperley},\ and\ \citenamefont {Senatore}}]{moroni2}%
  \BibitemOpen
  \bibfield  {author} {\bibinfo {author} {\bibfnamefont {S.}~\bibnamefont
  {Moroni}}, \bibinfo {author} {\bibfnamefont {D.~M.}\ \bibnamefont
  {Ceperley}}, \ and\ \bibinfo {author} {\bibfnamefont {G.}~\bibnamefont
  {Senatore}},\ }\bibfield  {title} {\enquote {\bibinfo {title} {Static
  response and local field factor of the electron gas},}\ }\href
  {http://link.aps.org/doi/10.1103/PhysRevLett.75.689} {\bibfield  {journal}
  {\bibinfo  {journal} {Phys. Rev. Lett}\ }\textbf {\bibinfo {volume} {75}},\
  \bibinfo {pages} {689} (\bibinfo {year} {1995})}\BibitemShut {NoStop}%
\bibitem [{\citenamefont {Ortiz}\ and\ \citenamefont
  {Ballone}(1994)}]{Ortiz_Ballone_PRB_1994}%
  \BibitemOpen
  \bibfield  {author} {\bibinfo {author} {\bibfnamefont {G.}~\bibnamefont
  {Ortiz}}\ and\ \bibinfo {author} {\bibfnamefont {P.}~\bibnamefont
  {Ballone}},\ }\bibfield  {title} {\enquote {\bibinfo {title} {Correlation
  energy, structure factor, radial distribution function, and momentum
  distribution of the spin-polarized uniform electron gas},}\ }\href {\doibase
  10.1103/PhysRevB.50.1391} {\bibfield  {journal} {\bibinfo  {journal} {Phys.
  Rev. B}\ }\textbf {\bibinfo {volume} {50}},\ \bibinfo {pages} {1391--1405}
  (\bibinfo {year} {1994})}\BibitemShut {NoStop}%
\bibitem [{\citenamefont {Ortiz}\ \emph {et~al.}(1999)\citenamefont {Ortiz},
  \citenamefont {Harris},\ and\ \citenamefont
  {Ballone}}]{Ortiz_Harris_Ballone_PRL_1999}%
  \BibitemOpen
  \bibfield  {author} {\bibinfo {author} {\bibfnamefont {G.}~\bibnamefont
  {Ortiz}}, \bibinfo {author} {\bibfnamefont {M.}~\bibnamefont {Harris}}, \
  and\ \bibinfo {author} {\bibfnamefont {P.}~\bibnamefont {Ballone}},\
  }\bibfield  {title} {\enquote {\bibinfo {title} {Zero temperature phases of
  the electron gas},}\ }\href {\doibase 10.1103/PhysRevLett.82.5317} {\bibfield
   {journal} {\bibinfo  {journal} {Phys. Rev. Lett.}\ }\textbf {\bibinfo
  {volume} {82}},\ \bibinfo {pages} {5317--5320} (\bibinfo {year}
  {1999})}\BibitemShut {NoStop}%
\bibitem [{\citenamefont {Zong}\ \emph {et~al.}(2002)\citenamefont {Zong},
  \citenamefont {Lin},\ and\ \citenamefont
  {Ceperley}}]{Zong_Lin_Ceperley_PRE_2002}%
  \BibitemOpen
  \bibfield  {author} {\bibinfo {author} {\bibfnamefont {F.~H.}\ \bibnamefont
  {Zong}}, \bibinfo {author} {\bibfnamefont {C.}~\bibnamefont {Lin}}, \ and\
  \bibinfo {author} {\bibfnamefont {D.~M.}\ \bibnamefont {Ceperley}},\
  }\bibfield  {title} {\enquote {\bibinfo {title} {Spin polarization of the
  low-density three-dimensional electron gas},}\ }\href {\doibase
  10.1103/PhysRevE.66.036703} {\bibfield  {journal} {\bibinfo  {journal} {Phys.
  Rev. E}\ }\textbf {\bibinfo {volume} {66}},\ \bibinfo {pages} {036703}
  (\bibinfo {year} {2002})}\BibitemShut {NoStop}%
\bibitem [{\citenamefont {Shepherd}\ \emph
  {et~al.}(2012{\natexlab{a}})\citenamefont {Shepherd}, \citenamefont {Booth},\
  and\ \citenamefont {Alavi}}]{Shepherd_UEG_2012}%
  \BibitemOpen
  \bibfield  {author} {\bibinfo {author} {\bibfnamefont {James~J.}\
  \bibnamefont {Shepherd}}, \bibinfo {author} {\bibfnamefont {George~H.}\
  \bibnamefont {Booth}}, \ and\ \bibinfo {author} {\bibfnamefont {Ali}\
  \bibnamefont {Alavi}},\ }\bibfield  {title} {\enquote {\bibinfo {title}
  {Investigation of the full configuration interaction quantum monte carlo
  method using homogeneous electron gas models},}\ }\href {\doibase
  10.1063/1.4720076} {\bibfield  {journal} {\bibinfo  {journal} {The Journal of
  Chemical Physics}\ }\textbf {\bibinfo {volume} {136}},\ \bibinfo {pages}
  {244101} (\bibinfo {year} {2012}{\natexlab{a}})},\ \Eprint
  {http://arxiv.org/abs/https://doi.org/10.1063/1.4720076}
  {https://doi.org/10.1063/1.4720076} \BibitemShut {NoStop}%
\bibitem [{\citenamefont {Shepherd}\ \emph
  {et~al.}(2012{\natexlab{b}})\citenamefont {Shepherd}, \citenamefont {Booth},
  \citenamefont {Gr\"uneis},\ and\ \citenamefont
  {Alavi}}]{Shepherd_UEG_PRB_2012}%
  \BibitemOpen
  \bibfield  {author} {\bibinfo {author} {\bibfnamefont {James~J.}\
  \bibnamefont {Shepherd}}, \bibinfo {author} {\bibfnamefont {George}\
  \bibnamefont {Booth}}, \bibinfo {author} {\bibfnamefont {Andreas}\
  \bibnamefont {Gr\"uneis}}, \ and\ \bibinfo {author} {\bibfnamefont {Ali}\
  \bibnamefont {Alavi}},\ }\bibfield  {title} {\enquote {\bibinfo {title} {Full
  configuration interaction perspective on the homogeneous electron gas},}\
  }\href {\doibase 10.1103/PhysRevB.85.081103} {\bibfield  {journal} {\bibinfo
  {journal} {Phys. Rev. B}\ }\textbf {\bibinfo {volume} {85}},\ \bibinfo
  {pages} {081103} (\bibinfo {year} {2012}{\natexlab{b}})}\BibitemShut
  {NoStop}%
\bibitem [{\citenamefont {Spink}\ \emph {et~al.}(2013)\citenamefont {Spink},
  \citenamefont {Needs},\ and\ \citenamefont {Drummond}}]{Spink_PRB_2013}%
  \BibitemOpen
  \bibfield  {author} {\bibinfo {author} {\bibfnamefont {G.~G.}\ \bibnamefont
  {Spink}}, \bibinfo {author} {\bibfnamefont {R.~J.}\ \bibnamefont {Needs}}, \
  and\ \bibinfo {author} {\bibfnamefont {N.~D.}\ \bibnamefont {Drummond}},\
  }\bibfield  {title} {\enquote {\bibinfo {title} {Quantum monte carlo study of
  the three-dimensional spin-polarized homogeneous electron gas},}\ }\href
  {\doibase 10.1103/PhysRevB.88.085121} {\bibfield  {journal} {\bibinfo
  {journal} {Phys. Rev. B}\ }\textbf {\bibinfo {volume} {88}},\ \bibinfo
  {pages} {085121} (\bibinfo {year} {2013})}\BibitemShut {NoStop}%
\bibitem [{\citenamefont {Drummond}\ \emph {et~al.}(2004)\citenamefont
  {Drummond}, \citenamefont {Radnai}, \citenamefont {Trail}, \citenamefont
  {Towler},\ and\ \citenamefont {Needs}}]{Drummond_Wigner_2004}%
  \BibitemOpen
  \bibfield  {author} {\bibinfo {author} {\bibfnamefont {N.~D.}\ \bibnamefont
  {Drummond}}, \bibinfo {author} {\bibfnamefont {Z.}~\bibnamefont {Radnai}},
  \bibinfo {author} {\bibfnamefont {J.~R.}\ \bibnamefont {Trail}}, \bibinfo
  {author} {\bibfnamefont {M.~D.}\ \bibnamefont {Towler}}, \ and\ \bibinfo
  {author} {\bibfnamefont {R.~J.}\ \bibnamefont {Needs}},\ }\bibfield  {title}
  {\enquote {\bibinfo {title} {Diffusion quantum monte carlo study of
  three-dimensional wigner crystals},}\ }\href {\doibase
  10.1103/PhysRevB.69.085116} {\bibfield  {journal} {\bibinfo  {journal} {Phys.
  Rev. B}\ }\textbf {\bibinfo {volume} {69}},\ \bibinfo {pages} {085116}
  (\bibinfo {year} {2004})}\BibitemShut {NoStop}%
\bibitem [{\citenamefont {Fraser}\ \emph {et~al.}(1996)\citenamefont {Fraser},
  \citenamefont {Foulkes}, \citenamefont {Rajagopal}, \citenamefont {Needs},
  \citenamefont {Kenny},\ and\ \citenamefont
  {Williamson}}]{Fraser_Foulkes_PRB_1996}%
  \BibitemOpen
  \bibfield  {author} {\bibinfo {author} {\bibfnamefont {Louisa~M.}\
  \bibnamefont {Fraser}}, \bibinfo {author} {\bibfnamefont {W.~M.~C.}\
  \bibnamefont {Foulkes}}, \bibinfo {author} {\bibfnamefont {G.}~\bibnamefont
  {Rajagopal}}, \bibinfo {author} {\bibfnamefont {R.~J.}\ \bibnamefont
  {Needs}}, \bibinfo {author} {\bibfnamefont {S.~D.}\ \bibnamefont {Kenny}}, \
  and\ \bibinfo {author} {\bibfnamefont {A.~J.}\ \bibnamefont {Williamson}},\
  }\bibfield  {title} {\enquote {\bibinfo {title} {Finite-size effects and
  coulomb interactions in quantum monte carlo calculations for homogeneous
  systems with periodic boundary conditions},}\ }\href {\doibase
  10.1103/PhysRevB.53.1814} {\bibfield  {journal} {\bibinfo  {journal} {Phys.
  Rev. B}\ }\textbf {\bibinfo {volume} {53}},\ \bibinfo {pages} {1814--1832}
  (\bibinfo {year} {1996})}\BibitemShut {NoStop}%
\bibitem [{\citenamefont {Perdew}\ and\ \citenamefont
  {Zunger}(1981)}]{Perdew_Zunger_PRB_1981}%
  \BibitemOpen
  \bibfield  {author} {\bibinfo {author} {\bibfnamefont {J.~P.}\ \bibnamefont
  {Perdew}}\ and\ \bibinfo {author} {\bibfnamefont {Alex}\ \bibnamefont
  {Zunger}},\ }\bibfield  {title} {\enquote {\bibinfo {title} {Self-interaction
  correction to density-functional approximations for many-electron systems},}\
  }\href {\doibase 10.1103/PhysRevB.23.5048} {\bibfield  {journal} {\bibinfo
  {journal} {Phys. Rev. B}\ }\textbf {\bibinfo {volume} {23}},\ \bibinfo
  {pages} {5048--5079} (\bibinfo {year} {1981})}\BibitemShut {NoStop}%
\bibitem [{\citenamefont {Perdew}\ and\ \citenamefont
  {Wang}(1992{\natexlab{a}})}]{Perdew_Wang_PRB_1992}%
  \BibitemOpen
  \bibfield  {author} {\bibinfo {author} {\bibfnamefont {John~P.}\ \bibnamefont
  {Perdew}}\ and\ \bibinfo {author} {\bibfnamefont {Yue}\ \bibnamefont
  {Wang}},\ }\bibfield  {title} {\enquote {\bibinfo {title} {Accurate and
  simple analytic representation of the electron-gas correlation energy},}\
  }\href {\doibase 10.1103/PhysRevB.45.13244} {\bibfield  {journal} {\bibinfo
  {journal} {Phys. Rev. B}\ }\textbf {\bibinfo {volume} {45}},\ \bibinfo
  {pages} {13244--13249} (\bibinfo {year} {1992}{\natexlab{a}})}\BibitemShut
  {NoStop}%
\bibitem [{\citenamefont {Perdew}\ and\ \citenamefont
  {Wang}(1992{\natexlab{b}})}]{Perdew_Wang_PDF_1992}%
  \BibitemOpen
  \bibfield  {author} {\bibinfo {author} {\bibfnamefont {John~P.}\ \bibnamefont
  {Perdew}}\ and\ \bibinfo {author} {\bibfnamefont {Yue}\ \bibnamefont
  {Wang}},\ }\bibfield  {title} {\enquote {\bibinfo {title} {Pair-distribution
  function and its coupling-constant average for the spin-polarized electron
  gas},}\ }\href {\doibase 10.1103/PhysRevB.46.12947} {\bibfield  {journal}
  {\bibinfo  {journal} {Phys. Rev. B}\ }\textbf {\bibinfo {volume} {46}},\
  \bibinfo {pages} {12947--12954} (\bibinfo {year}
  {1992}{\natexlab{b}})}\BibitemShut {NoStop}%
\bibitem [{\citenamefont {Vosko}\ \emph {et~al.}(1980)\citenamefont {Vosko},
  \citenamefont {Wilk},\ and\ \citenamefont {Nusair}}]{vwn}%
  \BibitemOpen
  \bibfield  {author} {\bibinfo {author} {\bibfnamefont {S.~H.}\ \bibnamefont
  {Vosko}}, \bibinfo {author} {\bibfnamefont {L.}~\bibnamefont {Wilk}}, \ and\
  \bibinfo {author} {\bibfnamefont {M.}~\bibnamefont {Nusair}},\ }\bibfield
  {title} {\enquote {\bibinfo {title} {Accurate spin-dependent electron liquid
  correlation energies for local spin density calculations: a critical
  analysis},}\ }\href {\doibase 10.1139/p80-159} {\bibfield  {journal}
  {\bibinfo  {journal} {Canadian Journal of Physics}\ }\textbf {\bibinfo
  {volume} {58}},\ \bibinfo {pages} {1200--1211} (\bibinfo {year} {1980})},\
  \Eprint {http://arxiv.org/abs/https://doi.org/10.1139/p80-159}
  {https://doi.org/10.1139/p80-159} \BibitemShut {NoStop}%
\bibitem [{\citenamefont {Gori-Giorgi}\ \emph {et~al.}(2000)\citenamefont
  {Gori-Giorgi}, \citenamefont {Sacchetti},\ and\ \citenamefont
  {Bachelet}}]{Gori_Giorgi_PRB_2000}%
  \BibitemOpen
  \bibfield  {author} {\bibinfo {author} {\bibfnamefont {Paola}\ \bibnamefont
  {Gori-Giorgi}}, \bibinfo {author} {\bibfnamefont {Francesco}\ \bibnamefont
  {Sacchetti}}, \ and\ \bibinfo {author} {\bibfnamefont {Giovanni~B.}\
  \bibnamefont {Bachelet}},\ }\bibfield  {title} {\enquote {\bibinfo {title}
  {Analytic static structure factors and pair-correlation functions for the
  unpolarized homogeneous electron gas},}\ }\href {\doibase
  10.1103/PhysRevB.61.7353} {\bibfield  {journal} {\bibinfo  {journal} {Phys.
  Rev. B}\ }\textbf {\bibinfo {volume} {61}},\ \bibinfo {pages} {7353--7363}
  (\bibinfo {year} {2000})}\BibitemShut {NoStop}%
\bibitem [{\citenamefont {Corradini}\ \emph {et~al.}(1998)\citenamefont
  {Corradini}, \citenamefont {Sole}, \citenamefont {Onida},\ and\ \citenamefont
  {Palummo}}]{cdop}%
  \BibitemOpen
  \bibfield  {author} {\bibinfo {author} {\bibfnamefont {M.}~\bibnamefont
  {Corradini}}, \bibinfo {author} {\bibfnamefont {R.~Del}\ \bibnamefont
  {Sole}}, \bibinfo {author} {\bibfnamefont {G.}~\bibnamefont {Onida}}, \ and\
  \bibinfo {author} {\bibfnamefont {M.}~\bibnamefont {Palummo}},\ }\bibfield
  {title} {\enquote {\bibinfo {title} {Analytical expressions for the
  local-field factor $g(q)$ and the exchange-correlation kernel
  ${K}_{\mathrm{xc}}(r)$ of the homogeneous electron gas},}\ }\href
  {http://link.aps.org/doi/10.1103/PhysRevB.57.14569} {\bibfield  {journal}
  {\bibinfo  {journal} {Phys. Rev. B}\ }\textbf {\bibinfo {volume} {57}},\
  \bibinfo {pages} {14569} (\bibinfo {year} {1998})}\BibitemShut {NoStop}%
\bibitem [{\citenamefont {Takada}(2016)}]{Takada_PRB_2016}%
  \BibitemOpen
  \bibfield  {author} {\bibinfo {author} {\bibfnamefont {Yasutami}\
  \bibnamefont {Takada}},\ }\bibfield  {title} {\enquote {\bibinfo {title}
  {Emergence of an excitonic collective mode in the dilute electron gas},}\
  }\href {\doibase 10.1103/PhysRevB.94.245106} {\bibfield  {journal} {\bibinfo
  {journal} {Phys. Rev. B}\ }\textbf {\bibinfo {volume} {94}},\ \bibinfo
  {pages} {245106} (\bibinfo {year} {2016})}\BibitemShut {NoStop}%
\bibitem [{\citenamefont {Perdew}\ \emph {et~al.}(1996)\citenamefont {Perdew},
  \citenamefont {Burke},\ and\ \citenamefont {Ernzerhof}}]{PBE_1996}%
  \BibitemOpen
  \bibfield  {author} {\bibinfo {author} {\bibfnamefont {John~P.}\ \bibnamefont
  {Perdew}}, \bibinfo {author} {\bibfnamefont {Kieron}\ \bibnamefont {Burke}},
  \ and\ \bibinfo {author} {\bibfnamefont {Matthias}\ \bibnamefont
  {Ernzerhof}},\ }\bibfield  {title} {\enquote {\bibinfo {title} {Generalized
  gradient approximation made simple},}\ }\href {\doibase
  10.1103/PhysRevLett.77.3865} {\bibfield  {journal} {\bibinfo  {journal}
  {Phys. Rev. Lett.}\ }\textbf {\bibinfo {volume} {77}},\ \bibinfo {pages}
  {3865--3868} (\bibinfo {year} {1996})}\BibitemShut {NoStop}%
\bibitem [{\citenamefont {Burke}(2012)}]{Burke_Perspective_JCP_2012}%
  \BibitemOpen
  \bibfield  {author} {\bibinfo {author} {\bibfnamefont {Kieron}\ \bibnamefont
  {Burke}},\ }\bibfield  {title} {\enquote {\bibinfo {title} {Perspective on
  density functional theory},}\ }\href {\doibase 10.1063/1.4704546} {\bibfield
  {journal} {\bibinfo  {journal} {The Journal of Chemical Physics}\ }\textbf
  {\bibinfo {volume} {136}},\ \bibinfo {pages} {150901} (\bibinfo {year}
  {2012})},\ \Eprint {http://arxiv.org/abs/https://doi.org/10.1063/1.4704546}
  {https://doi.org/10.1063/1.4704546} \BibitemShut {NoStop}%
\bibitem [{\citenamefont {Jones}(2015)}]{Jones_RevModPhys_2015}%
  \BibitemOpen
  \bibfield  {author} {\bibinfo {author} {\bibfnamefont {R.~O.}\ \bibnamefont
  {Jones}},\ }\bibfield  {title} {\enquote {\bibinfo {title} {Density
  functional theory: Its origins, rise to prominence, and future},}\ }\href
  {\doibase 10.1103/RevModPhys.87.897} {\bibfield  {journal} {\bibinfo
  {journal} {Rev. Mod. Phys.}\ }\textbf {\bibinfo {volume} {87}},\ \bibinfo
  {pages} {897--923} (\bibinfo {year} {2015})}\BibitemShut {NoStop}%
\bibitem [{\citenamefont {Saumon}\ \emph {et~al.}(1992)\citenamefont {Saumon},
  \citenamefont {Hubbard}, \citenamefont {Chabrier},\ and\ \citenamefont {van
  Horn}}]{saumon1}%
  \BibitemOpen
  \bibfield  {author} {\bibinfo {author} {\bibfnamefont {D.}~\bibnamefont
  {Saumon}}, \bibinfo {author} {\bibfnamefont {W.~B.}\ \bibnamefont {Hubbard}},
  \bibinfo {author} {\bibfnamefont {G.}~\bibnamefont {Chabrier}}, \ and\
  \bibinfo {author} {\bibfnamefont {H.~M.}\ \bibnamefont {van Horn}},\
  }\bibfield  {title} {\enquote {\bibinfo {title} {The role of the
  molecular-metallic transition of hydrogen in the evolution of jupiter,
  saturn, and brown dwarfs},}\ }\href
  {http://adsabs.harvard.edu/full/1992ApJ...391..827S} {\bibfield  {journal}
  {\bibinfo  {journal} {Astrophys. J}\ }\textbf {\bibinfo {volume} {391}},\
  \bibinfo {pages} {827--831} (\bibinfo {year} {1992})}\BibitemShut {NoStop}%
\bibitem [{\citenamefont {Militzer}\ \emph {et~al.}(2008)\citenamefont
  {Militzer}, \citenamefont {Hubbard}, \citenamefont {Vorberger}, \citenamefont
  {Tamblyn},\ and\ \citenamefont {Bonev}}]{Militzer_2008}%
  \BibitemOpen
  \bibfield  {author} {\bibinfo {author} {\bibfnamefont {B.}~\bibnamefont
  {Militzer}}, \bibinfo {author} {\bibfnamefont {W.~B.}\ \bibnamefont
  {Hubbard}}, \bibinfo {author} {\bibfnamefont {J.}~\bibnamefont {Vorberger}},
  \bibinfo {author} {\bibfnamefont {I.}~\bibnamefont {Tamblyn}}, \ and\
  \bibinfo {author} {\bibfnamefont {S.~A.}\ \bibnamefont {Bonev}},\ }\bibfield
  {title} {\enquote {\bibinfo {title} {A massive core in jupiter predicted from
  first-principles simulations},}\ }\href {\doibase 10.1086/594364} {\bibfield
  {journal} {\bibinfo  {journal} {The Astrophysical Journal}\ }\textbf
  {\bibinfo {volume} {688}},\ \bibinfo {pages} {L45--L48} (\bibinfo {year}
  {2008})}\BibitemShut {NoStop}%
\bibitem [{\citenamefont {Guillot}\ \emph {et~al.}(2018)\citenamefont
  {Guillot}, \citenamefont {Miguel}, \citenamefont {Militzer}, \citenamefont
  {Hubbard}, \citenamefont {Kaspi}, \citenamefont {Galanti}, \citenamefont
  {Cao}, \citenamefont {Helled}, \citenamefont {Wahl}, \citenamefont {Iess},
  \citenamefont {Folkner}, \citenamefont {Stevenson}, \citenamefont {Lunine},
  \citenamefont {Reese}, \citenamefont {Biekman}, \citenamefont {Parisi},
  \citenamefont {Durante}, \citenamefont {Connerney}, \citenamefont {Levin},\
  and\ \citenamefont {Bolton}}]{Guillot2018}%
  \BibitemOpen
  \bibfield  {author} {\bibinfo {author} {\bibfnamefont {T.}~\bibnamefont
  {Guillot}}, \bibinfo {author} {\bibfnamefont {Y.}~\bibnamefont {Miguel}},
  \bibinfo {author} {\bibfnamefont {B.}~\bibnamefont {Militzer}}, \bibinfo
  {author} {\bibfnamefont {W.~B.}\ \bibnamefont {Hubbard}}, \bibinfo {author}
  {\bibfnamefont {Y.}~\bibnamefont {Kaspi}}, \bibinfo {author} {\bibfnamefont
  {E.}~\bibnamefont {Galanti}}, \bibinfo {author} {\bibfnamefont
  {H.}~\bibnamefont {Cao}}, \bibinfo {author} {\bibfnamefont {R.}~\bibnamefont
  {Helled}}, \bibinfo {author} {\bibfnamefont {S.~M.}\ \bibnamefont {Wahl}},
  \bibinfo {author} {\bibfnamefont {L.}~\bibnamefont {Iess}}, \bibinfo {author}
  {\bibfnamefont {W.~M.}\ \bibnamefont {Folkner}}, \bibinfo {author}
  {\bibfnamefont {D.~J.}\ \bibnamefont {Stevenson}}, \bibinfo {author}
  {\bibfnamefont {J.~I.}\ \bibnamefont {Lunine}}, \bibinfo {author}
  {\bibfnamefont {D.~R.}\ \bibnamefont {Reese}}, \bibinfo {author}
  {\bibfnamefont {A.}~\bibnamefont {Biekman}}, \bibinfo {author} {\bibfnamefont
  {M.}~\bibnamefont {Parisi}}, \bibinfo {author} {\bibfnamefont
  {D.}~\bibnamefont {Durante}}, \bibinfo {author} {\bibfnamefont {J.~E.~P.}\
  \bibnamefont {Connerney}}, \bibinfo {author} {\bibfnamefont {S.~M.}\
  \bibnamefont {Levin}}, \ and\ \bibinfo {author} {\bibfnamefont {S.~J.}\
  \bibnamefont {Bolton}},\ }\bibfield  {title} {\enquote {\bibinfo {title} {A
  suppression of differential rotation in jupiter's deep interior},}\ }\href
  {\doibase 10.1038/nature25775} {\bibfield  {journal} {\bibinfo  {journal}
  {Nature}\ }\textbf {\bibinfo {volume} {555}},\ \bibinfo {pages} {227--230}
  (\bibinfo {year} {2018})}\BibitemShut {NoStop}%
\bibitem [{\citenamefont {Becker}\ \emph {et~al.}(2014)\citenamefont {Becker},
  \citenamefont {Lorenzen}, \citenamefont {Fortney}, \citenamefont
  {Nettelmann}, \citenamefont {Sch\"ottler},\ and\ \citenamefont
  {Redmer}}]{becker}%
  \BibitemOpen
  \bibfield  {author} {\bibinfo {author} {\bibfnamefont {A.}~\bibnamefont
  {Becker}}, \bibinfo {author} {\bibfnamefont {W.}~\bibnamefont {Lorenzen}},
  \bibinfo {author} {\bibfnamefont {J.~J.}\ \bibnamefont {Fortney}}, \bibinfo
  {author} {\bibfnamefont {N.}~\bibnamefont {Nettelmann}}, \bibinfo {author}
  {\bibfnamefont {M.}~\bibnamefont {Sch\"ottler}}, \ and\ \bibinfo {author}
  {\bibfnamefont {R.}~\bibnamefont {Redmer}},\ }\bibfield  {title} {\enquote
  {\bibinfo {title} {Ab initio equations of state for hydrogen (h-reos.3) and
  helium (he-reos.3) and their implications for the interior of brown
  dwarfs},}\ }\href
  {https://iopscience.iop.org/article/10.1088/0067-0049/215/2/21/meta}
  {\bibfield  {journal} {\bibinfo  {journal} {Astrophys. J. Suppl. Ser}\
  }\textbf {\bibinfo {volume} {215}},\ \bibinfo {pages} {21} (\bibinfo {year}
  {2014})}\BibitemShut {NoStop}%
\bibitem [{\citenamefont {Daligault}\ and\ \citenamefont
  {Gupta}(2009)}]{Daligault_2009}%
  \BibitemOpen
  \bibfield  {author} {\bibinfo {author} {\bibfnamefont {J.}~\bibnamefont
  {Daligault}}\ and\ \bibinfo {author} {\bibfnamefont {S.}~\bibnamefont
  {Gupta}},\ }\bibfield  {title} {\enquote {\bibinfo {title} {Electron-ion
  scattering in dense multi-component plasmas: application to the outer crust
  of an accreting star},}\ }\href {\doibase 10.1088/0004-637x/703/1/994}
  {\bibfield  {journal} {\bibinfo  {journal} {The Astrophysical Journal}\
  }\textbf {\bibinfo {volume} {703}},\ \bibinfo {pages} {994--1011} (\bibinfo
  {year} {2009})}\BibitemShut {NoStop}%
\bibitem [{\citenamefont {Hu}\ \emph {et~al.}(2011)\citenamefont {Hu},
  \citenamefont {Militzer}, \citenamefont {Goncharov},\ and\ \citenamefont
  {Skupsky}}]{hu_ICF}%
  \BibitemOpen
  \bibfield  {author} {\bibinfo {author} {\bibfnamefont {S.~X.}\ \bibnamefont
  {Hu}}, \bibinfo {author} {\bibfnamefont {B.}~\bibnamefont {Militzer}},
  \bibinfo {author} {\bibfnamefont {V.~N.}\ \bibnamefont {Goncharov}}, \ and\
  \bibinfo {author} {\bibfnamefont {S.}~\bibnamefont {Skupsky}},\ }\bibfield
  {title} {\enquote {\bibinfo {title} {First-principles equation-of-state table
  of deuterium for inertial confinement fusion applications},}\ }\href
  {https://journals.aps.org/prb/abstract/10.1103/PhysRevB.84.224109} {\bibfield
   {journal} {\bibinfo  {journal} {Phys. Rev. B}\ }\textbf {\bibinfo {volume}
  {84}},\ \bibinfo {pages} {224109} (\bibinfo {year} {2011})}\BibitemShut
  {NoStop}%
\bibitem [{\citenamefont {Brongersma}\ \emph {et~al.}(2015)\citenamefont
  {Brongersma}, \citenamefont {Halas},\ and\ \citenamefont
  {Nordlander}}]{Brongersma2015}%
  \BibitemOpen
  \bibfield  {author} {\bibinfo {author} {\bibfnamefont {Mark~L.}\ \bibnamefont
  {Brongersma}}, \bibinfo {author} {\bibfnamefont {Naomi~J.}\ \bibnamefont
  {Halas}}, \ and\ \bibinfo {author} {\bibfnamefont {Peter}\ \bibnamefont
  {Nordlander}},\ }\bibfield  {title} {\enquote {\bibinfo {title}
  {Plasmon-induced hot carrier science and technology},}\ }\href {\doibase
  10.1038/nnano.2014.311} {\bibfield  {journal} {\bibinfo  {journal} {Nature
  Nanotechnology}\ }\textbf {\bibinfo {volume} {10}},\ \bibinfo {pages}
  {25--34} (\bibinfo {year} {2015})}\BibitemShut {NoStop}%
\bibitem [{\citenamefont {Mukherjee}\ \emph {et~al.}(2013)\citenamefont
  {Mukherjee}, \citenamefont {Libisch}, \citenamefont {Large}, \citenamefont
  {Neumann}, \citenamefont {Brown}, \citenamefont {Cheng}, \citenamefont
  {Lassiter}, \citenamefont {Carter}, \citenamefont {Nordlander},\ and\
  \citenamefont {Halas}}]{Mukherjee2013}%
  \BibitemOpen
  \bibfield  {author} {\bibinfo {author} {\bibfnamefont {Shaunak}\ \bibnamefont
  {Mukherjee}}, \bibinfo {author} {\bibfnamefont {Florian}\ \bibnamefont
  {Libisch}}, \bibinfo {author} {\bibfnamefont {Nicolas}\ \bibnamefont
  {Large}}, \bibinfo {author} {\bibfnamefont {Oara}\ \bibnamefont {Neumann}},
  \bibinfo {author} {\bibfnamefont {Lisa~V.}\ \bibnamefont {Brown}}, \bibinfo
  {author} {\bibfnamefont {Jin}\ \bibnamefont {Cheng}}, \bibinfo {author}
  {\bibfnamefont {J.~Britt}\ \bibnamefont {Lassiter}}, \bibinfo {author}
  {\bibfnamefont {Emily~A.}\ \bibnamefont {Carter}}, \bibinfo {author}
  {\bibfnamefont {Peter}\ \bibnamefont {Nordlander}}, \ and\ \bibinfo {author}
  {\bibfnamefont {Naomi~J.}\ \bibnamefont {Halas}},\ }\bibfield  {title}
  {\enquote {\bibinfo {title} {Hot electrons do the impossible: Plasmon-induced
  dissociation of h2 on au},}\ }\href {\doibase 10.1021/nl303940z} {\bibfield
  {journal} {\bibinfo  {journal} {Nano Letters}\ }\textbf {\bibinfo {volume}
  {13}},\ \bibinfo {pages} {240--247} (\bibinfo {year} {2013})}\BibitemShut
  {NoStop}%
\bibitem [{\citenamefont {Falk}(2018)}]{falk_wdm}%
  \BibitemOpen
  \bibfield  {author} {\bibinfo {author} {\bibfnamefont {K.}~\bibnamefont
  {Falk}},\ }\bibfield  {title} {\enquote {\bibinfo {title} {Experimental
  methods for warm dense matter research},}\ }\href
  {https://www.cambridge.org/core/journals/high-power-laser-science-and-engineering/article/experimental-methods-for-warm-dense-matter-research/7205AE1029BEA0061044F84875F1CEDB}
  {\bibfield  {journal} {\bibinfo  {journal} {High Power Laser Sci. Eng}\
  }\textbf {\bibinfo {volume} {6}},\ \bibinfo {pages} {e59} (\bibinfo {year}
  {2018})}\BibitemShut {NoStop}%
\bibitem [{\citenamefont {Kraus}\ \emph {et~al.}(2016)\citenamefont {Kraus},
  \citenamefont {Ravasio}, \citenamefont {Gauthier}, \citenamefont {Gericke},
  \citenamefont {Vorberger}, \citenamefont {Frydrych}, \citenamefont
  {Helfrich}, \citenamefont {Fletcher}, \citenamefont {Schaumann},
  \citenamefont {Nagler}, \citenamefont {Barbrel}, \citenamefont {Bachmann},
  \citenamefont {Gamboa}, \citenamefont {G{\"o}de}, \citenamefont {Granados},
  \citenamefont {Gregori}, \citenamefont {Lee}, \citenamefont {Neumayer},
  \citenamefont {Schumaker}, \citenamefont {D{\"o}ppner}, \citenamefont
  {Falcone}, \citenamefont {Glenzer},\ and\ \citenamefont {Roth}}]{Kraus2016}%
  \BibitemOpen
  \bibfield  {author} {\bibinfo {author} {\bibfnamefont {D.}~\bibnamefont
  {Kraus}}, \bibinfo {author} {\bibfnamefont {A.}~\bibnamefont {Ravasio}},
  \bibinfo {author} {\bibfnamefont {M.}~\bibnamefont {Gauthier}}, \bibinfo
  {author} {\bibfnamefont {D.~O.}\ \bibnamefont {Gericke}}, \bibinfo {author}
  {\bibfnamefont {J.}~\bibnamefont {Vorberger}}, \bibinfo {author}
  {\bibfnamefont {S.}~\bibnamefont {Frydrych}}, \bibinfo {author}
  {\bibfnamefont {J.}~\bibnamefont {Helfrich}}, \bibinfo {author}
  {\bibfnamefont {L.~B.}\ \bibnamefont {Fletcher}}, \bibinfo {author}
  {\bibfnamefont {G.}~\bibnamefont {Schaumann}}, \bibinfo {author}
  {\bibfnamefont {B.}~\bibnamefont {Nagler}}, \bibinfo {author} {\bibfnamefont
  {B.}~\bibnamefont {Barbrel}}, \bibinfo {author} {\bibfnamefont
  {B.}~\bibnamefont {Bachmann}}, \bibinfo {author} {\bibfnamefont {E.~J.}\
  \bibnamefont {Gamboa}}, \bibinfo {author} {\bibfnamefont {S.}~\bibnamefont
  {G{\"o}de}}, \bibinfo {author} {\bibfnamefont {E.}~\bibnamefont {Granados}},
  \bibinfo {author} {\bibfnamefont {G.}~\bibnamefont {Gregori}}, \bibinfo
  {author} {\bibfnamefont {H.~J.}\ \bibnamefont {Lee}}, \bibinfo {author}
  {\bibfnamefont {P.}~\bibnamefont {Neumayer}}, \bibinfo {author}
  {\bibfnamefont {W.}~\bibnamefont {Schumaker}}, \bibinfo {author}
  {\bibfnamefont {T.}~\bibnamefont {D{\"o}ppner}}, \bibinfo {author}
  {\bibfnamefont {R.~W.}\ \bibnamefont {Falcone}}, \bibinfo {author}
  {\bibfnamefont {S.~H.}\ \bibnamefont {Glenzer}}, \ and\ \bibinfo {author}
  {\bibfnamefont {M.}~\bibnamefont {Roth}},\ }\bibfield  {title} {\enquote
  {\bibinfo {title} {Nanosecond formation of diamond and lonsdaleite by shock
  compression of graphite},}\ }\href {\doibase 10.1038/ncomms10970} {\bibfield
  {journal} {\bibinfo  {journal} {Nature Communications}\ }\textbf {\bibinfo
  {volume} {7}},\ \bibinfo {pages} {10970} (\bibinfo {year}
  {2016})}\BibitemShut {NoStop}%
\bibitem [{\citenamefont {Kraus}\ \emph {et~al.}(2017)\citenamefont {Kraus},
  \citenamefont {Vorberger}, \citenamefont {Pak}, \citenamefont {Hartley},
  \citenamefont {Fletcher}, \citenamefont {Frydrych}, \citenamefont {Galtier},
  \citenamefont {Gamboa}, \citenamefont {Gericke}, \citenamefont {Glenzer},
  \citenamefont {Granados}, \citenamefont {MacDonald}, \citenamefont
  {MacKinnon}, \citenamefont {McBride}, \citenamefont {Nam}, \citenamefont
  {Neumayer}, \citenamefont {Roth}, \citenamefont {Saunders}, \citenamefont
  {Schuster}, \citenamefont {Sun}, \citenamefont {van Driel}, \citenamefont
  {D{\"o}ppner},\ and\ \citenamefont {Falcone}}]{Kraus2017}%
  \BibitemOpen
  \bibfield  {author} {\bibinfo {author} {\bibfnamefont {D.}~\bibnamefont
  {Kraus}}, \bibinfo {author} {\bibfnamefont {J.}~\bibnamefont {Vorberger}},
  \bibinfo {author} {\bibfnamefont {A.}~\bibnamefont {Pak}}, \bibinfo {author}
  {\bibfnamefont {N.~J.}\ \bibnamefont {Hartley}}, \bibinfo {author}
  {\bibfnamefont {L.~B.}\ \bibnamefont {Fletcher}}, \bibinfo {author}
  {\bibfnamefont {S.}~\bibnamefont {Frydrych}}, \bibinfo {author}
  {\bibfnamefont {E.}~\bibnamefont {Galtier}}, \bibinfo {author} {\bibfnamefont
  {E.~J.}\ \bibnamefont {Gamboa}}, \bibinfo {author} {\bibfnamefont {D.~O.}\
  \bibnamefont {Gericke}}, \bibinfo {author} {\bibfnamefont {S.~H.}\
  \bibnamefont {Glenzer}}, \bibinfo {author} {\bibfnamefont {E.}~\bibnamefont
  {Granados}}, \bibinfo {author} {\bibfnamefont {M.~J.}\ \bibnamefont
  {MacDonald}}, \bibinfo {author} {\bibfnamefont {A.~J.}\ \bibnamefont
  {MacKinnon}}, \bibinfo {author} {\bibfnamefont {E.~E.}\ \bibnamefont
  {McBride}}, \bibinfo {author} {\bibfnamefont {I.}~\bibnamefont {Nam}},
  \bibinfo {author} {\bibfnamefont {P.}~\bibnamefont {Neumayer}}, \bibinfo
  {author} {\bibfnamefont {M.}~\bibnamefont {Roth}}, \bibinfo {author}
  {\bibfnamefont {A.~M.}\ \bibnamefont {Saunders}}, \bibinfo {author}
  {\bibfnamefont {A.~K.}\ \bibnamefont {Schuster}}, \bibinfo {author}
  {\bibfnamefont {P.}~\bibnamefont {Sun}}, \bibinfo {author} {\bibfnamefont
  {T.}~\bibnamefont {van Driel}}, \bibinfo {author} {\bibfnamefont
  {T.}~\bibnamefont {D{\"o}ppner}}, \ and\ \bibinfo {author} {\bibfnamefont
  {R.~W.}\ \bibnamefont {Falcone}},\ }\bibfield  {title} {\enquote {\bibinfo
  {title} {Formation of diamonds in laser-compressed hydrocarbons at planetary
  interior conditions},}\ }\href {\doibase 10.1038/s41550-017-0219-9}
  {\bibfield  {journal} {\bibinfo  {journal} {Nature Astronomy}\ }\textbf
  {\bibinfo {volume} {1}},\ \bibinfo {pages} {606--611} (\bibinfo {year}
  {2017})}\BibitemShut {NoStop}%
\bibitem [{\citenamefont {Sperling}\ \emph {et~al.}(2015)\citenamefont
  {Sperling}, \citenamefont {Gamboa}, \citenamefont {Lee}, \citenamefont
  {Chung}, \citenamefont {Galtier}, \citenamefont {Omarbakiyeva}, \citenamefont
  {Reinholz}, \citenamefont {R\"opke}, \citenamefont {Zastrau}, \citenamefont
  {Hastings}, \citenamefont {Fletcher},\ and\ \citenamefont
  {Glenzer}}]{Sperling_PRL_2015}%
  \BibitemOpen
  \bibfield  {author} {\bibinfo {author} {\bibfnamefont {P.}~\bibnamefont
  {Sperling}}, \bibinfo {author} {\bibfnamefont {E.~J.}\ \bibnamefont
  {Gamboa}}, \bibinfo {author} {\bibfnamefont {H.~J.}\ \bibnamefont {Lee}},
  \bibinfo {author} {\bibfnamefont {H.~K.}\ \bibnamefont {Chung}}, \bibinfo
  {author} {\bibfnamefont {E.}~\bibnamefont {Galtier}}, \bibinfo {author}
  {\bibfnamefont {Y.}~\bibnamefont {Omarbakiyeva}}, \bibinfo {author}
  {\bibfnamefont {H.}~\bibnamefont {Reinholz}}, \bibinfo {author}
  {\bibfnamefont {G.}~\bibnamefont {R\"opke}}, \bibinfo {author} {\bibfnamefont
  {U.}~\bibnamefont {Zastrau}}, \bibinfo {author} {\bibfnamefont
  {J.}~\bibnamefont {Hastings}}, \bibinfo {author} {\bibfnamefont {L.~B.}\
  \bibnamefont {Fletcher}}, \ and\ \bibinfo {author} {\bibfnamefont {S.~H.}\
  \bibnamefont {Glenzer}},\ }\bibfield  {title} {\enquote {\bibinfo {title}
  {Free-electron x-ray laser measurements of collisional-damped plasmons in
  isochorically heated warm dense matter},}\ }\href {\doibase
  10.1103/PhysRevLett.115.115001} {\bibfield  {journal} {\bibinfo  {journal}
  {Phys. Rev. Lett.}\ }\textbf {\bibinfo {volume} {115}},\ \bibinfo {pages}
  {115001} (\bibinfo {year} {2015})}\BibitemShut {NoStop}%
\bibitem [{\citenamefont {Graziani}\ \emph {et~al.}(2014)\citenamefont
  {Graziani}, \citenamefont {Desjarlais}, \citenamefont {Redmer},\ and\
  \citenamefont {Trickey}}]{wdm_book}%
  \BibitemOpen
  \bibinfo {editor} {\bibfnamefont {F.}~\bibnamefont {Graziani}}, \bibinfo
  {editor} {\bibfnamefont {M.~P.}\ \bibnamefont {Desjarlais}}, \bibinfo
  {editor} {\bibfnamefont {R.}~\bibnamefont {Redmer}}, \ and\ \bibinfo {editor}
  {\bibfnamefont {S.~B.}\ \bibnamefont {Trickey}},\ eds.,\ \href@noop {} {\emph
  {\bibinfo {title} {Frontiers and Challenges in Warm Dense Matter}}}\
  (\bibinfo  {publisher} {Springer},\ \bibinfo {address} {International
  Publishing},\ \bibinfo {year} {2014})\BibitemShut {NoStop}%
\bibitem [{\citenamefont {Bonitz}\ \emph {et~al.}(2020)\citenamefont {Bonitz},
  \citenamefont {Dornheim}, \citenamefont {Moldabekov}, \citenamefont {Zhang},
  \citenamefont {Hamann}, \citenamefont {Kählert}, \citenamefont {Filinov},
  \citenamefont {Ramakrishna},\ and\ \citenamefont {Vorberger}}]{new_POP}%
  \BibitemOpen
  \bibfield  {author} {\bibinfo {author} {\bibfnamefont {M.}~\bibnamefont
  {Bonitz}}, \bibinfo {author} {\bibfnamefont {T.}~\bibnamefont {Dornheim}},
  \bibinfo {author} {\bibfnamefont {Zh.~A.}\ \bibnamefont {Moldabekov}},
  \bibinfo {author} {\bibfnamefont {S.}~\bibnamefont {Zhang}}, \bibinfo
  {author} {\bibfnamefont {P.}~\bibnamefont {Hamann}}, \bibinfo {author}
  {\bibfnamefont {H.}~\bibnamefont {Kählert}}, \bibinfo {author}
  {\bibfnamefont {A.}~\bibnamefont {Filinov}}, \bibinfo {author} {\bibfnamefont
  {K.}~\bibnamefont {Ramakrishna}}, \ and\ \bibinfo {author} {\bibfnamefont
  {J.}~\bibnamefont {Vorberger}},\ }\bibfield  {title} {\enquote {\bibinfo
  {title} {Ab initio simulation of warm dense matter},}\ }\href {\doibase
  10.1063/1.5143225} {\bibfield  {journal} {\bibinfo  {journal} {Physics of
  Plasmas}\ }\textbf {\bibinfo {volume} {27}},\ \bibinfo {pages} {042710}
  (\bibinfo {year} {2020})},\ \Eprint
  {http://arxiv.org/abs/https://doi.org/10.1063/1.5143225}
  {https://doi.org/10.1063/1.5143225} \BibitemShut {NoStop}%
\bibitem [{\citenamefont {Ott}\ \emph {et~al.}(2018)\citenamefont {Ott},
  \citenamefont {Thomsen}, \citenamefont {Abraham}, \citenamefont {Dornheim},\
  and\ \citenamefont {Bonitz}}]{Ott2018}%
  \BibitemOpen
  \bibfield  {author} {\bibinfo {author} {\bibfnamefont {Torben}\ \bibnamefont
  {Ott}}, \bibinfo {author} {\bibfnamefont {Hauke}\ \bibnamefont {Thomsen}},
  \bibinfo {author} {\bibfnamefont {Jan~Willem}\ \bibnamefont {Abraham}},
  \bibinfo {author} {\bibfnamefont {Tobias}\ \bibnamefont {Dornheim}}, \ and\
  \bibinfo {author} {\bibfnamefont {Michael}\ \bibnamefont {Bonitz}},\
  }\bibfield  {title} {\enquote {\bibinfo {title} {Recent progress in the
  theory and simulation of strongly correlated plasmas: phase transitions,
  transport, quantum, and magnetic field effects},}\ }\href {\doibase
  10.1140/epjd/e2018-80385-7} {\bibfield  {journal} {\bibinfo  {journal} {The
  European Physical Journal D}\ }\textbf {\bibinfo {volume} {72}},\ \bibinfo
  {pages} {84} (\bibinfo {year} {2018})}\BibitemShut {NoStop}%
\bibitem [{\citenamefont {Mermin}(1965)}]{Mermin_DFT_1965}%
  \BibitemOpen
  \bibfield  {author} {\bibinfo {author} {\bibfnamefont {N.~David}\
  \bibnamefont {Mermin}},\ }\bibfield  {title} {\enquote {\bibinfo {title}
  {Thermal properties of the inhomogeneous electron gas},}\ }\href {\doibase
  10.1103/PhysRev.137.A1441} {\bibfield  {journal} {\bibinfo  {journal} {Phys.
  Rev.}\ }\textbf {\bibinfo {volume} {137}},\ \bibinfo {pages} {A1441--A1443}
  (\bibinfo {year} {1965})}\BibitemShut {NoStop}%
\bibitem [{\citenamefont {Ramakrishna}\ \emph
  {et~al.}(2020{\natexlab{a}})\citenamefont {Ramakrishna}, \citenamefont
  {Dornheim},\ and\ \citenamefont {Vorberger}}]{kushal}%
  \BibitemOpen
  \bibfield  {author} {\bibinfo {author} {\bibfnamefont {Kushal}\ \bibnamefont
  {Ramakrishna}}, \bibinfo {author} {\bibfnamefont {Tobias}\ \bibnamefont
  {Dornheim}}, \ and\ \bibinfo {author} {\bibfnamefont {Jan}\ \bibnamefont
  {Vorberger}},\ }\bibfield  {title} {\enquote {\bibinfo {title} {Influence of
  finite temperature exchange-correlation effects in hydrogen},}\ }\href
  {\doibase 10.1103/PhysRevB.101.195129} {\bibfield  {journal} {\bibinfo
  {journal} {Phys. Rev. B}\ }\textbf {\bibinfo {volume} {101}},\ \bibinfo
  {pages} {195129} (\bibinfo {year} {2020}{\natexlab{a}})}\BibitemShut
  {NoStop}%
\bibitem [{\citenamefont {Karasiev}\ \emph {et~al.}(2016)\citenamefont
  {Karasiev}, \citenamefont {Calderin},\ and\ \citenamefont
  {Trickey}}]{karasiev_importance}%
  \BibitemOpen
  \bibfield  {author} {\bibinfo {author} {\bibfnamefont {V.~V.}\ \bibnamefont
  {Karasiev}}, \bibinfo {author} {\bibfnamefont {L.}~\bibnamefont {Calderin}},
  \ and\ \bibinfo {author} {\bibfnamefont {S.~B.}\ \bibnamefont {Trickey}},\
  }\bibfield  {title} {\enquote {\bibinfo {title} {Importance of
  finite-temperature exchange correlation for warm dense matter
  calculations},}\ }\href
  {https://journals.aps.org/pre/abstract/10.1103/PhysRevE.93.063207} {\bibfield
   {journal} {\bibinfo  {journal} {Phys. Rev. E}\ }\textbf {\bibinfo {volume}
  {93}},\ \bibinfo {pages} {063207} (\bibinfo {year} {2016})}\BibitemShut
  {NoStop}%
\bibitem [{\citenamefont
  {Dharma-wardana}(2016)}]{Dharma-wardana_computation_2016}%
  \BibitemOpen
  \bibfield  {author} {\bibinfo {author} {\bibfnamefont {M.~W.~C.}\
  \bibnamefont {Dharma-wardana}},\ }\bibfield  {title} {\enquote {\bibinfo
  {title} {Current issues in finite-t density-functional theory and
  warm-correlated matter †},}\ }\href {\doibase 10.3390/computation4020016}
  {\bibfield  {journal} {\bibinfo  {journal} {Computation}\ }\textbf {\bibinfo
  {volume} {4}} (\bibinfo {year} {2016}),\
  10.3390/computation4020016}\BibitemShut {NoStop}%
\bibitem [{\citenamefont {Sjostrom}\ and\ \citenamefont
  {Daligault}(2014)}]{Sjostrom_PRB_2014}%
  \BibitemOpen
  \bibfield  {author} {\bibinfo {author} {\bibfnamefont {Travis}\ \bibnamefont
  {Sjostrom}}\ and\ \bibinfo {author} {\bibfnamefont {J\'er\^ome}\ \bibnamefont
  {Daligault}},\ }\bibfield  {title} {\enquote {\bibinfo {title} {Gradient
  corrections to the exchange-correlation free energy},}\ }\href {\doibase
  10.1103/PhysRevB.90.155109} {\bibfield  {journal} {\bibinfo  {journal} {Phys.
  Rev. B}\ }\textbf {\bibinfo {volume} {90}},\ \bibinfo {pages} {155109}
  (\bibinfo {year} {2014})}\BibitemShut {NoStop}%
\bibitem [{\citenamefont {Driver}\ and\ \citenamefont
  {Militzer}(2012)}]{Driver_Militzer_PRL_2012}%
  \BibitemOpen
  \bibfield  {author} {\bibinfo {author} {\bibfnamefont {K.~P.}\ \bibnamefont
  {Driver}}\ and\ \bibinfo {author} {\bibfnamefont {B.}~\bibnamefont
  {Militzer}},\ }\bibfield  {title} {\enquote {\bibinfo {title} {All-electron
  path integral monte carlo simulations of warm dense matter: Application to
  water and carbon plasmas},}\ }\href {\doibase 10.1103/PhysRevLett.108.115502}
  {\bibfield  {journal} {\bibinfo  {journal} {Phys. Rev. Lett.}\ }\textbf
  {\bibinfo {volume} {108}},\ \bibinfo {pages} {115502} (\bibinfo {year}
  {2012})}\BibitemShut {NoStop}%
\bibitem [{\citenamefont {Blunt}\ \emph {et~al.}(2014)\citenamefont {Blunt},
  \citenamefont {Rogers}, \citenamefont {Spencer},\ and\ \citenamefont
  {Foulkes}}]{Blunt_PRB_2014}%
  \BibitemOpen
  \bibfield  {author} {\bibinfo {author} {\bibfnamefont {N.~S.}\ \bibnamefont
  {Blunt}}, \bibinfo {author} {\bibfnamefont {T.~W.}\ \bibnamefont {Rogers}},
  \bibinfo {author} {\bibfnamefont {J.~S.}\ \bibnamefont {Spencer}}, \ and\
  \bibinfo {author} {\bibfnamefont {W.~M.~C.}\ \bibnamefont {Foulkes}},\
  }\bibfield  {title} {\enquote {\bibinfo {title} {Density-matrix quantum monte
  carlo method},}\ }\href {\doibase 10.1103/PhysRevB.89.245124} {\bibfield
  {journal} {\bibinfo  {journal} {Phys. Rev. B}\ }\textbf {\bibinfo {volume}
  {89}},\ \bibinfo {pages} {245124} (\bibinfo {year} {2014})}\BibitemShut
  {NoStop}%
\bibitem [{\citenamefont {Dornheim}\ \emph
  {et~al.}(2017{\natexlab{a}})\citenamefont {Dornheim}, \citenamefont {Groth},
  \citenamefont {Malone}, \citenamefont {Schoof}, \citenamefont {Sjostrom},
  \citenamefont {Foulkes},\ and\ \citenamefont {Bonitz}}]{dornheim_POP}%
  \BibitemOpen
  \bibfield  {author} {\bibinfo {author} {\bibfnamefont {Tobias}\ \bibnamefont
  {Dornheim}}, \bibinfo {author} {\bibfnamefont {Simon}\ \bibnamefont {Groth}},
  \bibinfo {author} {\bibfnamefont {Fionn~D.}\ \bibnamefont {Malone}}, \bibinfo
  {author} {\bibfnamefont {Tim}\ \bibnamefont {Schoof}}, \bibinfo {author}
  {\bibfnamefont {Travis}\ \bibnamefont {Sjostrom}}, \bibinfo {author}
  {\bibfnamefont {W.~M.~C.}\ \bibnamefont {Foulkes}}, \ and\ \bibinfo {author}
  {\bibfnamefont {Michael}\ \bibnamefont {Bonitz}},\ }\bibfield  {title}
  {\enquote {\bibinfo {title} {Ab initio quantum monte carlo simulation of the
  warm dense electron gas},}\ }\href {\doibase 10.1063/1.4977920} {\bibfield
  {journal} {\bibinfo  {journal} {Physics of Plasmas}\ }\textbf {\bibinfo
  {volume} {24}},\ \bibinfo {pages} {056303} (\bibinfo {year}
  {2017}{\natexlab{a}})},\ \Eprint
  {http://arxiv.org/abs/https://doi.org/10.1063/1.4977920}
  {https://doi.org/10.1063/1.4977920} \BibitemShut {NoStop}%
\bibitem [{\citenamefont {Brown}\ \emph {et~al.}(2013)\citenamefont {Brown},
  \citenamefont {Clark}, \citenamefont {DuBois},\ and\ \citenamefont
  {Ceperley}}]{Brown_PRL_2013}%
  \BibitemOpen
  \bibfield  {author} {\bibinfo {author} {\bibfnamefont {Ethan~W.}\
  \bibnamefont {Brown}}, \bibinfo {author} {\bibfnamefont {Bryan~K.}\
  \bibnamefont {Clark}}, \bibinfo {author} {\bibfnamefont {Jonathan~L.}\
  \bibnamefont {DuBois}}, \ and\ \bibinfo {author} {\bibfnamefont {David~M.}\
  \bibnamefont {Ceperley}},\ }\bibfield  {title} {\enquote {\bibinfo {title}
  {Path-integral monte carlo simulation of the warm dense homogeneous electron
  gas},}\ }\href {\doibase 10.1103/PhysRevLett.110.146405} {\bibfield
  {journal} {\bibinfo  {journal} {Phys. Rev. Lett.}\ }\textbf {\bibinfo
  {volume} {110}},\ \bibinfo {pages} {146405} (\bibinfo {year}
  {2013})}\BibitemShut {NoStop}%
\bibitem [{\citenamefont {Dornheim}\ \emph {et~al.}(2015)\citenamefont
  {Dornheim}, \citenamefont {Groth}, \citenamefont {Filinov},\ and\
  \citenamefont {Bonitz}}]{Dornheim_NJP_2015}%
  \BibitemOpen
  \bibfield  {author} {\bibinfo {author} {\bibfnamefont {Tobias}\ \bibnamefont
  {Dornheim}}, \bibinfo {author} {\bibfnamefont {Simon}\ \bibnamefont {Groth}},
  \bibinfo {author} {\bibfnamefont {Alexey}\ \bibnamefont {Filinov}}, \ and\
  \bibinfo {author} {\bibfnamefont {Michael}\ \bibnamefont {Bonitz}},\
  }\bibfield  {title} {\enquote {\bibinfo {title} {Permutation blocking path
  integral monte carlo: a highly efficient approach to the simulation of
  strongly degenerate non-ideal fermions},}\ }\href {\doibase
  10.1088/1367-2630/17/7/073017} {\bibfield  {journal} {\bibinfo  {journal}
  {New Journal of Physics}\ }\textbf {\bibinfo {volume} {17}},\ \bibinfo
  {pages} {073017} (\bibinfo {year} {2015})}\BibitemShut {NoStop}%
\bibitem [{\citenamefont {Schoof}\ \emph {et~al.}(2015)\citenamefont {Schoof},
  \citenamefont {Groth}, \citenamefont {Vorberger},\ and\ \citenamefont
  {Bonitz}}]{Schoof_PRL_2015}%
  \BibitemOpen
  \bibfield  {author} {\bibinfo {author} {\bibfnamefont {T.}~\bibnamefont
  {Schoof}}, \bibinfo {author} {\bibfnamefont {S.}~\bibnamefont {Groth}},
  \bibinfo {author} {\bibfnamefont {J.}~\bibnamefont {Vorberger}}, \ and\
  \bibinfo {author} {\bibfnamefont {M.}~\bibnamefont {Bonitz}},\ }\bibfield
  {title} {\enquote {\bibinfo {title} {Ab initio thermodynamic results for the
  degenerate electron gas at finite temperature},}\ }\href {\doibase
  10.1103/PhysRevLett.115.130402} {\bibfield  {journal} {\bibinfo  {journal}
  {Phys. Rev. Lett.}\ }\textbf {\bibinfo {volume} {115}},\ \bibinfo {pages}
  {130402} (\bibinfo {year} {2015})}\BibitemShut {NoStop}%
\bibitem [{\citenamefont {Malone}\ \emph {et~al.}(2015)\citenamefont {Malone},
  \citenamefont {Blunt}, \citenamefont {Shepherd}, \citenamefont {Lee},
  \citenamefont {Spencer},\ and\ \citenamefont {Foulkes}}]{Malone_JCP_2015}%
  \BibitemOpen
  \bibfield  {author} {\bibinfo {author} {\bibfnamefont {Fionn~D.}\
  \bibnamefont {Malone}}, \bibinfo {author} {\bibfnamefont {N.~S.}\
  \bibnamefont {Blunt}}, \bibinfo {author} {\bibfnamefont {James~J.}\
  \bibnamefont {Shepherd}}, \bibinfo {author} {\bibfnamefont {D.~K.~K.}\
  \bibnamefont {Lee}}, \bibinfo {author} {\bibfnamefont {J.~S.}\ \bibnamefont
  {Spencer}}, \ and\ \bibinfo {author} {\bibfnamefont {W.~M.~C.}\ \bibnamefont
  {Foulkes}},\ }\bibfield  {title} {\enquote {\bibinfo {title} {Interaction
  picture density matrix quantum monte carlo},}\ }\href {\doibase
  10.1063/1.4927434} {\bibfield  {journal} {\bibinfo  {journal} {The Journal of
  Chemical Physics}\ }\textbf {\bibinfo {volume} {143}},\ \bibinfo {pages}
  {044116} (\bibinfo {year} {2015})},\ \Eprint
  {http://arxiv.org/abs/https://doi.org/10.1063/1.4927434}
  {https://doi.org/10.1063/1.4927434} \BibitemShut {NoStop}%
\bibitem [{\citenamefont {Militzer}\ and\ \citenamefont
  {Driver}(2015)}]{Militzer_Driver_PRL_2015}%
  \BibitemOpen
  \bibfield  {author} {\bibinfo {author} {\bibfnamefont {Burkhard}\
  \bibnamefont {Militzer}}\ and\ \bibinfo {author} {\bibfnamefont {Kevin~P.}\
  \bibnamefont {Driver}},\ }\bibfield  {title} {\enquote {\bibinfo {title}
  {Development of path integral monte carlo simulations with localized nodal
  surfaces for second-row elements},}\ }\href {\doibase
  10.1103/PhysRevLett.115.176403} {\bibfield  {journal} {\bibinfo  {journal}
  {Phys. Rev. Lett.}\ }\textbf {\bibinfo {volume} {115}},\ \bibinfo {pages}
  {176403} (\bibinfo {year} {2015})}\BibitemShut {NoStop}%
\bibitem [{\citenamefont {Malone}\ \emph {et~al.}(2016)\citenamefont {Malone},
  \citenamefont {Blunt}, \citenamefont {Brown}, \citenamefont {Lee},
  \citenamefont {Spencer}, \citenamefont {Foulkes},\ and\ \citenamefont
  {Shepherd}}]{Malone_PRL_2016}%
  \BibitemOpen
  \bibfield  {author} {\bibinfo {author} {\bibfnamefont {Fionn~D.}\
  \bibnamefont {Malone}}, \bibinfo {author} {\bibfnamefont {N.~S.}\
  \bibnamefont {Blunt}}, \bibinfo {author} {\bibfnamefont {Ethan~W.}\
  \bibnamefont {Brown}}, \bibinfo {author} {\bibfnamefont {D.~K.~K.}\
  \bibnamefont {Lee}}, \bibinfo {author} {\bibfnamefont {J.~S.}\ \bibnamefont
  {Spencer}}, \bibinfo {author} {\bibfnamefont {W.~M.~C.}\ \bibnamefont
  {Foulkes}}, \ and\ \bibinfo {author} {\bibfnamefont {James~J.}\ \bibnamefont
  {Shepherd}},\ }\bibfield  {title} {\enquote {\bibinfo {title} {Accurate
  exchange-correlation energies for the warm dense electron gas},}\ }\href
  {\doibase 10.1103/PhysRevLett.117.115701} {\bibfield  {journal} {\bibinfo
  {journal} {Phys. Rev. Lett.}\ }\textbf {\bibinfo {volume} {117}},\ \bibinfo
  {pages} {115701} (\bibinfo {year} {2016})}\BibitemShut {NoStop}%
\bibitem [{\citenamefont {Dornheim}\ \emph {et~al.}(2016)\citenamefont
  {Dornheim}, \citenamefont {Groth}, \citenamefont {Sjostrom}, \citenamefont
  {Malone}, \citenamefont {Foulkes},\ and\ \citenamefont
  {Bonitz}}]{dornheim_prl}%
  \BibitemOpen
  \bibfield  {author} {\bibinfo {author} {\bibfnamefont {T.}~\bibnamefont
  {Dornheim}}, \bibinfo {author} {\bibfnamefont {S.}~\bibnamefont {Groth}},
  \bibinfo {author} {\bibfnamefont {T.}~\bibnamefont {Sjostrom}}, \bibinfo
  {author} {\bibfnamefont {F.~D.}\ \bibnamefont {Malone}}, \bibinfo {author}
  {\bibfnamefont {W.~M.~C.}\ \bibnamefont {Foulkes}}, \ and\ \bibinfo {author}
  {\bibfnamefont {M.}~\bibnamefont {Bonitz}},\ }\bibfield  {title} {\enquote
  {\bibinfo {title} {Ab initio quantum {M}onte {C}arlo simulation of the warm
  dense electron gas in the thermodynamic limit},}\ }\href
  {http://link.aps.org/doi/10.1103/PhysRevLett.117.156403} {\bibfield
  {journal} {\bibinfo  {journal} {Phys. Rev. Lett.}\ }\textbf {\bibinfo
  {volume} {117}},\ \bibinfo {pages} {156403} (\bibinfo {year}
  {2016})}\BibitemShut {NoStop}%
\bibitem [{\citenamefont {Dornheim}\ \emph
  {et~al.}(2017{\natexlab{b}})\citenamefont {Dornheim}, \citenamefont {Groth},\
  and\ \citenamefont {Bonitz}}]{dornheim_cpp}%
  \BibitemOpen
  \bibfield  {author} {\bibinfo {author} {\bibfnamefont {T.}~\bibnamefont
  {Dornheim}}, \bibinfo {author} {\bibfnamefont {S.}~\bibnamefont {Groth}}, \
  and\ \bibinfo {author} {\bibfnamefont {M.}~\bibnamefont {Bonitz}},\
  }\bibfield  {title} {\enquote {\bibinfo {title} {Ab initio results for the
  static structure factor of the warm dense electron gas},}\ }\href
  {https://onlinelibrary.wiley.com/doi/full/10.1002/ctpp.201700096} {\bibfield
  {journal} {\bibinfo  {journal} {Contrib. Plasma Phys}\ }\textbf {\bibinfo
  {volume} {57}},\ \bibinfo {pages} {468--478} (\bibinfo {year}
  {2017}{\natexlab{b}})}\BibitemShut {NoStop}%
\bibitem [{\citenamefont {Groth}\ \emph
  {et~al.}(2017{\natexlab{a}})\citenamefont {Groth}, \citenamefont {Dornheim},\
  and\ \citenamefont {Bonitz}}]{groth_jcp}%
  \BibitemOpen
  \bibfield  {author} {\bibinfo {author} {\bibfnamefont {S.}~\bibnamefont
  {Groth}}, \bibinfo {author} {\bibfnamefont {T.}~\bibnamefont {Dornheim}}, \
  and\ \bibinfo {author} {\bibfnamefont {M.}~\bibnamefont {Bonitz}},\
  }\bibfield  {title} {\enquote {\bibinfo {title} {Configuration path integral
  {M}onte {C}arlo approach to the static density response of the warm dense
  electron gas},}\ }\href {https://aip.scitation.org/doi/abs/10.1063/1.4999907}
  {\bibfield  {journal} {\bibinfo  {journal} {J. Chem. Phys}\ }\textbf
  {\bibinfo {volume} {147}},\ \bibinfo {pages} {164108} (\bibinfo {year}
  {2017}{\natexlab{a}})}\BibitemShut {NoStop}%
\bibitem [{\citenamefont {Dornheim}\ \emph
  {et~al.}(2017{\natexlab{c}})\citenamefont {Dornheim}, \citenamefont {Groth},
  \citenamefont {Vorberger},\ and\ \citenamefont {Bonitz}}]{dornheim_pre}%
  \BibitemOpen
  \bibfield  {author} {\bibinfo {author} {\bibfnamefont {T.}~\bibnamefont
  {Dornheim}}, \bibinfo {author} {\bibfnamefont {S.}~\bibnamefont {Groth}},
  \bibinfo {author} {\bibfnamefont {J.}~\bibnamefont {Vorberger}}, \ and\
  \bibinfo {author} {\bibfnamefont {M.}~\bibnamefont {Bonitz}},\ }\bibfield
  {title} {\enquote {\bibinfo {title} {Permutation blocking path integral
  {M}onte {C}arlo approach to the static density response of the warm dense
  electron gas},}\ }\href
  {https://journals.aps.org/pre/abstract/10.1103/PhysRevE.96.023203} {\bibfield
   {journal} {\bibinfo  {journal} {Phys. Rev. E}\ }\textbf {\bibinfo {volume}
  {96}},\ \bibinfo {pages} {023203} (\bibinfo {year}
  {2017}{\natexlab{c}})}\BibitemShut {NoStop}%
\bibitem [{\citenamefont {Driver}\ \emph {et~al.}(2018)\citenamefont {Driver},
  \citenamefont {Soubiran},\ and\ \citenamefont {Militzer}}]{Driver_PRE_2018}%
  \BibitemOpen
  \bibfield  {author} {\bibinfo {author} {\bibfnamefont {K.~P.}\ \bibnamefont
  {Driver}}, \bibinfo {author} {\bibfnamefont {F.}~\bibnamefont {Soubiran}}, \
  and\ \bibinfo {author} {\bibfnamefont {B.}~\bibnamefont {Militzer}},\
  }\bibfield  {title} {\enquote {\bibinfo {title} {Path integral monte carlo
  simulations of warm dense aluminum},}\ }\href {\doibase
  10.1103/PhysRevE.97.063207} {\bibfield  {journal} {\bibinfo  {journal} {Phys.
  Rev. E}\ }\textbf {\bibinfo {volume} {97}},\ \bibinfo {pages} {063207}
  (\bibinfo {year} {2018})}\BibitemShut {NoStop}%
\bibitem [{\citenamefont {Dornheim}\ \emph
  {et~al.}(2020{\natexlab{a}})\citenamefont {Dornheim}, \citenamefont
  {Vorberger},\ and\ \citenamefont {Bonitz}}]{Dornheim_PRL_2020}%
  \BibitemOpen
  \bibfield  {author} {\bibinfo {author} {\bibfnamefont {Tobias}\ \bibnamefont
  {Dornheim}}, \bibinfo {author} {\bibfnamefont {Jan}\ \bibnamefont
  {Vorberger}}, \ and\ \bibinfo {author} {\bibfnamefont {Michael}\ \bibnamefont
  {Bonitz}},\ }\bibfield  {title} {\enquote {\bibinfo {title} {Nonlinear
  electronic density response in warm dense matter},}\ }\href {\doibase
  10.1103/PhysRevLett.125.085001} {\bibfield  {journal} {\bibinfo  {journal}
  {Phys. Rev. Lett.}\ }\textbf {\bibinfo {volume} {125}},\ \bibinfo {pages}
  {085001} (\bibinfo {year} {2020}{\natexlab{a}})}\BibitemShut {NoStop}%
\bibitem [{\citenamefont {Dornheim}\ \emph
  {et~al.}(2020{\natexlab{b}})\citenamefont {Dornheim}, \citenamefont
  {Invernizzi}, \citenamefont {Vorberger},\ and\ \citenamefont
  {Hirshberg}}]{Dornheim_JCP_2020}%
  \BibitemOpen
  \bibfield  {author} {\bibinfo {author} {\bibfnamefont {Tobias}\ \bibnamefont
  {Dornheim}}, \bibinfo {author} {\bibfnamefont {Michele}\ \bibnamefont
  {Invernizzi}}, \bibinfo {author} {\bibfnamefont {Jan}\ \bibnamefont
  {Vorberger}}, \ and\ \bibinfo {author} {\bibfnamefont {Barak}\ \bibnamefont
  {Hirshberg}},\ }\bibfield  {title} {\enquote {\bibinfo {title} {Attenuating
  the fermion sign problem in path integral monte carlo simulations using the
  bogoliubov inequality and thermodynamic integration},}\ }\href {\doibase
  10.1063/5.0030760} {\bibfield  {journal} {\bibinfo  {journal} {The Journal of
  Chemical Physics}\ }\textbf {\bibinfo {volume} {153}},\ \bibinfo {pages}
  {234104} (\bibinfo {year} {2020}{\natexlab{b}})},\ \Eprint
  {http://arxiv.org/abs/https://doi.org/10.1063/5.0030760}
  {https://doi.org/10.1063/5.0030760} \BibitemShut {NoStop}%
\bibitem [{\citenamefont {Lee}\ \emph {et~al.}(2020)\citenamefont {Lee},
  \citenamefont {Morales},\ and\ \citenamefont {Malone}}]{lee2020phaseless}%
  \BibitemOpen
  \bibfield  {author} {\bibinfo {author} {\bibfnamefont {Joonho}\ \bibnamefont
  {Lee}}, \bibinfo {author} {\bibfnamefont {Miguel~A.}\ \bibnamefont
  {Morales}}, \ and\ \bibinfo {author} {\bibfnamefont {Fionn~D.}\ \bibnamefont
  {Malone}},\ }\href@noop {} {\enquote {\bibinfo {title} {A phaseless
  auxiliary-field quantum monte carlo perspective on the uniform electron gas
  at finite temperatures: Issues, observations, and benchmark study},}\ }
  (\bibinfo {year} {2020}),\ \Eprint {http://arxiv.org/abs/2012.12228}
  {arXiv:2012.12228 [physics.chem-ph]} \BibitemShut {NoStop}%
\bibitem [{\citenamefont {Liu}\ \emph {et~al.}(2018)\citenamefont {Liu},
  \citenamefont {Cho},\ and\ \citenamefont
  {Rubenstein}}]{Rubenstein_auxiliary_finite_T}%
  \BibitemOpen
  \bibfield  {author} {\bibinfo {author} {\bibfnamefont {Yuan}\ \bibnamefont
  {Liu}}, \bibinfo {author} {\bibfnamefont {Minsik}\ \bibnamefont {Cho}}, \
  and\ \bibinfo {author} {\bibfnamefont {Brenda}\ \bibnamefont {Rubenstein}},\
  }\bibfield  {title} {\enquote {\bibinfo {title} {Ab initio finite temperature
  auxiliary field quantum monte carlo},}\ }\href {\doibase
  10.1021/acs.jctc.8b00569} {\bibfield  {journal} {\bibinfo  {journal} {Journal
  of Chemical Theory and Computation}\ }\textbf {\bibinfo {volume} {14}},\
  \bibinfo {pages} {4722--4732} (\bibinfo {year} {2018})}\BibitemShut {NoStop}%
\bibitem [{\citenamefont {Yilmaz}\ \emph {et~al.}(2020)\citenamefont {Yilmaz},
  \citenamefont {Hunger}, \citenamefont {Dornheim}, \citenamefont {Groth},\
  and\ \citenamefont {Bonitz}}]{Yilmaz_JCP_2020}%
  \BibitemOpen
  \bibfield  {author} {\bibinfo {author} {\bibfnamefont {A.}~\bibnamefont
  {Yilmaz}}, \bibinfo {author} {\bibfnamefont {K.}~\bibnamefont {Hunger}},
  \bibinfo {author} {\bibfnamefont {T.}~\bibnamefont {Dornheim}}, \bibinfo
  {author} {\bibfnamefont {S.}~\bibnamefont {Groth}}, \ and\ \bibinfo {author}
  {\bibfnamefont {M.}~\bibnamefont {Bonitz}},\ }\bibfield  {title} {\enquote
  {\bibinfo {title} {Restricted configuration path integral monte carlo},}\
  }\href {\doibase 10.1063/5.0022800} {\bibfield  {journal} {\bibinfo
  {journal} {The Journal of Chemical Physics}\ }\textbf {\bibinfo {volume}
  {153}},\ \bibinfo {pages} {124114} (\bibinfo {year} {2020})},\ \Eprint
  {http://arxiv.org/abs/https://doi.org/10.1063/5.0022800}
  {https://doi.org/10.1063/5.0022800} \BibitemShut {NoStop}%
\bibitem [{\citenamefont {Groth}\ \emph
  {et~al.}(2017{\natexlab{b}})\citenamefont {Groth}, \citenamefont {Dornheim},
  \citenamefont {Sjostrom}, \citenamefont {Malone}, \citenamefont {Foulkes},\
  and\ \citenamefont {Bonitz}}]{groth_prl}%
  \BibitemOpen
  \bibfield  {author} {\bibinfo {author} {\bibfnamefont {S.}~\bibnamefont
  {Groth}}, \bibinfo {author} {\bibfnamefont {T.}~\bibnamefont {Dornheim}},
  \bibinfo {author} {\bibfnamefont {T.}~\bibnamefont {Sjostrom}}, \bibinfo
  {author} {\bibfnamefont {F.~D.}\ \bibnamefont {Malone}}, \bibinfo {author}
  {\bibfnamefont {W.~M.~C.}\ \bibnamefont {Foulkes}}, \ and\ \bibinfo {author}
  {\bibfnamefont {M.}~\bibnamefont {Bonitz}},\ }\bibfield  {title} {\enquote
  {\bibinfo {title} {Ab initio exchange--correlation free energy of the uniform
  electron gas at warm dense matter conditions},}\ }\href
  {https://journals.aps.org/prl/abstract/10.1103/PhysRevLett.119.135001}
  {\bibfield  {journal} {\bibinfo  {journal} {Phys. Rev. Lett.}\ }\textbf
  {\bibinfo {volume} {119}},\ \bibinfo {pages} {135001} (\bibinfo {year}
  {2017}{\natexlab{b}})}\BibitemShut {NoStop}%
\bibitem [{\citenamefont {Karasiev}\ \emph {et~al.}(2014)\citenamefont
  {Karasiev}, \citenamefont {Sjostrom}, \citenamefont {Dufty},\ and\
  \citenamefont {Trickey}}]{ksdt}%
  \BibitemOpen
  \bibfield  {author} {\bibinfo {author} {\bibfnamefont {Valentin~V.}\
  \bibnamefont {Karasiev}}, \bibinfo {author} {\bibfnamefont {Travis}\
  \bibnamefont {Sjostrom}}, \bibinfo {author} {\bibfnamefont {James}\
  \bibnamefont {Dufty}}, \ and\ \bibinfo {author} {\bibfnamefont {S.~B.}\
  \bibnamefont {Trickey}},\ }\bibfield  {title} {\enquote {\bibinfo {title}
  {Accurate homogeneous electron gas exchange-correlation free energy for local
  spin-density calculations},}\ }\href {\doibase
  10.1103/PhysRevLett.112.076403} {\bibfield  {journal} {\bibinfo  {journal}
  {Phys. Rev. Lett.}\ }\textbf {\bibinfo {volume} {112}},\ \bibinfo {pages}
  {076403} (\bibinfo {year} {2014})}\BibitemShut {NoStop}%
\bibitem [{\citenamefont {White}\ \emph {et~al.}(2013)\citenamefont {White},
  \citenamefont {Richardson}, \citenamefont {Crowley}, \citenamefont
  {Pattison}, \citenamefont {Harris},\ and\ \citenamefont
  {Gregori}}]{White_PRL_2013}%
  \BibitemOpen
  \bibfield  {author} {\bibinfo {author} {\bibfnamefont {T.~G.}\ \bibnamefont
  {White}}, \bibinfo {author} {\bibfnamefont {S.}~\bibnamefont {Richardson}},
  \bibinfo {author} {\bibfnamefont {B.~J.~B.}\ \bibnamefont {Crowley}},
  \bibinfo {author} {\bibfnamefont {L.~K.}\ \bibnamefont {Pattison}}, \bibinfo
  {author} {\bibfnamefont {J.~W.~O.}\ \bibnamefont {Harris}}, \ and\ \bibinfo
  {author} {\bibfnamefont {G.}~\bibnamefont {Gregori}},\ }\bibfield  {title}
  {\enquote {\bibinfo {title} {Orbital-free density-functional theory
  simulations of the dynamic structure factor of warm dense aluminum},}\ }\href
  {\doibase 10.1103/PhysRevLett.111.175002} {\bibfield  {journal} {\bibinfo
  {journal} {Phys. Rev. Lett.}\ }\textbf {\bibinfo {volume} {111}},\ \bibinfo
  {pages} {175002} (\bibinfo {year} {2013})}\BibitemShut {NoStop}%
\bibitem [{\citenamefont {Gao}\ \emph {et~al.}(2016)\citenamefont {Gao},
  \citenamefont {Zhang}, \citenamefont {Kang}, \citenamefont {Wang},
  \citenamefont {Zhang},\ and\ \citenamefont {He}}]{Gao_PRB_2016}%
  \BibitemOpen
  \bibfield  {author} {\bibinfo {author} {\bibfnamefont {Chang}\ \bibnamefont
  {Gao}}, \bibinfo {author} {\bibfnamefont {Shen}\ \bibnamefont {Zhang}},
  \bibinfo {author} {\bibfnamefont {Wei}\ \bibnamefont {Kang}}, \bibinfo
  {author} {\bibfnamefont {Cong}\ \bibnamefont {Wang}}, \bibinfo {author}
  {\bibfnamefont {Ping}\ \bibnamefont {Zhang}}, \ and\ \bibinfo {author}
  {\bibfnamefont {X.~T.}\ \bibnamefont {He}},\ }\bibfield  {title} {\enquote
  {\bibinfo {title} {Validity boundary of orbital-free molecular dynamics
  method corresponding to thermal ionization of shell structure},}\ }\href
  {\doibase 10.1103/PhysRevB.94.205115} {\bibfield  {journal} {\bibinfo
  {journal} {Phys. Rev. B}\ }\textbf {\bibinfo {volume} {94}},\ \bibinfo
  {pages} {205115} (\bibinfo {year} {2016})}\BibitemShut {NoStop}%
\bibitem [{\citenamefont {Zhang}\ \emph {et~al.}(2016)\citenamefont {Zhang},
  \citenamefont {Wang}, \citenamefont {Kang}, \citenamefont {Zhang},\ and\
  \citenamefont {He}}]{Zhang_POP_2016}%
  \BibitemOpen
  \bibfield  {author} {\bibinfo {author} {\bibfnamefont {Shen}\ \bibnamefont
  {Zhang}}, \bibinfo {author} {\bibfnamefont {Hongwei}\ \bibnamefont {Wang}},
  \bibinfo {author} {\bibfnamefont {Wei}\ \bibnamefont {Kang}}, \bibinfo
  {author} {\bibfnamefont {Ping}\ \bibnamefont {Zhang}}, \ and\ \bibinfo
  {author} {\bibfnamefont {X.~T.}\ \bibnamefont {He}},\ }\bibfield  {title}
  {\enquote {\bibinfo {title} {Extended application of kohn-sham
  first-principles molecular dynamics method with plane wave approximation at
  high energy—from cold materials to hot dense plasmas},}\ }\href {\doibase
  10.1063/1.4947212} {\bibfield  {journal} {\bibinfo  {journal} {Physics of
  Plasmas}\ }\textbf {\bibinfo {volume} {23}},\ \bibinfo {pages} {042707}
  (\bibinfo {year} {2016})},\ \Eprint
  {http://arxiv.org/abs/https://doi.org/10.1063/1.4947212}
  {https://doi.org/10.1063/1.4947212} \BibitemShut {NoStop}%
\bibitem [{\citenamefont {Ding}\ \emph {et~al.}(2018)\citenamefont {Ding},
  \citenamefont {White}, \citenamefont {Hu}, \citenamefont {Certik},\ and\
  \citenamefont {Collins}}]{Ding_PRL_2018}%
  \BibitemOpen
  \bibfield  {author} {\bibinfo {author} {\bibfnamefont {Y.~H.}\ \bibnamefont
  {Ding}}, \bibinfo {author} {\bibfnamefont {A.~J.}\ \bibnamefont {White}},
  \bibinfo {author} {\bibfnamefont {S.~X.}\ \bibnamefont {Hu}}, \bibinfo
  {author} {\bibfnamefont {O.}~\bibnamefont {Certik}}, \ and\ \bibinfo {author}
  {\bibfnamefont {L.~A.}\ \bibnamefont {Collins}},\ }\bibfield  {title}
  {\enquote {\bibinfo {title} {Ab initio studies on the stopping power of warm
  dense matter with time-dependent orbital-free density functional theory},}\
  }\href {\doibase 10.1103/PhysRevLett.121.145001} {\bibfield  {journal}
  {\bibinfo  {journal} {Phys. Rev. Lett.}\ }\textbf {\bibinfo {volume} {121}},\
  \bibinfo {pages} {145001} (\bibinfo {year} {2018})}\BibitemShut {NoStop}%
\bibitem [{\citenamefont {Sharma}\ \emph {et~al.}(2020)\citenamefont {Sharma},
  \citenamefont {Hamel}, \citenamefont {Bethkenhagen}, \citenamefont {Pask},\
  and\ \citenamefont {Suryanarayana}}]{Mandy_highT_DFT_JCP_2020}%
  \BibitemOpen
  \bibfield  {author} {\bibinfo {author} {\bibfnamefont {Abhiraj}\ \bibnamefont
  {Sharma}}, \bibinfo {author} {\bibfnamefont {Sebastien}\ \bibnamefont
  {Hamel}}, \bibinfo {author} {\bibfnamefont {Mandy}\ \bibnamefont
  {Bethkenhagen}}, \bibinfo {author} {\bibfnamefont {John~E.}\ \bibnamefont
  {Pask}}, \ and\ \bibinfo {author} {\bibfnamefont {Phanish}\ \bibnamefont
  {Suryanarayana}},\ }\bibfield  {title} {\enquote {\bibinfo {title}
  {Real-space formulation of the stress tensor for o(n) density functional
  theory: Application to high temperature calculations},}\ }\href {\doibase
  10.1063/5.0016783} {\bibfield  {journal} {\bibinfo  {journal} {The Journal of
  Chemical Physics}\ }\textbf {\bibinfo {volume} {153}},\ \bibinfo {pages}
  {034112} (\bibinfo {year} {2020})},\ \Eprint
  {http://arxiv.org/abs/https://doi.org/10.1063/5.0016783}
  {https://doi.org/10.1063/5.0016783} \BibitemShut {NoStop}%
\bibitem [{\citenamefont {Karasiev}\ \emph {et~al.}(2018)\citenamefont
  {Karasiev}, \citenamefont {Dufty},\ and\ \citenamefont
  {Trickey}}]{Karasiev_PRL_2018}%
  \BibitemOpen
  \bibfield  {author} {\bibinfo {author} {\bibfnamefont {Valentin~V.}\
  \bibnamefont {Karasiev}}, \bibinfo {author} {\bibfnamefont {James~W.}\
  \bibnamefont {Dufty}}, \ and\ \bibinfo {author} {\bibfnamefont {S.~B.}\
  \bibnamefont {Trickey}},\ }\bibfield  {title} {\enquote {\bibinfo {title}
  {Nonempirical semilocal free-energy density functional for matter under
  extreme conditions},}\ }\href {\doibase 10.1103/PhysRevLett.120.076401}
  {\bibfield  {journal} {\bibinfo  {journal} {Phys. Rev. Lett.}\ }\textbf
  {\bibinfo {volume} {120}},\ \bibinfo {pages} {076401} (\bibinfo {year}
  {2018})}\BibitemShut {NoStop}%
\bibitem [{\citenamefont {Glenzer}\ and\ \citenamefont
  {Redmer}(2009)}]{siegfried_review}%
  \BibitemOpen
  \bibfield  {author} {\bibinfo {author} {\bibfnamefont {S.~H.}\ \bibnamefont
  {Glenzer}}\ and\ \bibinfo {author} {\bibfnamefont {R.}~\bibnamefont
  {Redmer}},\ }\bibfield  {title} {\enquote {\bibinfo {title} {X-ray thomson
  scattering in high energy density plasmas},}\ }\href
  {https://journals.aps.org/rmp/abstract/10.1103/RevModPhys.81.1625} {\bibfield
   {journal} {\bibinfo  {journal} {Rev. Mod. Phys}\ }\textbf {\bibinfo {volume}
  {81}},\ \bibinfo {pages} {1625} (\bibinfo {year} {2009})}\BibitemShut
  {NoStop}%
\bibitem [{\citenamefont {Kraus}\ \emph {et~al.}(2019)\citenamefont {Kraus},
  \citenamefont {Bachmann}, \citenamefont {Barbrel}, \citenamefont {Falcone},
  \citenamefont {Fletcher}, \citenamefont {Frydrych}, \citenamefont {Gamboa},
  \citenamefont {Gauthier}, \citenamefont {Gericke}, \citenamefont {Glenzer},
  \citenamefont {G\"ode}, \citenamefont {Granados}, \citenamefont {Hartley},
  \citenamefont {Helfrich}, \citenamefont {Lee}, \citenamefont {Nagler},
  \citenamefont {Ravasio}, \citenamefont {Schumaker}, \citenamefont
  {Vorberger},\ and\ \citenamefont {D\"oppner}}]{kraus_xrts}%
  \BibitemOpen
  \bibfield  {author} {\bibinfo {author} {\bibfnamefont {D.}~\bibnamefont
  {Kraus}}, \bibinfo {author} {\bibfnamefont {B.}~\bibnamefont {Bachmann}},
  \bibinfo {author} {\bibfnamefont {B.}~\bibnamefont {Barbrel}}, \bibinfo
  {author} {\bibfnamefont {R.~W.}\ \bibnamefont {Falcone}}, \bibinfo {author}
  {\bibfnamefont {L.~B.}\ \bibnamefont {Fletcher}}, \bibinfo {author}
  {\bibfnamefont {S.}~\bibnamefont {Frydrych}}, \bibinfo {author}
  {\bibfnamefont {E.~J.}\ \bibnamefont {Gamboa}}, \bibinfo {author}
  {\bibfnamefont {M.}~\bibnamefont {Gauthier}}, \bibinfo {author}
  {\bibfnamefont {D.~O.}\ \bibnamefont {Gericke}}, \bibinfo {author}
  {\bibfnamefont {S.~H.}\ \bibnamefont {Glenzer}}, \bibinfo {author}
  {\bibfnamefont {S.}~\bibnamefont {G\"ode}}, \bibinfo {author} {\bibfnamefont
  {E.}~\bibnamefont {Granados}}, \bibinfo {author} {\bibfnamefont {N.~J.}\
  \bibnamefont {Hartley}}, \bibinfo {author} {\bibfnamefont {J.}~\bibnamefont
  {Helfrich}}, \bibinfo {author} {\bibfnamefont {H.~J.}\ \bibnamefont {Lee}},
  \bibinfo {author} {\bibfnamefont {B.}~\bibnamefont {Nagler}}, \bibinfo
  {author} {\bibfnamefont {A.}~\bibnamefont {Ravasio}}, \bibinfo {author}
  {\bibfnamefont {W.}~\bibnamefont {Schumaker}}, \bibinfo {author}
  {\bibfnamefont {J.}~\bibnamefont {Vorberger}}, \ and\ \bibinfo {author}
  {\bibfnamefont {T.}~\bibnamefont {D\"oppner}},\ }\bibfield  {title} {\enquote
  {\bibinfo {title} {Characterizing the ionization potential depression in
  dense carbon plasmas with high-precision spectrally resolved x-ray
  scattering},}\ }\href
  {https://iopscience.iop.org/article/10.1088/1361-6587/aadd6c/meta} {\bibfield
   {journal} {\bibinfo  {journal} {Plasma Phys. Control Fusion}\ }\textbf
  {\bibinfo {volume} {61}},\ \bibinfo {pages} {014015} (\bibinfo {year}
  {2019})}\BibitemShut {NoStop}%
\bibitem [{\citenamefont {Lu}(2014)}]{Lu_JCP_2014}%
  \BibitemOpen
  \bibfield  {author} {\bibinfo {author} {\bibfnamefont {Deyu}\ \bibnamefont
  {Lu}},\ }\bibfield  {title} {\enquote {\bibinfo {title} {Evaluation of model
  exchange-correlation kernels in the adiabatic connection
  fluctuation-dissipation theorem for inhomogeneous systems},}\ }\href
  {\doibase 10.1063/1.4867538} {\bibfield  {journal} {\bibinfo  {journal} {The
  Journal of Chemical Physics}\ }\textbf {\bibinfo {volume} {140}},\ \bibinfo
  {pages} {18A520} (\bibinfo {year} {2014})},\ \Eprint
  {http://arxiv.org/abs/https://doi.org/10.1063/1.4867538}
  {https://doi.org/10.1063/1.4867538} \BibitemShut {NoStop}%
\bibitem [{\citenamefont {Patrick}\ and\ \citenamefont
  {Thygesen}(2015)}]{Thygesen_JCP_2015}%
  \BibitemOpen
  \bibfield  {author} {\bibinfo {author} {\bibfnamefont {Christopher~E.}\
  \bibnamefont {Patrick}}\ and\ \bibinfo {author} {\bibfnamefont {Kristian~S.}\
  \bibnamefont {Thygesen}},\ }\bibfield  {title} {\enquote {\bibinfo {title}
  {Adiabatic-connection fluctuation-dissipation dft for the structural
  properties of solids—the renormalized alda and electron gas kernels},}\
  }\href {\doibase 10.1063/1.4919236} {\bibfield  {journal} {\bibinfo
  {journal} {The Journal of Chemical Physics}\ }\textbf {\bibinfo {volume}
  {143}},\ \bibinfo {pages} {102802} (\bibinfo {year} {2015})},\ \Eprint
  {http://arxiv.org/abs/https://doi.org/10.1063/1.4919236}
  {https://doi.org/10.1063/1.4919236} \BibitemShut {NoStop}%
\bibitem [{\citenamefont {G\"orling}(2019)}]{Goerling_PRB_2019}%
  \BibitemOpen
  \bibfield  {author} {\bibinfo {author} {\bibfnamefont {Andreas}\ \bibnamefont
  {G\"orling}},\ }\bibfield  {title} {\enquote {\bibinfo {title} {Hierarchies
  of methods towards the exact kohn-sham correlation energy based on the
  adiabatic-connection fluctuation-dissipation theorem},}\ }\href {\doibase
  10.1103/PhysRevB.99.235120} {\bibfield  {journal} {\bibinfo  {journal} {Phys.
  Rev. B}\ }\textbf {\bibinfo {volume} {99}},\ \bibinfo {pages} {235120}
  (\bibinfo {year} {2019})}\BibitemShut {NoStop}%
\bibitem [{\citenamefont {Pribram-Jones}\ \emph {et~al.}(2016)\citenamefont
  {Pribram-Jones}, \citenamefont {Grabowski},\ and\ \citenamefont
  {Burke}}]{pribram}%
  \BibitemOpen
  \bibfield  {author} {\bibinfo {author} {\bibfnamefont {A.}~\bibnamefont
  {Pribram-Jones}}, \bibinfo {author} {\bibfnamefont {P.~E.}\ \bibnamefont
  {Grabowski}}, \ and\ \bibinfo {author} {\bibfnamefont {K.}~\bibnamefont
  {Burke}},\ }\bibfield  {title} {\enquote {\bibinfo {title} {Thermal density
  functional theory: Time-dependent linear response and approximate functionals
  from the fluctuation-dissipation theorem},}\ }\href
  {https://journals.aps.org/prl/abstract/10.1103/PhysRevLett.116.233001}
  {\bibfield  {journal} {\bibinfo  {journal} {Phys. Rev. Lett}\ }\textbf
  {\bibinfo {volume} {116}},\ \bibinfo {pages} {233001} (\bibinfo {year}
  {2016})}\BibitemShut {NoStop}%
\bibitem [{\citenamefont {Gross}\ and\ \citenamefont {Kohn}(1985)}]{dynamic1}%
  \BibitemOpen
  \bibfield  {author} {\bibinfo {author} {\bibfnamefont {E.~K.~U.}\
  \bibnamefont {Gross}}\ and\ \bibinfo {author} {\bibfnamefont
  {W.}~\bibnamefont {Kohn}},\ }\bibfield  {title} {\enquote {\bibinfo {title}
  {Local density-functional theory of frequency-dependent linear response},}\
  }\href {https://journals.aps.org/prl/abstract/10.1103/PhysRevLett.55.2850}
  {\bibfield  {journal} {\bibinfo  {journal} {Phys. Rev. Lett}\ }\textbf
  {\bibinfo {volume} {55}},\ \bibinfo {pages} {2850} (\bibinfo {year}
  {1985})}\BibitemShut {NoStop}%
\bibitem [{\citenamefont {Baczewski}\ \emph {et~al.}(2016)\citenamefont
  {Baczewski}, \citenamefont {Shulenburger}, \citenamefont {Desjarlais},
  \citenamefont {Hansen},\ and\ \citenamefont {Magyar}}]{Baczewski_PRL_2016}%
  \BibitemOpen
  \bibfield  {author} {\bibinfo {author} {\bibfnamefont {A.~D.}\ \bibnamefont
  {Baczewski}}, \bibinfo {author} {\bibfnamefont {L.}~\bibnamefont
  {Shulenburger}}, \bibinfo {author} {\bibfnamefont {M.~P.}\ \bibnamefont
  {Desjarlais}}, \bibinfo {author} {\bibfnamefont {S.~B.}\ \bibnamefont
  {Hansen}}, \ and\ \bibinfo {author} {\bibfnamefont {R.~J.}\ \bibnamefont
  {Magyar}},\ }\bibfield  {title} {\enquote {\bibinfo {title} {X-ray thomson
  scattering in warm dense matter without the chihara decomposition},}\ }\href
  {https://journals.aps.org/prl/abstract/10.1103/PhysRevLett.116.115004}
  {\bibfield  {journal} {\bibinfo  {journal} {Phys. Rev. Lett}\ }\textbf
  {\bibinfo {volume} {116}},\ \bibinfo {pages} {115004} (\bibinfo {year}
  {2016})}\BibitemShut {NoStop}%
\bibitem [{\citenamefont {Moldabekov}\ \emph
  {et~al.}(2020{\natexlab{a}})\citenamefont {Moldabekov}, \citenamefont
  {Dornheim}, \citenamefont {Bonitz},\ and\ \citenamefont
  {Ramazanov}}]{Moldabekov_PRE_2020}%
  \BibitemOpen
  \bibfield  {author} {\bibinfo {author} {\bibfnamefont {Zh.~A.}\ \bibnamefont
  {Moldabekov}}, \bibinfo {author} {\bibfnamefont {T.}~\bibnamefont
  {Dornheim}}, \bibinfo {author} {\bibfnamefont {M.}~\bibnamefont {Bonitz}}, \
  and\ \bibinfo {author} {\bibfnamefont {T.~S.}\ \bibnamefont {Ramazanov}},\
  }\bibfield  {title} {\enquote {\bibinfo {title} {Ion energy-loss
  characteristics and friction in a free-electron gas at warm dense matter and
  nonideal dense plasma conditions},}\ }\href {\doibase
  10.1103/PhysRevE.101.053203} {\bibfield  {journal} {\bibinfo  {journal}
  {Phys. Rev. E}\ }\textbf {\bibinfo {volume} {101}},\ \bibinfo {pages}
  {053203} (\bibinfo {year} {2020}{\natexlab{a}})}\BibitemShut {NoStop}%
\bibitem [{\citenamefont {Senatore}\ \emph {et~al.}(1996)\citenamefont
  {Senatore}, \citenamefont {Moroni},\ and\ \citenamefont
  {Ceperley}}]{Ceperley_Potential_1996}%
  \BibitemOpen
  \bibfield  {author} {\bibinfo {author} {\bibfnamefont {G.}~\bibnamefont
  {Senatore}}, \bibinfo {author} {\bibfnamefont {S.}~\bibnamefont {Moroni}}, \
  and\ \bibinfo {author} {\bibfnamefont {D.M.}\ \bibnamefont {Ceperley}},\
  }\bibfield  {title} {\enquote {\bibinfo {title} {Local field factor and
  effective potentials in liquid metals},}\ }\href {\doibase
  https://doi.org/10.1016/S0022-3093(96)00316-X} {\bibfield  {journal}
  {\bibinfo  {journal} {Journal of Non-Crystalline Solids}\ }\textbf {\bibinfo
  {volume} {205-207}},\ \bibinfo {pages} {851 -- 854} (\bibinfo {year}
  {1996})}\BibitemShut {NoStop}%
\bibitem [{\citenamefont {Moldabekov}\ \emph
  {et~al.}(2017{\natexlab{a}})\citenamefont {Moldabekov}, \citenamefont
  {Groth}, \citenamefont {Dornheim}, \citenamefont {Bonitz},\ and\
  \citenamefont {Ramazanov}}]{Moldabekov_CPP_2017}%
  \BibitemOpen
  \bibfield  {author} {\bibinfo {author} {\bibfnamefont {Zh.A.}\ \bibnamefont
  {Moldabekov}}, \bibinfo {author} {\bibfnamefont {S.}~\bibnamefont {Groth}},
  \bibinfo {author} {\bibfnamefont {T.}~\bibnamefont {Dornheim}}, \bibinfo
  {author} {\bibfnamefont {M.}~\bibnamefont {Bonitz}}, \ and\ \bibinfo {author}
  {\bibfnamefont {T.S.}\ \bibnamefont {Ramazanov}},\ }\bibfield  {title}
  {\enquote {\bibinfo {title} {Ion potential in non-ideal dense quantum
  plasmas},}\ }\href {\doibase https://doi.org/10.1002/ctpp.201700109}
  {\bibfield  {journal} {\bibinfo  {journal} {Contributions to Plasma Physics}\
  }\textbf {\bibinfo {volume} {57}},\ \bibinfo {pages} {532--538} (\bibinfo
  {year} {2017}{\natexlab{a}})}\BibitemShut {NoStop}%
\bibitem [{\citenamefont {Moldabekov}\ \emph
  {et~al.}(2018{\natexlab{a}})\citenamefont {Moldabekov}, \citenamefont
  {Groth}, \citenamefont {Dornheim}, \citenamefont {K\"ahlert}, \citenamefont
  {Bonitz},\ and\ \citenamefont {Ramazanov}}]{zhandos1}%
  \BibitemOpen
  \bibfield  {author} {\bibinfo {author} {\bibfnamefont {Zh.A.}\ \bibnamefont
  {Moldabekov}}, \bibinfo {author} {\bibfnamefont {S.}~\bibnamefont {Groth}},
  \bibinfo {author} {\bibfnamefont {T.}~\bibnamefont {Dornheim}}, \bibinfo
  {author} {\bibfnamefont {H.}~\bibnamefont {K\"ahlert}}, \bibinfo {author}
  {\bibfnamefont {M.}~\bibnamefont {Bonitz}}, \ and\ \bibinfo {author}
  {\bibfnamefont {T.~S.}\ \bibnamefont {Ramazanov}},\ }\bibfield  {title}
  {\enquote {\bibinfo {title} {Structural characteristics of strongly coupled
  ions in a dense quantum plasma},}\ }\href
  {https://journals.aps.org/pre/abstract/10.1103/PhysRevE.98.023207} {\bibfield
   {journal} {\bibinfo  {journal} {Phys. Rev. E}\ }\textbf {\bibinfo {volume}
  {98}},\ \bibinfo {pages} {023207} (\bibinfo {year}
  {2018}{\natexlab{a}})}\BibitemShut {NoStop}%
\bibitem [{\citenamefont {Hamann}\ \emph
  {et~al.}(2020{\natexlab{a}})\citenamefont {Hamann}, \citenamefont {Dornheim},
  \citenamefont {Vorberger}, \citenamefont {Moldabekov},\ and\ \citenamefont
  {Bonitz}}]{Hamann_PRB_2020}%
  \BibitemOpen
  \bibfield  {author} {\bibinfo {author} {\bibfnamefont {Paul}\ \bibnamefont
  {Hamann}}, \bibinfo {author} {\bibfnamefont {Tobias}\ \bibnamefont
  {Dornheim}}, \bibinfo {author} {\bibfnamefont {Jan}\ \bibnamefont
  {Vorberger}}, \bibinfo {author} {\bibfnamefont {Zhandos~A.}\ \bibnamefont
  {Moldabekov}}, \ and\ \bibinfo {author} {\bibfnamefont {Michael}\
  \bibnamefont {Bonitz}},\ }\bibfield  {title} {\enquote {\bibinfo {title}
  {Dynamic properties of the warm dense electron gas based on $ab initio$ path
  integral monte carlo simulations},}\ }\href {\doibase
  10.1103/PhysRevB.102.125150} {\bibfield  {journal} {\bibinfo  {journal}
  {Phys. Rev. B}\ }\textbf {\bibinfo {volume} {102}},\ \bibinfo {pages}
  {125150} (\bibinfo {year} {2020}{\natexlab{a}})}\BibitemShut {NoStop}%
\bibitem [{\citenamefont {Diaw}\ and\ \citenamefont
  {Murillo}(2017)}]{Diaw2017}%
  \BibitemOpen
  \bibfield  {author} {\bibinfo {author} {\bibfnamefont {Abdourahmane}\
  \bibnamefont {Diaw}}\ and\ \bibinfo {author} {\bibfnamefont {Michael~S.}\
  \bibnamefont {Murillo}},\ }\bibfield  {title} {\enquote {\bibinfo {title} {A
  viscous quantum hydrodynamics model based on dynamic density functional
  theory},}\ }\href {\doibase 10.1038/s41598-017-14414-9} {\bibfield  {journal}
  {\bibinfo  {journal} {Scientific Reports}\ }\textbf {\bibinfo {volume} {7}},\
  \bibinfo {pages} {15352} (\bibinfo {year} {2017})}\BibitemShut {NoStop}%
\bibitem [{\citenamefont {Moldabekov}\ \emph
  {et~al.}(2018{\natexlab{b}})\citenamefont {Moldabekov}, \citenamefont
  {Bonitz},\ and\ \citenamefont {Ramazanov}}]{zhandos_QHD}%
  \BibitemOpen
  \bibfield  {author} {\bibinfo {author} {\bibfnamefont {Zh.~A.}\ \bibnamefont
  {Moldabekov}}, \bibinfo {author} {\bibfnamefont {M.}~\bibnamefont {Bonitz}},
  \ and\ \bibinfo {author} {\bibfnamefont {T.~S.}\ \bibnamefont {Ramazanov}},\
  }\bibfield  {title} {\enquote {\bibinfo {title} {Theoretical foundations of
  quantum hydrodynamics for plasmas},}\ }\href {\doibase 10.1063/1.5003910}
  {\bibfield  {journal} {\bibinfo  {journal} {Physics of Plasmas}\ }\textbf
  {\bibinfo {volume} {25}},\ \bibinfo {pages} {031903} (\bibinfo {year}
  {2018}{\natexlab{b}})}\BibitemShut {NoStop}%
\bibitem [{\citenamefont {Sterne}\ \emph {et~al.}(2007)\citenamefont {Sterne},
  \citenamefont {Hansen}, \citenamefont {Wilson},\ and\ \citenamefont
  {Isaacs}}]{Sterne_average_atom_HEDP_2007}%
  \BibitemOpen
  \bibfield  {author} {\bibinfo {author} {\bibfnamefont {P.~A.}\ \bibnamefont
  {Sterne}}, \bibinfo {author} {\bibfnamefont {S.~B.}\ \bibnamefont {Hansen}},
  \bibinfo {author} {\bibfnamefont {B.~G.}\ \bibnamefont {Wilson}}, \ and\
  \bibinfo {author} {\bibfnamefont {W.~A.}\ \bibnamefont {Isaacs}},\ }\bibfield
   {title} {\enquote {\bibinfo {title} {Equation of state, occupation
  probabilities and conductivities in the average atom purgatorio code},}\
  }\href {\doibase 10.1016/j.hedp.2007.02.037} {\bibfield  {journal} {\bibinfo
  {journal} {High Energy Density Physics}\ }\textbf {\bibinfo {volume} {3}},\
  \bibinfo {pages} {278--282} (\bibinfo {year} {2007})}\BibitemShut {NoStop}%
\bibitem [{\citenamefont {Ceperley}(1995)}]{cep}%
  \BibitemOpen
  \bibfield  {author} {\bibinfo {author} {\bibfnamefont {D.~M.}\ \bibnamefont
  {Ceperley}},\ }\bibfield  {title} {\enquote {\bibinfo {title} {Path integrals
  in the theory of condensed helium},}\ }\href
  {https://journals.aps.org/rmp/abstract/10.1103/RevModPhys.67.279} {\bibfield
  {journal} {\bibinfo  {journal} {Rev. Mod. Phys}\ }\textbf {\bibinfo {volume}
  {67}},\ \bibinfo {pages} {279} (\bibinfo {year} {1995})}\BibitemShut
  {NoStop}%
\bibitem [{\citenamefont {Dornheim}\ \emph {et~al.}(2019)\citenamefont
  {Dornheim}, \citenamefont {Vorberger}, \citenamefont {Groth}, \citenamefont
  {Hoffmann}, \citenamefont {Moldabekov},\ and\ \citenamefont
  {Bonitz}}]{dornheim_ML}%
  \BibitemOpen
  \bibfield  {author} {\bibinfo {author} {\bibfnamefont {T.}~\bibnamefont
  {Dornheim}}, \bibinfo {author} {\bibfnamefont {J.}~\bibnamefont {Vorberger}},
  \bibinfo {author} {\bibfnamefont {S.}~\bibnamefont {Groth}}, \bibinfo
  {author} {\bibfnamefont {N.}~\bibnamefont {Hoffmann}}, \bibinfo {author}
  {\bibfnamefont {Zh.A.}\ \bibnamefont {Moldabekov}}, \ and\ \bibinfo {author}
  {\bibfnamefont {M.}~\bibnamefont {Bonitz}},\ }\bibfield  {title} {\enquote
  {\bibinfo {title} {The static local field correction of the warm dense
  electron gas: An ab initio path integral {M}onte {C}arlo study and machine
  learning representation},}\ }\href
  {https://aip.scitation.org/doi/full/10.1063/1.5123013} {\bibfield  {journal}
  {\bibinfo  {journal} {J. Chem. Phys}\ }\textbf {\bibinfo {volume} {151}},\
  \bibinfo {pages} {194104} (\bibinfo {year} {2019})}\BibitemShut {NoStop}%
\bibitem [{\citenamefont {Groth}\ \emph {et~al.}(2019)\citenamefont {Groth},
  \citenamefont {Dornheim},\ and\ \citenamefont
  {Vorberger}}]{dynamic_folgepaper}%
  \BibitemOpen
  \bibfield  {author} {\bibinfo {author} {\bibfnamefont {S.}~\bibnamefont
  {Groth}}, \bibinfo {author} {\bibfnamefont {T.}~\bibnamefont {Dornheim}}, \
  and\ \bibinfo {author} {\bibfnamefont {J.}~\bibnamefont {Vorberger}},\
  }\bibfield  {title} {\enquote {\bibinfo {title} {Ab initio path integral
  {M}onte {C}arlo approach to the static and dynamic density response of the
  uniform electron gas},}\ }\href
  {https://link.aps.org/doi/10.1103/PhysRevB.99.235122} {\bibfield  {journal}
  {\bibinfo  {journal} {Phys. Rev. B}\ }\textbf {\bibinfo {volume} {99}},\
  \bibinfo {pages} {235122} (\bibinfo {year} {2019})}\BibitemShut {NoStop}%
\bibitem [{\citenamefont {Dornheim}\ \emph
  {et~al.}(2018{\natexlab{b}})\citenamefont {Dornheim}, \citenamefont {Groth},
  \citenamefont {Vorberger},\ and\ \citenamefont {Bonitz}}]{dornheim_dynamic}%
  \BibitemOpen
  \bibfield  {author} {\bibinfo {author} {\bibfnamefont {T.}~\bibnamefont
  {Dornheim}}, \bibinfo {author} {\bibfnamefont {S.}~\bibnamefont {Groth}},
  \bibinfo {author} {\bibfnamefont {J.}~\bibnamefont {Vorberger}}, \ and\
  \bibinfo {author} {\bibfnamefont {M.}~\bibnamefont {Bonitz}},\ }\bibfield
  {title} {\enquote {\bibinfo {title} {Ab initio path integral {M}onte {C}arlo
  results for the dynamic structure factor of correlated electrons: From the
  electron liquid to warm dense matter},}\ }\href
  {https://journals.aps.org/prl/abstract/10.1103/PhysRevLett.121.255001}
  {\bibfield  {journal} {\bibinfo  {journal} {Phys. Rev. Lett.}\ }\textbf
  {\bibinfo {volume} {121}},\ \bibinfo {pages} {255001} (\bibinfo {year}
  {2018}{\natexlab{b}})}\BibitemShut {NoStop}%
\bibitem [{\citenamefont {Dornheim}\ and\ \citenamefont
  {Vorberger}(2020)}]{Dornheim_PRE_2020}%
  \BibitemOpen
  \bibfield  {author} {\bibinfo {author} {\bibfnamefont {Tobias}\ \bibnamefont
  {Dornheim}}\ and\ \bibinfo {author} {\bibfnamefont {Jan}\ \bibnamefont
  {Vorberger}},\ }\bibfield  {title} {\enquote {\bibinfo {title} {Finite-size
  effects in the reconstruction of dynamic properties from ab initio path
  integral monte carlo simulations},}\ }\href {\doibase
  10.1103/PhysRevE.102.063301} {\bibfield  {journal} {\bibinfo  {journal}
  {Phys. Rev. E}\ }\textbf {\bibinfo {volume} {102}},\ \bibinfo {pages}
  {063301} (\bibinfo {year} {2020})}\BibitemShut {NoStop}%
\bibitem [{\citenamefont {Hamann}\ \emph
  {et~al.}(2020{\natexlab{b}})\citenamefont {Hamann}, \citenamefont
  {Vorberger}, \citenamefont {Dornheim}, \citenamefont {Moldabekov},\ and\
  \citenamefont {Bonitz}}]{Hamann_CPP_2020}%
  \BibitemOpen
  \bibfield  {author} {\bibinfo {author} {\bibfnamefont {Paul}\ \bibnamefont
  {Hamann}}, \bibinfo {author} {\bibfnamefont {Jan}\ \bibnamefont {Vorberger}},
  \bibinfo {author} {\bibfnamefont {Tobias}\ \bibnamefont {Dornheim}}, \bibinfo
  {author} {\bibfnamefont {Zhandos~A.}\ \bibnamefont {Moldabekov}}, \ and\
  \bibinfo {author} {\bibfnamefont {Michael}\ \bibnamefont {Bonitz}},\
  }\bibfield  {title} {\enquote {\bibinfo {title} {Ab initio results for the
  plasmon dispersion and damping of the warm dense electron gas},}\ }\href
  {\doibase https://doi.org/10.1002/ctpp.202000147} {\bibfield  {journal}
  {\bibinfo  {journal} {Contributions to Plasma Physics}\ }\textbf {\bibinfo
  {volume} {60}},\ \bibinfo {pages} {e202000147} (\bibinfo {year}
  {2020}{\natexlab{b}})}\BibitemShut {NoStop}%
\bibitem [{\citenamefont {Dornheim}\ \emph
  {et~al.}(2020{\natexlab{c}})\citenamefont {Dornheim}, \citenamefont
  {Moldabekov}, \citenamefont {Vorberger},\ and\ \citenamefont
  {Groth}}]{dornheim_HEDP}%
  \BibitemOpen
  \bibfield  {author} {\bibinfo {author} {\bibfnamefont {Tobias}\ \bibnamefont
  {Dornheim}}, \bibinfo {author} {\bibfnamefont {Zhandos~A}\ \bibnamefont
  {Moldabekov}}, \bibinfo {author} {\bibfnamefont {Jan}\ \bibnamefont
  {Vorberger}}, \ and\ \bibinfo {author} {\bibfnamefont {Simon}\ \bibnamefont
  {Groth}},\ }\bibfield  {title} {\enquote {\bibinfo {title} {Ab initio path
  integral monte carlo simulation of the uniform electron gas in the high
  energy density regime},}\ }\href {\doibase 10.1088/1361-6587/ab8bb4}
  {\bibfield  {journal} {\bibinfo  {journal} {Plasma Physics and Controlled
  Fusion}\ }\textbf {\bibinfo {volume} {62}},\ \bibinfo {pages} {075003}
  (\bibinfo {year} {2020}{\natexlab{c}})}\BibitemShut {NoStop}%
\bibitem [{\citenamefont {Dornheim}\ \emph
  {et~al.}(2020{\natexlab{d}})\citenamefont {Dornheim}, \citenamefont {Cangi},
  \citenamefont {Ramakrishna}, \citenamefont {B\"ohme}, \citenamefont
  {Tanaka},\ and\ \citenamefont {Vorberger}}]{dornheim_PRL_ESA_2020}%
  \BibitemOpen
  \bibfield  {author} {\bibinfo {author} {\bibfnamefont {Tobias}\ \bibnamefont
  {Dornheim}}, \bibinfo {author} {\bibfnamefont {Attila}\ \bibnamefont
  {Cangi}}, \bibinfo {author} {\bibfnamefont {Kushal}\ \bibnamefont
  {Ramakrishna}}, \bibinfo {author} {\bibfnamefont {Maximilian}\ \bibnamefont
  {B\"ohme}}, \bibinfo {author} {\bibfnamefont {Shigenori}\ \bibnamefont
  {Tanaka}}, \ and\ \bibinfo {author} {\bibfnamefont {Jan}\ \bibnamefont
  {Vorberger}},\ }\bibfield  {title} {\enquote {\bibinfo {title} {Effective
  static approximation: A fast and reliable tool for warm-dense matter
  theory},}\ }\href {\doibase 10.1103/PhysRevLett.125.235001} {\bibfield
  {journal} {\bibinfo  {journal} {Phys. Rev. Lett.}\ }\textbf {\bibinfo
  {volume} {125}},\ \bibinfo {pages} {235001} (\bibinfo {year}
  {2020}{\natexlab{d}})}\BibitemShut {NoStop}%
\bibitem [{\citenamefont {Hunger}\ \emph {et~al.}(2021)\citenamefont {Hunger},
  \citenamefont {Schoof}, \citenamefont {Dornheim}, \citenamefont {Bonitz},\
  and\ \citenamefont {Filinov}}]{Hunger_PRE_2021}%
  \BibitemOpen
  \bibfield  {author} {\bibinfo {author} {\bibfnamefont {Kai}\ \bibnamefont
  {Hunger}}, \bibinfo {author} {\bibfnamefont {Tim}\ \bibnamefont {Schoof}},
  \bibinfo {author} {\bibfnamefont {Tobias}\ \bibnamefont {Dornheim}}, \bibinfo
  {author} {\bibfnamefont {Michael}\ \bibnamefont {Bonitz}}, \ and\ \bibinfo
  {author} {\bibfnamefont {Alexey}\ \bibnamefont {Filinov}},\ }\href@noop {}
  {\enquote {\bibinfo {title} {Momentum distribution function and short-range
  correlations of the warm dense electron gas -- ab initio quantum monte carlo
  results},}\ } (\bibinfo {year} {2021}),\ \Eprint
  {http://arxiv.org/abs/2101.00842} {arXiv:2101.00842 [physics.plasm-ph]}
  \BibitemShut {NoStop}%
\bibitem [{cod()}]{code}%
  \BibitemOpen
  \href@noop {} {}\bibinfo {note} {A link to the repository will be given upon
  publication.}\BibitemShut {Stop}%
\bibitem [{\citenamefont {Kugler}(1975)}]{kugler1}%
  \BibitemOpen
  \bibfield  {author} {\bibinfo {author} {\bibfnamefont {A.~A.}\ \bibnamefont
  {Kugler}},\ }\bibfield  {title} {\enquote {\bibinfo {title} {Theory of the
  local field correction in an electron gas},}\ }\href
  {http://link.springer.com/article/10.1007/BF01024183} {\bibfield  {journal}
  {\bibinfo  {journal} {J. Stat. Phys}\ }\textbf {\bibinfo {volume} {12}},\
  \bibinfo {pages} {35} (\bibinfo {year} {1975})}\BibitemShut {NoStop}%
\bibitem [{\citenamefont {Dornheim}\ \emph
  {et~al.}(2020{\natexlab{e}})\citenamefont {Dornheim}, \citenamefont
  {Sjostrom}, \citenamefont {Tanaka},\ and\ \citenamefont
  {Vorberger}}]{dornheim_electron_liquid}%
  \BibitemOpen
  \bibfield  {author} {\bibinfo {author} {\bibfnamefont {Tobias}\ \bibnamefont
  {Dornheim}}, \bibinfo {author} {\bibfnamefont {Travis}\ \bibnamefont
  {Sjostrom}}, \bibinfo {author} {\bibfnamefont {Shigenori}\ \bibnamefont
  {Tanaka}}, \ and\ \bibinfo {author} {\bibfnamefont {Jan}\ \bibnamefont
  {Vorberger}},\ }\bibfield  {title} {\enquote {\bibinfo {title} {Strongly
  coupled electron liquid: Ab initio path integral monte carlo simulations and
  dielectric theories},}\ }\href {\doibase 10.1103/PhysRevB.101.045129}
  {\bibfield  {journal} {\bibinfo  {journal} {Phys. Rev. B}\ }\textbf {\bibinfo
  {volume} {101}},\ \bibinfo {pages} {045129} (\bibinfo {year}
  {2020}{\natexlab{e}})}\BibitemShut {NoStop}%
\bibitem [{\citenamefont {Tanaka}\ and\ \citenamefont {Ichimaru}(1986)}]{stls}%
  \BibitemOpen
  \bibfield  {author} {\bibinfo {author} {\bibfnamefont {S.}~\bibnamefont
  {Tanaka}}\ and\ \bibinfo {author} {\bibfnamefont {S.}~\bibnamefont
  {Ichimaru}},\ }\bibfield  {title} {\enquote {\bibinfo {title} {Thermodynamics
  and correlational properties of finite-temperature electron liquids in the
  {S}ingwi-{T}osi-{Land}-{S}j\"olander approximation},}\ }\href
  {http://journals.jps.jp/doi/abs/10.1143/JPSJ.55.2278} {\bibfield  {journal}
  {\bibinfo  {journal} {J. Phys. Soc. Jpn}\ }\textbf {\bibinfo {volume} {55}},\
  \bibinfo {pages} {2278--2289} (\bibinfo {year} {1986})}\BibitemShut {NoStop}%
\bibitem [{\citenamefont {Holas}(1987)}]{holas_limit}%
  \BibitemOpen
  \bibfield  {author} {\bibinfo {author} {\bibfnamefont {A.}~\bibnamefont
  {Holas}},\ }\bibfield  {title} {\enquote {\bibinfo {title} {Exact asymptotic
  expression for the static dielectric function of a uniform electron liquid at
  large wave vector},}\ }in\ \href@noop {} {\emph {\bibinfo {booktitle}
  {Strongly Coupled Plasma Physics}}},\ \bibinfo {editor} {edited by\ \bibinfo
  {editor} {\bibfnamefont {F.J.}\ \bibnamefont {Rogers}}\ and\ \bibinfo
  {editor} {\bibfnamefont {H.E.}\ \bibnamefont {DeWitt}}}\ (\bibinfo
  {publisher} {Plenum},\ \bibinfo {address} {New York},\ \bibinfo {year}
  {1987})\BibitemShut {NoStop}%
\bibitem [{\citenamefont {Farid}\ \emph {et~al.}(1993)\citenamefont {Farid},
  \citenamefont {Heine}, \citenamefont {Engel},\ and\ \citenamefont
  {Robertson}}]{farid}%
  \BibitemOpen
  \bibfield  {author} {\bibinfo {author} {\bibfnamefont {B.}~\bibnamefont
  {Farid}}, \bibinfo {author} {\bibfnamefont {V.}~\bibnamefont {Heine}},
  \bibinfo {author} {\bibfnamefont {G.~E.}\ \bibnamefont {Engel}}, \ and\
  \bibinfo {author} {\bibfnamefont {I.~J.}\ \bibnamefont {Robertson}},\
  }\bibfield  {title} {\enquote {\bibinfo {title} {Extremal properties of the
  harris-foulkes functional and an improved screening calculation for the
  electron gas},}\ }\href {http://link.aps.org/doi/10.1103/PhysRevB.48.11602}
  {\bibfield  {journal} {\bibinfo  {journal} {Phys. Rev. B}\ }\textbf {\bibinfo
  {volume} {48}},\ \bibinfo {pages} {11602} (\bibinfo {year}
  {1993})}\BibitemShut {NoStop}%
\bibitem [{\citenamefont {Sjostrom}\ and\ \citenamefont {Dufty}(2013)}]{stls2}%
  \BibitemOpen
  \bibfield  {author} {\bibinfo {author} {\bibfnamefont {T.}~\bibnamefont
  {Sjostrom}}\ and\ \bibinfo {author} {\bibfnamefont {J.}~\bibnamefont
  {Dufty}},\ }\bibfield  {title} {\enquote {\bibinfo {title} {Uniform electron
  gas at finite temperatures},}\ }\href
  {http://link.aps.org/doi/10.1103/PhysRevB.88.115123} {\bibfield  {journal}
  {\bibinfo  {journal} {Phys. Rev. B}\ }\textbf {\bibinfo {volume} {88}},\
  \bibinfo {pages} {115123} (\bibinfo {year} {2013})}\BibitemShut {NoStop}%
\bibitem [{\citenamefont {Militzer}\ and\ \citenamefont
  {Pollock}(2002)}]{Militzer_PRL_2002}%
  \BibitemOpen
  \bibfield  {author} {\bibinfo {author} {\bibfnamefont {B.}~\bibnamefont
  {Militzer}}\ and\ \bibinfo {author} {\bibfnamefont {E.~L.}\ \bibnamefont
  {Pollock}},\ }\bibfield  {title} {\enquote {\bibinfo {title} {Lowering of the
  kinetic energy in interacting quantum systems},}\ }\href {\doibase
  10.1103/PhysRevLett.89.280401} {\bibfield  {journal} {\bibinfo  {journal}
  {Phys. Rev. Lett.}\ }\textbf {\bibinfo {volume} {89}},\ \bibinfo {pages}
  {280401} (\bibinfo {year} {2002})}\BibitemShut {NoStop}%
\bibitem [{\citenamefont {Singwi}\ \emph {et~al.}(1968)\citenamefont {Singwi},
  \citenamefont {Tosi}, \citenamefont {Land},\ and\ \citenamefont
  {Sj\"olander}}]{stls_original}%
  \BibitemOpen
  \bibfield  {author} {\bibinfo {author} {\bibfnamefont {K.~S.}\ \bibnamefont
  {Singwi}}, \bibinfo {author} {\bibfnamefont {M.~P.}\ \bibnamefont {Tosi}},
  \bibinfo {author} {\bibfnamefont {R.~H.}\ \bibnamefont {Land}}, \ and\
  \bibinfo {author} {\bibfnamefont {A.}~\bibnamefont {Sj\"olander}},\
  }\bibfield  {title} {\enquote {\bibinfo {title} {Electron correlations at
  metallic densities},}\ }\href
  {http://link.aps.org/doi/10.1103/PhysRev.176.589} {\bibfield  {journal}
  {\bibinfo  {journal} {Phys. Rev}\ }\textbf {\bibinfo {volume} {176}},\
  \bibinfo {pages} {589} (\bibinfo {year} {1968})}\BibitemShut {NoStop}%
\bibitem [{\citenamefont {Segal}\ \emph {et~al.}(2010)\citenamefont {Segal},
  \citenamefont {Millis},\ and\ \citenamefont {Reichman}}]{Segal_PRB_2010}%
  \BibitemOpen
  \bibfield  {author} {\bibinfo {author} {\bibfnamefont {Dvira}\ \bibnamefont
  {Segal}}, \bibinfo {author} {\bibfnamefont {Andrew~J.}\ \bibnamefont
  {Millis}}, \ and\ \bibinfo {author} {\bibfnamefont {David~R.}\ \bibnamefont
  {Reichman}},\ }\bibfield  {title} {\enquote {\bibinfo {title} {Numerically
  exact path-integral simulation of nonequilibrium quantum transport and
  dissipation},}\ }\href {\doibase 10.1103/PhysRevB.82.205323} {\bibfield
  {journal} {\bibinfo  {journal} {Phys. Rev. B}\ }\textbf {\bibinfo {volume}
  {82}},\ \bibinfo {pages} {205323} (\bibinfo {year} {2010})}\BibitemShut
  {NoStop}%
\bibitem [{\citenamefont {Dornheim}(2019)}]{dornheim_sign_problem}%
  \BibitemOpen
  \bibfield  {author} {\bibinfo {author} {\bibfnamefont {T.}~\bibnamefont
  {Dornheim}},\ }\bibfield  {title} {\enquote {\bibinfo {title} {Fermion sign
  problem in path integral {M}onte {C}arlo simulations: Quantum dots, ultracold
  atoms, and warm dense matter},}\ }\href
  {https://journals.aps.org/pre/abstract/10.1103/PhysRevE.100.023307}
  {\bibfield  {journal} {\bibinfo  {journal} {Phys. Rev. E}\ }\textbf {\bibinfo
  {volume} {100}},\ \bibinfo {pages} {023307} (\bibinfo {year}
  {2019})}\BibitemShut {NoStop}%
\bibitem [{\citenamefont {Kugler}(1970)}]{kugler_bounds}%
  \BibitemOpen
  \bibfield  {author} {\bibinfo {author} {\bibfnamefont {A.~A.}\ \bibnamefont
  {Kugler}},\ }\bibfield  {title} {\enquote {\bibinfo {title} {Bounds for some
  equilibrium properties of an electron gas},}\ }\href
  {https://journals.aps.org/pra/abstract/10.1103/PhysRevA.1.1688} {\bibfield
  {journal} {\bibinfo  {journal} {Phys. Rev. A}\ }\textbf {\bibinfo {volume}
  {1}},\ \bibinfo {pages} {1688} (\bibinfo {year} {1970})}\BibitemShut
  {NoStop}%
\bibitem [{\citenamefont {Mazevet}\ \emph {et~al.}(2005)\citenamefont
  {Mazevet}, \citenamefont {Desjarlais}, \citenamefont {Collins}, \citenamefont
  {Kress},\ and\ \citenamefont {Magee}}]{low_density1}%
  \BibitemOpen
  \bibfield  {author} {\bibinfo {author} {\bibfnamefont {S.}~\bibnamefont
  {Mazevet}}, \bibinfo {author} {\bibfnamefont {M.~P.}\ \bibnamefont
  {Desjarlais}}, \bibinfo {author} {\bibfnamefont {L.~A.}\ \bibnamefont
  {Collins}}, \bibinfo {author} {\bibfnamefont {J.~D.}\ \bibnamefont {Kress}},
  \ and\ \bibinfo {author} {\bibfnamefont {N.~H.}\ \bibnamefont {Magee}},\
  }\bibfield  {title} {\enquote {\bibinfo {title} {Simulations of the optical
  properties of warm dense aluminum},}\ }\href
  {https://journals.aps.org/pre/abstract/10.1103/PhysRevE.71.016409} {\bibfield
   {journal} {\bibinfo  {journal} {Phys. Rev. E}\ }\textbf {\bibinfo {volume}
  {71}},\ \bibinfo {pages} {016409} (\bibinfo {year} {2005})}\BibitemShut
  {NoStop}%
\bibitem [{\citenamefont {Zastrau}\ \emph {et~al.}(2014)\citenamefont
  {Zastrau}, \citenamefont {Sperling}, \citenamefont {Harmand}, \citenamefont
  {Becker}, \citenamefont {Bornath}, \citenamefont {Bredow}, \citenamefont
  {Dziarzhytski}, \citenamefont {Fennel}, \citenamefont {Fletcher},
  \citenamefont {F{"o}rster}, \citenamefont {G{"o}de}, \citenamefont {Gregori},
  \citenamefont {Hilbert}, \citenamefont {Hochhaus}, \citenamefont {Holst},
  \citenamefont {Laarmann}, \citenamefont {Lee}, \citenamefont {Ma},
  \citenamefont {Mithen}, \citenamefont {Mitzner}, \citenamefont {Murphy},
  \citenamefont {Nakatsutsumi}, \citenamefont {Neumayer}, \citenamefont
  {Przystawik}, \citenamefont {Roling}, \citenamefont {Schulz}, \citenamefont
  {Siemer}, \citenamefont {Skruszewicz}, \citenamefont {Tiggesb{"a}umker},
  \citenamefont {Toleikis}, \citenamefont {Tschentscher}, \citenamefont
  {White}, \citenamefont {W{"o}stmann}, \citenamefont {Zacharias},
  \citenamefont {D{"o}ppner}, \citenamefont {Glenzer},\ and\ \citenamefont
  {Redmer}}]{Zastrau}%
  \BibitemOpen
  \bibfield  {author} {\bibinfo {author} {\bibfnamefont {U.}~\bibnamefont
  {Zastrau}}, \bibinfo {author} {\bibfnamefont {P.}~\bibnamefont {Sperling}},
  \bibinfo {author} {\bibfnamefont {M.}~\bibnamefont {Harmand}}, \bibinfo
  {author} {\bibfnamefont {A.}~\bibnamefont {Becker}}, \bibinfo {author}
  {\bibfnamefont {T.}~\bibnamefont {Bornath}}, \bibinfo {author} {\bibfnamefont
  {R.}~\bibnamefont {Bredow}}, \bibinfo {author} {\bibfnamefont
  {S.}~\bibnamefont {Dziarzhytski}}, \bibinfo {author} {\bibfnamefont
  {T.}~\bibnamefont {Fennel}}, \bibinfo {author} {\bibfnamefont {L.~B.}\
  \bibnamefont {Fletcher}}, \bibinfo {author} {\bibfnamefont {E.}~\bibnamefont
  {F{"o}rster}}, \bibinfo {author} {\bibfnamefont {S.}~\bibnamefont {G{"o}de}},
  \bibinfo {author} {\bibfnamefont {G.}~\bibnamefont {Gregori}}, \bibinfo
  {author} {\bibfnamefont {V.}~\bibnamefont {Hilbert}}, \bibinfo {author}
  {\bibfnamefont {D.}~\bibnamefont {Hochhaus}}, \bibinfo {author}
  {\bibfnamefont {B.}~\bibnamefont {Holst}}, \bibinfo {author} {\bibfnamefont
  {T.}~\bibnamefont {Laarmann}}, \bibinfo {author} {\bibfnamefont {H.~J.}\
  \bibnamefont {Lee}}, \bibinfo {author} {\bibfnamefont {T.}~\bibnamefont
  {Ma}}, \bibinfo {author} {\bibfnamefont {J.~P.}\ \bibnamefont {Mithen}},
  \bibinfo {author} {\bibfnamefont {R.}~\bibnamefont {Mitzner}}, \bibinfo
  {author} {\bibfnamefont {C.~D.}\ \bibnamefont {Murphy}}, \bibinfo {author}
  {\bibfnamefont {M.}~\bibnamefont {Nakatsutsumi}}, \bibinfo {author}
  {\bibfnamefont {P.}~\bibnamefont {Neumayer}}, \bibinfo {author}
  {\bibfnamefont {A.}~\bibnamefont {Przystawik}}, \bibinfo {author}
  {\bibfnamefont {S.}~\bibnamefont {Roling}}, \bibinfo {author} {\bibfnamefont
  {M.}~\bibnamefont {Schulz}}, \bibinfo {author} {\bibfnamefont
  {B.}~\bibnamefont {Siemer}}, \bibinfo {author} {\bibfnamefont
  {S.}~\bibnamefont {Skruszewicz}}, \bibinfo {author} {\bibfnamefont
  {J.}~\bibnamefont {Tiggesb{"a}umker}}, \bibinfo {author} {\bibfnamefont
  {S.}~\bibnamefont {Toleikis}}, \bibinfo {author} {\bibfnamefont
  {T.}~\bibnamefont {Tschentscher}}, \bibinfo {author} {\bibfnamefont
  {T.}~\bibnamefont {White}}, \bibinfo {author} {\bibfnamefont
  {M.}~\bibnamefont {W{"o}stmann}}, \bibinfo {author} {\bibfnamefont
  {H.}~\bibnamefont {Zacharias}}, \bibinfo {author} {\bibfnamefont
  {T.}~\bibnamefont {D{"o}ppner}}, \bibinfo {author} {\bibfnamefont {S.~H.}\
  \bibnamefont {Glenzer}}, \ and\ \bibinfo {author} {\bibfnamefont
  {R.}~\bibnamefont {Redmer}},\ }\bibfield  {title} {\enquote {\bibinfo {title}
  {Resolving ultrafast heating of dense cryogenic hydrogen},}\ }\href
  {https://journals.aps.org/prl/abstract/10.1103/PhysRevLett.112.105002}
  {\bibfield  {journal} {\bibinfo  {journal} {Phys. Rev. Lett}\ }\textbf
  {\bibinfo {volume} {112}},\ \bibinfo {pages} {105002} (\bibinfo {year}
  {2014})}\BibitemShut {NoStop}%
\bibitem [{\citenamefont {Benage}\ \emph {et~al.}(1999)\citenamefont {Benage},
  \citenamefont {Shanahan},\ and\ \citenamefont {Murillo}}]{benage}%
  \BibitemOpen
  \bibfield  {author} {\bibinfo {author} {\bibfnamefont {J.~F.}\ \bibnamefont
  {Benage}}, \bibinfo {author} {\bibfnamefont {W.~R.}\ \bibnamefont
  {Shanahan}}, \ and\ \bibinfo {author} {\bibfnamefont {M.~S.}\ \bibnamefont
  {Murillo}},\ }\bibfield  {title} {\enquote {\bibinfo {title} {Electrical
  resistivity measurements of hot dense aluminum},}\ }\href
  {https://journals.aps.org/prl/abstract/10.1103/PhysRevLett.83.2953}
  {\bibfield  {journal} {\bibinfo  {journal} {Phys. Rev. Lett}\ }\textbf
  {\bibinfo {volume} {83}},\ \bibinfo {pages} {2953} (\bibinfo {year}
  {1999})}\BibitemShut {NoStop}%
\bibitem [{\citenamefont {Desjarlais}\ \emph {et~al.}(2002)\citenamefont
  {Desjarlais}, \citenamefont {Kress},\ and\ \citenamefont
  {Collins}}]{low_density2}%
  \BibitemOpen
  \bibfield  {author} {\bibinfo {author} {\bibfnamefont {M.~P.}\ \bibnamefont
  {Desjarlais}}, \bibinfo {author} {\bibfnamefont {J.~D.}\ \bibnamefont
  {Kress}}, \ and\ \bibinfo {author} {\bibfnamefont {L.~A.}\ \bibnamefont
  {Collins}},\ }\bibfield  {title} {\enquote {\bibinfo {title} {Electrical
  conductivity for warm, dense aluminum plasmas and liquids},}\ }\href
  {https://journals.aps.org/pre/abstract/10.1103/PhysRevE.66.025401} {\bibfield
   {journal} {\bibinfo  {journal} {Phys. Rev. E}\ }\textbf {\bibinfo {volume}
  {66}},\ \bibinfo {pages} {025401(R)} (\bibinfo {year} {2002})}\BibitemShut
  {NoStop}%
\bibitem [{\citenamefont {Ramakrishna}\ \emph
  {et~al.}(2020{\natexlab{b}})\citenamefont {Ramakrishna}, \citenamefont
  {Cangi}, \citenamefont {Dornheim},\ and\ \citenamefont
  {Vorberger}}]{Ramakrishna_PRB_2021}%
  \BibitemOpen
  \bibfield  {author} {\bibinfo {author} {\bibfnamefont {Kushal}\ \bibnamefont
  {Ramakrishna}}, \bibinfo {author} {\bibfnamefont {Attila}\ \bibnamefont
  {Cangi}}, \bibinfo {author} {\bibfnamefont {Tobias}\ \bibnamefont
  {Dornheim}}, \ and\ \bibinfo {author} {\bibfnamefont {Jan}\ \bibnamefont
  {Vorberger}},\ }\href@noop {} {\enquote {\bibinfo {title} {First-principles
  modeling of plasmons in aluminum under ambient and extreme conditions},}\ }
  (\bibinfo {year} {2020}{\natexlab{b}}),\ \Eprint
  {http://arxiv.org/abs/2009.12163} {arXiv:2009.12163 [cond-mat.mtrl-sci]}
  \BibitemShut {NoStop}%
\bibitem [{\citenamefont {Ichimaru}\ \emph {et~al.}(1987)\citenamefont
  {Ichimaru}, \citenamefont {Iyetomi},\ and\ \citenamefont
  {Tanaka}}]{ICHIMARU198791}%
  \BibitemOpen
  \bibfield  {author} {\bibinfo {author} {\bibfnamefont {Setsuo}\ \bibnamefont
  {Ichimaru}}, \bibinfo {author} {\bibfnamefont {Hiroshi}\ \bibnamefont
  {Iyetomi}}, \ and\ \bibinfo {author} {\bibfnamefont {Shigenori}\ \bibnamefont
  {Tanaka}},\ }\bibfield  {title} {\enquote {\bibinfo {title} {Statistical
  physics of dense plasmas: Thermodynamics, transport coefficients and dynamic
  correlations},}\ }\href {\doibase
  https://doi.org/10.1016/0370-1573(87)90125-6} {\bibfield  {journal} {\bibinfo
   {journal} {Physics Reports}\ }\textbf {\bibinfo {volume} {149}},\ \bibinfo
  {pages} {91 -- 205} (\bibinfo {year} {1987})}\BibitemShut {NoStop}%
\bibitem [{\citenamefont {Mithen}\ \emph {et~al.}(2012)\citenamefont {Mithen},
  \citenamefont {Daligault},\ and\ \citenamefont {Gregori}}]{Mithen_PRE_2012}%
  \BibitemOpen
  \bibfield  {author} {\bibinfo {author} {\bibfnamefont {James~P.}\
  \bibnamefont {Mithen}}, \bibinfo {author} {\bibfnamefont {J\'er\^ome}\
  \bibnamefont {Daligault}}, \ and\ \bibinfo {author} {\bibfnamefont
  {Gianluca}\ \bibnamefont {Gregori}},\ }\bibfield  {title} {\enquote {\bibinfo
  {title} {Comparative merits of the memory function and dynamic local-field
  correction of the classical one-component plasma},}\ }\href {\doibase
  10.1103/PhysRevE.85.056407} {\bibfield  {journal} {\bibinfo  {journal} {Phys.
  Rev. E}\ }\textbf {\bibinfo {volume} {85}},\ \bibinfo {pages} {056407}
  (\bibinfo {year} {2012})}\BibitemShut {NoStop}%
\bibitem [{\citenamefont {Bonitz}(2016)}]{bonitz_book}%
  \BibitemOpen
  \bibfield  {author} {\bibinfo {author} {\bibfnamefont {M.}~\bibnamefont
  {Bonitz}},\ }\href@noop {} {\emph {\bibinfo {title} {Quantum kinetic
  theory}}}\ (\bibinfo  {publisher} {Springer},\ \bibinfo {address}
  {Heidelberg},\ \bibinfo {year} {2016})\BibitemShut {NoStop}%
\bibitem [{\citenamefont {Alexandrov}\ \emph {et~al.}(1984)\citenamefont
  {Alexandrov}, \citenamefont {Bogdankievich},\ and\ \citenamefont
  {Rukhadse}}]{alexandrov}%
  \BibitemOpen
  \bibfield  {author} {\bibinfo {author} {\bibnamefont {Alexandrov}}, \bibinfo
  {author} {\bibnamefont {Bogdankievich}}, \ and\ \bibinfo {author}
  {\bibnamefont {Rukhadse}},\ }\href@noop {} {\emph {\bibinfo {title}
  {Principles of Plasma Electrodynamics}}}\ (\bibinfo  {publisher} {Springer},\
  \bibinfo {address} {Heidelberg, Germany},\ \bibinfo {year}
  {1984})\BibitemShut {NoStop}%
\bibitem [{\citenamefont {Vashishta}\ and\ \citenamefont
  {Singwi}(1972)}]{vs_original}%
  \BibitemOpen
  \bibfield  {author} {\bibinfo {author} {\bibfnamefont {P.}~\bibnamefont
  {Vashishta}}\ and\ \bibinfo {author} {\bibfnamefont {K.~S.}\ \bibnamefont
  {Singwi}},\ }\bibfield  {title} {\enquote {\bibinfo {title} {Electron
  correlations at metallic densities v},}\ }\href
  {http://link.aps.org/doi/10.1103/PhysRevB.6.875} {\bibfield  {journal}
  {\bibinfo  {journal} {Phys. Rev. B}\ }\textbf {\bibinfo {volume} {6}},\
  \bibinfo {pages} {875} (\bibinfo {year} {1972})}\BibitemShut {NoStop}%
\bibitem [{\citenamefont {Stolzmann}\ and\ \citenamefont
  {R\"osler}(2001)}]{stolzmann}%
  \BibitemOpen
  \bibfield  {author} {\bibinfo {author} {\bibfnamefont {W.}~\bibnamefont
  {Stolzmann}}\ and\ \bibinfo {author} {\bibfnamefont {M.}~\bibnamefont
  {R\"osler}},\ }\bibfield  {title} {\enquote {\bibinfo {title} {Static
  local-field corrected dielectric and thermodynamic functions},}\ }\href
  {http://onlinelibrary.wiley.com/doi/10.1002/1521-3986(200103)41:2/3<203::AID-CTPP203>3.0.CO;2-S/abstract}
  {\bibfield  {journal} {\bibinfo  {journal} {Contrib. Plasma Phys}\ }\textbf
  {\bibinfo {volume} {41}},\ \bibinfo {pages} {203} (\bibinfo {year}
  {2001})}\BibitemShut {NoStop}%
\bibitem [{\citenamefont {Galam}\ and\ \citenamefont {Hansen}(1976)}]{Hansen}%
  \BibitemOpen
  \bibfield  {author} {\bibinfo {author} {\bibfnamefont {Serge}\ \bibnamefont
  {Galam}}\ and\ \bibinfo {author} {\bibfnamefont {Jean-Pierre}\ \bibnamefont
  {Hansen}},\ }\bibfield  {title} {\enquote {\bibinfo {title} {Statistical
  mechanics of dense ionized matter. vi. electron screening corrections to the
  thermodynamic properties of the one-component plasma},}\ }\href {\doibase
  10.1103/PhysRevA.14.816} {\bibfield  {journal} {\bibinfo  {journal} {Phys.
  Rev. A}\ }\textbf {\bibinfo {volume} {14}},\ \bibinfo {pages} {816--832}
  (\bibinfo {year} {1976})}\BibitemShut {NoStop}%
\bibitem [{\citenamefont {Moldabekov}\ \emph
  {et~al.}(2020{\natexlab{b}})\citenamefont {Moldabekov}, \citenamefont
  {Dornheim},\ and\ \citenamefont {Bonitz}}]{zhandos_cpp21}%
  \BibitemOpen
  \bibfield  {author} {\bibinfo {author} {\bibfnamefont {Zh.A.}\ \bibnamefont
  {Moldabekov}}, \bibinfo {author} {\bibfnamefont {T.}~\bibnamefont
  {Dornheim}}, \ and\ \bibinfo {author} {\bibfnamefont {M.}~\bibnamefont
  {Bonitz}},\ }\bibfield  {title} {\enquote {\bibinfo {title} {Screening of a
  test charge in a free-electron gas at warm dense matter and dense non-ideal
  plasma conditions},}\ }\href@noop {} {\bibfield  {journal} {\bibinfo
  {journal} {accepted for publication in Contrib. Plasma
  Phys.(arXiv:2009.09180)}\ } (\bibinfo {year}
  {2020}{\natexlab{b}})}\BibitemShut {NoStop}%
\bibitem [{\citenamefont {Moldabekov}\ \emph
  {et~al.}(2017{\natexlab{b}})\citenamefont {Moldabekov}, \citenamefont
  {Groth}, \citenamefont {Dornheim}, \citenamefont {Bonitz},\ and\
  \citenamefont {Ramazanov}}]{zhandos_cpp17}%
  \BibitemOpen
  \bibfield  {author} {\bibinfo {author} {\bibfnamefont {Zh.A.}\ \bibnamefont
  {Moldabekov}}, \bibinfo {author} {\bibfnamefont {S.}~\bibnamefont {Groth}},
  \bibinfo {author} {\bibfnamefont {T.}~\bibnamefont {Dornheim}}, \bibinfo
  {author} {\bibfnamefont {M.}~\bibnamefont {Bonitz}}, \ and\ \bibinfo {author}
  {\bibfnamefont {T.S.}\ \bibnamefont {Ramazanov}},\ }\bibfield  {title}
  {\enquote {\bibinfo {title} {Ion potential in non‐ideal dense quantum
  plasmas},}\ }\href {\doibase 10.1002/ctpp.201700109} {\bibfield  {journal}
  {\bibinfo  {journal} {Contrib. Plasma Phys.}\ }\textbf {\bibinfo {volume}
  {57}},\ \bibinfo {pages} {532--538} (\bibinfo {year}
  {2017}{\natexlab{b}})}\BibitemShut {NoStop}%
\bibitem [{\citenamefont {Grabowski}\ \emph {et~al.}(2020)\citenamefont
  {Grabowski}, \citenamefont {Hansen}, \citenamefont {Murillo}, \citenamefont
  {Stanton}, \citenamefont {Graziani}, \citenamefont {Zylstra}, \citenamefont
  {Baalrud}, \citenamefont {Arnault}, \citenamefont {Baczewski}, \citenamefont
  {Benedict}, \citenamefont {Blancard}, \citenamefont {Čertík}, \citenamefont
  {Clérouin}, \citenamefont {Collins}, \citenamefont {Copeland}, \citenamefont
  {Correa}, \citenamefont {Dai}, \citenamefont {Daligault}, \citenamefont
  {Desjarlais}, \citenamefont {Dharma-wardana}, \citenamefont {Faussurier},
  \citenamefont {Haack}, \citenamefont {Haxhimali}, \citenamefont
  {Hayes-Sterbenz}, \citenamefont {Hou}, \citenamefont {Hu}, \citenamefont
  {Jensen}, \citenamefont {Jungman}, \citenamefont {Kagan}, \citenamefont
  {Kang}, \citenamefont {Kress}, \citenamefont {Ma}, \citenamefont {Marciante},
  \citenamefont {Meyer}, \citenamefont {Rudd}, \citenamefont {Saumon},
  \citenamefont {Shulenburger}, \citenamefont {Singleton}, \citenamefont
  {Sjostrom}, \citenamefont {Stanek}, \citenamefont {Starrett}, \citenamefont
  {Ticknor}, \citenamefont {Valaitis}, \citenamefont {Venzke},\ and\
  \citenamefont {White}}]{GRABOWSKI2020100905}%
  \BibitemOpen
  \bibfield  {author} {\bibinfo {author} {\bibfnamefont {P.E.}\ \bibnamefont
  {Grabowski}}, \bibinfo {author} {\bibfnamefont {S.B.}\ \bibnamefont
  {Hansen}}, \bibinfo {author} {\bibfnamefont {M.S.}\ \bibnamefont {Murillo}},
  \bibinfo {author} {\bibfnamefont {L.G.}\ \bibnamefont {Stanton}}, \bibinfo
  {author} {\bibfnamefont {F.R.}\ \bibnamefont {Graziani}}, \bibinfo {author}
  {\bibfnamefont {A.B.}\ \bibnamefont {Zylstra}}, \bibinfo {author}
  {\bibfnamefont {S.D.}\ \bibnamefont {Baalrud}}, \bibinfo {author}
  {\bibfnamefont {P.}~\bibnamefont {Arnault}}, \bibinfo {author} {\bibfnamefont
  {A.D.}\ \bibnamefont {Baczewski}}, \bibinfo {author} {\bibfnamefont {L.X.}\
  \bibnamefont {Benedict}}, \bibinfo {author} {\bibfnamefont {C.}~\bibnamefont
  {Blancard}}, \bibinfo {author} {\bibfnamefont {O.}~\bibnamefont {Čertík}},
  \bibinfo {author} {\bibfnamefont {J.}~\bibnamefont {Clérouin}}, \bibinfo
  {author} {\bibfnamefont {L.A.}\ \bibnamefont {Collins}}, \bibinfo {author}
  {\bibfnamefont {S.}~\bibnamefont {Copeland}}, \bibinfo {author}
  {\bibfnamefont {A.A.}\ \bibnamefont {Correa}}, \bibinfo {author}
  {\bibfnamefont {J.}~\bibnamefont {Dai}}, \bibinfo {author} {\bibfnamefont
  {J.}~\bibnamefont {Daligault}}, \bibinfo {author} {\bibfnamefont {M.P.}\
  \bibnamefont {Desjarlais}}, \bibinfo {author} {\bibfnamefont {M.W.C.}\
  \bibnamefont {Dharma-wardana}}, \bibinfo {author} {\bibfnamefont
  {G.}~\bibnamefont {Faussurier}}, \bibinfo {author} {\bibfnamefont
  {J.}~\bibnamefont {Haack}}, \bibinfo {author} {\bibfnamefont
  {T.}~\bibnamefont {Haxhimali}}, \bibinfo {author} {\bibfnamefont
  {A.}~\bibnamefont {Hayes-Sterbenz}}, \bibinfo {author} {\bibfnamefont
  {Y.}~\bibnamefont {Hou}}, \bibinfo {author} {\bibfnamefont {S.X.}\
  \bibnamefont {Hu}}, \bibinfo {author} {\bibfnamefont {D.}~\bibnamefont
  {Jensen}}, \bibinfo {author} {\bibfnamefont {G.}~\bibnamefont {Jungman}},
  \bibinfo {author} {\bibfnamefont {G.}~\bibnamefont {Kagan}}, \bibinfo
  {author} {\bibfnamefont {D.}~\bibnamefont {Kang}}, \bibinfo {author}
  {\bibfnamefont {J.D.}\ \bibnamefont {Kress}}, \bibinfo {author}
  {\bibfnamefont {Q.}~\bibnamefont {Ma}}, \bibinfo {author} {\bibfnamefont
  {M.}~\bibnamefont {Marciante}}, \bibinfo {author} {\bibfnamefont
  {E.}~\bibnamefont {Meyer}}, \bibinfo {author} {\bibfnamefont {R.E.}\
  \bibnamefont {Rudd}}, \bibinfo {author} {\bibfnamefont {D.}~\bibnamefont
  {Saumon}}, \bibinfo {author} {\bibfnamefont {L.}~\bibnamefont
  {Shulenburger}}, \bibinfo {author} {\bibfnamefont {R.L.}\ \bibnamefont
  {Singleton}}, \bibinfo {author} {\bibfnamefont {T.}~\bibnamefont {Sjostrom}},
  \bibinfo {author} {\bibfnamefont {L.J.}\ \bibnamefont {Stanek}}, \bibinfo
  {author} {\bibfnamefont {C.E.}\ \bibnamefont {Starrett}}, \bibinfo {author}
  {\bibfnamefont {C.}~\bibnamefont {Ticknor}}, \bibinfo {author} {\bibfnamefont
  {S.}~\bibnamefont {Valaitis}}, \bibinfo {author} {\bibfnamefont
  {J.}~\bibnamefont {Venzke}}, \ and\ \bibinfo {author} {\bibfnamefont
  {A.}~\bibnamefont {White}},\ }\bibfield  {title} {\enquote {\bibinfo {title}
  {Review of the first charged-particle transport coefficient comparison
  workshop},}\ }\href {\doibase https://doi.org/10.1016/j.hedp.2020.100905}
  {\bibfield  {journal} {\bibinfo  {journal} {High Energy Density Physics}\
  }\textbf {\bibinfo {volume} {37}},\ \bibinfo {pages} {100905} (\bibinfo
  {year} {2020})}\BibitemShut {NoStop}%
\bibitem [{\citenamefont {Kodanova}\ \emph {et~al.}(2018)\citenamefont
  {Kodanova}, \citenamefont {Issanova}, \citenamefont {Amirov}, \citenamefont
  {Ramazanov}, \citenamefont {Tikhonov},\ and\ \citenamefont
  {Moldabekov}}]{MRE2018}%
  \BibitemOpen
  \bibfield  {author} {\bibinfo {author} {\bibfnamefont {S.K.}\ \bibnamefont
  {Kodanova}}, \bibinfo {author} {\bibfnamefont {M.K.}\ \bibnamefont
  {Issanova}}, \bibinfo {author} {\bibfnamefont {S.M.}\ \bibnamefont {Amirov}},
  \bibinfo {author} {\bibfnamefont {T.S.}\ \bibnamefont {Ramazanov}}, \bibinfo
  {author} {\bibfnamefont {A.}~\bibnamefont {Tikhonov}}, \ and\ \bibinfo
  {author} {\bibfnamefont {Zh.A.}\ \bibnamefont {Moldabekov}},\ }\bibfield
  {title} {\enquote {\bibinfo {title} {Relaxation of non-isothermal hot dense
  plasma parameters},}\ }\href {\doibase 10.1016/j.mre.2017.07.005} {\bibfield
  {journal} {\bibinfo  {journal} {Matter and Radiation at Extremes}\ }\textbf
  {\bibinfo {volume} {3}},\ \bibinfo {pages} {40--49} (\bibinfo {year}
  {2018})}\BibitemShut {NoStop}%
\bibitem [{\citenamefont {Arista}\ and\ \citenamefont
  {Brandt}(1981)}]{PhysRevA.23.1898}%
  \BibitemOpen
  \bibfield  {author} {\bibinfo {author} {\bibfnamefont {N\'estor~R.}\
  \bibnamefont {Arista}}\ and\ \bibinfo {author} {\bibfnamefont {Werner}\
  \bibnamefont {Brandt}},\ }\bibfield  {title} {\enquote {\bibinfo {title}
  {Energy loss and straggling of charged particles in plasmas of all
  degeneracies},}\ }\href {\doibase 10.1103/PhysRevA.23.1898} {\bibfield
  {journal} {\bibinfo  {journal} {Phys. Rev. A}\ }\textbf {\bibinfo {volume}
  {23}},\ \bibinfo {pages} {1898--1905} (\bibinfo {year} {1981})}\BibitemShut
  {NoStop}%
\bibitem [{\citenamefont {Zwicknagel}\ \emph {et~al.}(1999)\citenamefont
  {Zwicknagel}, \citenamefont {Toepffer},\ and\ \citenamefont
  {Reinhard}}]{Zwicknagel}%
  \BibitemOpen
  \bibfield  {author} {\bibinfo {author} {\bibfnamefont {G.}~\bibnamefont
  {Zwicknagel}}, \bibinfo {author} {\bibfnamefont {C.}~\bibnamefont
  {Toepffer}}, \ and\ \bibinfo {author} {\bibfnamefont {P.-G.}\ \bibnamefont
  {Reinhard}},\ }\bibfield  {title} {\enquote {\bibinfo {title} {Stopping of
  heavy ions in plasmas at strong coupling},}\ }\href {\doibase
  https://doi.org/10.1016/S0370-1573(98)00056-8} {\bibfield  {journal}
  {\bibinfo  {journal} {Physics Reports}\ }\textbf {\bibinfo {volume} {309}},\
  \bibinfo {pages} {117 -- 208} (\bibinfo {year} {1999})}\BibitemShut {NoStop}%
\bibitem [{\citenamefont {Higuchi}\ and\ \citenamefont
  {Yasuhara}(2000)}]{Higuchi_Japan_2000}%
  \BibitemOpen
  \bibfield  {author} {\bibinfo {author} {\bibfnamefont {Masahiko}\
  \bibnamefont {Higuchi}}\ and\ \bibinfo {author} {\bibfnamefont {Hiroshi}\
  \bibnamefont {Yasuhara}},\ }\bibfield  {title} {\enquote {\bibinfo {title}
  {Kleinman's dielectric function and interband optical absorption strength of
  simple metals},}\ }\href {\doibase 10.1143/JPSJ.69.2099} {\bibfield
  {journal} {\bibinfo  {journal} {Journal of the Physical Society of Japan}\
  }\textbf {\bibinfo {volume} {69}},\ \bibinfo {pages} {2099--2106} (\bibinfo
  {year} {2000})},\ \Eprint
  {http://arxiv.org/abs/https://doi.org/10.1143/JPSJ.69.2099}
  {https://doi.org/10.1143/JPSJ.69.2099} \BibitemShut {NoStop}%
\bibitem [{\citenamefont {Chihara}(1987)}]{Chihara_1987}%
  \BibitemOpen
  \bibfield  {author} {\bibinfo {author} {\bibfnamefont {J}~\bibnamefont
  {Chihara}},\ }\bibfield  {title} {\enquote {\bibinfo {title} {Difference in
  x-ray scattering between metallic and non-metallic liquids due to conduction
  electrons},}\ }\href {\doibase 10.1088/0305-4608/17/2/002} {\bibfield
  {journal} {\bibinfo  {journal} {Journal of Physics F: Metal Physics}\
  }\textbf {\bibinfo {volume} {17}},\ \bibinfo {pages} {295--304} (\bibinfo
  {year} {1987})}\BibitemShut {NoStop}%
\bibitem [{\citenamefont {Moldabekov}\ \emph
  {et~al.}(2015{\natexlab{a}})\citenamefont {Moldabekov}, \citenamefont
  {Schoof}, \citenamefont {Ludwig}, \citenamefont {Bonitz},\ and\ \citenamefont
  {Ramazanov}}]{zhandos_POP15}%
  \BibitemOpen
  \bibfield  {author} {\bibinfo {author} {\bibfnamefont {Zhandos}\ \bibnamefont
  {Moldabekov}}, \bibinfo {author} {\bibfnamefont {Tim}\ \bibnamefont
  {Schoof}}, \bibinfo {author} {\bibfnamefont {Patrick}\ \bibnamefont
  {Ludwig}}, \bibinfo {author} {\bibfnamefont {Michael}\ \bibnamefont
  {Bonitz}}, \ and\ \bibinfo {author} {\bibfnamefont {Tlekkabul}\ \bibnamefont
  {Ramazanov}},\ }\bibfield  {title} {\enquote {\bibinfo {title} {Statically
  screened ion potential and bohm potential in a quantum plasma},}\ }\href
  {\doibase 10.1063/1.4932051} {\bibfield  {journal} {\bibinfo  {journal}
  {Physics of Plasmas}\ }\textbf {\bibinfo {volume} {22}},\ \bibinfo {pages}
  {102104} (\bibinfo {year} {2015}{\natexlab{a}})},\ \Eprint
  {http://arxiv.org/abs/https://doi.org/10.1063/1.4932051}
  {https://doi.org/10.1063/1.4932051} \BibitemShut {NoStop}%
\bibitem [{\citenamefont {Moldabekov}\ \emph
  {et~al.}(2015{\natexlab{b}})\citenamefont {Moldabekov}, \citenamefont
  {Ludwig}, \citenamefont {Bonitz},\ and\ \citenamefont
  {Ramazanov}}]{PhysRevE.91.023102}%
  \BibitemOpen
  \bibfield  {author} {\bibinfo {author} {\bibfnamefont {Zhandos}\ \bibnamefont
  {Moldabekov}}, \bibinfo {author} {\bibfnamefont {Patrick}\ \bibnamefont
  {Ludwig}}, \bibinfo {author} {\bibfnamefont {Michael}\ \bibnamefont
  {Bonitz}}, \ and\ \bibinfo {author} {\bibfnamefont {Tlekkabul}\ \bibnamefont
  {Ramazanov}},\ }\bibfield  {title} {\enquote {\bibinfo {title} {Ion potential
  in warm dense matter: Wake effects due to streaming degenerate electrons},}\
  }\href {\doibase 10.1103/PhysRevE.91.023102} {\bibfield  {journal} {\bibinfo
  {journal} {Phys. Rev. E}\ }\textbf {\bibinfo {volume} {91}},\ \bibinfo
  {pages} {023102} (\bibinfo {year} {2015}{\natexlab{b}})}\BibitemShut
  {NoStop}%
\bibitem [{\citenamefont {Moldabekov}\ \emph {et~al.}(2016)\citenamefont
  {Moldabekov}, \citenamefont {Ludwig}, \citenamefont {Bonitz},\ and\
  \citenamefont {Ramazanov}}]{ctpp.201500137}%
  \BibitemOpen
  \bibfield  {author} {\bibinfo {author} {\bibfnamefont {Zh.~A.}\ \bibnamefont
  {Moldabekov}}, \bibinfo {author} {\bibfnamefont {P.}~\bibnamefont {Ludwig}},
  \bibinfo {author} {\bibfnamefont {M.}~\bibnamefont {Bonitz}}, \ and\ \bibinfo
  {author} {\bibfnamefont {T.~S.}\ \bibnamefont {Ramazanov}},\ }\bibfield
  {title} {\enquote {\bibinfo {title} {Notes on anomalous quantum wake
  effects},}\ }\href {\doibase https://doi.org/10.1002/ctpp.201500137}
  {\bibfield  {journal} {\bibinfo  {journal} {Contributions to Plasma Physics}\
  }\textbf {\bibinfo {volume} {56}},\ \bibinfo {pages} {442--447} (\bibinfo
  {year} {2016})},\ \Eprint
  {http://arxiv.org/abs/https://onlinelibrary.wiley.com/doi/pdf/10.1002/ctpp.201500137}
  {https://onlinelibrary.wiley.com/doi/pdf/10.1002/ctpp.201500137} \BibitemShut
  {NoStop}%
\bibitem [{\citenamefont {Valtierra~Rodriguez}\ \emph
  {et~al.}(2019)\citenamefont {Valtierra~Rodriguez}, \citenamefont {Wang},
  \citenamefont {Ofori-Opoku}, \citenamefont {Provatas},\ and\ \citenamefont
  {Bevan}}]{PhysRevB.100.235116}%
  \BibitemOpen
  \bibfield  {author} {\bibinfo {author} {\bibfnamefont {Salvador}\
  \bibnamefont {Valtierra~Rodriguez}}, \bibinfo {author} {\bibfnamefont {Nan}\
  \bibnamefont {Wang}}, \bibinfo {author} {\bibfnamefont {Nana}\ \bibnamefont
  {Ofori-Opoku}}, \bibinfo {author} {\bibfnamefont {Nikolas}\ \bibnamefont
  {Provatas}}, \ and\ \bibinfo {author} {\bibfnamefont {Kirk~H.}\ \bibnamefont
  {Bevan}},\ }\bibfield  {title} {\enquote {\bibinfo {title} {Capturing the
  dynamics of wigner crystals within the phase-field crystal method},}\ }\href
  {\doibase 10.1103/PhysRevB.100.235116} {\bibfield  {journal} {\bibinfo
  {journal} {Phys. Rev. B}\ }\textbf {\bibinfo {volume} {100}},\ \bibinfo
  {pages} {235116} (\bibinfo {year} {2019})}\BibitemShut {NoStop}%
\bibitem [{\citenamefont {Vorberger}\ \emph {et~al.}(2010)\citenamefont
  {Vorberger}, \citenamefont {Gericke}, \citenamefont {Bornath},\ and\
  \citenamefont {Schlanges}}]{transfer1}%
  \BibitemOpen
  \bibfield  {author} {\bibinfo {author} {\bibfnamefont {J.}~\bibnamefont
  {Vorberger}}, \bibinfo {author} {\bibfnamefont {D.~O.}\ \bibnamefont
  {Gericke}}, \bibinfo {author} {\bibfnamefont {Th.}\ \bibnamefont {Bornath}},
  \ and\ \bibinfo {author} {\bibfnamefont {M.}~\bibnamefont {Schlanges}},\
  }\bibfield  {title} {\enquote {\bibinfo {title} {Energy relaxation in dense,
  strongly coupled two-temperature plasmas},}\ }\href
  {https://journals.aps.org/pre/abstract/10.1103/PhysRevE.96.023203} {\bibfield
   {journal} {\bibinfo  {journal} {Phys. Rev. E}\ }\textbf {\bibinfo {volume}
  {81}},\ \bibinfo {pages} {046404} (\bibinfo {year} {2010})}\BibitemShut
  {NoStop}%
\bibitem [{\citenamefont {Benedict}\ \emph {et~al.}(2017)\citenamefont
  {Benedict}, \citenamefont {Surh}, \citenamefont {Stanton}, \citenamefont
  {Scullard}, \citenamefont {Correa}, \citenamefont {Castor}, \citenamefont
  {Graziani}, \citenamefont {Collins}, \citenamefont {Certík}, \citenamefont
  {Kress},\ and\ \citenamefont {Murillo}}]{transfer2}%
  \BibitemOpen
  \bibfield  {author} {\bibinfo {author} {\bibfnamefont {L.~X.}\ \bibnamefont
  {Benedict}}, \bibinfo {author} {\bibfnamefont {M.~P.}\ \bibnamefont {Surh}},
  \bibinfo {author} {\bibfnamefont {L.~G.}\ \bibnamefont {Stanton}}, \bibinfo
  {author} {\bibfnamefont {C.~R.}\ \bibnamefont {Scullard}}, \bibinfo {author}
  {\bibfnamefont {A.~A.}\ \bibnamefont {Correa}}, \bibinfo {author}
  {\bibfnamefont {J.~I.}\ \bibnamefont {Castor}}, \bibinfo {author}
  {\bibfnamefont {F.~R.}\ \bibnamefont {Graziani}}, \bibinfo {author}
  {\bibfnamefont {L.~A.}\ \bibnamefont {Collins}}, \bibinfo {author}
  {\bibfnamefont {O.}~\bibnamefont {Certík}}, \bibinfo {author} {\bibfnamefont
  {J.~D.}\ \bibnamefont {Kress}}, \ and\ \bibinfo {author} {\bibfnamefont
  {M.~S.}\ \bibnamefont {Murillo}},\ }\bibfield  {title} {\enquote {\bibinfo
  {title} {Molecular dynamics studies of electron-ion temperature equilibration
  in hydrogen plasmas within the coupled-mode regime},}\ }\href
  {https://journals.aps.org/pre/abstract/10.1103/PhysRevE.95.043202} {\bibfield
   {journal} {\bibinfo  {journal} {Phys. Rev. E}\ }\textbf {\bibinfo {volume}
  {95}},\ \bibinfo {pages} {043202} (\bibinfo {year} {2017})}\BibitemShut
  {NoStop}%
\bibitem [{\citenamefont {Scullard}\ \emph {et~al.}(2018)\citenamefont
  {Scullard}, \citenamefont {Serna}, \citenamefont {Benedict}, \citenamefont
  {Ellison},\ and\ \citenamefont {Graziani}}]{PhysRevE.97.013205}%
  \BibitemOpen
  \bibfield  {author} {\bibinfo {author} {\bibfnamefont {Christian~R.}\
  \bibnamefont {Scullard}}, \bibinfo {author} {\bibfnamefont {Susana}\
  \bibnamefont {Serna}}, \bibinfo {author} {\bibfnamefont {Lorin~X.}\
  \bibnamefont {Benedict}}, \bibinfo {author} {\bibfnamefont {C.~Leland}\
  \bibnamefont {Ellison}}, \ and\ \bibinfo {author} {\bibfnamefont {Frank~R.}\
  \bibnamefont {Graziani}},\ }\bibfield  {title} {\enquote {\bibinfo {title}
  {Analytic expressions for electron-ion temperature equilibration rates from
  the {L}enard-{B}alescu equation},}\ }\href {\doibase
  10.1103/PhysRevE.97.013205} {\bibfield  {journal} {\bibinfo  {journal} {Phys.
  Rev. E}\ }\textbf {\bibinfo {volume} {97}},\ \bibinfo {pages} {013205}
  (\bibinfo {year} {2018})}\BibitemShut {NoStop}%
\bibitem [{\citenamefont {Tanaka}(2016)}]{tanaka_hnc}%
  \BibitemOpen
  \bibfield  {author} {\bibinfo {author} {\bibfnamefont {S.}~\bibnamefont
  {Tanaka}},\ }\bibfield  {title} {\enquote {\bibinfo {title} {Correlational
  and thermodynamic properties of finite-temperature electron liquids in the
  hypernetted-chain approximation},}\ }\href
  {https://aip.scitation.org/doi/abs/10.1063/1.4969071} {\bibfield  {journal}
  {\bibinfo  {journal} {J. Chem. Phys}\ }\textbf {\bibinfo {volume} {145}},\
  \bibinfo {pages} {214104} (\bibinfo {year} {2016})}\BibitemShut {NoStop}%
\bibitem [{\citenamefont {Tanaka}(2017)}]{Tanaka_CPP_2017}%
  \BibitemOpen
  \bibfield  {author} {\bibinfo {author} {\bibfnamefont {Shigenori}\
  \bibnamefont {Tanaka}},\ }\bibfield  {title} {\enquote {\bibinfo {title}
  {Improved equation of state for finite-temperature spin-polarized electron
  liquids on the basis of singwi–tosi–land–sjölander approximation},}\
  }\href {\doibase https://doi.org/10.1002/ctpp.201600096} {\bibfield
  {journal} {\bibinfo  {journal} {Contributions to Plasma Physics}\ }\textbf
  {\bibinfo {volume} {57}},\ \bibinfo {pages} {126--136} (\bibinfo {year}
  {2017})}\BibitemShut {NoStop}%
\bibitem [{\citenamefont {Panholzer}\ \emph {et~al.}(2018)\citenamefont
  {Panholzer}, \citenamefont {Gatti},\ and\ \citenamefont
  {Reining}}]{Panholzer_PRL_2018}%
  \BibitemOpen
  \bibfield  {author} {\bibinfo {author} {\bibfnamefont {Martin}\ \bibnamefont
  {Panholzer}}, \bibinfo {author} {\bibfnamefont {Matteo}\ \bibnamefont
  {Gatti}}, \ and\ \bibinfo {author} {\bibfnamefont {Lucia}\ \bibnamefont
  {Reining}},\ }\bibfield  {title} {\enquote {\bibinfo {title} {Nonlocal and
  nonadiabatic effects in the charge-density response of solids: A
  time-dependent density-functional approach},}\ }\href {\doibase
  10.1103/PhysRevLett.120.166402} {\bibfield  {journal} {\bibinfo  {journal}
  {Phys. Rev. Lett.}\ }\textbf {\bibinfo {volume} {120}},\ \bibinfo {pages}
  {166402} (\bibinfo {year} {2018})}\BibitemShut {NoStop}%
\bibitem [{\citenamefont {Arora}\ \emph {et~al.}(2017)\citenamefont {Arora},
  \citenamefont {Kumar},\ and\ \citenamefont {Moudgil}}]{arora}%
  \BibitemOpen
  \bibfield  {author} {\bibinfo {author} {\bibfnamefont {P.}~\bibnamefont
  {Arora}}, \bibinfo {author} {\bibfnamefont {K.}~\bibnamefont {Kumar}}, \ and\
  \bibinfo {author} {\bibfnamefont {R.~K.}\ \bibnamefont {Moudgil}},\
  }\bibfield  {title} {\enquote {\bibinfo {title} {Spin-resolved correlations
  in the warm-dense homogeneous electron gas},}\ }\href
  {https://link.springer.com/article/10.1140/epjb/e2017-70532-y} {\bibfield
  {journal} {\bibinfo  {journal} {Eur. Phys. J. B}\ }\textbf {\bibinfo {volume}
  {90}},\ \bibinfo {pages} {76} (\bibinfo {year} {2017})}\BibitemShut {NoStop}%
\bibitem [{\citenamefont {Schweng}\ and\ \citenamefont
  {B\"ohm}(1993)}]{schweng}%
  \BibitemOpen
  \bibfield  {author} {\bibinfo {author} {\bibfnamefont {H.~K.}\ \bibnamefont
  {Schweng}}\ and\ \bibinfo {author} {\bibfnamefont {H.~M.}\ \bibnamefont
  {B\"ohm}},\ }\bibfield  {title} {\enquote {\bibinfo {title}
  {Finite-temperature electron correlations in the framework of a dynamic
  local-field correction},}\ }\href
  {https://journals.aps.org/prb/abstract/10.1103/PhysRevB.48.2037} {\bibfield
  {journal} {\bibinfo  {journal} {Phys. Rev. B}\ }\textbf {\bibinfo {volume}
  {48}},\ \bibinfo {pages} {2037} (\bibinfo {year} {1993})}\BibitemShut
  {NoStop}%
\bibitem [{\citenamefont {Holas}\ and\ \citenamefont
  {Rahman}(1987)}]{dynamic_ii}%
  \BibitemOpen
  \bibfield  {author} {\bibinfo {author} {\bibfnamefont {A.}~\bibnamefont
  {Holas}}\ and\ \bibinfo {author} {\bibfnamefont {S.}~\bibnamefont {Rahman}},\
  }\bibfield  {title} {\enquote {\bibinfo {title} {Dynamic local-field factor
  of an electron liquid in the quantum versions of the
  {S}ingwi-{T}osi-{L}and-{S}j\"olander and {V}ashishta-{S}ingwi theories},}\
  }\href {https://journals.aps.org/prb/abstract/10.1103/PhysRevB.35.2720}
  {\bibfield  {journal} {\bibinfo  {journal} {Phys. Rev. B}\ }\textbf {\bibinfo
  {volume} {35}},\ \bibinfo {pages} {2720} (\bibinfo {year}
  {1987})}\BibitemShut {NoStop}%
\end{thebibliography}%
\end{document}